\documentclass{aa}

\usepackage{caption}
\usepackage{graphicx}
\usepackage{txfonts}
\usepackage{rotating} 
\usepackage{placeins}           

\usepackage{hyperref}

\begin{document}

\title{Singular spectrum analysis of {\it Fermi}-LAT blazar light curves:\\ 
A systematic search for periodicity and trends in the time domain}
\titlerunning{SSA of {\it Fermi}-LAT Blazar LCs}  

\author{Alba Rico\inst{1,2}\thanks{E-mail: aricoro@clemson.edu}
    \and A. Domínguez\inst{1}\thanks{E-mail: alberto.d@ucm.es}
    \and P. Pe\~nil\inst{2}\thanks{E-mail: ppenil@clemson.edu}
    \and M. Ajello\inst{2}
    \and S. Buson\inst{3}
    \and S. Adhikari\inst{2}
    \and M. Movahedifar\inst{4}}
\authorrunning{A. Rico et al.}
\institute{IPARCOS and Department of EMFTEL, Universidad Complutense de Madrid, E-28040 Madrid, Spain
  \and Department of Physics and Astronomy, Clemson University,	Kinard Lab of Physics, Clemson, SC 29634-0978, USA
  \and Julius-Maximilians-Universität, 97070, Würzburg, Germany
  \and Institute for Statistics, University of Bremen, 28334, Bremen, Germany
    }
\date{Received 4 October 2024 / Accepted 9 March 2025}
\abstract
   {A majority of blazars exhibit variable emission across the entire electromagnetic spectrum, observed over various timescales. In particular, discernible periodic patterns are detected in the $\gamma$-ray light curves (LCs) of a few blazars, such as PG 1553+113, S5 1044+71, and PKS 0426-380. The presence of trends, flares, and noise complicates the detection of periodicity, requiring careful analysis to determine whether these patterns are related to emission mechanisms within the source or occur by chance.}
   {We employ singular spectrum analysis (SSA) for the first time on data from the Large Area Telescope (LAT) aboard the \textit{Fermi} Gamma-ray Space Telescope to systematically search for periodicity in the time domain, using 28-day binned LCs. Our aim is to isolate any potential periodic nature of the emission from trends and noise, thereby reducing uncertainties in revealing periodicity. Additionally, we aim to characterize long-term trends and develop a forecasting algorithm based on SSA, enabling accurate predictions of future emission behavior.}
   {We apply SSA to analyze 494 sources detected by \textit{Fermi}-LAT, focusing on identifying and isolating oscillatory components from trends and noise in their $\gamma$-ray LCs. We calculate the Lomb-Scargle periodogram (LSP) for the oscillatory components extracted by SSA to determine the most significant periods. The local and global significance of these periods is then assessed to validate their authenticity.}
   {Our analysis identifies 46 blazars as potential candidates for quasi-periodic $\gamma$-ray emissions, each with a local significance level $\geq 2\sigma$. Notably, 33 of these candidates exhibit a local significance of $\geq 4\sigma$ (corresponding to a global significance of $\geq 2.2\sigma$). Our findings introduce 25 new $\gamma$-ray candidates, effectively doubling the number of potentially periodic sources. This study provides a foundation for future investigations by identifying promising candidates and highlighting their potential significance within the context of blazar variability.}
   {}


\maketitle

\section{Introduction}
Active galactic nuclei (AGN) are among the most luminous and distant persistent objects in the Universe. These are supermassive black holes (SMBH) at the center of some galaxies that accrete matter from surrounding disks of gas and dust \citep[e.g.,][]{Urry_1996,AGN_book}. These SMBHs can produce highly collimated jets of particles, accelerated to relativistic velocities \citep[e.g.,][]{Blandford_2019}. When the jet is closely aligned with our line of sight, these objects are referred to as blazars.

Blazars exhibit variability across the entire electromagnetic spectrum, with timescales ranging from seconds to years \citep[e.g.,][]{urry11}. While this variability is typically considered stochastic, it can also reveal characteristic patterns \citep[e.g.,][]{radio_optical,Sandrinelli_2,P20,Pablo22}. These patterns provide insight into the internal structure of blazars and their emission processes \citep[e.g.,][]{Bhatta_2019}. Periodicity in blazar emissions can result from several physical mechanisms, including (1) jet precession \citep[e.g.,][]{lighthouse}, (2) modulations in the accretion flow \citep[e.g.,][]{modulation}, or (3) the presence of binary SMBH systems \citep[e.g.,][]{sobachi,celoria}.  Binary SMBHs, which typically result from galaxy mergers and play a fundamental role in galaxy evolution \citep[e.g.,][]{jiang2022ticktock}, are a key focus of this work. Theoretical studies exploring periodic variability linked to binary SMBHs \citep[e.g.,][]{sobachi,celoria} highlight the relevance of searching for year-scale periodicities to investigate their dynamic evolution in the gas-driven regime and role in galaxy evolution \citep[e.g.,][]{penil_mwl_pg1553,penil24}.

In addition to periodic signals, long-term trends can also be observed in blazars \citep[e.g.,][]{Marscher_trend,rani2018radio_trend,Valverde_2020_trend}. The variability timescales in SMBH systems reflect their physical properties and dynamic processes. For single SMBH systems, variability spans from minutes to years, with short-term changes often caused by instabilities in the inner accretion disk or jet phenomena, while long-term changes may result from variations in the accretion rate or disk precession \citep[e.g.,][]{peterson01,mchardy06}. In binary SMBH systems, additional periodicities from the orbital motion of the black holes are superimposed on the variability from individual accretion disks \citep[e.g.,][]{graham15,sobacchi17}. During SMBH mergers, tidal interactions and dynamic evolution lead to complex variability patterns, including quasi-periodic oscillations (QPOs) with timescales from weeks to years \citep[e.g.,][]{merritt05,valtonen08}.  While linking periodic behavior to specific physical scenarios is an essential goal, achieving this will require more detailed analyses, such as multiwavelength modeling and theoretical simulations, which are beyond the scope of this work.

Identifying QPOs and long-term trends poses an observational challenge due to the need for continuous monitoring over extended periods. In the $\gamma$-ray domain, the Large Area Telescope (LAT) aboard the \textit{Fermi} Gamma-ray Space Telescope, launched in June 2008, monitors the entire sky with high sensitivity. Using this data, \citet{P20} conducted the first systematic periodicity study based on the third release of the \textit{Fermi}-LAT catalog \citep[3FGL,][]{3FGL}, analyzing 9 years of LAT data and 10 different techniques. \citet{P20} identified 11 blazars with periodic signals showing local significance $\geq 2.0 \sigma$.

However, these studies primarily relied on Fourier-domain techniques, which face inherent limitations in addressing nonstationary signals typical in astrophysical observations \citep[e.g.,][]{prokhorov,seyfert1,Sandrinelli,B21520,quasi,P20,Pablo22,s51044,QPO,pks1510}. The complexity of blazar emissions, including noise structures \citep[e.g.,][]{vaughan_criticism}, long-term trends, and flares, can be better addressed using time-domain methods that can effectively isolate periodic components from the overall signal \citep[see also, e.g.,][]{rueda22}.

Singular spectrum analysis (SSA) emerges as a promising solution \citep[e.g.,][]{SSA_paper,SSA}. This statistical method, designed for time series data, decomposes a signal into its trend, oscillatory components, and noise structure. By separating these components, SSA provides a clearer picture of periodic patterns and helps in developing more accurate predictive models. SSA has been successfully applied across various scientific domains, including economics \citep[e.g.,][]{HASSANI2009103}, climate science \citep[e.g.,][]{SSA_meteo}, and biomedicine \citep[e.g.,][]{SSA_biomedics}.

In this study, we performed a systematic search for periodic emissions in a sample of AGN, including blazars, radio galaxies, and narrow-line Seyfert 1 galaxies (see Sect. \ref{sample}), detected by \textit{Fermi}-LAT in the second release of the fourth \textit{Fermi}-LAT source catalog \citep[4FGL-DR2,][]{Abdollahi_2020,DR2}.

The paper is organized as follows: Sect. \ref{sample} presents the blazar sample and data reduction methodology. Sect. \ref{method} describes the SSA method, including its application to periodicity, trends, and forecasts. The results are discussed in Sect. \ref{results}, and a summary and conclusion are provided in Sect. \ref{sec:conclusions}.

\section{Data reduction and sample selection}\label{sample}
In this section, we describe the $\gamma$-ray data reduction and sample selection.

\subsection{Analysis of Gamma-ray Data}
{\it Fermi}-LAT is a $\gamma$-ray telescope and pair-conversion detector, capable of measuring energies from below 20 MeV to over 300 GeV \citep{Atwood_2009}. Thanks to its wide field of view (2.4 sr), it continuously scans the entire sky with the possibility of observing the entire sky every 3 hours. Data collected by LAT are made publicly available through the {\it Fermi} Science Support Center\footnote{\url{https://fermi.gsfc.nasa.gov/ssc/}} (FSSC). The \textit{Fermi}-LAT data used in this study are processed with \texttt{Fermipy}, a \texttt{Python} package, following the procedure outlined by \cite{wood2017fermipy}. We specifically use the Pass 8 SOURCE class, from the \texttt{P8R3$\_$SOURCE$\_$V2} instrument response function (IRF), to select the photons \citep{atwood2013pass}. Our region of interest (RoI) is a $15^{\circ} \times 15^{\circ}$ square, centered at the target. To mitigate contamination from $\gamma$-rays produced in the Earth's upper atmosphere, we apply a zenith angle cut of $\theta < 90^{\circ}$. Standard quality cuts are applied (\texttt{DATA QUAL > 0)}, (\texttt{LAT CONFIG == 1}), and time periods coinciding with solar flares and $\gamma$-ray bursts are excluded. The standard LAT analysis of a RoI includes sources located within $20^{\circ}$ from the RoI center, using the 4FGL catalog \citep{Abdollahi_2020}, which includes the Galactic and isotropic diffuse emissions (\texttt{gll$\_$iem$\_$v07.fits} and \texttt{iso$\_$P8R3$\_$SOURCE$\_$V6$\_$v06.txt}). 

We conduct a binned analysis, using 10 energy bins per decade from 0.1 to 800 GeV with 0.1$^{\circ}$ spatial bins, which produces a maximum likelihood analysis over the entire observation period (2008 Aug 04 15:43:36 UTC to 2020 Dec 10 00:01:26 UTC, approximately 12 years). Using the parameters and spectral shapes from the 4FGL catalog, we model the sources within the RoI. The normalization and spectral index of the target source, as well as sources within a $3^{\circ}$ radius centered at the target position, are allowed to vary, along with the two diffuse backgrounds.  We construct a test statistic (TS) map for searching for newly detected sources relative to the 4FGL catalog. The TS is defined as $2\log(L/L_0)$ \citep{mattox}, where $L$ is the likelihood of the model with a point source at a given position and $L_0$ is the likelihood without the point source. This statistic quantifies how significantly a source emerges from the background \citep{Abdollahi_2020}. A power-law spectrum is used to model new test sources, where only the normalization is allowed to vary during the fitting process, while the photon index is fixed at 2. We look for significant peaks, retaining only sources with TS $> 25$ and separated by at least $0.5^{\circ}$ from existing sources in the model. We then add a new point source at the position of the most significant peak. The model for the RoI is readjusted, and a new TS map is produced. This iterative process continues until no further significant excesses are found, generally leading to the addition of two point sources. Any $\gamma$-ray excess not listed in the DR2 catalog but with TS > 25 in the full-time interval analysis is included in the RoI as part of our analysis strategy.

To generate light curves (LCs), we split the data for each source into 28-day bins and perform a full likelihood fit for each bin. The best-fit RoI model from the time-interval analysis is adopted. We first focus on the target source and all nearby sources within the RoI (i.e.,~within $3^{\circ}$), including the diffuse components. During the fitting process, only the normalization parameters are allowed to vary. If the fitting does not converge, we gradually reduce the number of free parameters until convergence is achieved.

An iterative approach is used, starting by fixing sources with low detection significance (TS < 4) within the RoI. Next, we address sources with TS < 9, and then fix sources within a $1^{\circ}$ radius of the RoI center if their TS < 25. Finally, we leave only the normalization of the target source free, fixing all other parameters. Note these LCs are the same as those used by \citet{Pablo22}, \citet{penil_mwl_pg1553}, and \citet{penil24}.

\subsection{Sample selection}\label{sample_selection}
Our sample is drawn from the 4FGL-DR2 catalog, which includes data from approximately 3500 blazars \citep{Ajello_2021}. We filtered the 4FGL-DR2 sample by selecting only sources that show variability in their emissions. Specifically, we used a variability index threshold, requiring the variability index to be greater than 18.48 \citep[][]{Abdollahi_2020}. This threshold indicates that the probability of the source remaining steady is less than 1\% \citep[see Table~12 of][]{Abdollahi_2020}. We used 28-day binned LCs to be compatible with our data reduction, which is optimized for detecting periodicities on the order of 1 year, as this provides a practical balance between computational efficiency and sensitivity to long-term variations.

In cases of non-detection, we established an upper limit for the flux in all time bins where the TS is below 4, making our sample more conservative. The choice of using upper limits for <50\% of the data was tested by \cite{P20}, who found that LCs with more upper limits were too incomplete for reliable analysis. We followed the same approach as \cite{Pablo22} for handling upper limits. To address potential inaccuracies from weak signals or non-detections (TS < 1), we replaced flux values with those that maximize the likelihood function for each time bin. This approach was applied to about 8\% of the data. For the sources listed in Table \ref{table results}, the average percentage of time bins with upper limits (TS < 4) is approximately 16\%.

\section{Methodology} \label{method}
\subsection{Theoretical background on singular spectrum analysis}\label{SSA}

SSA is a statistical technique for time series analysis that integrates elements of classical time series analysis, dynamical systems, multivariate statistics, multivariate geometry, and signal processing \citep[e.g.,~][]{SSA_paper,SSA}. The core idea of SSA is to decompose the original signal into a sum of components, such as a smooth trend, an oscillatory component, and a noise structure.

The SSA technique consists of two main stages: decomposition and reconstruction. Both stages include two separate steps. In the first stage, the series is decomposed to enable signal extraction and noise reduction. In the second stage, the original series is reconstructed, and the less noisy reconstructed series is utilized for forecasting new data points. Below, we provide a concise explanation of the SSA methodology.

\begin{enumerate}
  \item {\it Embedding.}
 The starting point of SSA is to embed a one-dimensional time series $Y_N=(y_1, \dots, y_N)$ of length $N$ into a multidimensional series $X_1,...,X_K$ with vectors $X_i =\left(y_i,..., y_{i+L-1}\right)\in\mathbb{R}^{L}$, where $K = N - L +1$. The single parameter for embedding is the window length $L$, an integer such that~$2\leq L \leq N/2$. The result of this step is the trajectory matrix $\textbf{X} =\left[X_1,...,X_K\right]$:
\begin{equation}
  \label{eq.traje.SSA}
                \mathbf{\textbf{X}} = \left(
                  \begin{array}{ccccc}
                      y_1 & y_2 & y_3 &\cdots & y_K     \\
                   y_2 & y_3 & y_4 &\cdots & y_{K+1}   \\
                   \vdots  & \vdots  & \vdots &\ddots & \vdots \\
                   y_L & y_{L+1} & y_{L+2} &\cdots & y_N   \\
                  \end{array}
                \right),
\end{equation}
where the trajectory matrix $\textbf{X}$ is a Hankel matrix, meaning that all elements along the diagonal $i + j = const$ are equal. This step breaks down the LC into overlapping time windows, similar to a sliding window approach, to analyze trends and oscillations in smaller sections.\\

\item {\it Decomposition.}
 The trajectory matrix $\textbf{X}$ is then decomposed using singular value decomposition (SVD):
\begin{align}
\label{eq.SVD M}
  {\textbf{X}}={\textbf{U}}\boldsymbol{\Sigma}{ \textbf{V}}^T,
\end{align}
where $\boldsymbol{\Sigma}= \mathrm{diag}(\sqrt{\lambda_1}\geq\sqrt{\lambda_2}\geq...\geq\sqrt{\lambda_L})
\in\mathbb{R}^{L\times L}$, $\textbf{V}=[V_1,V_2,...,V_L]\in\mathbb{R}^{K\times L}$ and $\textbf{U}=[U_1,U_2,...,U_L]\in\mathbb{R}^{L\times L}$. Here, $\lambda_1,\lambda_2,...,\lambda_L $ are the eigenvalues of $\textbf{XX}^T$ in decreasing order $(\lambda_1\geq\lambda_2\geq...\geq\lambda_L\geq0)$, and $U_1,U_2,...,U_L$ are the corresponding eigenvectors. Set $d$ as the maximum $i$, such that $\lambda_i>0$, representing the rank of $\textbf{X}$. If we denote $V_i=\textbf{X}^TU_i/\sqrt{\lambda_i}$, the SVD of the trajectory matrix can be written as:
\begin{align}
\label{eq.x}
 {\textbf{X}}={\textbf{X}}_1+{\textbf{X}}_2+...+{\textbf{X}}_d,
\end{align}
where $\textbf{X}_i=\sqrt{\lambda_i}U_iV_i^T \quad (i=1,2,...,d)$. The matrices $\textbf{X}_i$ have rank 1. SVD (\ref{eq.x}) is optimal in the sense that, among all matrices of rank $r < d$, the matrix $\textbf{X}^{(r)} =\sum_{i=1}^{r} \textbf{X}_i$ provides the best approximation to the trajectory matrix $\textbf{X}$, minimizing $\|\textbf{X}-\textbf{X}^{(r)}\|_F$, where $\|.\|_F$ is the Frobenius norm. Note that $\| \textbf{X} \|^{2} =\sum_{i=1}^{d}\lambda_i$ and $\| \textbf{X}_i \|^{2}=\lambda_i$ for $i=1,...,d$. The ratio ${\lambda_i}/{\sum_{i=1}^{d}\lambda_i}$ quantifies the contribution of matrix $X_i$ to the expansion (\ref{eq.x}). Consequently, ${\sum_{i=1}^{r}\lambda_i}/{\sum_{i=1}^{d}\lambda_i}$ is a measure of the approximation of the trajectory matrix by matrices of rank $r$. This step separates the LC into its main components, such as long-term trends, oscillatory behaviors, and random noise, allowing us to analyze each aspect separately.\\

\item {\it Grouping.}
The grouping step involves splitting the elementary matrices $\textbf{X}_i$ into several groups and adding the matrices within each group. The grouping procedure partitions the set of indices $\{1, ...,L\}$ into $m$ disjoint subsets $\{I_1, ... , I_m\}$ (eigentriple grouping), where each group $I_c$ contains a set of principal components $\{i_1, ..., i_p\}$ representing specific components of the signal (trend, oscillations, harmonic components, etc.). For example, if the periodic component can be obtained using $ I_{period} = \{1,2\}$, then $\textbf{X}_{1} + \textbf{X}_{2} = \textbf{X}_{I_{period}}$. For more details, see Section 2.2.2 in \cite{SSA_Saeid}. Let $I_{c} = \{ i_1,..., i_p\}$ be a group of indices. Then the matrix $\textbf{X}_{I_c}$ corresponding to group $I_c$ is defined as $\textbf{X}_{I_c} = \textbf{X}_{i_1} +... + \textbf{X}_{i_p}$. These matrices are computed for $I_{c} = I_1, ..., I_m$, and the SVD expansion leads to the decomposition $\textbf{X} = \textbf{X}_{I_1} +... + \textbf{X}_{I_m}$. Each $\textbf{X}_{I_c}$ represents a set of eigentriples describing a specific component of the original time series. In short, after separating the different components of the LC, we group those that represent similar behaviors. For example, if two components both show a periodic pattern, we group them together to extract the full periodic signal. This helps us isolate different types of behavior in the data, such as long-term trends or periodic oscillations.\\
 
\item {\it Reconstruction.} 
The final step in basic SSA is to determine the components corresponding to the original series. This involves deriving $m$ new time series $\widetilde{Y}^{(c)}$ of length $N$, where each time series corresponds to a specific component of the signal (e.g., trend, oscillations). Diagonal averaging of each $\textbf{X}_{I_c}$ provides the $c$-th component of the series $Y_N$. The $n$-th sample is obtained by averaging over the cross-diagonal $i + j = n + 1 = const$ of $\textbf{X}_{I_c}$, since each $\textbf{X}_{I_c}$ is expected to be a Hankel matrix for the corresponding embedded component series. This procedure is called diagonal averaging or Hankelization of a matrix. Applying Hankelization to all matrix components of $\textbf{X}_{I_1} +... + \textbf{X}_{I_m}$ yields another expansion:
\begin{equation}
\textbf{X} = \widetilde{\textbf{X}}_{I_1} + \dots + \widetilde{\textbf{X}}_{I_m}. \label{eq:2.3}
\end{equation}
 This is equivalent to decomposing the initial series $Y_N=(y_1, \dots, y_N)$ into a sum of $m$ series:
 
\begin{equation}\label{reconstructed series}
y_t = \sum_{c=1}^{m} \widetilde{y}^{(c)}_t 
\end{equation}
where $\widetilde{Y}^{(c)}_N = (\widetilde{y}^{(c)}_1, \dots, \widetilde{y}^{(c)}_N)$ corresponds to the matrix $X_{I_c}$.

In summary, after decomposition and grouping, we combine the components we are interested in, such as the periodic signal, while filtering out noise.
\end{enumerate}

For more details on SSA and its application, see \citet{SSA_paper} and \citet{Movahedifar2018}.

\subsection{Further notes on SSA: Separability and grouping selection}\label{Further notes on SSA}

As mentioned earlier, SSA decomposes a time series into various components, which can be categorized as trend, oscillatory components, and noise. The window length $L$ is the only parameter in the decomposition stage. If the time series contains a periodic component with an integer period, it is recommended to set the window length proportional to that period to enhance the separability of the periodic component. A periodic component generates two eigentriples with closely matched singular values, making it beneficial to check the breaks obtained from the eigenvalue spectra. Additionally, a pure noise series typically produces a slowly decreasing sequence of singular values. 

To assess the quality of the decomposition of the observed data into signal and noise components, we determine separability. To measure this separability, we use the  w-correlation matrix, which is useful for finding correlations and identifying groups. This matrix consists of weighted cosines of angles between the reconstructed time series components. Specifically, we analyze the matrix of absolute values of the $w$-correlations. If the absolute value of the $w$-correlations is small, then the corresponding series (here, signal and noise) are almost $w$-orthogonal. However, if the $w$-correlation is large, then the two series are far from being $w$-orthogonal and are thus poorly separable. If two reconstructed components have zero $w$-correlations, they are separable. Large $w$-correlations between reconstructed components suggest they should be grouped together and correspond to the same component in SSA decomposition.

\begin{figure*}
    \centering
    \includegraphics[width=\columnwidth]{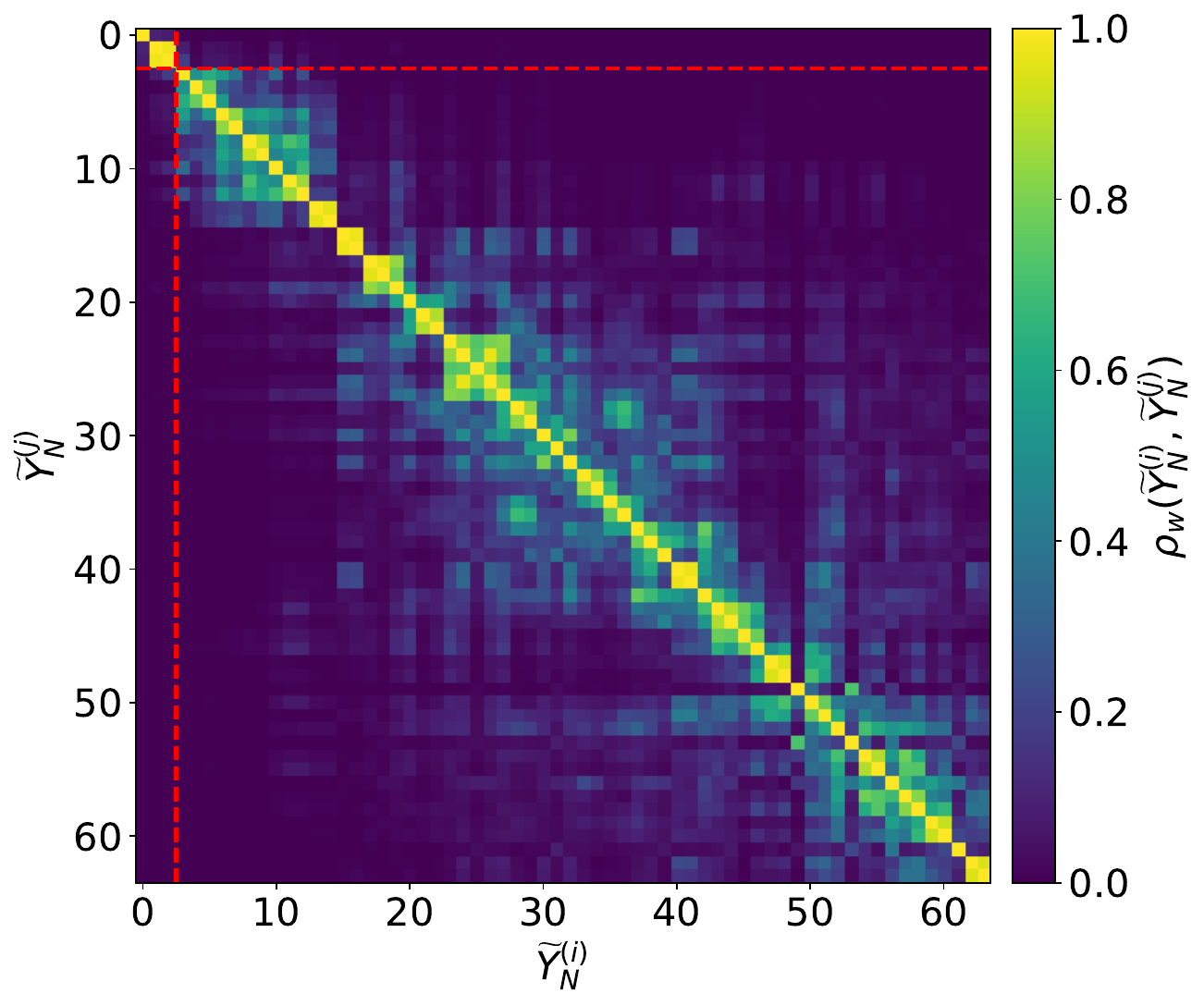}
    \includegraphics[width=\columnwidth]{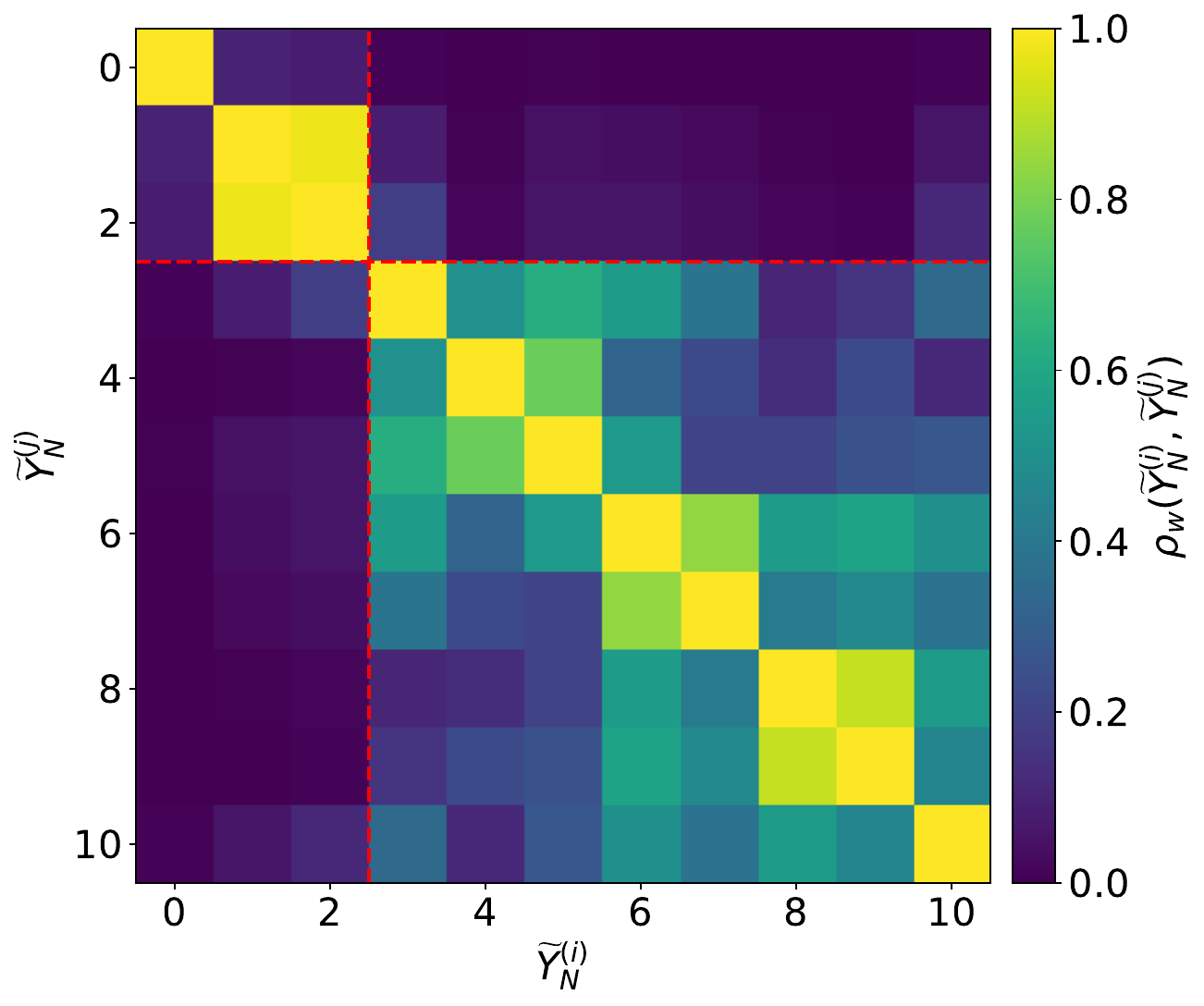}
    \caption{$w$-correlation matrices from SSA decomposition of S5 1044+71 LC. Dashed red lines separate the top-left region, containing the trend and periodic components, from the rest of the matrix, where noise components are more dominant. ({\it Left panel}) General $w$-correlation matrix, where each point $\left(\widetilde{Y}^{(i)}_N, \widetilde{Y}^{(j)}_N\right)$ represents the correlation between the \(i\)-th and \(j\)-th component. The color scale indicates the strength of the correlation, with yellow representing high correlation and dark purple representing low correlation. The diagonal represents self-correlation. ({\it Right panel}) Zoomed-in view focusing on the first 10 components, providing a detailed examination of the correlations between the initial components that typically capture key trends and periodic signals.}
\label{fig:w_correlation}
    \label{fig:correlations}
\end{figure*}

 Let us consider a time series $Y_N$, where $Y_N = Y_N^{(1)} + Y_N^{(2)}$, with $Y_N^{(1)}$ representing the main signal (trend + periodic components), and $Y_N^{(2)}$ representing the noise. If $Y_N^{(1)}$ can be separated from $Y_N$, it means there exists a partition into groups at the grouping stage such that $\widetilde{Y}_N^{(m)} = Y_N^{(m)}$.

We apply the separability formalism defined in Section~2.1.3 of \cite{SSA_R}, summarized as follows. Let $\mathbf{X}^{(m)}$ be the trajectory matrix of the series, with their SVDs, $\mathbf{X}^{(m)}= \sum ^{d_m}_{i=1}\sqrt{\lambda_{m,i}}U_{m,i}V_{m,i}^{\mathrm{T}}$. The column and row spaces of the trajectory matrix are called column and row trajectory spaces, respectively. The separability conditions are:

\begin{enumerate}
\item Two series $Y_N^{(1)}$ and $Y_N^{(2)}$ are weakly separable if their column and row trajectory spaces are orthogonal.
\item Two series are strongly separable if they are weakly separable and the sets of singular values of their $L$-trajectory matrices are disjoint, $\lambda_{1,i}\neq \lambda_{2,j}$, for any $i$ and $j$.
\end{enumerate}

Let $w_n \quad (n=1, \ldots, N)$ be the number of times the series element $x_n$ appears in the trajectory matrix. Consider two series of length $N$, $Y_N^{(1)}$ and $Y_N^{(2)}$. Set $(Y_N^{(1)}, Y_N^{(2)})_w = \sum_{n=1}^Nw_ny_n^{(1)}y_n^{(2)}$, the $w$-scalar product of the time series. The $w$-correlation matrix is defined as:
\begin{equation}
\rho_w(Y_N^{(1)}, Y_N^{(2)})=\frac{\left(Y_N^{(1)}, Y_N^{(2)}\right)}{(\|Y_N^{(1)}\|_w \|Y_N^{(2)}\|_w)}.
\end{equation}

The well-separated components in Equation~\ref{reconstructed series} are weakly correlated, while poorly separated components have high correlation. By examining the matrix of $w$-correlations between elementary reconstructed series $\widetilde{Y}^{(i)}_N$ and $\widetilde{Y}^{(j)}_N$, it is possible to identify groups of correlated series components. It is important to avoid grouping highly correlated components separately. For more details, see \cite{SSA_paper}.

 An example of the $w$-correlation matrices is shown in Figure~\ref{fig:correlations}. The matrix plots different components of the decomposed LC along both the $x$-axis and $y$-axis. Each point $\left(\widetilde{Y}^{(i)}_N, \widetilde{Y}^{(j)}_N\right)$ represents the degree of correlation between the \(i\)-th and \(j\)-th component. The firsts components are usually the trend and periodicity, whereas the rest of the components can be characterized as noise. The color represents the level of correlation, where yellow indicates a high correlation (close to 1) and dark purple shows near-zero correlation. A correlation of 1 means the components are almost identical, while a correlation close to 0 suggests the components are well-separated (orthogonal). If the components are well-separated (low correlation), we can confidently treat them as distinct parts of the signal, such as separating periodic behavior from noise. The diagonal elements (always 1) show the self-correlation of each component. Typically, a value greater than 0.8 means that the components are significantly correlated, and we consider them from the same group, that is, either trend, periodicity, or noise. For instance, in this example plotted in Figure~\ref{fig:correlations}, the components 1 $\left(\widetilde{Y}^{(1)}_N\right)$  and 2 $\left(\widetilde{Y}^{(2)}_N\right)$ are considered correlated and they both store the periodicity information, while the component 0 $\left(\widetilde{Y}^{(0)}_N\right)$ is characterized as trend. The dashed red lines separate the trend and periodicity components from the noise. Figure~\ref{fig:correlations} (right) shows a zoom of the first ten components.

\subsection{Challenges and limitations in the search for periodicity}\label{periodicity}

Several limitations are inherent in any periodicity analysis, including the presence of different noise types. Noise can be characterized by an exponent $\beta$, which represents the power-law index of the power spectral density (PSD). The PSD is crucial for characterizing variability emissions and quantifies the variability power as a function of the frequency $\nu$, with PSD$(\nu) \propto  \nu ^{-\beta}$, where $\beta = 0$ for white noise (spurious background independent of frequency), $\beta = 1$ for pink noise, and $\beta = 2$ for red or Brownian noise \citep{galaxies7010028}. These noise types can distort the signal and make the detection of periodic patterns more challenging \citep{vanderplas}, potentially leading to false detections \citep{vaughan_criticism}.

Underlying trends and/or unpredictable stochastic emissions, such as flares, can also distort the periodicity analysis. SSA effectively mitigates these challenges by decomposing the signal, allowing us to search for inherent periodic patterns by isolating the oscillatory component from the underlying trend and noise structure.

In Section~\ref{Further notes on SSA}, we discussed the significance of the window length as a key parameter in our analysis. There are no strict guidelines for selecting a specific window length in SSA, as long as it falls within the range $2 \leq L \leq N/2$, where $N$ is the data length. However, larger window lengths (up to 30–45\% of the length of the time series) are sometimes necessary to adequately separate periodicities from the overall trend. For more details on window length selection, see \cite{SSA}.

We find that a window length of 40\% of the data length is optimal for our analysis. When the window length is lower, stochastic events tend to dominate the signal, whereas when the window length is larger, there are too many components for adequate grouping.

We search for long-term periodicities in the range of 1 to 6 years, as our 12-year LCs allow for the detection of at most two cycles within this timeframe. This range is particularly relevant for binary SMBH candidates in the gas-driven regime \citep[see, e.g.,][]{penil_mwl_pg1553}, where systems with parsec to sub-parsec separations between the black holes, which cannot generally be spatially resolved, are expected to show periodicities. Focusing on this range also minimizes the impact of red noise contamination, which primarily affects longer periodicities, while shorter periodicities are less relevant to this study.

\subsection{Periodicity search pipeline}
We applied SSA to the LCs of the blazars selected in our sample, as detailed in Section~\ref{sample_selection}.

To calculate the period, we employed a widely recognized technique for identifying periodicity in time series data, the Lomb-Scargle periodogram \citep[LSP,][]{Lomb, Scargle}. We applied the LSP to the periodicity component obtained after applying the SSA code to the original signal. The period corresponded to the highest peak in the periodogram.

\subsubsection{Systematic search for periodicity candidates}\label{systematic search}
\begin{figure}
    \centering
    \includegraphics[width=\columnwidth]{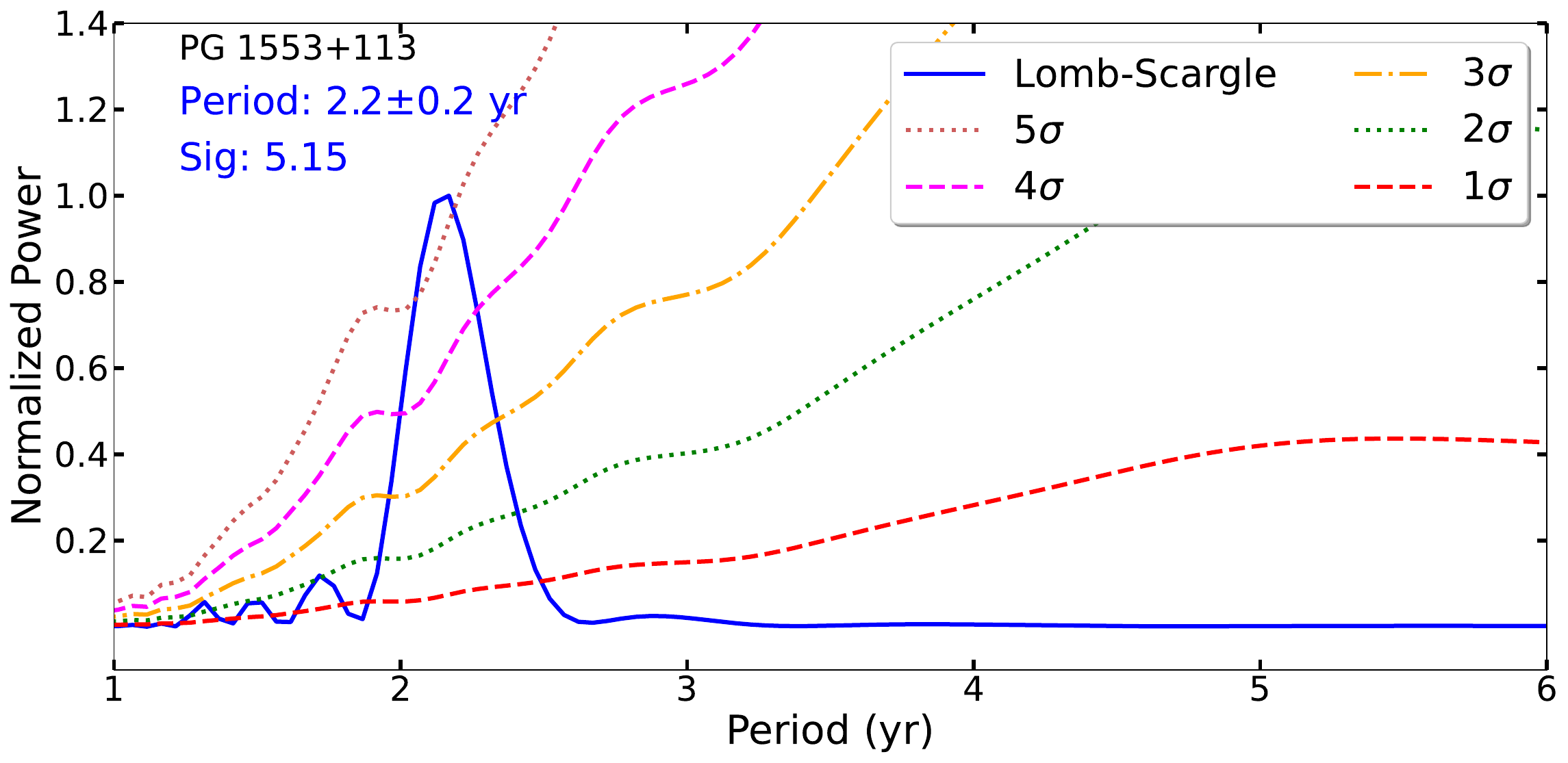}
    \caption{Lomb-Scargle periodogram of PG 1553$+$113. 
    A significant period peak is detected at $\sim 2.2$ yr. The different dashed lines represent the significance levels computed by simulating artificial LCs (see Section \ref{systematic search}).}
    \label{fig:levels}
\end{figure}

We performed a systematic search for periodicity in our sample to select the most promising blazars. The LSP was computed for simulated LCs to determine their power values. These simulated LCs were generated using the best-fitting parameters derived from the PSD and the Probability Density Function (PDF) of the original LC. We applied the Emmanoulopoulos technique\footnote{\url{https://github.com/samconnolly/DELightcurveSimulation}\label{emma}} to generate artificial LCs and evaluated the significance of the periods \citep{Emmanoulopoulos}. The mean power value within each period bin was multiplied by various percentiles corresponding to confidence significance levels. The significance of the detected period was assessed by comparing the periodicity peak derived from the LSP for the periodicity component of the studied blazar with the computed significance levels. As an example, Figure~\ref{fig:levels} shows the significance levels for PG 1553+113. We selected all blazars where the significance of the peak in the LSP is $\geq 2.0 \sigma$.

\subsubsection{Local significance}\label{simulated}
For this subsample, we refined the analysis and calculated a more robust significance. Specifically, we generated simulated LCs\textsuperscript{\ref{emma}} to estimate the significance of the identified periods by computing the local significance, as explained below. We used the same procedure outlined in Appendix A of \citet{O’Neill_2022}.  Note that this procedure already accounts for the uncertainty arising from the fact that the signal frequency is initially unknown, necessitating a systematic search over a range of possible periods.

 First, we calculate the period significance, defined as the probability of identifying a periodic signal with the same period, \(p_{\mathrm{peak}}\), and equal or greater power, \(P_{\mathrm{peak}}\), at the LSP. This involves evaluating how often the observed period of the AGN is replicated in the simulated LC. Consequently, we derive a $p$-value, denoted as \(p\)-value\(_{\mathrm{peak}}\). The local significance is obtained by applying the same procedure with the periods obtained in the different simulated LCs.

     In summary, the steps of the significance procedure are as follows: 
    \begin{itemize}
        \item[1.] For each simulated LC, we apply the SSA decomposition and obtain the periodicity component.
        \item[2.] We compute the LSP for each periodicity component, obtaining the period, $p_{\mathrm{sim}}$, and the power of the period peak, $P_{\mathrm{sim}}$.
        \item[3.] For each simulated LC, the period significance is calculated, $p$-value$_{\mathrm{sim}}$. We count all simulations where the period $p_{\mathrm{sim}}$ is detected with power $\geq P_{\mathrm{sim}}$.
        \item[4.] To estimate the local significance, we compute the probability that $p$-value$_{\mathrm{sim}} \leq p$-value$_{\mathrm{peak}}$. 
    \end{itemize}

\subsubsection{Global Significance}\label{sec:global}
In periodicity analyses, we must correct the local significance for the look-elsewhere effect \citep[e.g.,~][]{quasi}.  This occurs in situations where a signal is searched for only in a specific region of parameter space \citep{look-else}. In our case, we are searching for periodicity signals within a sample of sources. The look-elsewhere effect describes the ratio between the probability of observing the excess at a fixed value and the probability of observing it anywhere in the range. It is quantified in terms of a specific number of trials \citep{look-else_2, Pablo22} as follows,
\begin{equation}
    p_{\mathrm{global}} = 1 - (1 - p_{\mathrm{local}})^N,
\end{equation}

\noindent  where $p_{\mathrm{local}}$ is the $p$-value associated with the local significance, and $N$ is the trial factor, computed from the total number of blazars in our sample ($N = 494$).

Since our local significance was limited by the total number of simulations, we performed a large number of artificial LC simulations to obtain a meaningful significance. One million simulations were conducted for each AGN in our sample, with the highest significance obtained being 4.8$\sigma$, 1 event as significant as the one observed in 1 million simulated LCs assuming no real periodic signal. After applying the trial factor corrections, we obtained the global significances listed in Table~\ref{tab:trials}.

\vspace{6mm}
\begin{table}
\caption{\label{tab:trials}Local to Global significance}
\centering
\begin{tabular}{c|c}
    \hline\hline
    Local significance ($\sigma$) &  Global significance ($\sigma$)  \\
    \hline
    $ 4.8$ & $ 3.4$\\
    $ 4.5$ & $ 3.0$\\
    $ 4.0$ & $ 2.2$\\
    $ 3.5$ & $ 1.2$\\
    $<3.0$ & $ <1.0$\\
\end{tabular}
\tablefoot{Global significance corresponding to local significance after trial corrections (see Section~\ref{sec:global}).}\label{tab:trials}
\end{table}

\subsection{Evaluation of the method}\label{test}
In this section, we perform tests to evaluate the performance of SSA under different conditions.

\subsubsection{Tests against noise}\label{Tests against noise}
Our evaluation involves a purely periodic signal, resembling a sinusoidal waveform with a predefined period, using the same time sampling as the original LCs. To simulate realistic conditions, we introduced random noise to the signal, specifically white, pink, and red noise, following \citet{timmer}. This process was repeated multiple times to observe how often SSA successfully detects the predetermined period in the presence of noise.

In this study, we used the \citet{timmer} method via the \texttt{colorednoise} \texttt{Python} package to simulate red, pink, and white noise components. This approach evaluates the robustness of SSA in detecting periodic signals under realistic conditions for blazar emissions.

We note that this methodology is consistent with Monte Carlo SSA techniques \citep{MC_SSA}, which combine SSA with Monte Carlo simulations using AR(1) noise models (autoregressive processes where each value depends on the previous one) to assess the statistical significance of detected components.

The periodic signal is computed as:
\begin{equation}
    \phi(t) = A\sin{\left(\frac{2\pi t}{T} + \theta\right)} + O,
\end{equation}
where $A$ is the amplitude, $T$ the period, $\theta$ the phase, and $O$ the offset of the signal. These parameters are randomly selected within a range of values. The ranges were computed using the Markov Chain Monte Carlo sinusoidal fitting method from the blazars in Table \ref{table results}, as explained in Section~3.1.6 of \cite{P20}. The period range is selected from the periods found in this work,
\begin{gather*}
    A = [1-20] \; \times 10^{-6} \; \mathrm{MeV \; cm^{-2}s^{-1}},\\
    O = [0-150] \; \times 10^{-6} \; \mathrm{MeV \; cm^{-2}s^{-1}},\\
    T = [1.0-4.5] \; \mathrm{yr},\\
    \theta = [0,2\pi] \; \mathrm{rad}.
\end{gather*}

We generated the noise structure using the \texttt{Python} package \texttt{colorednoise}\footnote{\url{https://github.com/felixpatzelt/colorednoise}}, based on \citet{timmer}, and considered red, pink, and white noise.

\begin{figure}
    \centering
    \includegraphics[width=\columnwidth]{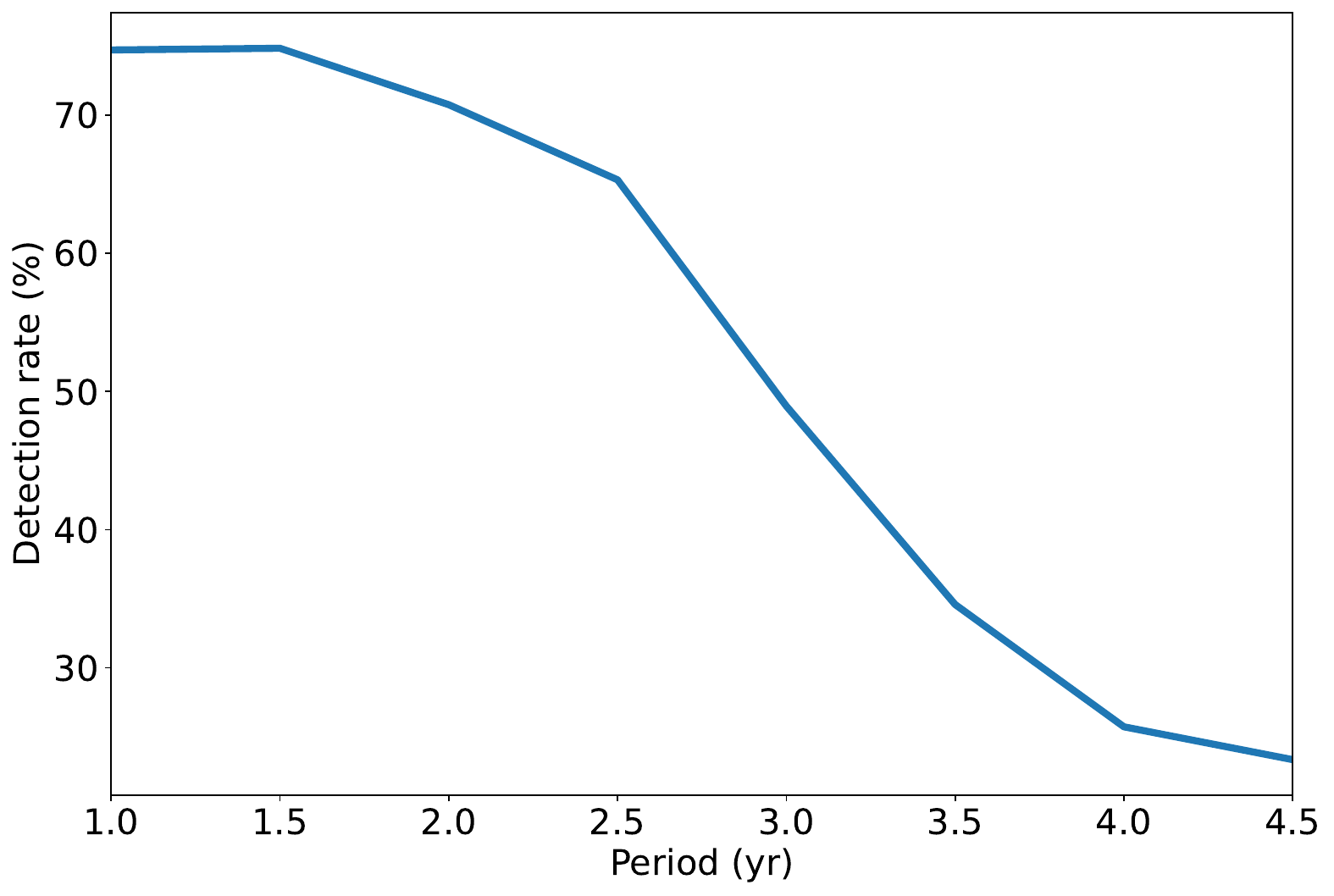}
    \caption{Detection rates as a function of period.
    Rates are computed using simulated LCs with added noise (see Section \ref{Tests against noise}).}
    \label{fig:detection_rates}
\end{figure}

We simulated 100,000 LCs by adding random noise with the same data length and amplitude to the sinusoidal time series for each period. The detection rate depends on the period, as shown in Figure~\ref{fig:detection_rates}.

As expected, the detection rate decreases as the signal period increases. This is a logical outcome since longer periods result in fewer cycles within the analyzed time interval. The methods utilized in \cite{Pablo22} report a detection range spanning approximately 12\% to 65\%, depending on the period range under analysis. For periods in the [1-2.5] year range, SSA exhibits a detection rate of 78\%, improving the methods in \cite{Pablo22} by 18\%. For longer periods in the [3.5-4.5] year range, SSA shows a 50\% enhancement in detection compared to the methods in \cite{Pablo22}. These findings highlight the advantages of SSA in detecting periods in signals contaminated by stochastic noise.

\subsubsection{False positive rate}
We also assessed whether the algorithm used to obtain the period component produces genuine results. For this, we generated pure red, pink, and white noise LCs \citep[using][]{Emmanoulopoulos} with the same time sampling as the original LCs. We applied our algorithm to determine how often we obtained significant results. This allows us to quantify the probability of detecting a period with high significance that has occurred stochastically. We performed 500,000 simulations and find that $0.13\%$ have a significance $\geq 3 \sigma$, and $2.27\%$ have a significance $\geq 2 \sigma$. A false positive rate of 2.27\% implies that, out of 494 stochastic LCs, you would expect 13 to show periodic signals at $2.0 \sigma$.

\subsection{Long-term trends}\label{trend}
Identifying periodic patterns in time series data can be challenging when there is a prominent trend. Applying the LSP to analyze signals with both periodicity and trend components often amplifies the periodogram's peaks. This complicates the detection of underlying periodicity and increases uncertainty in the derived period. It is recommended to detrend the signal as an initial step in periodicity analysis \citep[e.g.,][]{detrend_welsh}, since trends can impact the low-frequency components of the PSD. Such trends may falsely suggest periodic patterns within the time series, leading to erroneous detection of false periodicity \citep[][]{mcquillan_trend_fake_detection}. Therefore, separating the periodic and trend components is essential for accurate periodicity estimates. Additionally, understanding the general condition of the blazar in which periodicity is detected is crucial, as it can significantly impact emissions and the observed periodicity.

In some cases, the trend can dominate the periodic structure, such as when the periodicity and amplitude of the signal increase or decrease as the trend rises or falls. Since the first components obtained through SSA decomposition represent the trend of the time series, we analyze these components to characterize the trend.

It is common to apply statistical time series tests such as the F-test to assess trends. The null hypothesis of the F-test assumes that the trend can be represented as a constant. Applying the F-test to our analysis indicated that all blazars exhibited a trend. However, note that this test assesses the entire time series signal, while we focus solely on the trend reported by SSA. The F-test concludes that the trend is not constant because SSA does not report a perfectly consistent line; instead, it reveals small-scale variations around the overall blazar emission state. Since we are interested in long-term trends, the F-test is unsuitable for this analysis, as it consistently indicates a trend due to small-scale fluctuations. To address this, we propose an ad-hoc conservative criterion to select blazars with trends. A linear regression of the trend component was performed to find the best-fit slope. We considered the trend as increasing when the slope was $>0.02$, decreasing when the slope was $<-0.02$, and constant when the slope was in the range $[-0.02, 0.02]$.

In Section~\ref{trends_results}, we characterize the trends of our periodicity candidates.

\subsection{Forecasts}\label{forecast_methodology}
Predicting future states of $\gamma$-ray blazars is important, especially for planning observational proposals. In the context of periodicity, understanding and characterizing underlying trends and periodic patterns enables the development of predictive models for future source behavior. To achieve this, we applied SSA, specifically recurrent SSA forecasting \citep[SSA-R\footnote{\url{https://github.com/AndrewSukhobok95/ssa}},][]{recurrent_forecast}.

Our method forecasts under the assumption that the time series satisfies a Linear Recurrent Relation (LRR). The time series $(y_1, \ldots, y_N)$ satisfies an LRR of order $d$ if there are coefficients $a_1, \ldots, a_d$ such that:
\begin{equation}
    y_{i+d} = \displaystyle\sum_{k=1}^d a_k y_{i+d-k}, \quad 1\leq i \leq N-d, \quad a_d \neq 0, \quad d < N.
\end{equation}
The coefficients $(a_1,\ldots, a_d)$ are the LRR coefficients.

Let $I$ be the set of eigentriples determined at the grouping step of SSA, as described in Section~\ref{SSA}. The first $r$ eigentriples are selected to obtain forecasts of the reconstructed series. In our case, we considered only the group containing the trend and the periodicity component. If $i \in I$, and the eigenvectors of the chosen eigentriples are $U_i \in \mathbf{R}^L$, denoted by $\underline{U_i} \in \mathbf{R}^{L-1}$ (the vector consisting of the first $L-1$ components of $U_i$, where $L$ is the window length referenced earlier), and $\pi_i$ the last component of $U_i$, then $v^2 = \sum_{i\in I}\pi_i^2$. Let $\mathcal{L} \subset \mathbf{R}^L$ be the linear space spanned by the $U_i$ vectors, where $\mathcal{L} = \text{span}(U_i, i \in I)$. Assuming $\mathcal{L}$ is a non-vertical space (i.e., $e_L \notin \mathcal{L}$, with $e_L = (0, 0, \ldots, 1)^\mathrm{T} \in \mathbf{R}^L$), then $v^2 < 1$.

We define the vector $R=(a_{L-1},\dots, a_1)^\mathrm{T}$ as:
\begin{equation}
    R = \frac{1}{1-v^2}\sum_{i\in I}\pi_i \underline{U_i},
\end{equation}
and the SSA-R forecasting algorithm can be summarized as follows:\\

The $i$-th element of the time series $Y_{N+h} = \{y_1, \ldots, y_{N+h}\}$ is defined as:
\begin{equation}\label{forecast_eq}
y_i = \begin{cases}
    \widetilde{y_i} & \text{for } i = 1, \dots, N \\
    \sum_{j=1}^{L-1} a_j y_{i-j} & \text{for } i = N+1, \dots, N+h,
\end{cases}
\end{equation}
where $\widetilde{y_i}$ ($i=1,\dots, N$) are the reconstructed series. The values $y_{N+1},\dots, y_{N+h}$ are the $h$-step-ahead recurrent forecasts \citep[for further details, see Section~3 in][]{SSA_R}.

The confidence intervals are obtained by considering the residuals, that is, the components not used for forecasting (in particular, noise). The procedure is as follows.

Set the $Y_N$ time series, which can be described as a sum of components $Y_N^{(1)} + Y_N^{(2)}$, where $Y_N^{(1)}$ is governed by an LRR of relatively small dimension, and the residual series $Y_N^{(2)}$, $Y_N^{(2)} = Y_N - Y_N^{(1)}$, is approximately strongly separable from $Y_N^{(1)}$ at some window length $L$.

To establish confidence bounds for the forecast, we assume that $Y^{(2)}$ is a finite random noise series perturbing the signal $Y^{(1)}$.

Consider a bootstrap method to construct the confidence bounds for the signal $Y_N^{(1)}$ with $\mathrm{rank} \; r < L$. By applying SSA and reconstructing the signal, we obtain $Y_N = \Tilde Y_N^{(1)} + \Tilde Y_N^{(2)}$. In our case, $\Tilde Y_N^{(1)}$ is the reconstructed component related to the trend and periodicity of the studied source. The next step is to calculate the empirical distribution of the residuals $\Tilde Y_N^{(2)}$ and simulate $Q$ independent copies $\Tilde Y_{N, q}^{(2)}$ of the $Y_N^{(2)}$ series. We apply the forecasting procedure to $Q$ independent time series defined as $\Tilde{Y}_{N,q}=  \Tilde Y_N^{(1)} +  \Tilde Y_{N,q}^{(2)} \quad (q = 1, \ldots, Q)$. With an $M$ step-ahead forecast, we obtain the elements $\Tilde{y}^{(1)}_{N+M,q} \quad (1\leq q \leq Q)$. Afterward, we calculate the (empirical) lower and upper quantiles at some level (in our case, $95\%$) and obtain the corresponding confidence interval for the forecast, known as bootstrap confidence intervals.

\subsection{Software}

We used \texttt{Python} to perform our analysis, documented in a Jupyter Notebook that is made public at \url{www.ucm.es/blazars/ssa} for clarity and reproducibility. The underlying SSA analysis relies on commonly used libraries such as \texttt{numpy} for matrix operations (e.g., SVD using \texttt{numpy.linalg.svd}).

Established \texttt{Python} libraries such as \texttt{pyrssa}\footnote{\url{https://pypi.org/project/pyrssa/}} and \texttt{pyts.decomposition}\footnote{\url{https://pyts.readthedocs.io/en/stable/generated/pyts.decomposition.SingularSpectrumAnalysis.html}} can be used to reproduce the results presented in this paper. A tutorial available on Kaggle\footnote{\url{https://www.kaggle.com/code/jdarcy/introducing-ssa-for-time-series-decomposition}} offers a helpful introduction to SSA.

To compute LSP, the \texttt{scipy.signal.lombscargle} function was utilized. For forecasting, SSA-R forecasting techniques were employed to predict the future behavior of the sources. \texttt{Python} packages such as \texttt{pyrssa}, \texttt{MSSA}\footnote{\url{https://pypi.org/project/py-ssa-lib/}}, and the implementation available on GitHub\footnote{\url{https://github.com/AndrewSukhobok95/ssa}} provide robust tools for reproducing forecasting results.

\section{Results and discussion} \label{results}
In this section, we provide a comprehensive discussion of the results derived in this work, listing periodicity candidates and conducting an extensive literature search to compare our findings with prior research. In Section \ref{test}, we evaluated SSA's detection capabilities in the presence of random noise and assessed the likelihood of obtaining high-significance candidates that may correspond to purely stochastic emission sources.

\subsection{Periodicity candidates}

From our systematic periodicity search pipeline, we identified 46 AGN with evidence of periodic $\gamma$-ray emissions, including 24 flat spectrum radio quasars (FSRQs), 18 BL Lacertae objects (BL Lacs), 3 blazars of uncertain type (BCUs), and 1 Narrow-Line Seyfert 1 (NLSy1) galaxy. Table~\ref{table results} lists these candidates, along with their periods and local significance calculated with $10^6$ simulations.

\begin{figure}
    \centering
    \includegraphics[width=\columnwidth]{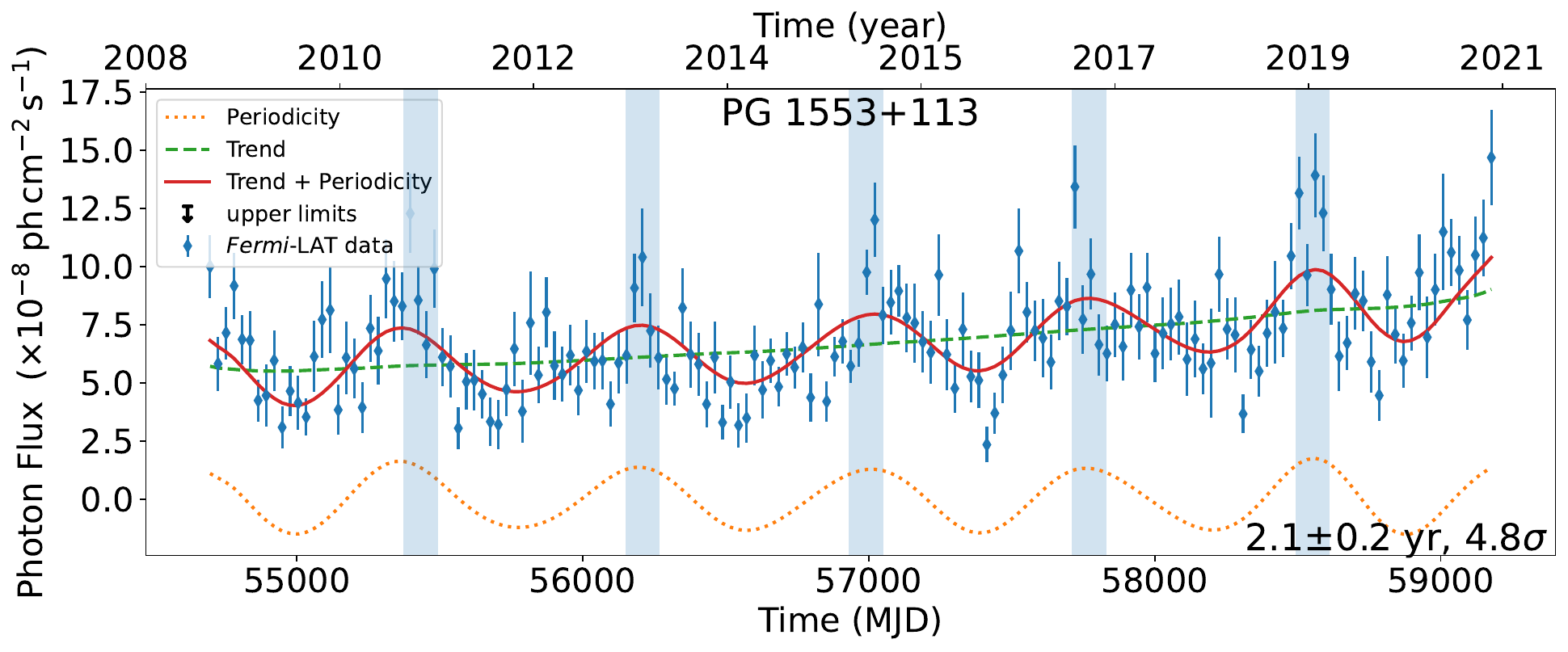}
    \caption{SSA applied to PG 1553+113. Blue dots represent the original signal with error bars. Orange, green, and red lines correspond to the periodicity component, the trend, and the sum of the trend and the periodicity reported by the SSA algorithm, respectively. The spacing between the blue bars indicates the period computed through the LSP (2.2 yr). The width of the blue bars is the period uncertainty (0.2 yr) computed from the FWHM of the period peak in the LSP.}
    \label{fig:SSA}
\end{figure}

Figure~\ref{fig:SSA} shows the results obtained using SSA for the blazar PG 1553+113. This blazar has gained significant attention due to the detection of a 2.2-year period with notable significance \citep{Ackermann_2015, prokhorov, Pablo22}. Our analysis yielded a $2.2 \pm 0.2$ yr period,  which is consistent with previous studies, supported by a local significance of $\sim 4 \sigma$ (global significance $\sim 2.2 \sigma$), as reported in table \ref{table results}.  Detailed plots corresponding to the candidates presented in this paper can be found in Appendix~\ref{plots}.

\begin{figure}
    \centering
    \includegraphics[width=\columnwidth]{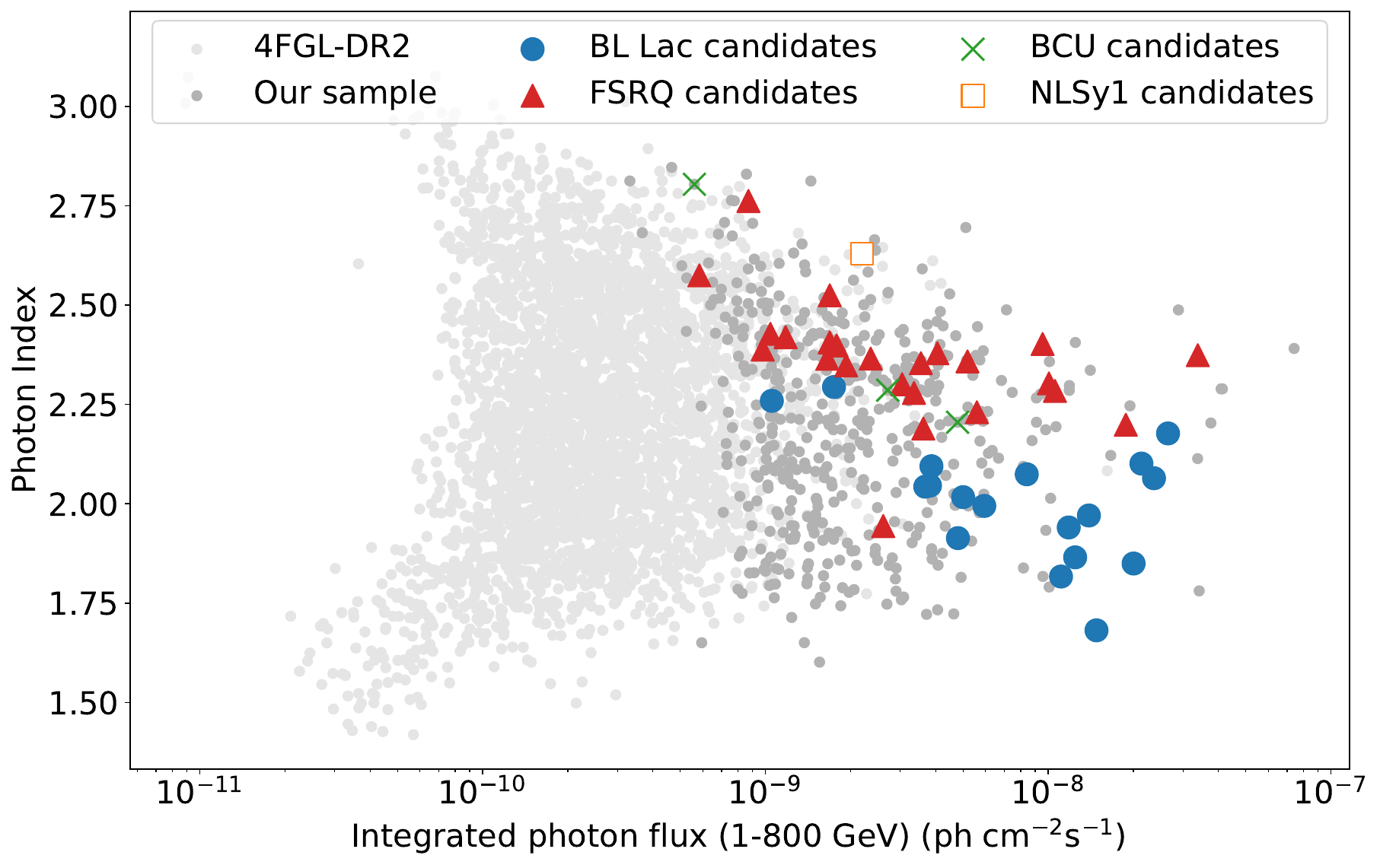}
    \caption{Photon spectral index as a function of integrated photon flux. We show all the blazars detected by \textit{Fermi}-LAT (gray circles) and blazars selected for periodicity analysis with a variability index above 18.48 (dark gray circles). Our periodicity candidates are shown according to their AGN type: BL Lacs (blue circles), FSRQs (red triangles), NLSy1 galaxies (orange square), and BCUs (green crosses).}
    \label{fig:blazars}
\end{figure}

Figure~\ref{fig:blazars} shows the relationship between the photon index and integrated photon flux of all the blazars detected by \textit{Fermi}-LAT and our periodicity candidates. Our periodicity candidates are typically detected for bright sources with integrated flux $\gtrsim 10^{-9} \mathrm{ph \; cm^{-2}s^{-1}}$.

\begin{figure}
    \centering
    \includegraphics[width=\columnwidth]{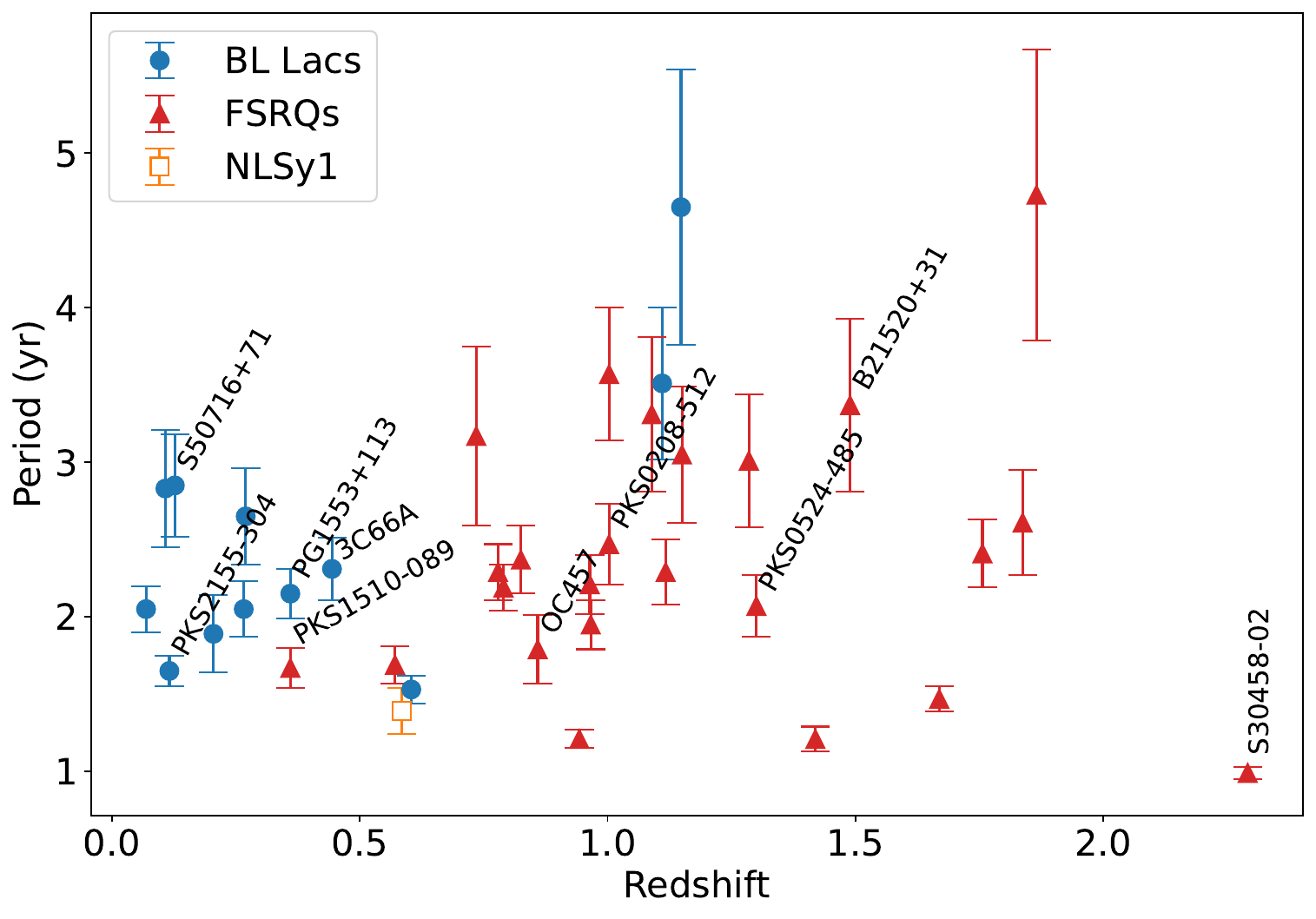}
    \caption{ Period in years as a function of redshift, with sources classified according to AGN type. PG 1553+113 and other candidates discussed in Section~4.1.1 and Section~4.1.2 are labeled. }
    \label{fig:redshift}
\end{figure}

\begin{figure}
    \centering
    \includegraphics[width=\columnwidth]{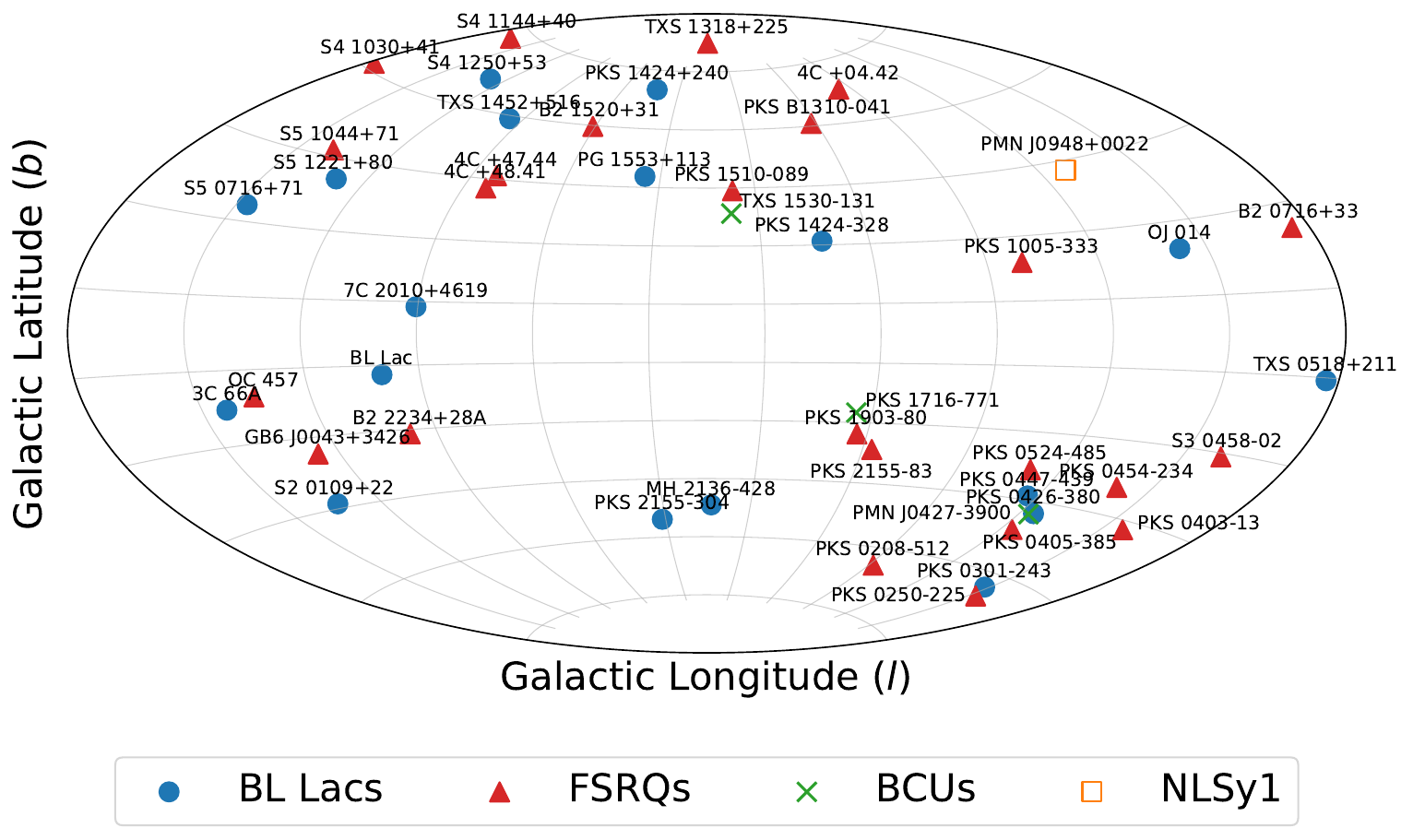}
    \caption{Sky map of periodicity candidates classified by AGN type.}
    \label{fig:skymap}
\end{figure}

Furthermore,  Figure~\ref{fig:redshift}  shows the distribution of redshifts for our candidates according to the detected periodicities. This distribution aligns with expectations, as BL Lac objects tend to be found at lower redshifts and FSRQs at higher redshifts \citep{4LAC}. Candidates with longer periods tend to exhibit greater uncertainty, as red noise has more impact at lower frequencies, leading to broader peaks in the LSP and fewer cycles in the data. Figure~\ref{fig:skymap} shows the sky map of the periodicity candidates.

We categorized the selected sources into two groups: those for which previous studies have reported periodicity in $\gamma$-ray emissions and those that are newly identified periodicity candidates presented in this work.

\subsubsection{New $\gamma$-ray periodicity candidates}
Twenty-five sources are reported as new $\gamma$-ray periodicity candidates in our analysis with a local significance $\gtrsim 2 \sigma$. Of these, 14 are classified as FSRQs, 8 as BL Lac-type blazars, and 3 as BCUs. BCUs have a broad-band SED resembling that of blazars but lack spectroscopic confirmation. Below are examples of new candidates, with their periods and local significance.  Note that a local significance of 4.8$\sigma$ is the maximum significance achievable given the simulations we conducted.

\begin{itemize}
    \item OC 457. We find a $1.8 \pm 0.2$ yr $(4.8 \sigma)$ period. The LC is shown in Appendix \ref{plots}. This blazar is a case where the trend decreases, causing the periodicity to disappear toward the end. Consecutive upper limits and low emission also impact the analysis, but this is the first time OC 457 has been identified as a periodicity candidate.
    
    \item PKS 0524$-$485. A $2.1 \pm 0.2$ yr $(4.8 \sigma)$ period was detected. Unlike OC 457, this blazar shows a clear periodic pattern over time, with a consistent and steady increase in the long-term trend, suggesting a strong connection between the trend and the observed periodicity.

    \item TXS 1530$-$131. A $1.4 \pm 0.1$ yr $(4.8 \sigma)$ period was detected. The trend is highly consistent, with the periodic cycles maintaining both their frequency and amplitude, indicating regularity in their behavior.
    
    \item PKS 2155$-$83. A $4.7 \pm 0.9$ yr $(4.8 \sigma)$ period was found. This is the longest-period candidate, but the uncertainty is higher due to red noise contamination at lower frequencies, leading to a broader LSP peak.
    
    \item 3C 66A. Previous studies report quasi-periodic behavior in the optical band but not in $\gamma$-rays, with a period of $\sim 2.3 \pm 0.3$ yr in the \textit{V}-band \citep{3c66a_1, 3c66a_2}. We find a $2.3 \pm 0.2$ yr $(\sim 3.8 \sigma)$ period in $\gamma$-rays, indicating a connection between the optical and $\gamma$-ray emission mechanisms.
\end{itemize}

\subsubsection{Periodicity candidates in literature}
Twenty-one of the 46 blazars with significant periodic emission in this work have a period reported in the literature across multiple bands. Table~\ref{literature} lists these blazars, ordered by highest to lowest local significance according to our results. Many of the results align with those in previous works. Below, we discuss some specific cases.

\begin{itemize}

    \item S3 0458$-$02. \cite{Pablo22} find a $\sim 3.8 \pm 0.9$ yr period, while we find a $\sim 1.00 \pm 0.04$ yr $(4.8 \sigma)$ period. Although SSA can detect both periods, the significance of the 3.8-year period is lower than 2$\sigma$. The 1-year period is more distinct in the LC, demonstrating SSA's ability to uncover underlying periodicities more effectively.

    \item B2 1520+31. We detect a $3.4 \pm 0.6$ yr $(4.8 \sigma)$ period. The literature reports a periodicity of a few days \citep{B21520, QPO}, but no long-term period has been reported.

    \item S5 0716+71. \cite{Pablo22} find a $\sim 2.7 \pm 0.4$ yr period, compatible with our result of $2.9 \pm 0.3$ yr $(\sim 4.5 \sigma)$. \cite{QPO} and \cite{Pablo22} also report a secondary period of $\sim 0.9$ yr, which we also detect with SSA.

    \item PKS 0208$-$512. \cite{P20} reported a $\sim 2.7 \pm 0.1$ yr period, while \cite{Pablo22} reported a $\sim 3.8 \pm 0.5$ yr period. We detect a $2.5 \pm 0.3$ yr $(\sim 4.5\sigma)$ period. The discrepancy arises from a flare in 2020 that distorts the periodicity, leading to a detected 4-year period. SSA more accurately detects the underlying $\sim 2.5$ yr period.

    \item PKS 1510$-$089. \cite{pks1510} find a period of $\sim 3.6 \pm 0.2$ yr, while we detect a period of $1.7 \pm 0.1$ yr $(\sim 2.5 \sigma)$. Although no other studies report long-term periodicity, there have been studies reporting transient periodicities of a few days \citep{pks1510_2}.
\end{itemize}

\begin{figure}
    \centering
    \includegraphics[width=\hsize]{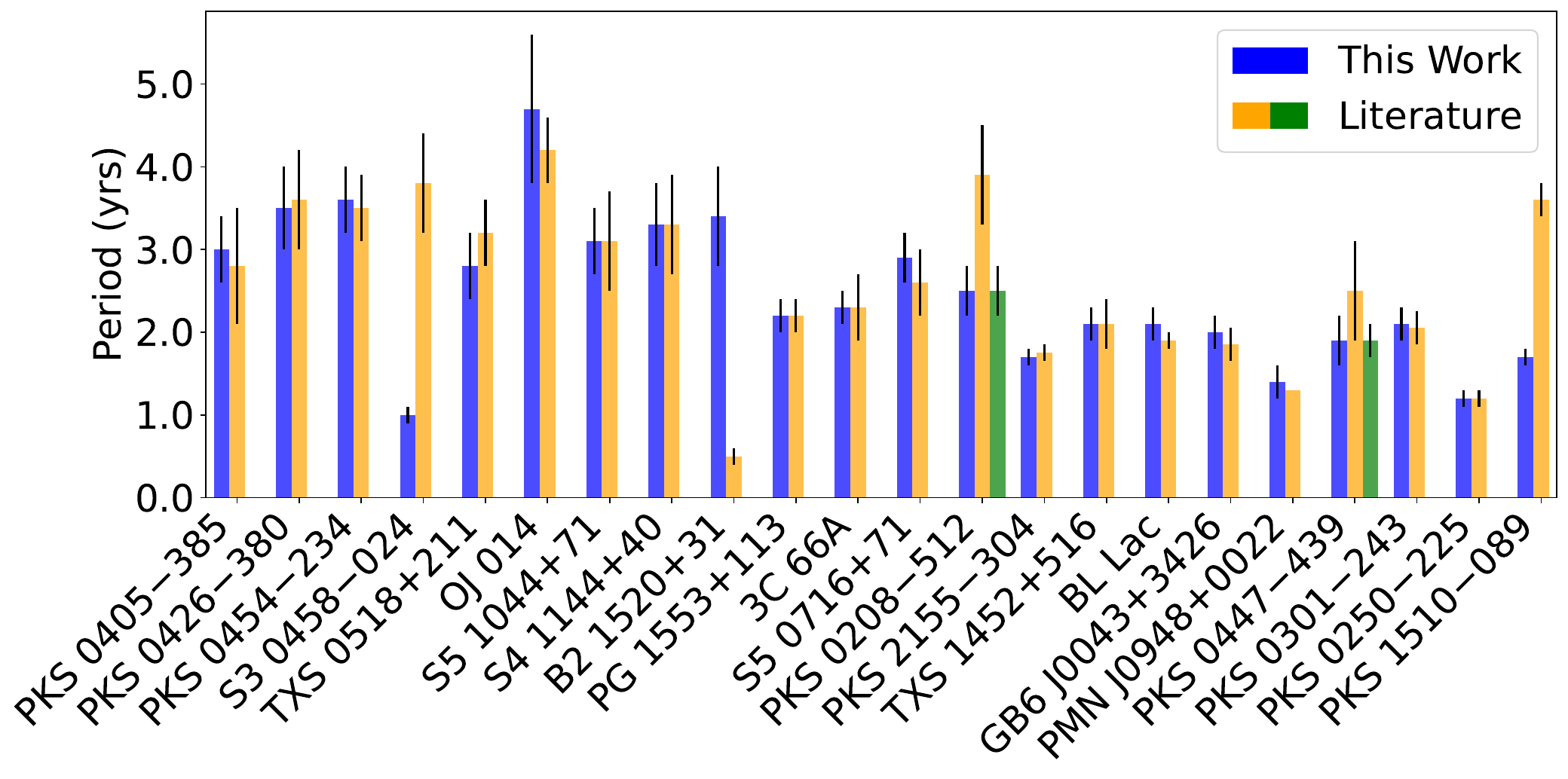}
    \caption{Periodicity results using SSA compared with those reported in the literature, as summarized in Table~\ref{literature}. Blue bars represent SSA results from this work, while orange and green bars represent periodicity values reported in the literature. Orange bars indicate periodicity values from individual studies, while green bars are added when other studies report additional periodicity values for the same blazar. Error bars indicate the uncertainties associated with each period.}
    \label{fig:comparison}
\end{figure}
Figure ~\ref{fig:comparison} presents a comparison between the periods detected using SSA and those reported in the literature, as summarized in Table ~\ref{literature}. 

\subsection{Long-term trends}\label{trends_results}
We conducted a comprehensive trend analysis, following the methods outlined in Section \ref{trend}, for all the periodicity candidates in this study. The results are presented in Table~\ref{table4}, showing that 21 candidates exhibit a constant long-term trend, 13 show increasing trends, and 12 show decreasing trends.

An example is the NLSy1 PKS 0250$-$225, where the trend is decreasing, and the periodicity disappears over time, following the trend (see LC in Appendix \ref{plots}). The opposite case is the blazar 4FGL MH 2136$-$428, where the periodicity becomes more evident and the amplitude increases as the trend increases. SSA is particularly effective in such cases for detecting significant periods.

\subsection{Forecast}\label{forecast_results}
For all candidates, we used the algorithm described in Section \ref{forecast_methodology} to predict the $\sim$10-year upcoming gamma-ray emission peaks from 2021. Table~\ref{table4} details the next four emission cycles. The uncertainty in peak prediction was based on the period uncertainty calculated with the LSP. In some cases, there were no values for the second peak prediction because the next emission occurs beyond the time interval used for forecasting. For the NLSy1 galaxy PMN J0948+0022, predictions were not possible because the amplitude trends toward zero, leading to the disappearance of periodicity.

In cases where periodicity disappears, the computed confidence interval becomes a rectangle, where the periodic time structure fades, and successive cycles cannot be predicted reliably. To determine significant forecasts and calculate the dates of future emission peaks, we applied a methodology that fits the upper bound of the confidence interval to a linear regression. We considered candidates with an $R^2$ parameter of less than 50\%, a moderate value according to \citet{hair_r2_2011}. Forecasting is inadequate for blazars where periodicity was detected in more advanced SSA components, such as those associated with flares (e.g., PKS 0208-512). Our forecasting models are designed to work with AGN both with and without long-term trends. When a source lacks a long-term trend, the model adapts to the data without introducing an artificial trend. As a result, the absence of long-term trends does not negatively impact the overall results, as the models are flexible and reflect the true characteristics of each source. 

\begin{figure}
    \centering
    \includegraphics[width=\columnwidth]{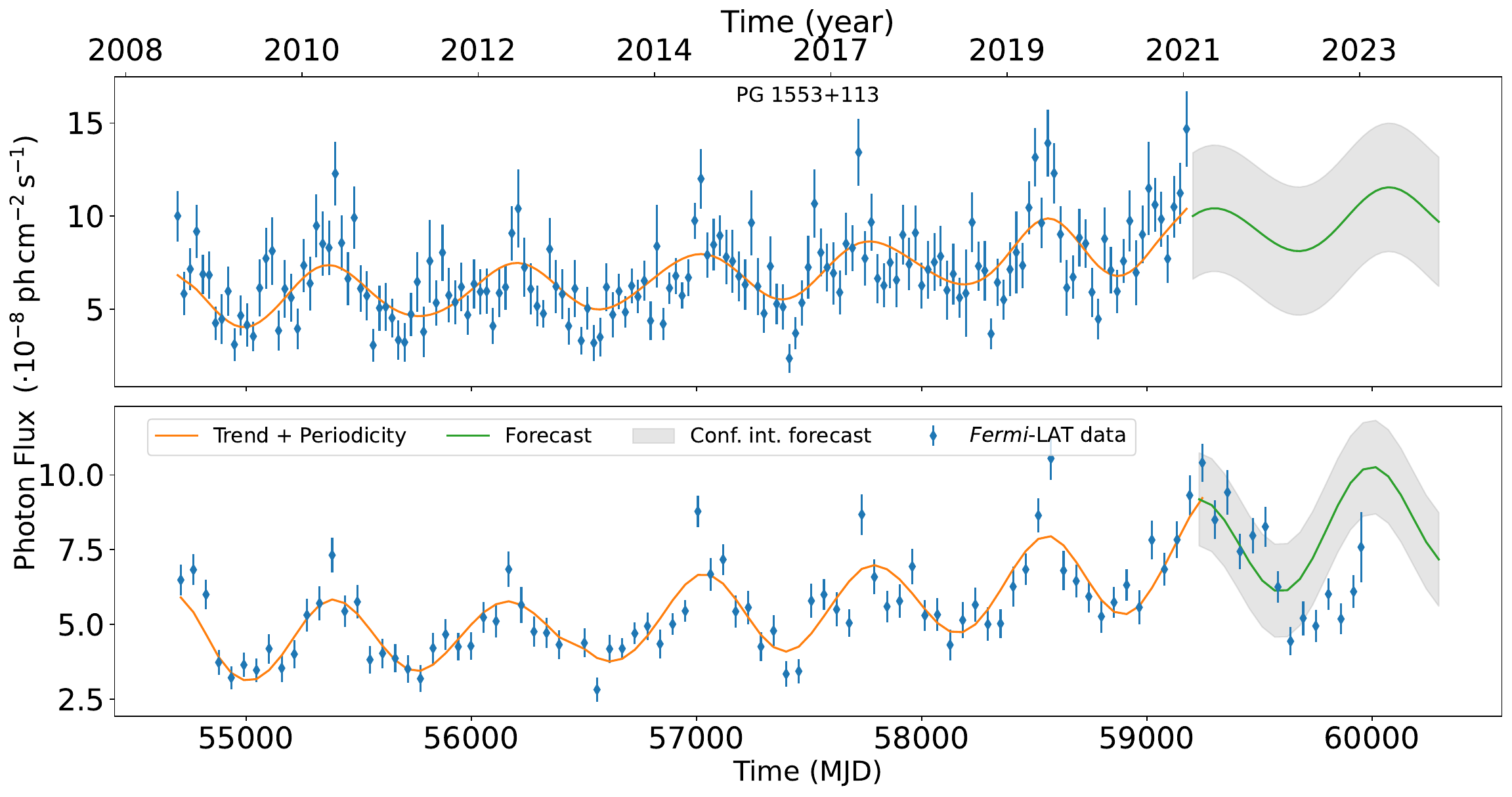}
    \caption{SSA-R forecasting models for PG 1553$+$113.
    (\textit{Top panel}) SSA-R forecasting model for PG 1553+113 using 28-day time-bin data. 
    (\textit{Bottom panel}) SSA-R forecasting model for PG 1553+113 using 56-day time-bin data until 2023.}
    \label{fig:forecast}
\end{figure}

We successfully predicted the next four $\gamma$-ray emission cycles for 28 candidates. Figure~\ref{fig:forecast} shows the forecasting model for PG 1553+113. We used the 28-day time-bin data from 2008 to 2021 to predict the next two cycles. Appendix \ref{plots_forecast} contains forecasting plots for all candidates, except those where the methodology was inadequate.

As a proof of concept, we evaluated the model's efficacy by forecasting the 56-day time-bin data of PG 1553+113 up to early 2023. We initiated the forecast around the end of 2020 and predicted the next two cycles. The 56-day time-bin data is shown as blue points in Figure~\ref{fig:forecast}. The results demonstrate a good fit between the forecasted and observed data,  as can be visually observed in Figure~\ref{fig:forecast} (bottom).

When comparing our predicted dates with updated data from the \textit{Fermi}-LAT Light Curve Repository\footnote{\url{https://fermi.gsfc.nasa.gov/ssc/data/access/lat/LightCurveRepository/}}, we observe that the next cycle occurred in February 2021, compared to our initial prediction of March 2021. The uncertainties arise from the period uncertainty. The subsequent cycle occurred on May 3, 2023, while our prediction was May 8, 2023. These results for PG 1553+113 show a good agreement with \textit{Fermi}-LAT data, considering the uncertainties in cycle period determination.

\section{Summary and conclusions}\label{sec:conclusions}
In this study, we employed SSA to conduct a systematic search for periodic patterns in AGN within the time domain. This marks the first application of this powerful statistical technique in this field. Our analysis includes 494 gamma-ray sources with well-sampled LCs. SSA effectively isolates underlying oscillatory components within the signal, detecting periodicities that may remain undetected by conventional methods. We employed a robust methodology to determine the significance of the detected periodicities by generating artificial LCs using Monte Carlo methods. SSA also facilitates long-term trend characterization and the development of forecasting models capable of predicting future emission peaks.

The main results of our study are as follows:
\begin{enumerate}
    \item We identified 46 $\gamma$-ray periodicity candidates with a local significance of $\geq 2\sigma$, including 21 known candidates and 25 newly discovered in this work, effectively doubling the number of known candidates.
    \item Long-term trends were characterized for 25 of these AGN.
    \item Forecasting models were provided for 28 of the AGN.
\end{enumerate}

While this study successfully identifies $\gamma$-ray periodicities in a sample of blazars, the task of linking these periodicities to specific physical mechanisms remains unresolved. The periodicities observed on year-long timescales could plausibly be attributed to processes such as jet precession \citep[e.g.,][]{lighthouse,Britzen18}, accretion flow instabilities such as warped disks or quasi-periodic oscillations \citep[e.g.,][]{modulation, Mishra_2022}, or SMBBH orbital motion \citep[e.g.,][]{graham15,sobacchi17}. However, distinguishing between these scenarios necessitates detailed multiwavelength observational campaigns to investigate correlations across different parts of the electromagnetic spectrum \citep[e.g.,][]{penil_mwl_pg1553}, the study of the time-dependence of the QPOs to examine their stability or evolution \citep[e.g.,~][]{2024ApJ...976..203A}, and theoretical simulations \citep[e.g.,~][]{2022PhRvD.106j3010W}. Such analyses would enable a more comprehensive understanding of the physical origins of the observed periodic behavior but fall outside the scope of this work. This study provides a foundation for future investigations by identifying promising candidates and highlighting their potential significance within the context of blazar variability.


\begin{acknowledgements}
The authors thank Sarah Wagner and the anonymous referee for their valuable comments, which improved the manuscript.

A.R. acknowledges the support of an Investigo grant funded by the European Union, Next Generation EU. A.D. is thankful for the support of Proyecto PID2021-126536OA-I00 funded by MCIN / AEI / 10.13039/501100011033.  A.R. and M.A. are grateful for the support of the {\it Fermi} Guest Investigator Cycle 17 program (proposal number 171063).

The \textit{Fermi}-LAT Collaboration acknowledges generous ongoing support from a number of agencies and institutes that have supported both the development and the operation of the LAT as well as scientific data analysis. These include the National Aeronautics and Space Administration and the Department of Energy in the United States, the Commissariat \`a l'Energie Atomique and the Centre National de la Recherche Scientifique / Institut National de Physique Nucl\'eaire et de Physique des Particules in France, the Agenzia Spaziale Italiana and the Istituto Nazionale di Fisica Nucleare in Italy, the Ministry of Education,Culture, Sports, Science and Technology (MEXT), High Energy Accelerator Research Organization (KEK) and Japan Aerospace Exploration Agency (JAXA) in Japan, and the K.~A.~Wallenberg Foundation, the Swedish Research Council and the Swedish National Space Board in Sweden.

Additional support for science analysis during the operations phase is gratefully acknowledged from the Istituto Nazionale di Astrofisica in Italy and the Centre National d'\'Etudes Spatiales in France. This work was performed in part under DOE Contract DE-AC02-76SF00515.

\end{acknowledgements}



\bibliographystyle{aa} 
\bibliography{ssa} 

\begin{thebibliography}{77}
\expandafter\ifx\csname natexlab\endcsname\relax\def\natexlab#1{#1}\fi

\bibitem[{Abdollahi {et~al.}(2020)Abdollahi, Acero, Ackermann, Ajello, Atwood, Axelsson, Baldini, Ballet, Barbiellini, Bastieri, Gonzalez, Bellazzini, Berretta, Bissaldi, Blandford, Bloom, Bonino, Bottacini, Brandt, Bregeon, Bruel, Buehler, Burnett, Buson, Cameron, Caputo, Caraveo, Casandjian, Castro, Cavazzuti, Charles, Chaty, Chen, Cheung, Chiaro, Ciprini, Cohen-Tanugi, Cominsky, Coronado-Bl{\'{a} }zquez, Costantin, Cuoco, Cutini, D'Ammando, DeKlotz, de~la Torre~Luque, de~Palma, Desai, Digel, Lalla, Mauro, Venere, Dom{\'{\i}}nguez, Dumora, Dirirsa, Fegan, Ferrara, Franckowiak, Fukazawa, Funk, Fusco, Gargano, Gasparrini, Giglietto, Giommi, Giordano, Giroletti, Glanzman, Green, Grenier, Griffin, Grondin, Grove, Guiriec, Harding, Hayashi, Hays, Hewitt, Horan, J{\'{o}}hannesson, Johnson, Kamae, Kerr, Kocevski, Kovac'evic', Kuss, Landriu, Larsson, Latronico, Lemoine-Goumard, Li, Liodakis, Longo, Loparco, Lott, Lovellette, Lubrano, Madejski, Maldera, Malyshev, Manfreda, Marchesini, Marcotulli,
  Mart{\'{\i}}-Devesa, Martin, Massaro, Mazziotta, McEnery, Mereu, Meyer, Michelson, Mirabal, Mizuno, Monzani, Morselli, Moskalenko, Negro, Nuss, Ojha, Omodei, Orienti, Orlando, Ormes, Palatiello, Paliya, Paneque, Pei, Pe{\~{n}}a-Herazo, Perkins, Persic, Pesce-Rollins, Petrosian, Petrov, Piron, Poon, Porter, Principe, Rain{\`{o}}, Rando, Razzano, Razzaque, Reimer, Reimer, Remy, Reposeur, Romani, Parkinson, Schinzel, Serini, Sgr{\`{o}}, Siskind, Smith, Spandre, Spinelli, Strong, Suson, Tajima, Takahashi, Tak, Thayer, Thompson, Tibaldo, Torres, Torresi, Valverde, Klaveren, van Zyl, Wood, Yassine, \& Zaharijas}]{Abdollahi_2020}
Abdollahi, S., Acero, F., Ackermann, M., {et~al.} 2020, ApJS, 247, 33

\bibitem[{{Abdollahi} {et~al.}(2024){Abdollahi}, {Baldini}, {Barbiellini}, {Bellazzini}, {Berenji}, {Bissaldi}, {Blandford}, {Bonino}, {Bruel}, {Buson}, {Cameron}, {Caraveo}, {Casaburo}, {Cavazzuti}, {Cheung}, {Chiaro}, {Ciprini}, {Cozzolongo}, {Cristarella Orestano}, {Cutini}, {D'Ammando}, {Di Lalla}, {Dirirsa}, {Di Venere}, {Dom{\'\i}nguez}, {Fegan}, {Ferrara}, {Fiori}, {Fukazawa}, {Funk}, {Fusco}, {Gargano}, {Garrappa}, {Gasparrini}, {Germani}, {Giglietto}, {Giordano}, {Giroletti}, {Green}, {Grenier}, {Guiriec}, {Hays}, {Horan}, {Kuss}, {Larsson}, {Laurenti}, {Li}, {Liodakis}, {Longo}, {Loparco}, {Lott}, {Lovellette}, {Lubrano}, {Maldera}, {Malyshev}, {Manfreda}, {Marcotulli}, {Mart{\'\i}-Devesa}, {Mazziotta}, {Mereu}, {Michelson}, {Mitthumsiri}, {Mizuno}, {Monzani}, {Morselli}, {Moskalenko}, {Negro}, {Omodei}, {Orienti}, {Orlando}, {Ormes}, {Paneque}, {Perri}, {Persic}, {Pesce-Rollins}, {Porter}, {Principe}, {Rain{\`o}}, {Rando}, {Rani}, {Razzano}, {Reimer}, {Reimer}, {Saz Parkinson}, {Scotton}, {Serini},
  {Sesana}, {Sgr{\`o}}, {Siskind}, {Spandre}, {Spinelli}, {Suson}, {Tajima}, {Takahashi}, {Tak}, {Thayer}, {Thompson}, {Torres}, {Valverde}, {Verrecchia}, \& {Zaharijas}}]{2024ApJ...976..203A}
{Abdollahi}, S., {Baldini}, L., {Barbiellini}, G., {et~al.} 2024, \apj, 976, 203

\bibitem[{{Acero} {et~al.}(2015){Acero}, {Ackermann}, {Ajello}, {Albert}, {Atwood}, {Axelsson}, {Baldini}, {Ballet}, {Barbiellini}, {Bastieri}, {Belfiore}, {Bellazzini}, {Bissaldi}, {Blandford}, {Bloom}, {Bogart}, {Bonino}, {Bottacini}, {Bregeon}, {Britto}, {Bruel}, {Buehler}, {Burnett}, {Buson}, {Caliandro}, {Cameron}, {Caputo}, {Caragiulo}, {Caraveo}, {Casandjian}, {Cavazzuti}, {Charles}, {Chaves}, {Chekhtman}, {Cheung}, {Chiang}, {Chiaro}, {Ciprini}, {Claus}, {Cohen-Tanugi}, {Cominsky}, {Conrad}, {Cutini}, {D'Ammando}, {de Angelis}, {DeKlotz}, {de Palma}, {Desiante}, {Digel}, {Di Venere}, {Drell}, {Dubois}, {Dumora}, {Favuzzi}, {Fegan}, {Ferrara}, {Finke}, {Franckowiak}, {Fukazawa}, {Funk}, {Fusco}, {Gargano}, {Gasparrini}, {Giebels}, {Giglietto}, {Giommi}, {Giordano}, {Giroletti}, {Glanzman}, {Godfrey}, {Grenier}, {Grondin}, {Grove}, {Guillemot}, {Guiriec}, {Hadasch}, {Harding}, {Hays}, {Hewitt}, {Hill}, {Horan}, {Iafrate}, {Jogler}, {J{\'o}hannesson}, {Johnson}, {Johnson}, {Johnson}, {Johnson}, {Kamae},
  {Kataoka}, {Katsuta}, {Kuss}, {La Mura}, {Landriu}, {Larsson}, {Latronico}, {Lemoine-Goumard}, {Li}, {Li}, {Longo}, {Loparco}, {Lott}, {Lovellette}, {Lubrano}, {Madejski}, {Massaro}, {Mayer}, {Mazziotta}, {McEnery}, {Michelson}, {Mirabal}, {Mizuno}, {Moiseev}, {Mongelli}, {Monzani}, {Morselli}, {Moskalenko}, {Murgia}, {Nuss}, {Ohno}, {Ohsugi}, {Omodei}, {Orienti}, {Orlando}, {Ormes}, {Paneque}, {Panetta}, {Perkins}, {Pesce-Rollins}, {Piron}, {Pivato}, {Porter}, {Racusin}, {Rando}, {Razzano}, {Razzaque}, {Reimer}, {Reimer}, {Reposeur}, {Rochester}, {Romani}, {Salvetti}, {S{\'a}nchez-Conde}, {Saz Parkinson}, {Schulz}, {Siskind}, {Smith}, {Spada}, {Spandre}, {Spinelli}, {Stephens}, {Strong}, {Suson}, {Takahashi}, {Takahashi}, {Tanaka}, {Thayer}, {Thayer}, {Thompson}, {Tibaldo}, {Tibolla}, {Torres}, {Torresi}, {Tosti}, {Troja}, {Van Klaveren}, {Vianello}, {Winer}, {Wood}, {Wood}, {Zimmer}, \& {Fermi-LAT Collaboration}}]{3FGL}
{Acero}, F., {Ackermann}, M., {Ajello}, M., {et~al.} 2015, \apjs, 218, 23

\bibitem[{Ackermann {et~al.}(2015)Ackermann, Ajello, Albert, Atwood, Baldini, Ballet, Barbiellini, Bastieri, Gonzalez, Bellazzini, Bissaldi, Blandford, Bloom, Bonino, Bottacini, Bregeon, Bruel, Buehler, Buson, Caliandro, Cameron, Caputo, Caragiulo, Caraveo, Cavazzuti, Cecchi, Chekhtman, Chiang, Chiaro, Ciprini, Cohen-Tanugi, Conrad, Cutini, D’Ammando, de~Angelis, de~Palma, Desiante, Venere, Domi´nguez, Drell, Favuzzi, Fegan, Ferrara, Focke, Fuhrmann, Fukazawa, Fusco, Gargano, Gasparrini, Giglietto, Giommi, Giordano, Giroletti, Godfrey, Green, Grenier, Grove, Guiriec, Harding, Hays, Hewitt, Hill, Horan, Jogler, Jóhannesson, Johnson, Kamae, Kuss, Larsson, Latronico, Li, Li, Longo, Loparco, Lott, Lovellette, Lubrano, Magill, Maldera, Manfreda, Max-Moerbeck, Mayer, Mazziotta, McEnery, Michelson, Mizuno, Monzani, Morselli, Moskalenko, Murgia, Nuss, Ohno, Ohsugi, Ojha, Omodei, Orlando, Ormes, Paneque, Pearson, Perkins, Perri, Pesce-Rollins, Petrosian, Piron, Pivato, Porter, Rainò, Rando, Razzano, Readhead,
  Reimer, Reimer, Schulz, Sgrò, Siskind, Spada, Spandre, Spinelli, Suson, Takahashi, Thayer, Thompson, Tibaldo, Torres, Tosti, Troja, Uchiyama, Vianello, Wood, Wood, Zimmer, Berdyugin, Corbet, Hovatta, Lindfors, Nilsson, Reinthal, Sillanpää, Stamerra, Takalo, \& Valtonen}]{Ackermann_2015}
Ackermann, M., Ajello, M., Albert, A., {et~al.} 2015, ApJL, 813, L41

\bibitem[{{Ait Benkhali} {et~al.}(2020){Ait Benkhali}, {Hofmann}, {Rieger}, \& {Chakraborty}}]{quasi}
{Ait Benkhali}, F., {Hofmann}, W., {Rieger}, F.~M., \& {Chakraborty}, N. 2020, \aap, 634, A120

\bibitem[{Ajello {et~al.}(2020)Ajello, Angioni, Axelsson, Ballet, Barbiellini, Bastieri, Gonzalez, Bellazzini, Bissaldi, Bloom, Bonino, Bottacini, Bruel, Buson, Cafardo, Cameron, Cavazzuti, Chen, Cheung, Ciprini, Costantin, Cutini, D’Ammando, de~la Torre~Luque, de~Menezes, de~Palma, Desai, Lalla, Venere, Domínguez, Dirirsa, Ferrara, Finke, Franckowiak, Fukazawa, Funk, Fusco, Gargano, Garrappa, Gasparrini, Giglietto, Giordano, Giroletti, Green, Grenier, Guiriec, Harita, Hays, Horan, Itoh, Jóhannesson, Kovac’evic’, Krauss, Kreter, Kuss, Larsson, Leto, Li, Liodakis, Longo, Loparco, Lott, Lovellette, Lubrano, Madejski, Maldera, Manfreda, Martí-Devesa, Massaro, Mazziotta, Mereu, Meyer, Migliori, Mirabal, Mizuno, Monzani, Morselli, Moskalenko, Negro, Nemmen, Nuss, Ojha, Ojha, Omodei, Orienti, Orlando, Ormes, Paliya, Pei, Peña-Herazo, Persic, Pesce-Rollins, Petrov, Piron, Poon, Principe, Rainò, Rando, Rani, Razzano, Razzaque, Reimer, Reimer, Schinzel, Serini, Sgrò, Siskind, Spandre, Spinelli, Suson,
  Tachibana, Thompson, Torres, Torresi, Troja, Valverde, van Zyl, \& Yassine}]{4LAC}
Ajello, M., Angioni, R., Axelsson, M., {et~al.} 2020, ApJ, 892, 105

\bibitem[{Ajello {et~al.}(2021)Ajello, Atwood, Axelsson, Bagagli, Bagni, Baldini, Bastieri, Bellardi, Bellazzini, Bissaldi, Bloom, Bonino, Bregeon, Brez, Bruel, Buehler, Buson, Cameron, Caraveo, Cavazzuti, Ceccanti, Chen, Cheung, Ciprini, Cognard, Cohen-Tanugi, Cutini, D’Ammando, de~la Torre~Luque, de~Palma, Digel, Dirirsa, Lalla, Venere, Domínguez, Fabiani, Ferrara, Fiori, Foglia, Fukazawa, Fusco, Gargano, Gasparrini, Giroletti, Glanzman, Green, Griffin, Grondin, Grove, Guillemot, Guiriec, Gustafsson, Hays, Horan, Jóhannesson, Johnson, Kamae, Kerr, Kuss, Larsson, Latronico, Lemoine-Goumard, Li, Liodakis, Longo, Loparco, Lovellette, Lubrano, Maldera, Manfreda, Martí-Devesa, Mazziotta, Menon, Mereu, Meyer, Michelson, Minuti, Mitthumsiri, Mizuno, Mongelli, Monzani, Moskalenko, Negro, Nuss, Ojha, Orienti, Orlando, Paccagnella, Paliya, Paneque, Pei, Perkins, Pesce-Rollins, Pinchera, Piron, Poon, Porter, Primavera, Principe, Racusin, Rainò, Rando, Rani, Rapposelli, Razzano, Razzaque, Reimer, Reimer, Russell,
  Saggini, Parkinson, Scolieri, Serini, Sgrò, Siskind, Smith, Spandre, Spinelli, Suson, Tajima, Thayer, Thompson, Tibaldo, Torres, Tosti, Valverde, Vigiani, \& Zaharijas}]{Ajello_2021}
Ajello, M., Atwood, W.~B., Axelsson, M., {et~al.} 2021, ApJS, 256, 12

\bibitem[{Allen \& Smith(1996)}]{MC_SSA}
Allen, M.~R. \& Smith, L.~A. 1996, Journal of Climate, 9, 3373

\bibitem[{Atwood {et~al.}(2013)Atwood, Albert, Baldini, Tinivella, Bregeon, Pesce-Rollins, Sgrò, Bruel, Charles, Drlica-Wagner, Franckowiak, Jogler, Rochester, Usher, Wood, Cohen-Tanugi, \& Zimmer}]{atwood2013pass}
Atwood, W., Albert, A., Baldini, L., {et~al.} 2013, Pass 8: Toward the Full Realization of the Fermi-LAT Scientific Potential

\bibitem[{Atwood {et~al.}(2009)Atwood, Abdo, Ackermann, Althouse, Anderson, Axelsson, Baldini, Ballet, Band, Barbiellini, Bartelt, Bastieri, Baughman, Bechtol, Bédérède, Bellardi, Bellazzini, Berenji, Bignami, Bisello, Bissaldi, Blandford, Bloom, Bogart, Bonamente, Bonnell, Borgland, Bouvier, Bregeon, Brez, Brigida, Bruel, Burnett, Busetto, Caliandro, Cameron, Caraveo, Carius, Carlson, Casandjian, Cavazzuti, Ceccanti, Cecchi, Charles, Chekhtman, Cheung, Chiang, Chipaux, Cillis, Ciprini, Claus, Cohen-Tanugi, Condamoor, Conrad, Corbet, Corucci, Costamante, Cutini, Davis, Decotigny, DeKlotz, Dermer, de~Angelis, Digel, do~Couto~e Silva, Drell, Dubois, Dumora, Edmonds, Fabiani, Farnier, Favuzzi, Flath, Fleury, Focke, Funk, Fusco, Gargano, Gasparrini, Gehrels, Gentit, Germani, Giebels, Giglietto, Giommi, Giordano, Glanzman, Godfrey, Grenier, Grondin, Grove, Guillemot, Guiriec, Haller, Harding, Hart, Hays, Healey, Hirayama, Hjalmarsdotter, Horn, Hughes, Jóhannesson, Johansson, Johnson, Johnson, Johnson, Johnson,
  Kamae, Katagiri, Kataoka, Kavelaars, Kawai, Kelly, Kerr, Klamra, Knödlseder, Kocian, Komin, Kuehn, Kuss, Landriu, Latronico, Lee, Lee, Lemoine-Goumard, Lionetto, Longo, Loparco, Lott, Lovellette, Lubrano, Madejski, Makeev, Marangelli, Massai, Mazziotta, McEnery, Menon, Meurer, Michelson, Minuti, Mirizzi, Mitthumsiri, Mizuno, Moiseev, Monte, Monzani, Moretti, Morselli, Moskalenko, Murgia, Nakamori, Nishino, Nolan, Norris, Nuss, Ohno, Ohsugi, Omodei, Orlando, Ormes, Paccagnella, Paneque, Panetta, Parent, Pearce, Pepe, Perazzo, Pesce-Rollins, Picozza, Pieri, Pinchera, Piron, Porter, Poupard, Rainò, Rando, Rapposelli, Razzano, Reimer, Reimer, Reposeur, Reyes, Ritz, Rochester, Rodriguez, Romani, Roth, Russell, Ryde, Sabatini, Sadrozinski, Sanchez, Sander, Sapozhnikov, Parkinson, Scargle, Schalk, Scolieri, Sgrò, Share, Shaw, Shimokawabe, Shrader, Sierpowska-Bartosik, Siskind, Smith, Smith, Spandre, Spinelli, Starck, Stephens, Strickman, Strong, Suson, Tajima, Takahashi, Takahashi, Tanaka, Tenze, Tether,
  Thayer, Thayer, Thompson, Tibaldo, Tibolla, Torres, Tosti, Tramacere, Turri, Usher, Vilchez, Vitale, Wang, Watters, Winer, Wood, Ylinen, \& Ziegler}]{Atwood_2009}
Atwood, W.~B., Abdo, A.~A., Ackermann, M., {et~al.} 2009, ApJ, 697, 1071

\bibitem[{Ballet {et~al.}(2020)Ballet, Burnett, Digel, \& Lott}]{DR2}
Ballet, J., Burnett, T.~H., Digel, S.~W., \& Lott, B. 2020, ArXiv e-prints [\eprint[arXiv]{2005.11208}]

\bibitem[{Bhatta(2019)}]{Bhatta_2019}
Bhatta, G. 2019, in Recent Progress in Relativistic Astrophysics ({MDPI})

\bibitem[{Blandford {et~al.}(2019)Blandford, Meier, \& Readhead}]{Blandford_2019}
Blandford, R., Meier, D., \& Readhead, A. 2019, Annual Review of Astronomy and Astrophysics, 57, 467

\bibitem[{{Britzen, S.} {et~al.}(2019){Britzen, S.}, {Fendt, C.}, {Böttcher, M.}, {Zajaček, M.}, {Jaron, F.}, {Pashchenko, I. N.}, {Araudo, A.}, {Karas, V.}, \& {Kurtanidze, O.}}]{Britzen18}
{Britzen, S.}, {Fendt, C.}, {Böttcher, M.}, {et~al.} 2019, A\&A, 630, A103

\bibitem[{{Camenzind} \& {Krockenberger}(1992)}]{lighthouse}
{Camenzind}, M. \& {Krockenberger}, M. 1992, \aap, 255, 59

\bibitem[{Celoria {et~al.}(2018)Celoria, Oliveri, Sesana, \& Mapelli}]{celoria}
Celoria, M., Oliveri, R., Sesana, A., \& Mapelli, M. 2018, Lecture notes on black hole binary astrophysics

\bibitem[{{Cheng} {et~al.}(2022){Cheng}, {Liu}, {Sun}, \& {Dong}}]{3c66a_1}
{Cheng}, Y., {Liu}, F., {Sun}, Z.-n., \& {Dong}, F.-t. 2022, \caa, 46, 204

\bibitem[{Combes(2021)}]{AGN_book}
Combes, F. 2021, in Active Galactic Nuclei, 2514-3433 (IOP Publishing), 1--1 to 1--25

\bibitem[{Dabbakuti \& G(2019)}]{SSA_meteo}
Dabbakuti, J. R. K.~K. \& G, B.~L. 2019, IEEE Journal of Selected Topics in Applied Earth Observations and Remote Sensing, 12, 5101

\bibitem[{{Dorigo Jones} {et~al.}(2022){Dorigo Jones}, {Johnson}, {Muzahid}, {Charlton}, {Chen}, {Narayanan}, {Sameer}, {Schaye}, \& {Wijers}}]{DorigoJones2022}
{Dorigo Jones}, J., {Johnson}, S.~D., {Muzahid}, S., {et~al.} 2022, \mnras, 509, 4330

\bibitem[{{Emmanoulopoulos} {et~al.}(2013){Emmanoulopoulos}, {McHardy}, \& {Papadakis}}]{Emmanoulopoulos}
{Emmanoulopoulos}, D., {McHardy}, I.~M., \& {Papadakis}, I.~E. 2013, \mnras, 433, 907

\bibitem[{Golyandina {et~al.}(2018)Golyandina, Korobeynikov, \& Zhigljavsky}]{SSA_R}
Golyandina, N., Korobeynikov, A., \& Zhigljavsky, A. 2018, Singular spectrum analysis with {R} (Springer-Verlag Berlin Heidelberg)

\bibitem[{{Gong} {et~al.}(2022){Gong}, {Zhou}, {Yuan}, {Zhang}, {Yi}, \& {Fang}}]{pks0405}
{Gong}, Y., {Zhou}, L., {Yuan}, M., {et~al.} 2022, \apj, 931, 168

\bibitem[{Gracia {et~al.}(2003)Gracia, Peitz, Keller, \& Camenzind}]{modulation}
Gracia, J., Peitz, J., Keller, C., \& Camenzind, M. 2003, MNRAS, 344, 468

\bibitem[{Graham {et~al.}(2015)Graham, Djorgovski, \& Stern}]{graham15}
Graham, M.~J., Djorgovski, S.~G., \& Stern, D. e.~a. 2015, Nature, 518, 74

\bibitem[{Gross \& Vitells(2010)}]{look-else_2}
Gross, E. \& Vitells, O. 2010, The European Physical Journal C, 70, 525

\bibitem[{{Gupta} {et~al.}(2019){Gupta}, {Tripathi}, {Wiita}, {Kushwaha}, {Zhang}, \& {Bambi}}]{B21520}
{Gupta}, A.~C., {Tripathi}, A., {Wiita}, P.~J., {et~al.} 2019, \mnras, 484, 5785

\bibitem[{Hair {et~al.}(2011)Hair, Ringle, \& Sarstedt}]{hair_r2_2011}
Hair, J.~F., Ringle, C.~M., \& Sarstedt, M. 2011, Journal of Marketing Theory and Practice, 19, 139

\bibitem[{Hassani(2007)}]{SSA_paper}
Hassani, H. 2007, University Library of Munich, Germany, MPRA Paper, 5

\bibitem[{Hassani {et~al.}(2009)Hassani, Heravi, \& Zhigljavsky}]{HASSANI2009103}
Hassani, H., Heravi, S., \& Zhigljavsky, A. 2009, International Journal of Forecasting, 25, 103

\bibitem[{Jiang {et~al.}(2022)Jiang, Yang, Wang, Zhu, Lyu, Dou, Wang, Wang, Pan, Liu, Shu, \& Zheng}]{jiang2022ticktock}
Jiang, N., Yang, H., Wang, T., {et~al.} 2022, Tick-Tock: The Imminent Merger of a Supermassive Black Hole Binary

\bibitem[{{Johnson} {et~al.}(2019){Johnson}, {Mulchaey}, {Chen}, {Wijers}, {Connor}, {Muzahid}, {Schaye}, {Cen}, {Carlsten}, {Charlton}, {Drout}, {Goulding}, {Hansen}, \& {Walth}}]{Johnson2019}
{Johnson}, S.~D., {Mulchaey}, J.~S., {Chen}, H.-W., {et~al.} 2019, \apjl, 884, L31

\bibitem[{{Kalantari} {et~al.}(2020){Kalantari}, {Hassani}, \& {Sirimal Silva}}]{recurrent_forecast}
{Kalantari}, M., {Hassani}, H., \& {Sirimal Silva}, E. 2020, Fluctuation and Noise Letters, 19, 2050010

\bibitem[{{Li} {et~al.}(2023){Li}, {Cai}, {Yang}, {L{\"a}hteenm{\"a}ki}, {Tornikoski}, {Tammi}, {Suutarinen}, {Yang}, {Luo}, \& {Wang}}]{pks1510}
{Li}, X.-P., {Cai}, Y., {Yang}, H.-Y., {et~al.} 2023, \mnras, 519, 4893

\bibitem[{{Lomb}(1976)}]{Lomb}
{Lomb}, N.~R. 1976, \apss, 39, 447

\bibitem[{{Marscher} {et~al.}(2011){Marscher}, {Jorstad}, {Larionov}, {Aller}, \& {L{\"a}hteenm{\"a}ki}}]{Marscher_trend}
{Marscher}, A., {Jorstad}, S.~G., {Larionov}, V.~M., {Aller}, M.~F., \& {L{\"a}hteenm{\"a}ki}, A. 2011, JA\&A, 32, 233

\bibitem[{{Mattox} {et~al.}(1996){Mattox}, {Bertsch}, {Chiang}, {Dingus}, {Digel}, {Esposito}, {Fierro}, {Hartman}, {Hunter}, {Kanbach}, {Kniffen}, {Lin}, {Macomb}, {Mayer-Hasselwander}, {Michelson}, {von Montigny}, {Mukherjee}, {Nolan}, {Ramanamurthy}, {Schneid}, {Sreekumar}, {Thompson}, \& {Willis}}]{mattox}
{Mattox}, J.~R., {Bertsch}, D.~L., {Chiang}, J., {et~al.} 1996, \apj, 461, 396

\bibitem[{McHardy {et~al.}(2006)McHardy, Koerding, Knigge, Uttley, \& Fender}]{mchardy06}
McHardy, I.~M., Koerding, E., Knigge, C., Uttley, P., \& Fender, R.~P. 2006, Nature, 444, 730

\bibitem[{{McQuillan} {et~al.}(2013){McQuillan}, {Aigrain}, \& {Mazeh}}]{mcquillan_trend_fake_detection}
{McQuillan}, A., {Aigrain}, S., \& {Mazeh}, T. 2013, \mnras, 432, 1203

\bibitem[{Merritt \& Milosavljevi{\'c}(2005)}]{merritt05}
Merritt, D. \& Milosavljevi{\'c}, M. 2005, Living Reviews in Relativity, 8, 8

\bibitem[{Mishra {et~al.}(2022)Mishra, Fragile, Anderson, Blankenship, Li, \& Nalewajko}]{Mishra_2022}
Mishra, B., Fragile, P.~C., Anderson, J., {et~al.} 2022, ApJ, 939, 31

\bibitem[{Movahedifar {et~al.}(2018)Movahedifar, Yarmohammadi, \& Hassani}]{Movahedifar2018}
Movahedifar, M., Yarmohammadi, M., \& Hassani, H. 2018, Mathematical Biosciences, 303, 52

\bibitem[{Nina~Golyandina(2020)}]{SSA}
Nina~Golyandina, A.~Z. 2020, Singular Spectrum Analysis for Time Series, 2nd edn., S (Springer Berlin, Heidelberg)

\bibitem[{{Otero-Santos} {et~al.}(2020){Otero-Santos}, {Acosta-Pulido}, {Becerra Gonz{\'a}lez}, {Raiteri}, {Larionov}, {Pe{\~n}il}, {Smith}, {Ballester Niebla}, {Borman}, {Carnerero}, {Castro Segura}, {Grishina}, {Kopatskaya}, {Larionova}, {Morozova}, {Nikiforova}, {Savchenko}, {Troitskaya}, {Troitsky}, {Vasilyev}, \& {Villata}}]{3c66a_2}
{Otero-Santos}, J., {Acosta-Pulido}, J.~A., {Becerra Gonz{\'a}lez}, J., {et~al.} 2020, \mnras, 492, 5524

\bibitem[{O’Neill {et~al.}(2022)O’Neill, Kiehlmann, Readhead, Aller, Blandford, Liodakis, Lister, Mróz, O’Dea, Pearson, Ravi, Vallisneri, Cleary, Graham, Grainge, Hodges, Hovatta, Lähteenmäki, Lamb, Lazio, Max-Moerbeck, Pavlidou, Prince, Reeves, Tornikoski, de~la Parra, \& Zensus}]{O’Neill_2022}
O’Neill, S., Kiehlmann, S., Readhead, A. C.~S., {et~al.} 2022, ApJL, 926, L35

\bibitem[{{Pe{\~n}il} {et~al.}(2020){Pe{\~n}il}, {Dom{\'\i}nguez}, {Buson}, {Ajello}, {Otero-Santos}, {Barrio}, {Nemmen}, {Cutini}, {Rani}, {Franckowiak}, \& {Cavazzuti}}]{P20}
{Pe{\~n}il}, P., {Dom{\'\i}nguez}, A., {Buson}, S., {et~al.} 2020, \apj, 896, 134

\bibitem[{{Pe{\~n}il} {et~al.}(2024{\natexlab{a}}){Pe{\~n}il}, {Otero-Santos}, {Ajello}, {Buson}, {Dom{\'\i}nguez}, {Marcotulli}, {Torres-Alb{\`a}}, {Becerra Gonz{\'a}lez}, \& {Acosta-Pulido}}]{penil24}
{Pe{\~n}il}, P., {Otero-Santos}, J., {Ajello}, M., {et~al.} 2024{\natexlab{a}}, \mnras, 529, 1365

\bibitem[{{Pe{\~n}il} {et~al.}(2024{\natexlab{b}}){Pe{\~n}il}, {Westernacher-Schneider}, {Ajello}, {Dom{\'\i}nguez}, {Buson}, {Otero-Santos}, {Marcotulli}, {Torres-Alb{\`a}}, \& {Zrake}}]{penil_mwl_pg1553}
{Pe{\~n}il}, P., {Westernacher-Schneider}, J.~R., {Ajello}, M., {et~al.} 2024{\natexlab{b}}, \mnras, 527, 10168

\bibitem[{Peterson(2001)}]{peterson01}
Peterson, B.~M. 2001, Variability of Active Galactic Nuclei (World Scientific), 3–68

\bibitem[{Peñil {et~al.}(2022)Peñil, Ajello, Buson, Domínguez, Westernacher-Schneider, \& Zrake}]{Pablo22}
Peñil, P., Ajello, M., Buson, S., {et~al.} 2022, ArXiv e-prints [\eprint[arXiv]{2211.01894}]

\bibitem[{Prokhorov \& Moraghan(2017)}]{prokhorov}
Prokhorov, D.~A. \& Moraghan, A. 2017, MNRAS, 471, 3036

\bibitem[{{Raiteri} {et~al.}(2001){Raiteri}, {Villata}, {Aller}, {Aller}, {Heidt}, {Kurtanidze}, {Lanteri}, {Maesano}, {Massaro}, {Montagni}, {Nesci}, {Nilsson}, {Nikolashvili}, {Nurmi}, {Ostorero}, {Pursimo}, {Rekola}, {Sillanp{\"a}{\"a}}, {Takalo}, {Ter{\"a}sranta}, {Tosti}, {Balonek}, {Feldt}, {Heines}, {Heisler}, {Hu}, {Kidger}, {Mattox}, {McGrath}, {Pati}, {Robb}, {Sadun}, {Shastri}, {Wagner}, {Wei}, \& {Wu}}]{radio_optical}
{Raiteri}, C.~M., {Villata}, M., {Aller}, H.~D., {et~al.} 2001, \aap, 377, 396

\bibitem[{Rani(2018)}]{rani2018radio_trend}
Rani, B. 2018, ArXiv e-prints [\eprint[arXiv]{1811.00567}]

\bibitem[{{Ren} {et~al.}(2023){Ren}, {Cerruti}, \& {Sahakyan}}]{QPO}
{Ren}, H.~X., {Cerruti}, M., \& {Sahakyan}, N. 2023, \aap, 672, A86

\bibitem[{Rieger(2019)}]{galaxies7010028}
Rieger, F.~M. 2019, Galaxies, 7

\bibitem[{{Roy} {et~al.}(2022){Roy}, {Sarkar}, {Chatterjee}, {Gupta}, {Chitnis}, \& {Wiita}}]{pks1510_2}
{Roy}, A., {Sarkar}, A., {Chatterjee}, A., {et~al.} 2022, \mnras, 510, 3641

\bibitem[{{Rueda} {et~al.}(2022){Rueda}, {Glicenstein}, \& {Brun}}]{rueda22}
{Rueda}, H., {Glicenstein}, J.-F., \& {Brun}, F. 2022, \apj, 934, 6

\bibitem[{Saeid~Sanei(2015)}]{SSA_Saeid}
Saeid~Sanei, H.~H. 2015, Singular Spectrum Analysis of Biomedical Signals (CRC Press)

\bibitem[{{Sandrinelli} {et~al.}(2016){Sandrinelli}, {Covino}, {Dotti}, \& {Treves}}]{Sandrinelli_2}
{Sandrinelli}, A., {Covino}, S., {Dotti}, M., \& {Treves}, A. 2016, \aj, 151, 54

\bibitem[{{Sandrinelli} {et~al.}(2018){Sandrinelli}, {Covino}, {Treves}, {Holgado}, {Sesana}, {Lindfors}, \& {Ramazani}}]{Sandrinelli}
{Sandrinelli}, A., {Covino}, S., {Treves}, A., {et~al.} 2018, \aap, 615, A118

\bibitem[{Sanei \& Hassani(2015)}]{SSA_biomedics}
Sanei, S. \& Hassani, H. 2015, Singular spectrum analysis of biomedical signals (CRC press)

\bibitem[{{Scargle}(1982)}]{Scargle}
{Scargle}, J.~D. 1982, \apj, 263, 835

\bibitem[{Sobacchi {et~al.}(2016)Sobacchi, Sormani, \& Stamerra}]{sobachi}
Sobacchi, E., Sormani, M.~C., \& Stamerra, A. 2016, MNRAS, 465, 161

\bibitem[{Sobacchi {et~al.}(2017)Sobacchi, Sormani, \& Stamerra}]{sobacchi17}
Sobacchi, E., Sormani, M.~C., \& Stamerra, A. 2017, MNRAS Lett., 465, L26

\bibitem[{{Timmer} \& {Koenig}(1995)}]{timmer}
{Timmer}, J. \& {Koenig}, M. 1995, \aap, 300, 707

\bibitem[{Urry(1996)}]{Urry_1996}
Urry, C.~M. 1996, ArXiv e-prints [\eprint[arXiv]{astro-ph/9609023}]

\bibitem[{{Urry}(2011)}]{urry11}
{Urry}, M. 2011, JA\&A, 32, 139

\bibitem[{Valtonen {et~al.}(2008)Valtonen, Lehto, \& Sillanpää}]{valtonen08}
Valtonen, M.~J., Lehto, H.~J., \& Sillanpää, A. e.~a. 2008, Nature, 452, 851

\bibitem[{Valverde {et~al.}(2020)Valverde, Horan, Bernard, Fegan, Collaboration), Abeysekara, Archer, Benbow, Bird, Brill, Brose, Buchovecky, Buckley, Christiansen, Cui, Falcone, Feng, Finley, Fortson, Furniss, Gent, Gillanders, Giuri, Gueta, Hanna, Hassan, Hervet, Holder, Hughes, Humensky, Kaaret, Kelley-Hoskins, Kertzman, Kieda, Krause, Krennrich, Lang, Maier, Moriarty, Mukherjee, Nieto, Nievas-Rosillo, O’Brien, Ong, Otte, Park, Petrashyk, Pfrang, Pichel, Pohl, Prado, Pueschel, Quinn, Ragan, Reynolds, Ribeiro, Richards, Roache, Sadeh, Santander, Scott, Sembroski, Shahinyan, Shang, Sushch, Vassiliev, Weinstein, Wells, Wilcox, Wilhelm, Williams, Williamson, (VERITAS Collaboration), Noto, Edwards, Piner, Ramazani, Hovatta, Jormanainen, Lindfors, Nilsson, Takalo, Kovalev, Lister, Pushkarev, Savolainen, Kiehlmann, Max-Moerbeck, Readhead, Lähteenmäki, \& Tornikoski}]{Valverde_2020_trend}
Valverde, J., Horan, D., Bernard, D., {et~al.} 2020, ApJ, 891, 170

\bibitem[{{VanderPlas}(2018)}]{vanderplas}
{VanderPlas}, J.~T. 2018, \apjs, 236, 16

\bibitem[{{Vaughan} {et~al.}(2016){Vaughan}, {Uttley}, {Markowitz}, {Huppenkothen}, {Middleton}, {Alston}, {Scargle}, \& {Farr}}]{vaughan_criticism}
{Vaughan}, S., {Uttley}, P., {Markowitz}, A.~G., {et~al.} 2016, \mnras, 461, 3145

\bibitem[{Vitells(2011)}]{look-else}
Vitells, O. 2011, CERN Document Server, 183

\bibitem[{{Wang} {et~al.}(2022){Wang}, {Cai}, \& {Fan}}]{s51044}
{Wang}, G.~G., {Cai}, J.~T., \& {Fan}, J.~H. 2022, \apj, 929, 130

\bibitem[{{Welsh}(1999)}]{detrend_welsh}
{Welsh}, W.~F. 1999, \pasp, 111, 1347

\bibitem[{{Westernacher-Schneider} {et~al.}(2022){Westernacher-Schneider}, {Zrake}, {MacFadyen}, \& {Haiman}}]{2022PhRvD.106j3010W}
{Westernacher-Schneider}, J.~R., {Zrake}, J., {MacFadyen}, A., \& {Haiman}, Z. 2022, \prd, 106, 103010

\bibitem[{Wood {et~al.}(2017)Wood, Caputo, Charles, Mauro, Magill, \& Perkins}]{wood2017fermipy}
Wood, M., Caputo, R., Charles, E., {et~al.} 2017, ArXiv e-prints [\eprint[arXiv]{1707.09551}]

\bibitem[{{Zhang} {et~al.}(2017){Zhang}, {Zhang}, {Zhu}, {Yi}, {Yao}, {Lu}, \& {Liang}}]{seyfert1}
{Zhang}, J., {Zhang}, H.-M., {Zhu}, Y.-K., {et~al.} 2017, \apj, 849, 42

\end{thebibliography}



\newpage
\onecolumn

\begin{appendix}

\section{Tables}

\FloatBarrier
\begin{table*}[h!]
     \caption{\label{table results}Periodicity candidates}

    \centering
    \begin{tabular}{lcccccccc}
    \hline\hline
	4FGL &    RA(J2000) &  Dec(J2000) & Association &  Type &  Redshift &  Period (yr) &    Local & Global\\

	Source Name &  & & Name & &   &   &    Significance & Significance\\
    \hline

        J0137.0+4751*  & 24.260 & 47.864 &  OC 457        &   fsrq  &      0.859  &  1.79 $\pm$0.22  &        4.8$\sigma$ & 3.3$\sigma$ \\
        J0407.0$-$3826  & 61.763 & -38.439 &  PKS 0405$-$385  &   fsrq  &      1.285  &  3.01 $\pm$0.43  &        4.8$\sigma$ & 3.3$\sigma$\\
        J0428.6$-$3756  & 67.173 & -37.94 &  PKS 0426$-$380  &    bll  &      1.110  &  3.51 $\pm$0.49  &        4.8$\sigma$ & 3.3$\sigma$\\
        J0457.0$-$2324  & 74.261 & -23.415 &  PKS 0454$-$234  &   fsrq  &      1.003  &  3.57 $\pm$0.43  &        4.8$\sigma$ & 3.3$\sigma$ \\
        J0501.2$-$0158  & 75.302 & -1.975 &  S3 0458$-$02    &   fsrq  &      2.291  &  1.00 $\pm$0.04  &        4.8$\sigma$ & 3.3$\sigma$ \\
        J0521.7+2112  & 80.445 & 21.213 &  TXS 0518+211  &    bll  &      0.108  &  2.83 $\pm$0.38  &        4.8$\sigma$ & 3.3$\sigma$\\
        J0526.2$-$4830*  & 81.571 & -48.515 &  PKS 0524$-$485  &   fsrq  &      1.300  &   2.07 $\pm$0.20  &        4.8$\sigma$ & 3.3$\sigma$ \\
        J0811.4+0146  & 122.861 & 1.776 &  OJ 014        &    bll  &      1.148  &  4.65 $\pm$0.89  &        4.8$\sigma$ & 3.3$\sigma$ \\
        J1007.6$-$3332*  & 151.912 & -33.543 &  PKS 1005$-$333  &   fsrq  &      1.837  &  2.61 $\pm$0.34  &        4.8$\sigma$ & 3.3$\sigma$\\
        J1048.4+7143  & 162.107 & 71.730 &  S5 1044+71    &   fsrq  &      1.150  &  3.05 $\pm$0.44  &        4.8$\sigma$ & 3.3$\sigma$\\
        J1146.9+3958  & 176.740 & 39.978 &  S4 1144+40    &   fsrq  &      1.089  &   3.31 $\pm$0.50  &        4.8$\sigma$ & 3.3$\sigma$\\
        J1522.1+3144  & 230.545 & 31.740 &  B2 1520+31    &   fsrq  &      1.489  &  3.37 $\pm$0.56  &        4.8$\sigma$ & 3.3$\sigma$\\
        J1532.7$-$1319*  & 233.197 & -13.326 &  TXS 1530$-$131  &   bcu   &       -  &  1.37 $\pm$0.07  &        4.8$\sigma$ & 3.3$\sigma$\\ 
        J1555.7+1111  & 238.931 & 11.188 &  PG 1553+113   &    bll  &      0.433  &  2.15 $\pm$0.16  &        4.8$\sigma$ & 3.3$\sigma$\\
        J1637.7+4717*  & 249.434 & 47.291 &  4C +47.44     &   fsrq  &      0.735  &  3.17 $\pm$0.58  &        4.8$\sigma$ & 3.3$\sigma$\\
        J1657.7+4808*  & 254.438 & 48.137 &  4C +48.41     &   fsrq  &      1.669  &  1.47 $\pm$0.08  &        4.8$\sigma$ & 3.3$\sigma$\\
        J1913.0$-$8009*  & 288.27 & -80.157 &  PKS 1903$-$80   &   fsrq  &      1.756  &  2.41 $\pm$0.22  &        4.8$\sigma$ & 3.3$\sigma$\\
        J2201.5$-$8339*  & 330.379 & -83.663 &  PKS 2155$-$83   &   fsrq  &      1.865  &  4.73 $\pm$0.94  &        4.8$\sigma$ & 3.3$\sigma$\\
        J2139.4$-$4235*  & 324.855 & -42.590 &  MH 2136$-$428   &    bll  &       -  &  1.81 $\pm$0.18  &        4.8$\sigma$ & 3.3$\sigma$\\
        J0222.6+4302*  & 35.670 & 43.036 &  3C 66A        &    bll  &      0.444  &   2.31 $\pm$0.20  &        4.6$\sigma$ & 3.1$\sigma$ \\
        J1312.8$-$0425*  & 198.217 & -4.420 &  PKS B1310$-$041 &   fsrq  &      0.825  &  2.37 $\pm$0.22  &        4.6$\sigma$ & 3.1$\sigma$ \\
        J0721.9+7120  & 110.488 & 71.340 &  S5 0716+71    &    bll  &      0.127  &  2.85 $\pm$0.33  &        4.5$\sigma$  & 3.0$\sigma$\\
        J0210.7$-$5101  & 32.695 & -51.022 &  PKS 0208$-$512  &   fsrq  &      1.003  &  2.47 $\pm$0.26  &        4.5$\sigma$ & 3.0$\sigma$\\
        J2158.8$-$3013  & 329.714 & -30.225 &  PKS 2155$-$304  &    bll  &      0.116  &   1.65 $\pm$0.10  &        4.5$\sigma$ & 3.0$\sigma$ \\
        J1427.0+2348*  & 216.756 & 23.801 &  PKS 1424+240  &    bll  &      0.604  &  1.53 $\pm$0.09  &        4.5$\sigma$ & 3.0$\sigma$ \\
        J1454.4+5124  & 223.625 & 51.409 &  TXS 1452+516  &    bll  &       -  &   2.07 $\pm$0.20  &        4.2$\sigma$ & 2.5$\sigma$ \\
        J2012.0+4629*  & 303.020 & 46.488 &  7C 2010+4619  &    bll  &       -  &  3.23 $\pm$0.46  &        4.2$\sigma$ & 2.5$\sigma$\\
        J2202.7+4216  & 330.695 & 42.282 &  BL Lac        &    bll  &      0.069  &  2.05 $\pm$0.15  &        4.1$\sigma$ & 2.3$\sigma$ \\
        J2236.3+2828*  & 339.096 & 28.483 &  B2 2234+28A   &   fsrq  &      0.790  &  2.19 $\pm$0.15  &        4.1$\sigma$ & 2.3$\sigma$  \\
        J0043.8+3425  & 10.972 & 34.432 &  GB6 J0043+3426               &   fsrq  &      0.966  &  1.95 $\pm$0.16  &    4.1$\sigma$ & 2.3$\sigma$ \\
        J0112.1+2245*  & 18.029 & 22.751 &  S2 0109+22    &    bll  &      0.27  &  2.65 $\pm$0.31  &        4.1$\sigma$ & 2.3$\sigma$ \\
        J0948.9+0022  & 147.244 & 0.372 &  PMN J0948+0022               &  nlsy1  &      0.585  &  1.39 $\pm$0.15  &        4.0$\sigma$ & 2.2$\sigma$  \\
        J0449.4$-$4350  & 72.358 & -43.835 &  PKS 0447$-$439  &    bll  &      0.205  &  1.89 $\pm$0.25  &        4.0$\sigma$ & 2.2$\sigma$  \\
        J0427.3$-$3900*  & 66.826 & -39.01 &  PMN J0427$-$3900               &   bcu   &       -  &  2.79 $\pm$0.31  &        3.9$\sigma$ & 2.0$\sigma$   \\
        J1321.1+2216*  & 200.296 & 22.281 &  TXS 1318+225  &   fsrq  &      0.943  &  1.21 $\pm$0.06  &        3.8$\sigma$ & 1.8$\sigma$\\
        J1223.8+8039*  & 185.971 & 80.660 &  S5 1221+80    &    bll  &       -  &  2.67 $\pm$0.27  &        3.5$\sigma$ & 1.2$\sigma$\\
        J0303.4$-$2407  & 45.862 & -24.122 &  PKS 0301$-$243  &    bll  &      0.266  &  2.05 $\pm$0.18  &        3.3$\sigma$ & 1.0$\sigma$\\
        J1723.6$-$7714*  & 260.922 & -77.238 &  PKS 1716$-$771  &   bcu   &       -  &   2.27 $\pm$0.20  &        2.9$\sigma$  & $<$1.0$\sigma$\\
        J1427.6$-$3305*  & 216.913 & -33.094 &  PKS 1424$-$328  &    bll  &       -  &  1.27 $\pm$0.08  &        2.9$\sigma$ & $<$1.0$\sigma$\\
        J1222.5+0414*  & 185.627 & 4.239 &  4C +04.42     &   fsrq  &      0.964  &  2.21 $\pm$0.19  &        2.7$\sigma$ & $<$1.0$\sigma$ \\
        J0252.8$-$2219  & 43.201 & -22.320 &  PKS 0250$-$225  &   fsrq  &      1.419  &  1.21 $\pm$0.08  &        2.7$\sigma$ & $<$1.0$\sigma$ \\
        J1253.2+5301*  & 193.307 & 53.017 &  S4 1250+53    &    bll  &       -  &  2.31$\pm$0.21  &       2.6$\sigma$ & $<$1.0$\sigma$\\        
        J1512.8$-$0906  & 228.215 & -9.106 &  PKS 1510$-$089  &   fsrq  &      0.360  &  1.67 $\pm$0.13  &        2.5$\sigma$ & $<$1.0$\sigma$\\
        J0405.6$-$1308*  & 61.419 & -13.144 &   PKS 0403$-$13  &   fsrq  &      0.571  &  1.69 $\pm$0.12  &        2.3$\sigma$ & $<$1.0$\sigma$ \\
        J0719.3+3307*  & 109.840 & 33.123 &  B2 0716+33    &   fsrq  &      0.779  &  2.29 $\pm$0.18  &        2.2$\sigma$ & $<$1.0$\sigma$\\
        J1033.1+4115*  & 158.275 & 41.262 &  S4 1030+41    &   fsrq  &      1.117  &  2.29 $\pm$0.21  &        2.0$\sigma$ & $<$1.0$\sigma$\\  
\hline  
    \end{tabular}
    \tablefoot{Periodicity candidates with a local significance $\geq$ 2$\sigma$, sorted from largest to smallest. We use the \href{https://fermi.gsfc.nasa.gov/ssc/data/access/lat/10yr_catalog/}{4FGL-DR2} catalog for the coordinate and redshift. The redshift of PG 1553$+$113 was taken from \citet{Johnson2019} and \citet{DorigoJones2022}. The `*' denotes blazars without previous evidence of periodicity in $\gamma$-rays.}
\end{table*}

\FloatBarrier

\begin{table*}[h!]
\caption{\label{literature}Periodicity candidates previously reported in the literature}

\centering
\begin{tabular}{lcccc}
\hline\hline
Association &  Period & Local & Period in & References\\
Name & (yr) & Significance & literature (yr) & \\  
\hline

PKS 0405$-$385 & $ 3.0 \pm 0.4$ & $4.8 \sigma$ & $\sim 2.8$ & \cite{pks0405}\\

PKS 0426$-$380 & $ 3.5 \pm 0.5$ & $4.8 \sigma$ & $\sim 3.6, \sim 3.4$  & \cite{quasi}, \cite{Pablo22}\\

PKS 0454$-$234 & $ 3.6 \pm 0.4$ & $4.8 \sigma$ & $\sim 3.6$ & \cite{Pablo22}\\

S3 0458$-$024 & $ 1.0 \pm 0.1$ & $4.8 \sigma$ & $\sim 3.8$ & \cite{Pablo22}\\

TXS 0518$+$211 & $ 2.8 \pm 0.4$ & $4.8 \sigma$ & $\sim 3.1$  & \cite{Pablo22}\\

OJ 014 & $ 4.7 \pm 0.9$ & $4.8 \sigma$ & $\sim 4.1$  & \cite{Pablo22}\\

S5 1044$+$71 & $ 3.1 \pm 0.4 $ & $4.8 \sigma$ &  $\sim 3.1$ & \cite{s51044}, \cite{QPO}\\

S4 1144$+$40 & $ 3.3 \pm 0.5 $ & $4.8 \sigma$ &  $\sim 3.3$ & \cite{Pablo22}\\

B2 1520$+$31 & $ 3.4 \pm 0.6$ & $4.8 \sigma$ &  $\sim 0.2$ & \cite{B21520}, \cite{QPO} \\

PG 1553$+$113 & $ 2.2 \pm 0.2$ & $4.8 \sigma$ &  $\sim 2.2$ & \cite{prokhorov}, \cite{Sandrinelli}\\ 
& & & & \cite{Pablo22}, \cite{rueda22}\\

3C 66A & $ 2.3 \pm 0.2$ & $4.6 \sigma$ & $\sim 2.3$ (\textit{V}$-$band) &  \cite{3c66a_2}, \cite{3c66a_1}\\

S5 0716$+$71 & $ 2.9 \pm 0.3$ & $4.5 \sigma$ & $\sim 2.7$  & \cite{Pablo22}\\

PKS 0208$-$512 & $ 2.5 \pm 0.3$ & $4.5 \sigma$ & $\sim 3.8$, $\sim 2.6$  & \cite{P20}, \cite{Pablo22}, \cite{rueda22} \\

PKS 2155$-$304 & $ 1.7 \pm 0.1$ & $4.5 \sigma$ & $\sim 1.7, \sim 1.8$  &  \cite{prokhorov}, \cite{Sandrinelli}\\
& & & &\cite{Pablo22}, \cite{rueda22}\\

TXS 1452$+$516 & $ 2.1 \pm 0.2$ & $4.2 \sigma$ & $\sim 2.1$ & \cite{Pablo22}, \cite{rueda22}\\

BL Lac & $ 2.1 \pm 0.2$ & $4.1 \sigma$ & $\sim 1.9$  & \cite{prokhorov}, \cite{Sandrinelli},\\
& & & &\cite{rueda22}\\

GB6 J0043$+$3426 & $ 2.0 \pm 0.2$ & $4.1 \sigma$ & $\sim 1.9$, $\sim 1.8$  & \cite{Pablo22}, \cite{rueda22}\\

PMN J0948$+$0022 & $ 1.4 \pm 0.2$ & $4.0 \sigma$ & $\sim 1.3$  & \cite{seyfert1}\\

PKS 0447$-$439 & $ 1.9 \pm 0.3$ & $4.0 \sigma$ & $\sim 2.5, \sim 1.9$, $\sim 2.0$  & \cite{quasi}, \cite{Pablo22}, \cite{rueda22}\\

PKS 0301$-$243 & $ 2.1 \pm 0.2$ & $3.3 \sigma$ & $\sim 2.1$, $\sim 2.0$  & \cite{quasi}, \cite{Pablo22}, \cite{rueda22}\\

PKS 0250$-$225 & $ 1.2 \pm 0.1$ & $2.7 \sigma$ & $\sim 1.2$  & \cite{Pablo22}, \cite{rueda22}\\

PKS 1510$-$089 & $ 1.7 \pm 0.1$ & $2.5 \sigma$ & $\sim 3.6$  & \cite{pks1510}\\

\hline
\end{tabular}

\tablefoot{Periodicity candidates already discussed in the literature, sorted from larger to smaller local significance from our analysis. Note that the periods in the literature are typically given without uncertainties and also as local significance.}
\end{table*}

\FloatBarrier
\begin{table*}
\caption{\label{table4} Long-term trend characterization}
    \begin{tabular}{cccccc}
    \hline\hline
        4FGL Source Name & Long-term Trend & Predicted cycle 1 & Predicted cycle 2 & Predicted cycle 3 & Predicted cycle 4  \\
        \hline

OC 457& Decreasing & -- & -- & -- & -- \\
PKS 0405$-$385 & Increasing & 2022-03-14 & 2025-04-07 & 2028-05-01 & -- \\
PKS 0426$-$380 & Decreasing & -- & -- & -- & -- \\
PKS 0454$-$234 & Decreasing & 2024-01-15 & 2028-04-03 & -- & -- \\
S3 0458$-$02 & Increasing & -- & -- & -- & -- \\
TXS 0518+211 & -- & -- & -- & -- & -- \\
PKS 0524$-$485 & Increasing & 2022-10-24 & 2024-11-18 & 2026-11-16 & 2028-12-11 \\
OJ 014 & Increasing & 2024-08-26 & 2029-01-08 & -- & -- \\
PKS 1005$-$333 & -- & -- & -- & -- & --\\
S5 1044+71 & Increasing & 2023-01-16 & 2026-02-09 & 2029-02-05 & -- \\
S4 1144+40 & -- & 2021-05-10 & 2024-10-21 & 2028-04-03 & -- \\
B2 1520+31 & Decreasing & -- & -- & -- & -- \\
TXS 1530$-$131 & -- & 2021-03-15 & 2022-08-29 & 2024-01-15 & 2025-06-02 \\
PG 1553+113 & Increasing & 2021-03-15 & 2023-05-08 & 2025-06-02 & 2027-07-26 \\
4C +47.44 & Decreasing & -- & -- & -- & -- \\
4C +48.41 & -- & -- & -- & -- & -- \\
PKS 1903$-$80 & -- & 2022-08-01 & 2025-09-22 & 2028-11-13 & -- \\
PKS 2155$-$83 & -- & 2024-05-06 & 2029-08-20 & -- & -- \\
MH 2136$-$428 & Increasing & 2021-08-30 & 2023-07-03 & 2025-06-02 & 2027-04-05 \\
3C 66A & Decreasing & 2022-01-17 & 2024-05-06 & 2026-09-21 & 2029-01-08 \\
PKS B1310$-$041 & Increasing & -- & -- & -- & -- \\
S5 0716+71 & -- & 2023-09-25 & 2026-07-27 & 2029-05-28 & -- \\
PKS 0208$-$512 & Increasing & -- & -- & -- & -- \\
PKS 2155$-$304 & -- & 2022-08-29 & 2024-05-06 & 2026-01-12 & 2027-09-20 \\
PKS 1424+240 & Decreasing & 2021-07-05 & 2023-01-16 & 2024-07-29 & 2026-02-09 \\
TXS 1452+516 & Increasing & 2022-10-24 & 2025-03-10 & 2027-05-03 & 2029-08-20 \\
7C 2010+4619 & Increasing & -- & -- & -- & -- \\
BL Lac & Increasing & 2022-07-04 & -- & -- & -- \\
B2 2234+28A & Increasing & 2021-02-15 & 2022-08-29 & 2024-03-11 & 2025-10-20 \\
GB6 J0043+3426 & -- & 2021-04-12 & 2021-07-05 & 2021-09-27 & 2021-12-20 \\
S2 0109+22 & -- & 2022-08-29 & 2025-03-10 & 2027-08-23 & 2030-03-04 \\
PMN J0948+0022 & Decreasing & -- & -- & -- & -- \\
PKS 0447$-$439 & Increasing & 2022-09-26 & 2024-08-26 & 2026-08-24 & 2028-07-24 \\
PMN J0427$-$3900 & -- & -- & -- & -- & -- \\
TXS 1318+225 & -- & -- & -- & -- & --\\
S5 1221+80 & -- & 2023-05-08 & 2025-12-15 & 2028-07-24 & -- \\
PKS 0301$-$243 & -- & -- & -- & -- & -- \\
PKS 1716$-$771 & -- & 2023-02-13 & 2025-09-22 & 2028-03-06 & 2030-08-19 \\
PKS 1424$-$328 & -- & -- & -- & -- & -- \\
4C +04.42 & Decreasing  & 2023-02-13 & 2025-04-07 & 2027-05-31 & -- \\
PKS 0250$-$225 & Decreasing & -- & -- & -- & -- \\
S4 1250+53 & -- & 2021-04-12 & 2023-07-31 & 2025-11-17 & 2028-03-06 \\
PKS 1510$-$089 & Decreasing & 2021-12-20 & 2023-07-31 & 2025-05-05 & 2026-12-14 \\
PKS 0403$-$13 & -- & 2021-07-05 & 2023-03-13 & 2024-11-18 & 2026-07-27 \\
B2 0716+33 & Decreasing & -- & -- & -- & -- \\
S4 1030+41 & -- & 2021-01-18 & 2023-04-10 & 2025-07-28 & -- \\
        \hline
        \end{tabular}
  \tablefoot{Our periodicity candidates sorter according to the \textit{local significance} (same as Table~\ref{table results}) with their 4FGL source name, long-term trend, and the dates for the next four gamma ray cycles. The \textit{--} in the Long-term Trend column means that no trend was inferred and in the other columns mean that there is no predicted cycle.}
\end{table*}

\FloatBarrier
\section{Singular spectrum analysis plots}\label{plots}
This section presents the LCs of all our candidates. The blue dots represent the $\gamma$-ray data from \textit{Fermi}-LAT. Black arrows indicate the upper limits, which are considered when the flux value has a test statistic lower than 1. The upper-limit flux value is the flux that maximizes the likelihood function for the corresponding time-bin. The orange, green, and red solid lines represent the periodicity, trend, and the sum of the trend and periodicity from SSA, respectively. The distance between the blue rectangles represents the source period, and their width indicates the period uncertainty.

\begin{figure*}[h!]
	\centering
         \includegraphics[scale=0.262]{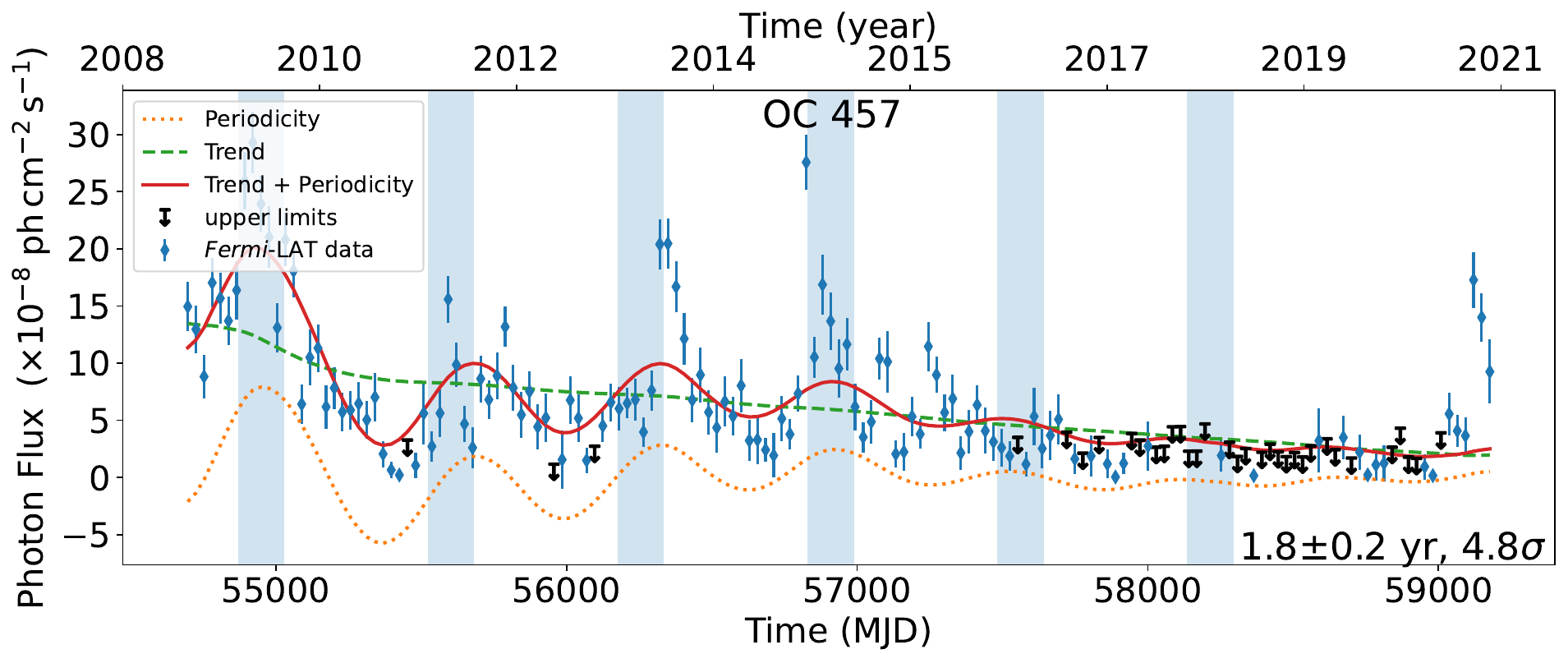}
         \includegraphics[scale=0.262]{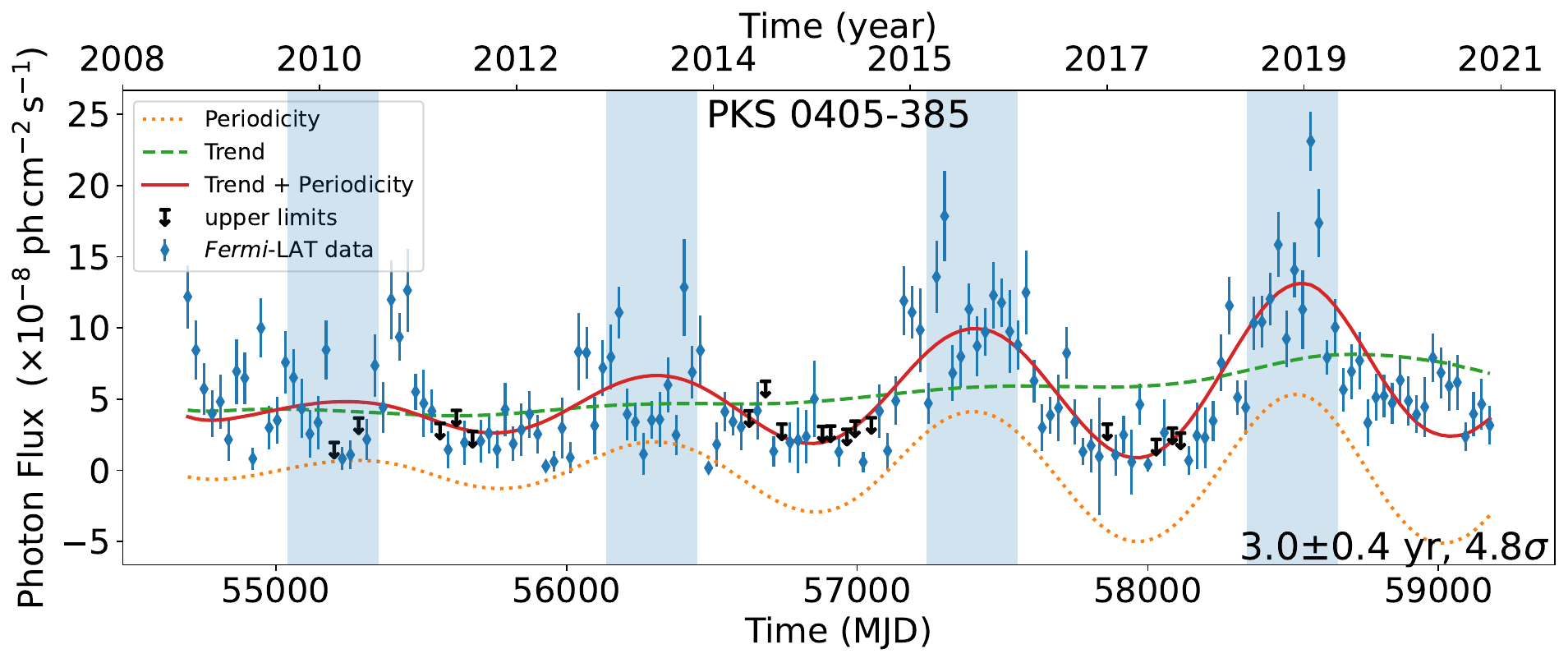}
         
         \includegraphics[scale=0.262]{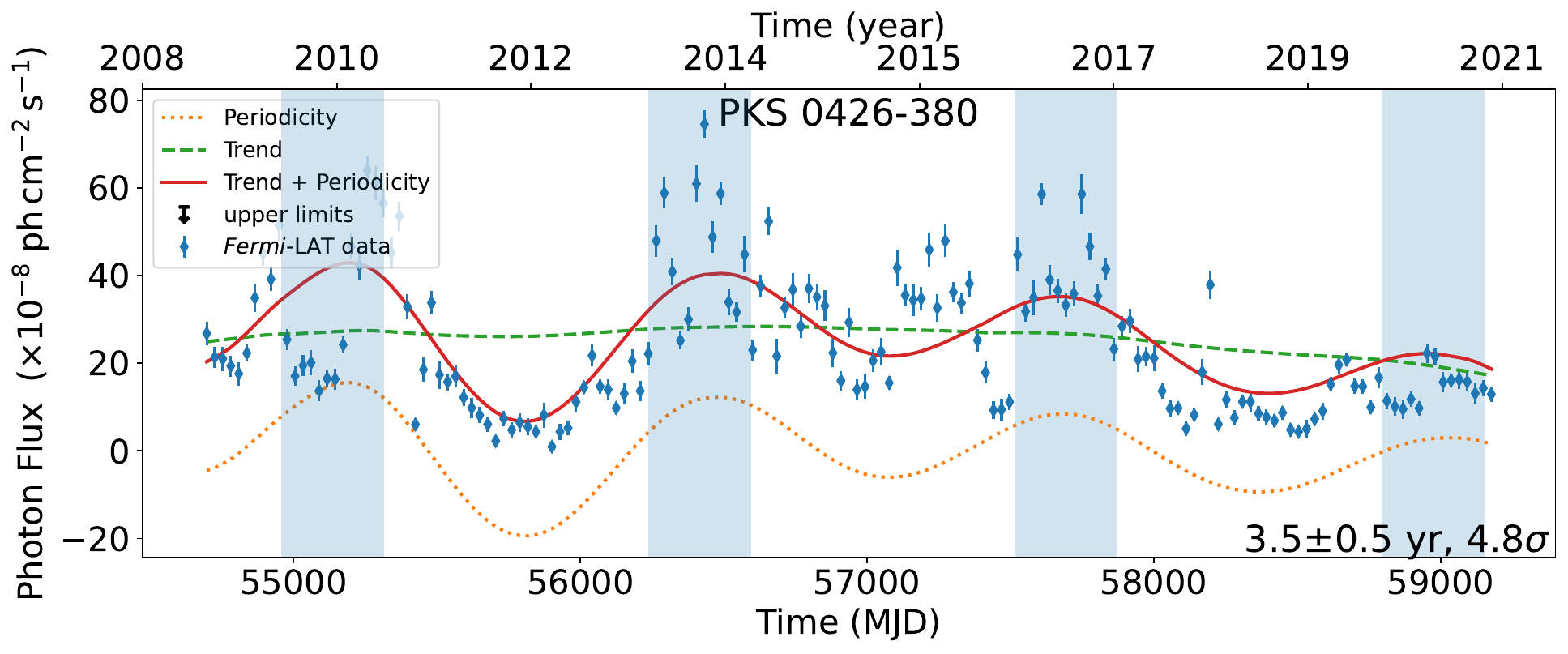}
         \includegraphics[scale=0.262]{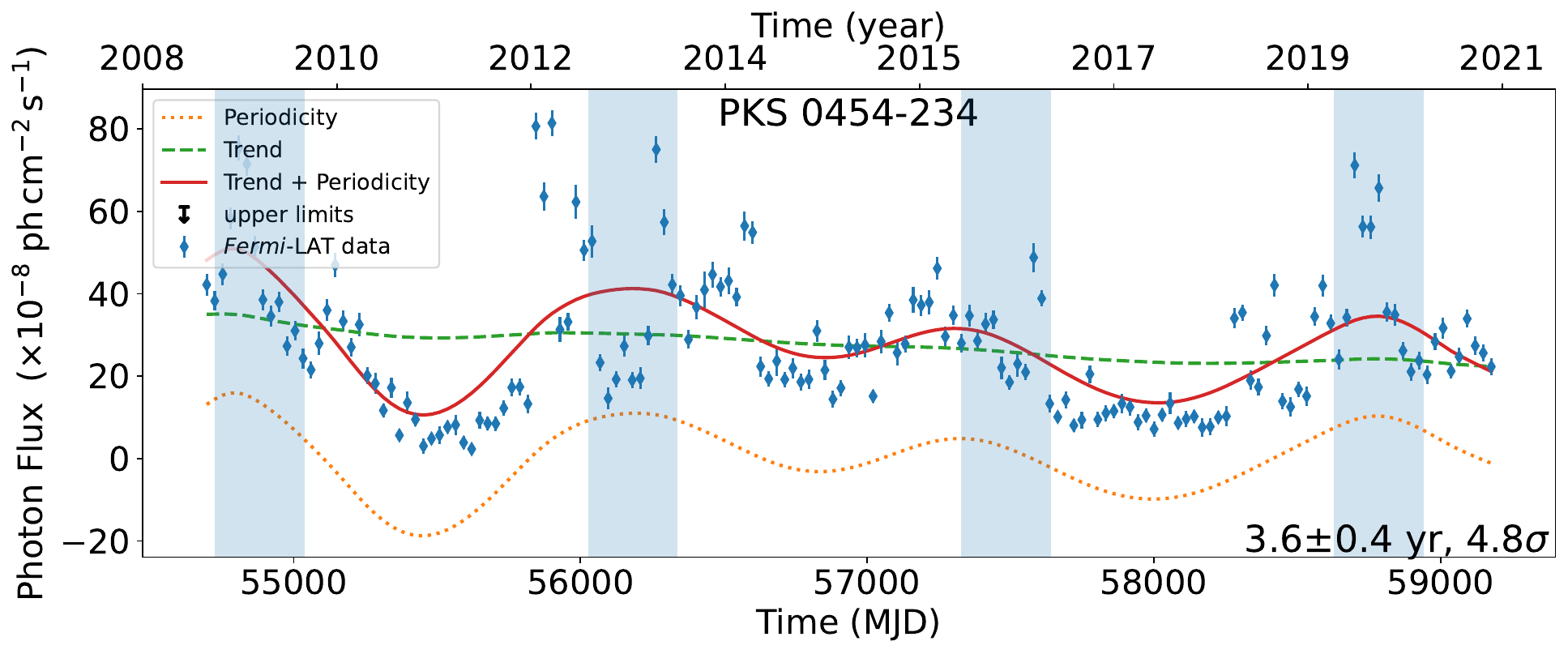} 
         
         \includegraphics[scale=0.262]{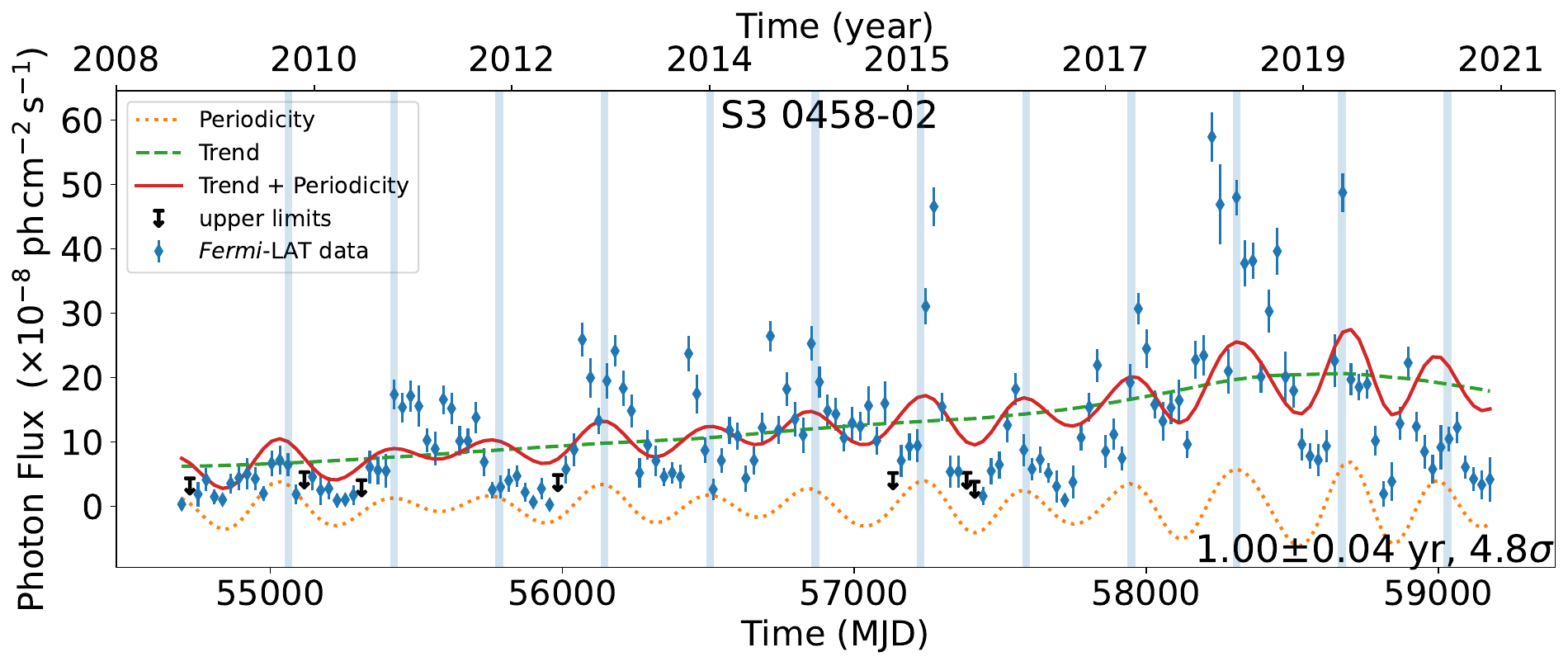}
         \includegraphics[scale=0.262]{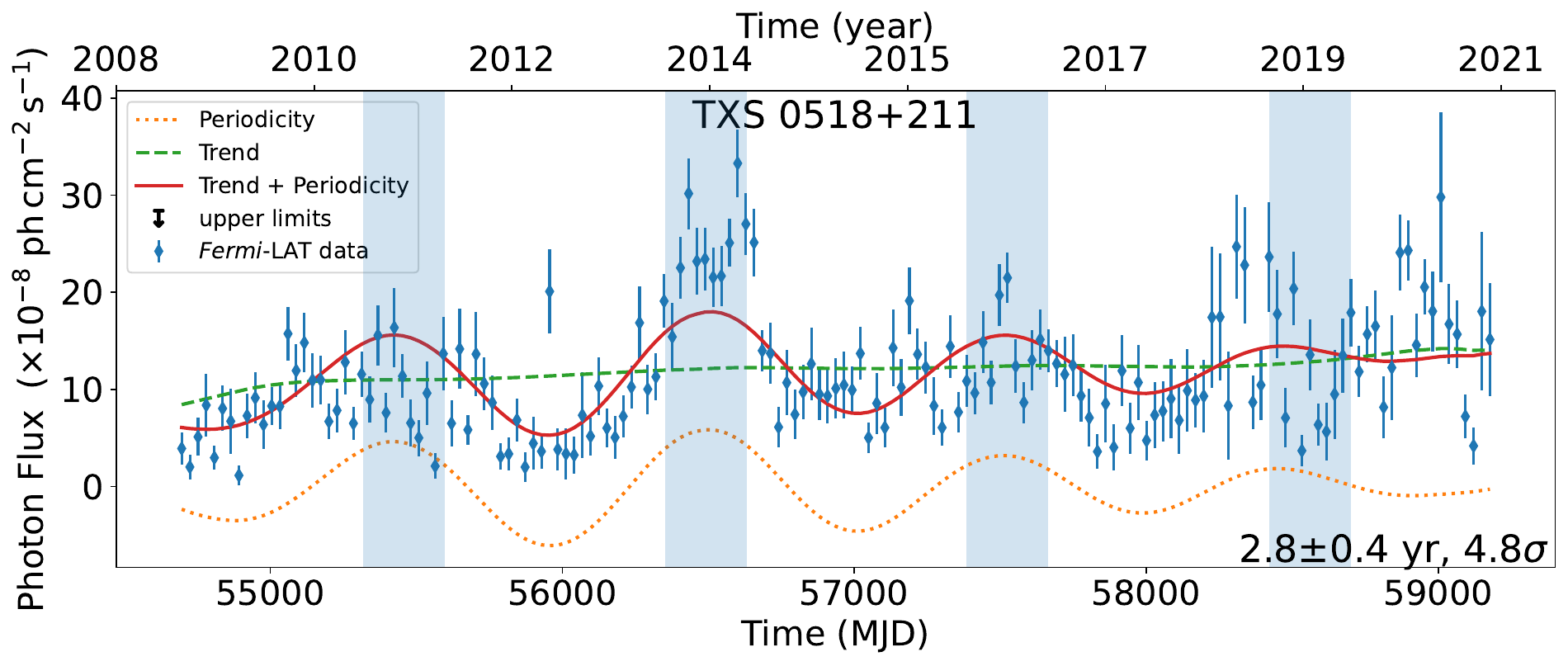}
         
         \includegraphics[scale=0.262]{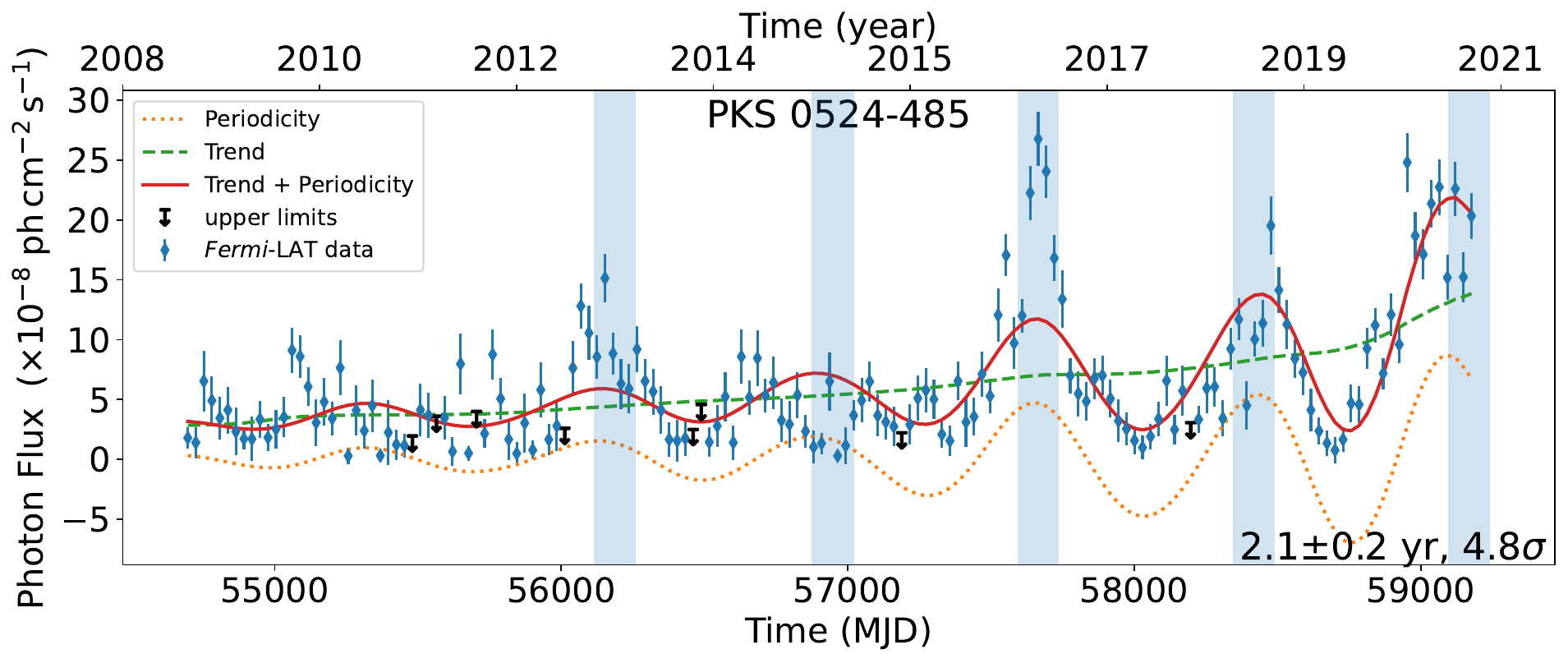}
         \includegraphics[scale=0.262]{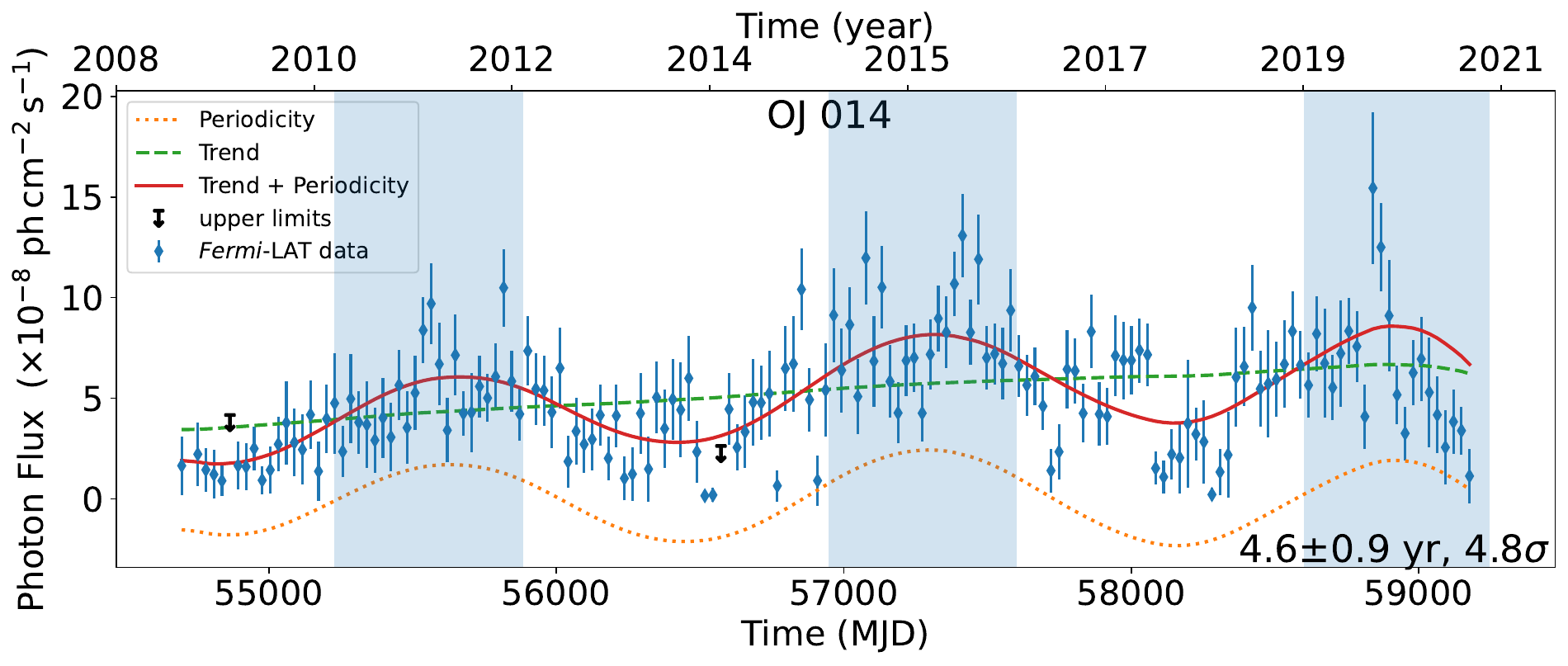}

	\caption{Light curves of quasi-periodic candidates presented in Table \ref{table results}. }
	\label{fig: plots}
\end{figure*}

\begin{figure*}
        \centering
        \ContinuedFloat
        
        \includegraphics[scale=0.262]{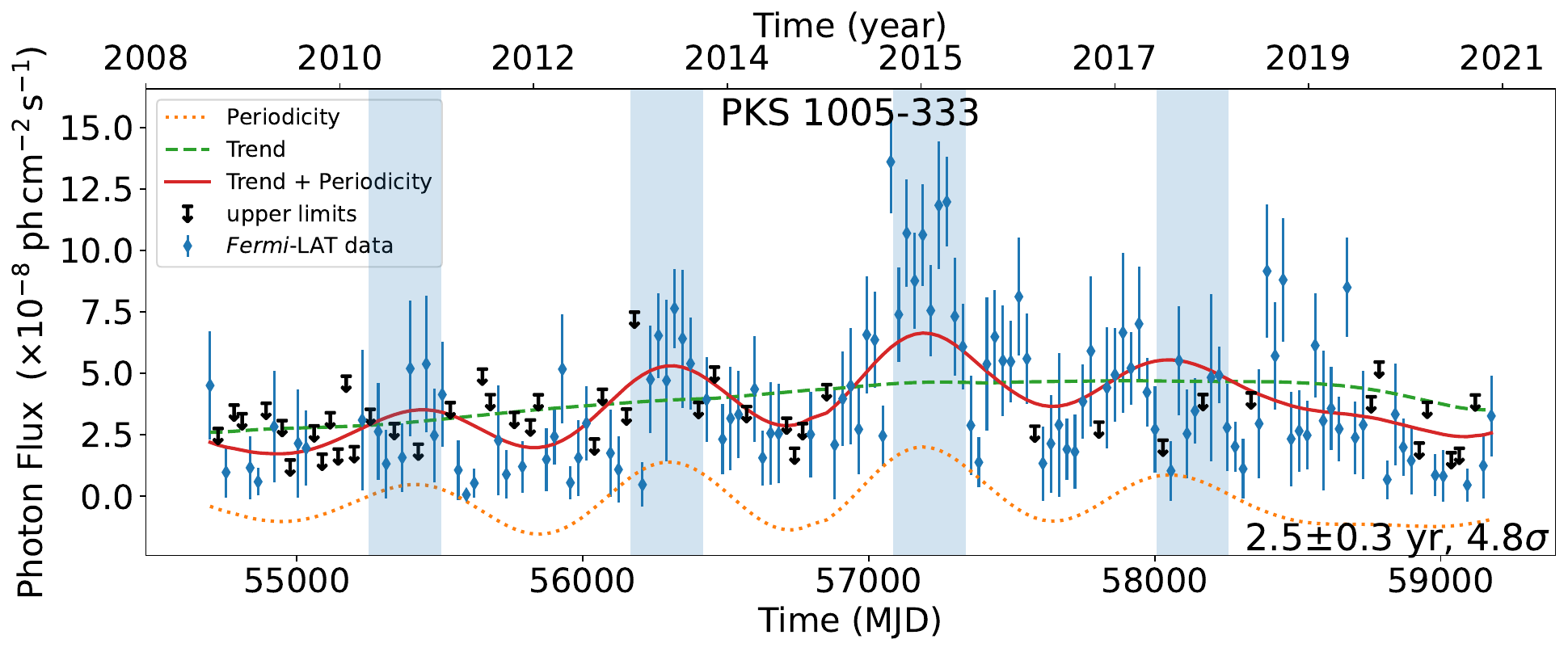}
        \includegraphics[scale=0.262]{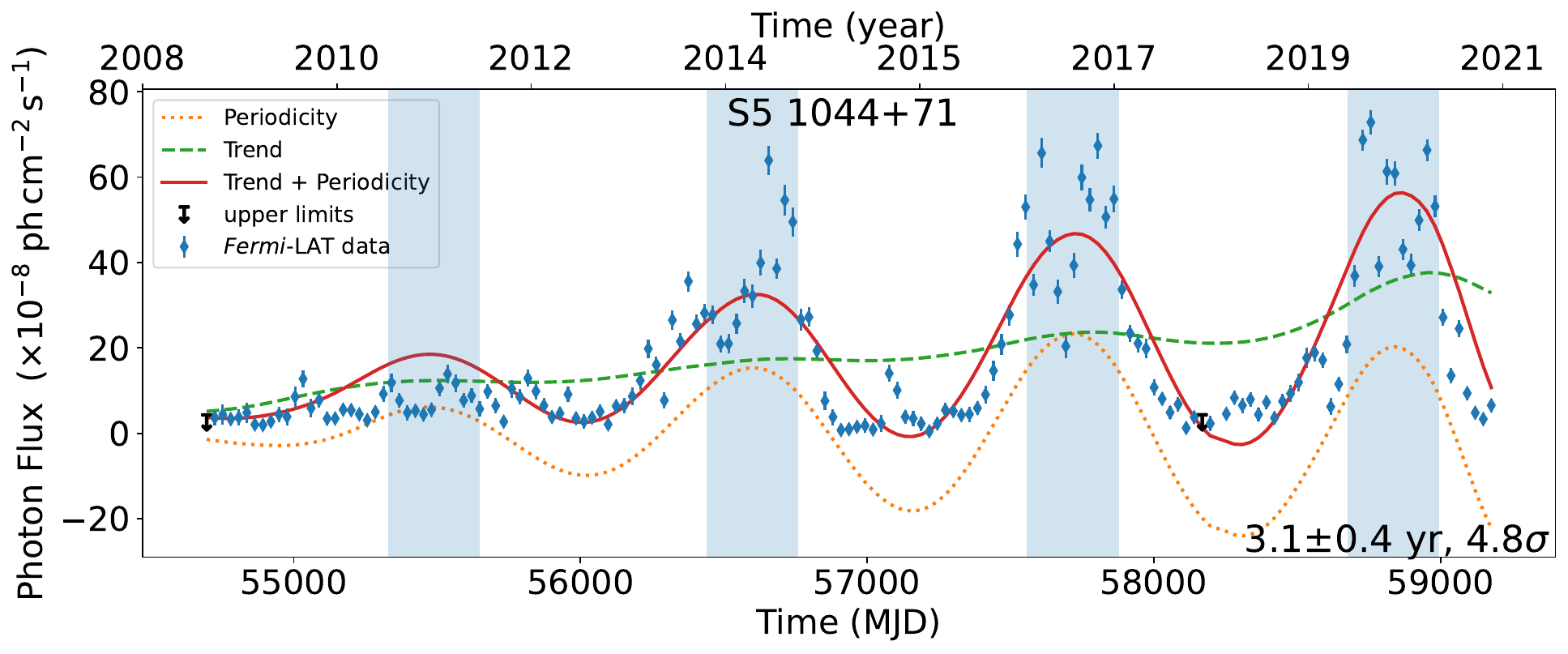}
  
        \includegraphics[scale=0.262]{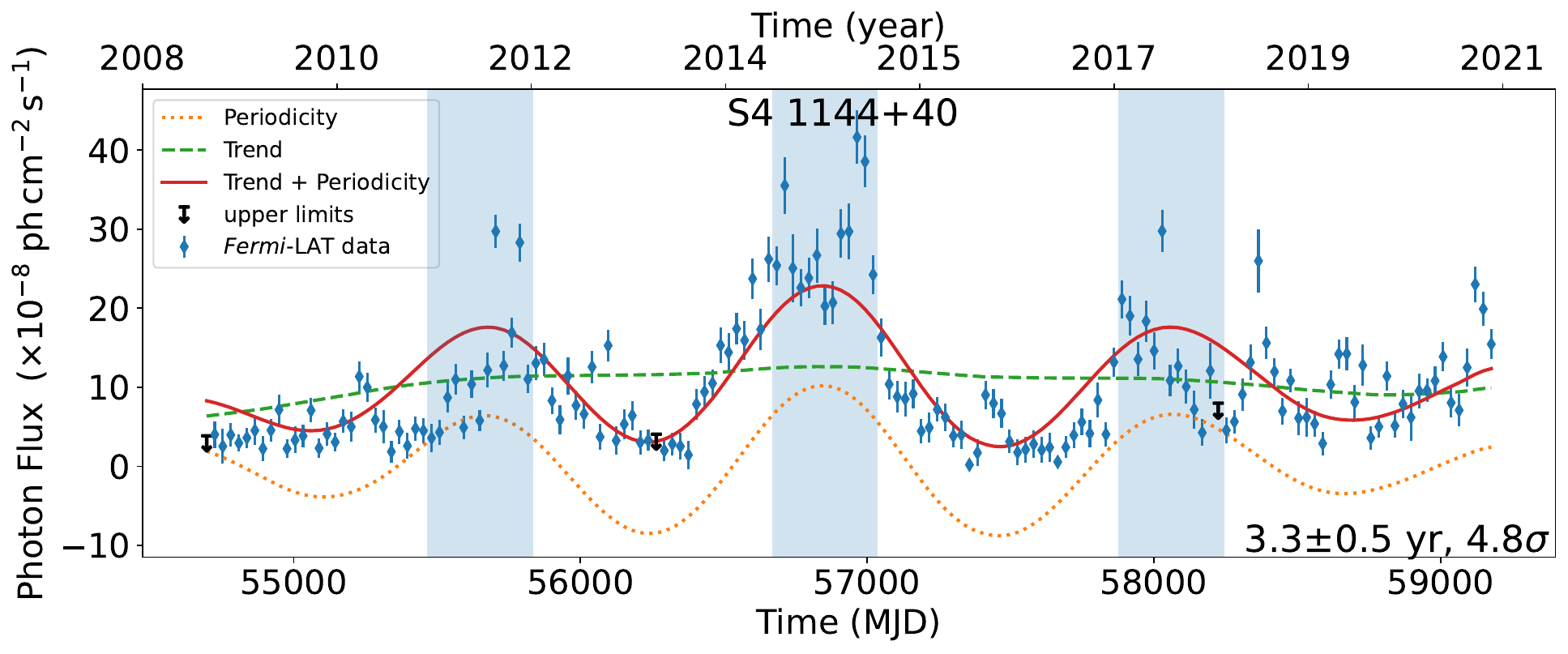}
        \includegraphics[scale=0.262]{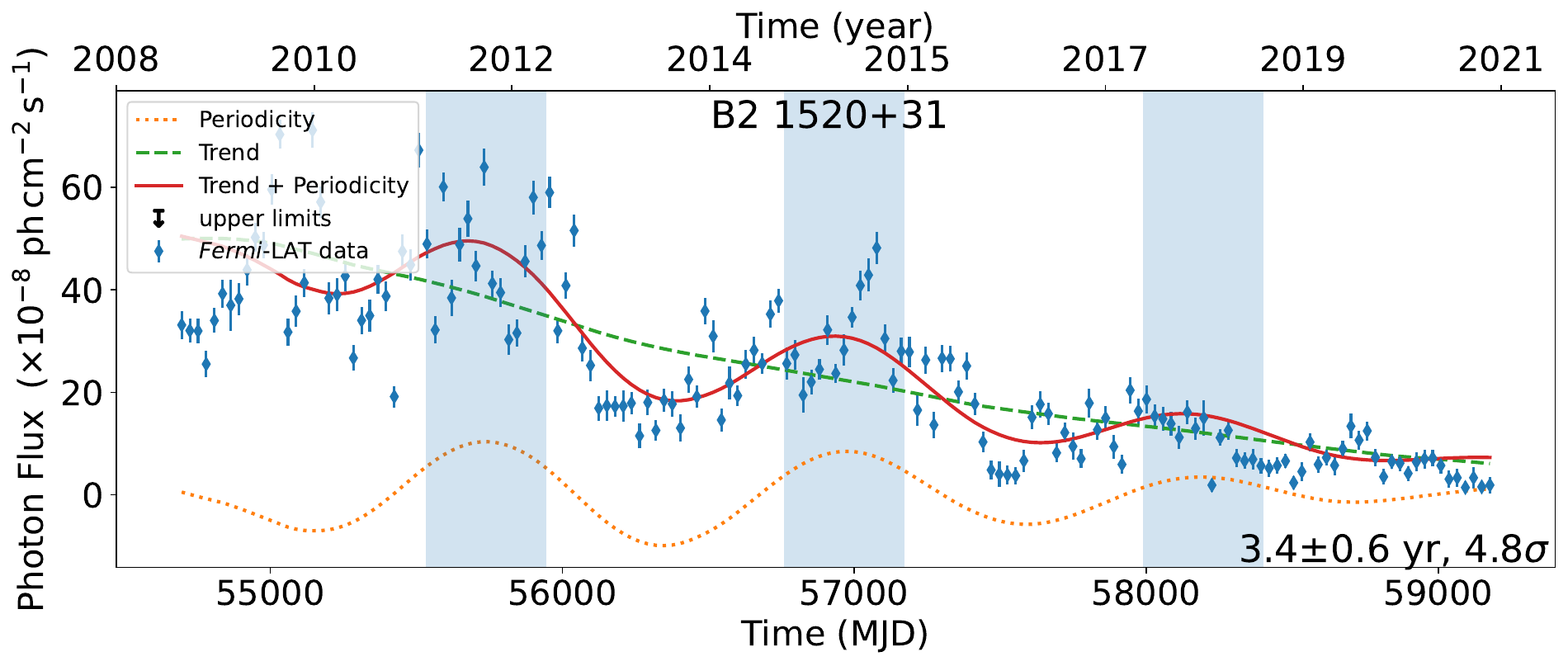}
         
        \includegraphics[scale=0.262]{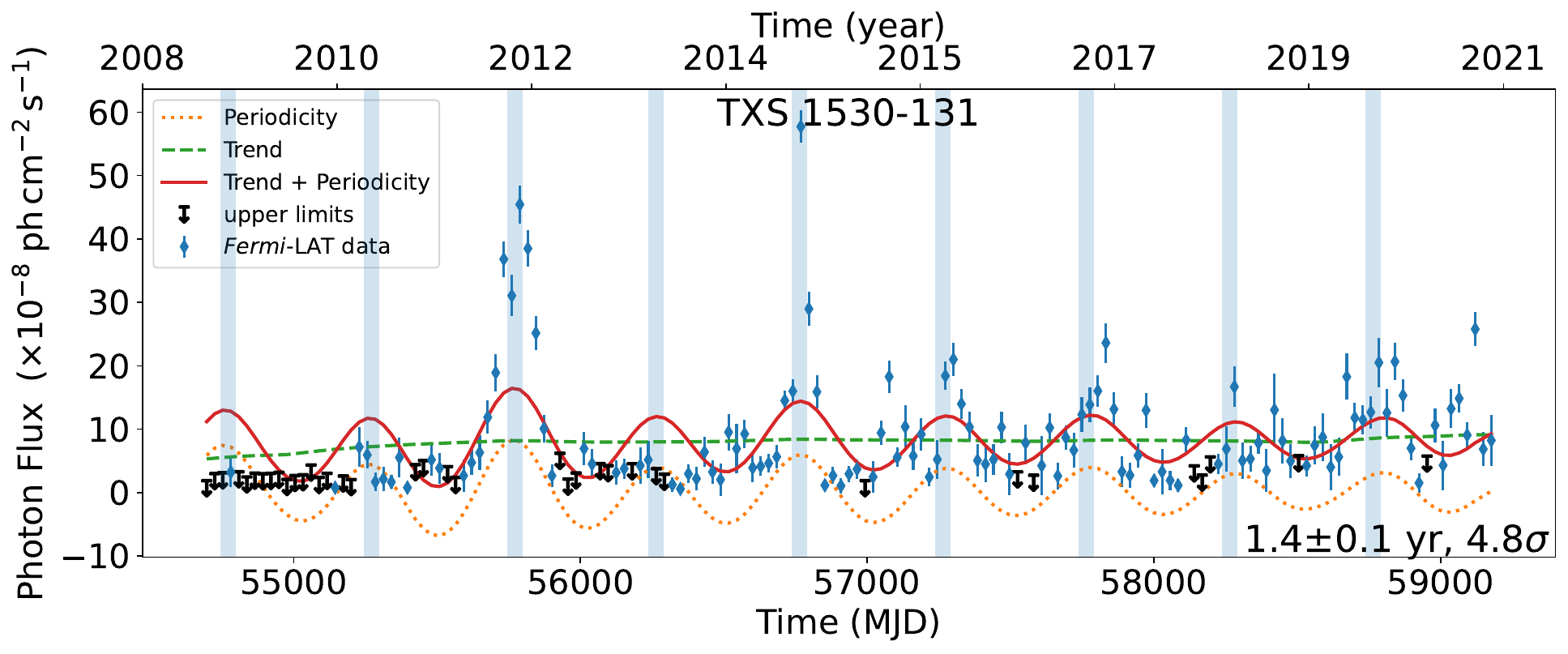}
        \includegraphics[scale=0.262]{4FGLJ1555.7+1111.pdf}
         
        \includegraphics[scale=0.262]{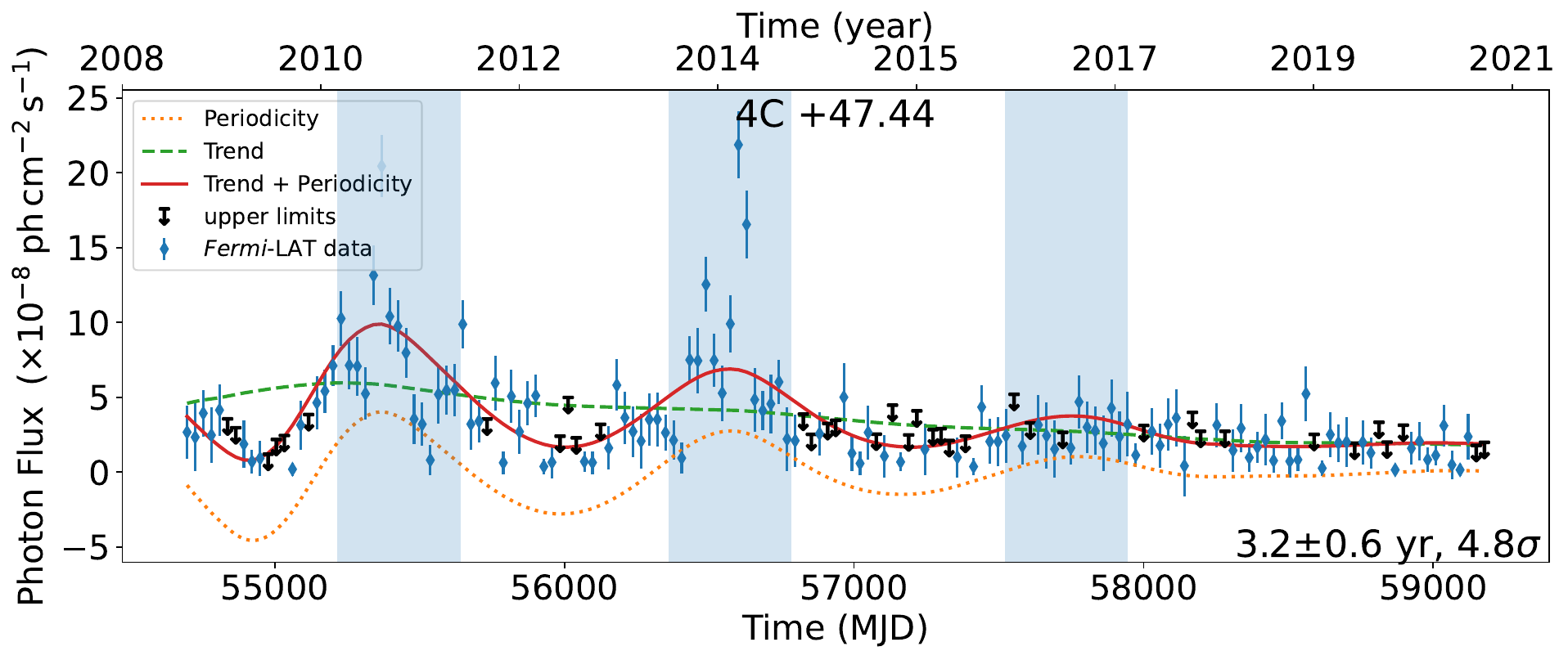}
        \includegraphics[scale=0.262]{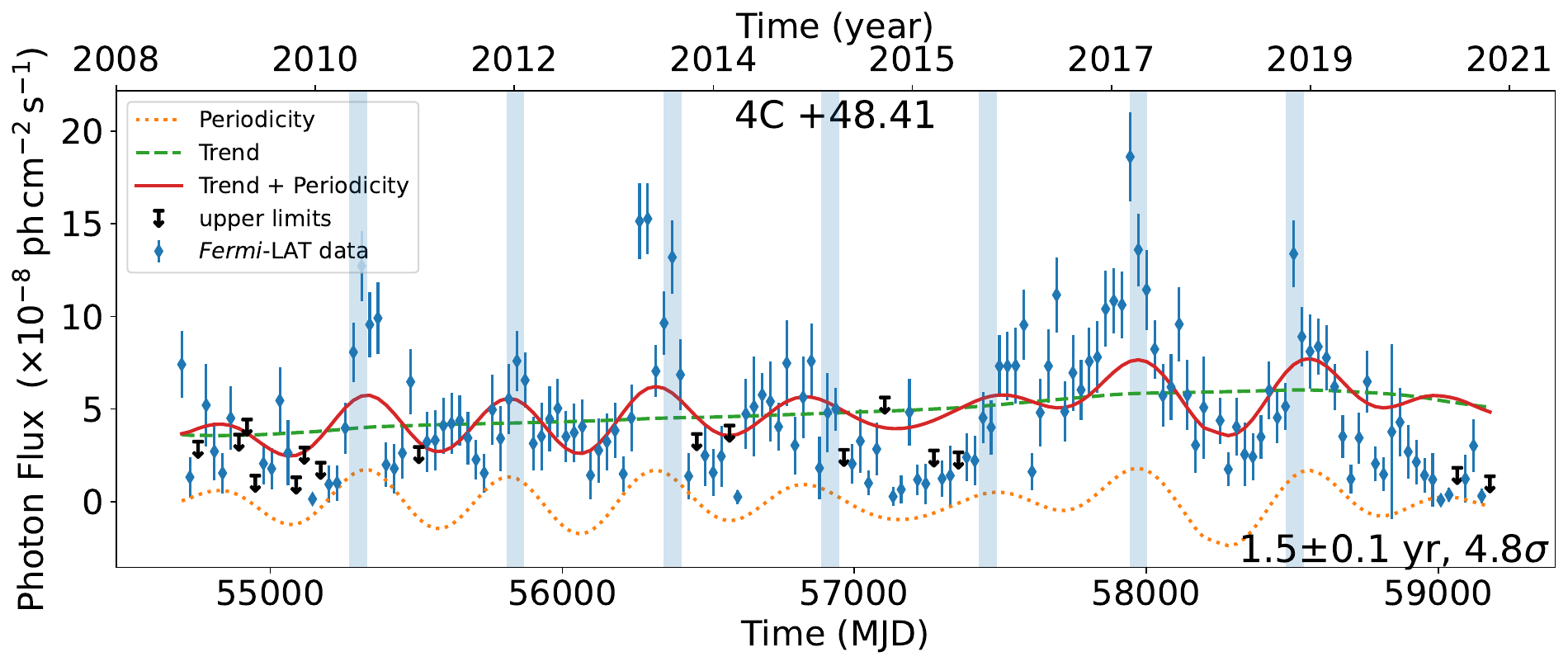}

        \includegraphics[scale=0.262]{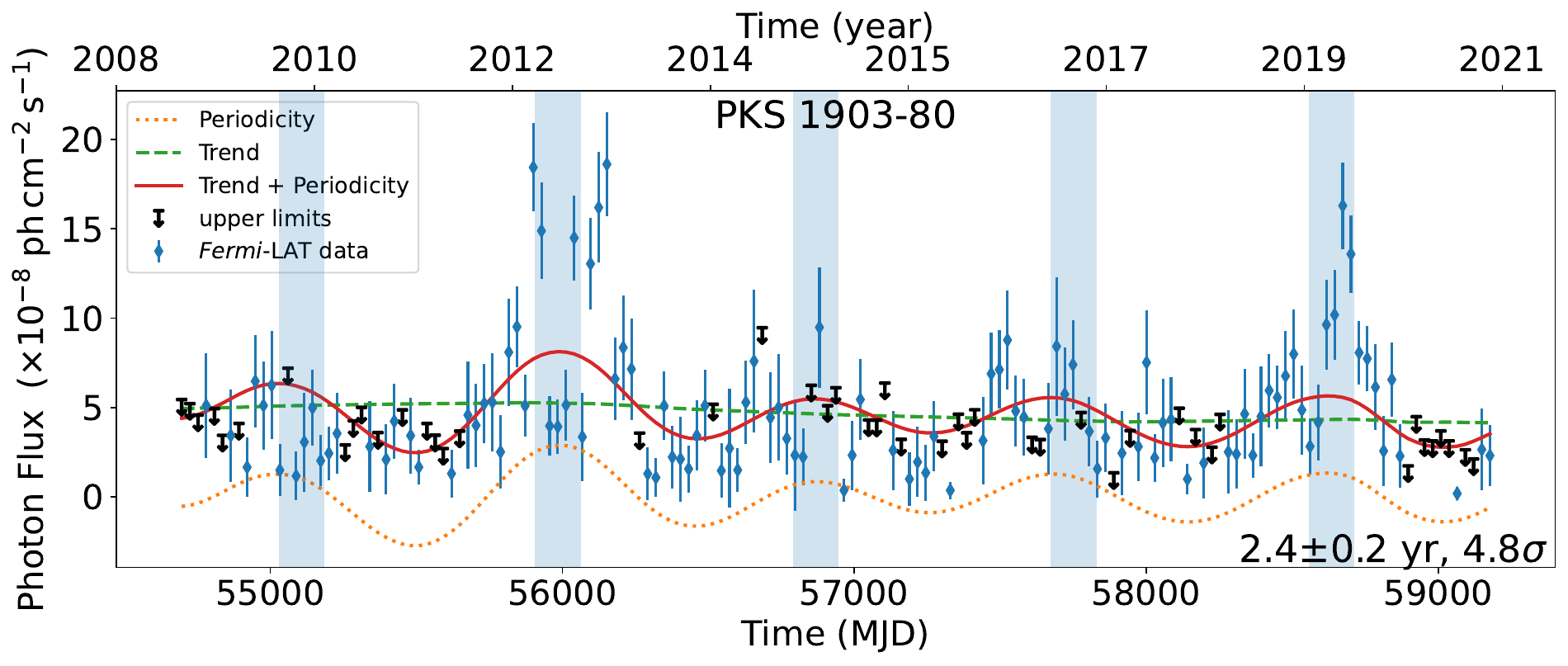}
        \includegraphics[scale=0.262]{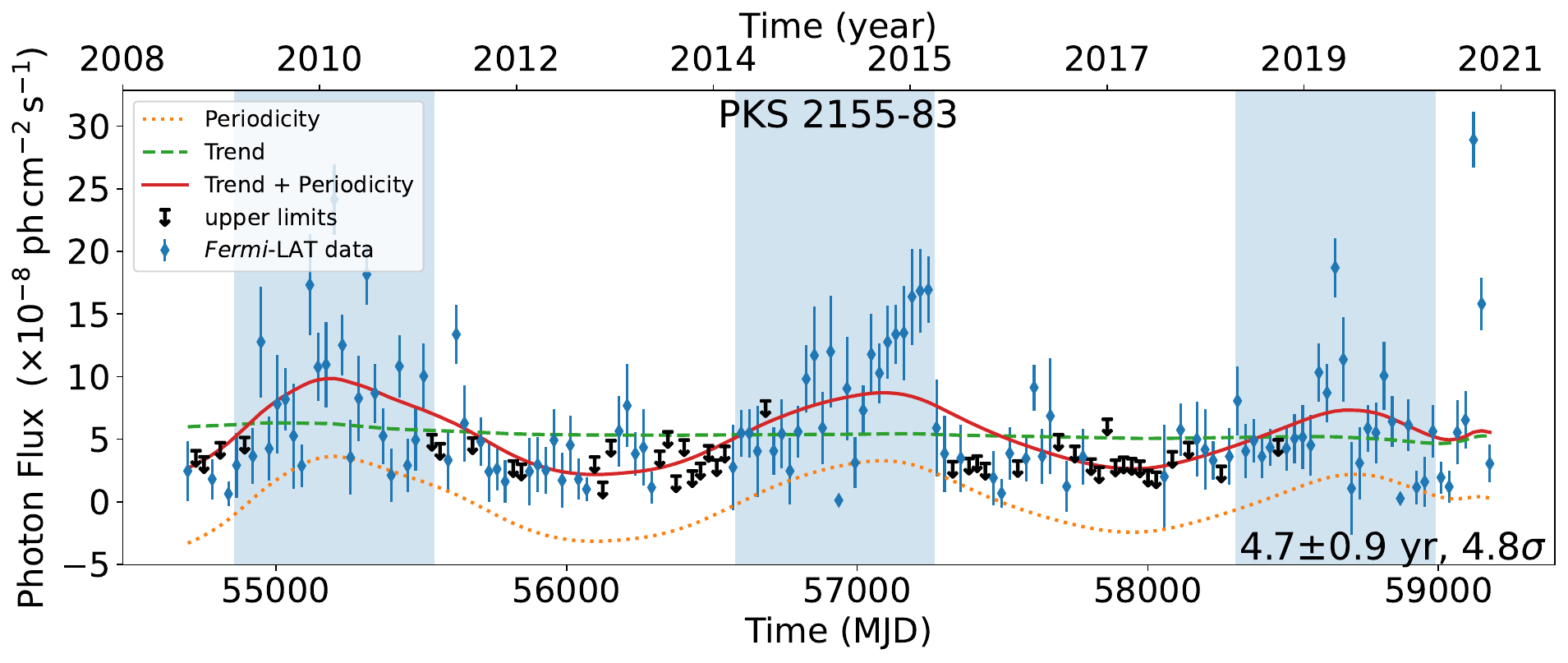}

         \caption{(Continued).}
    
\end{figure*}

\begin{figure*}
        \centering
        \ContinuedFloat

         \includegraphics[scale=0.262]{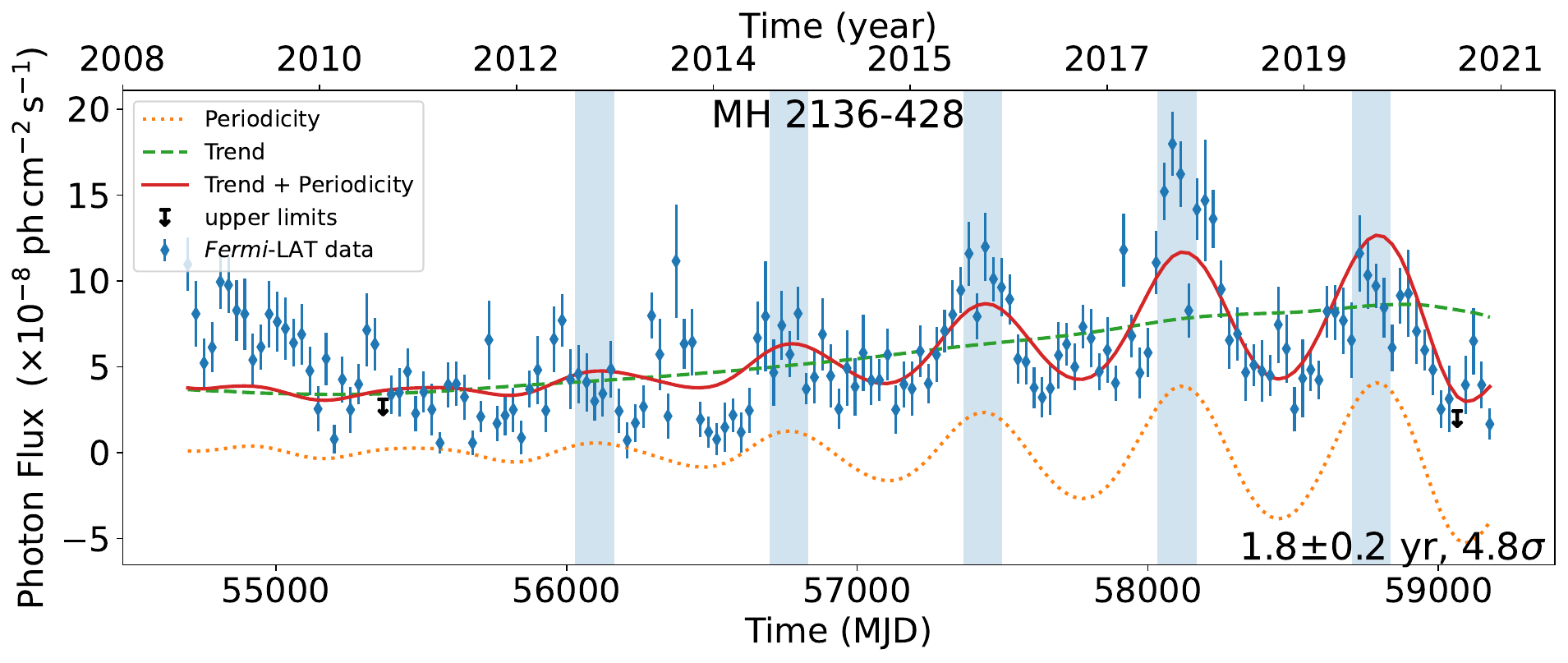}
         \includegraphics[scale=0.262]{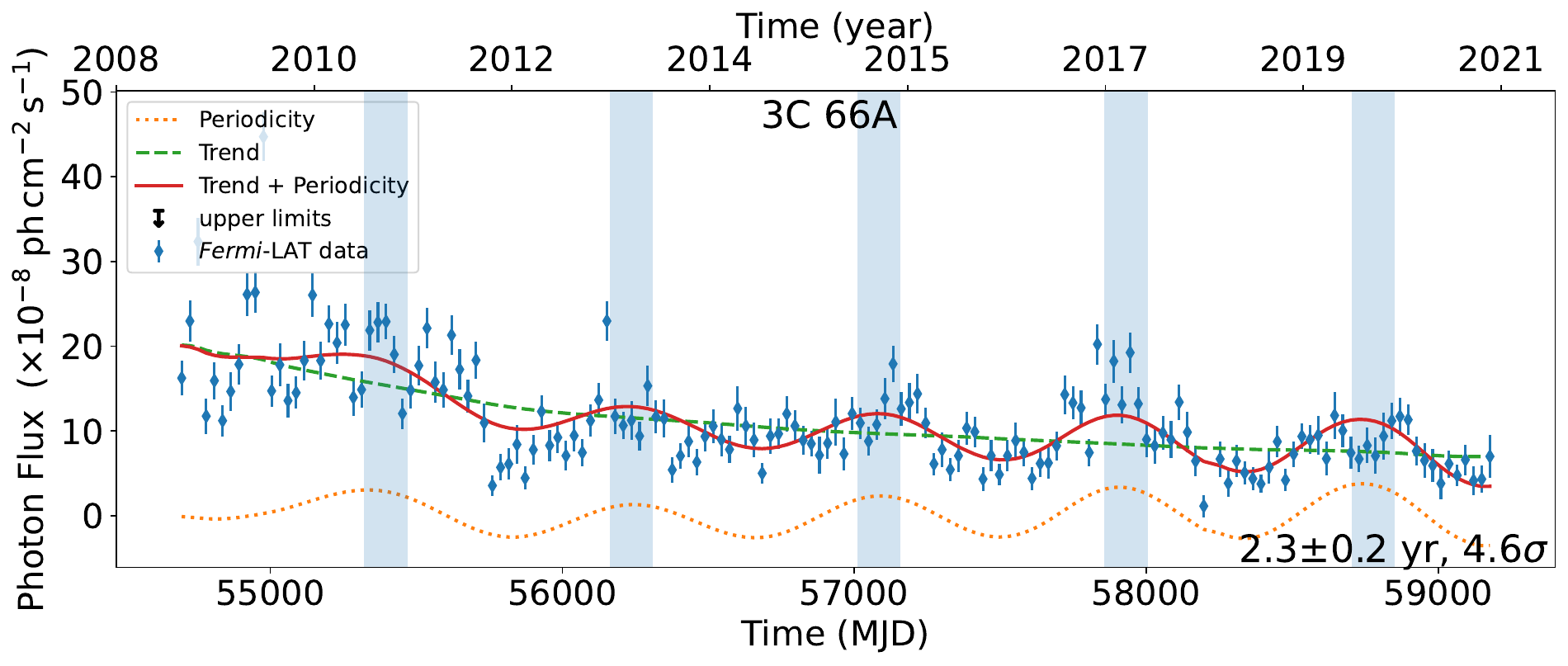}
         
         \includegraphics[scale=0.262]{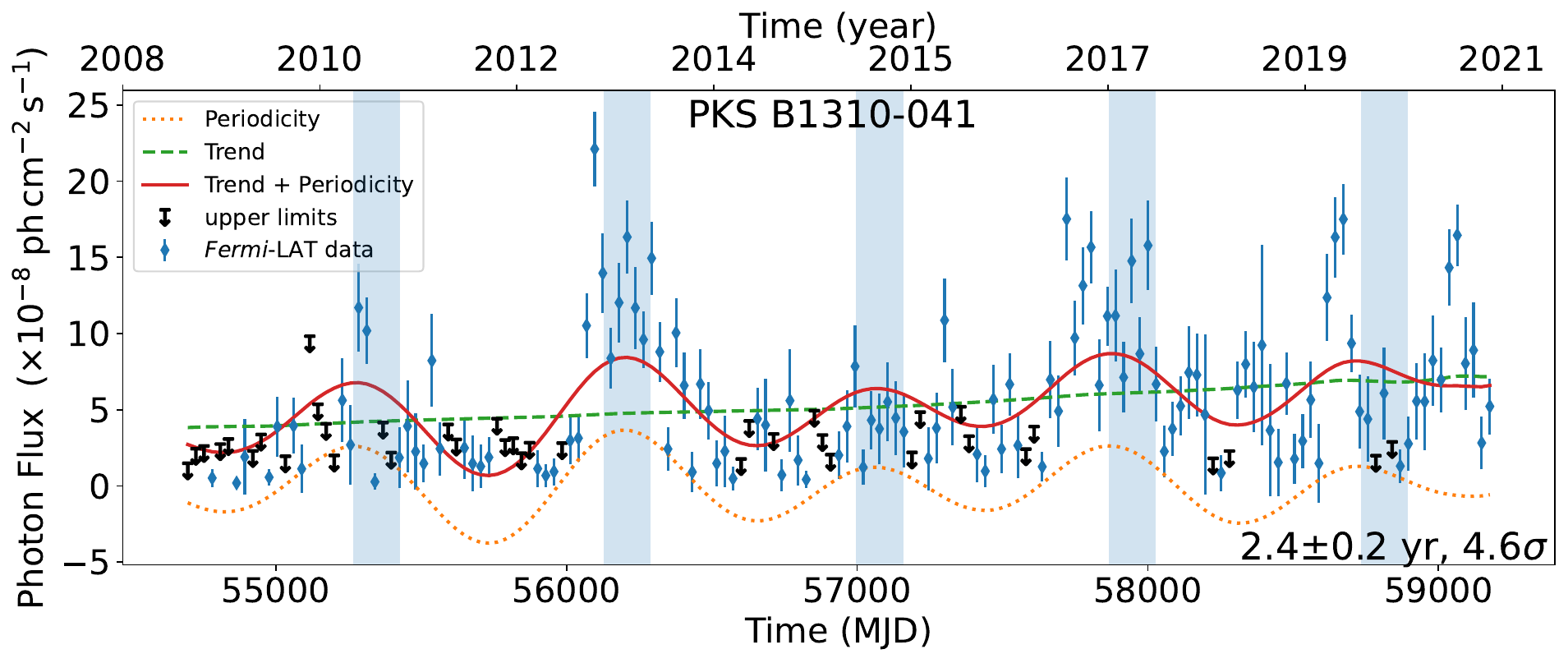}
         \includegraphics[scale=0.262]{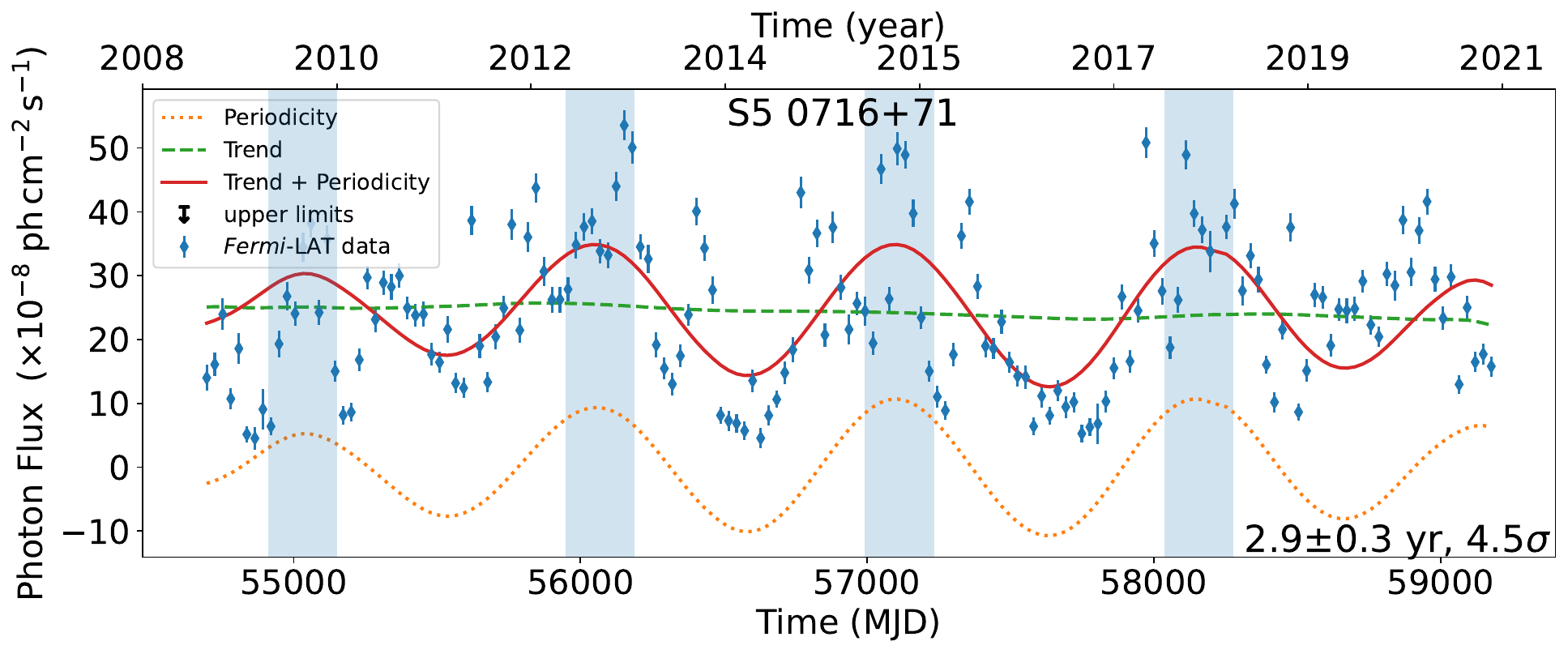}

         \includegraphics[scale=0.262]{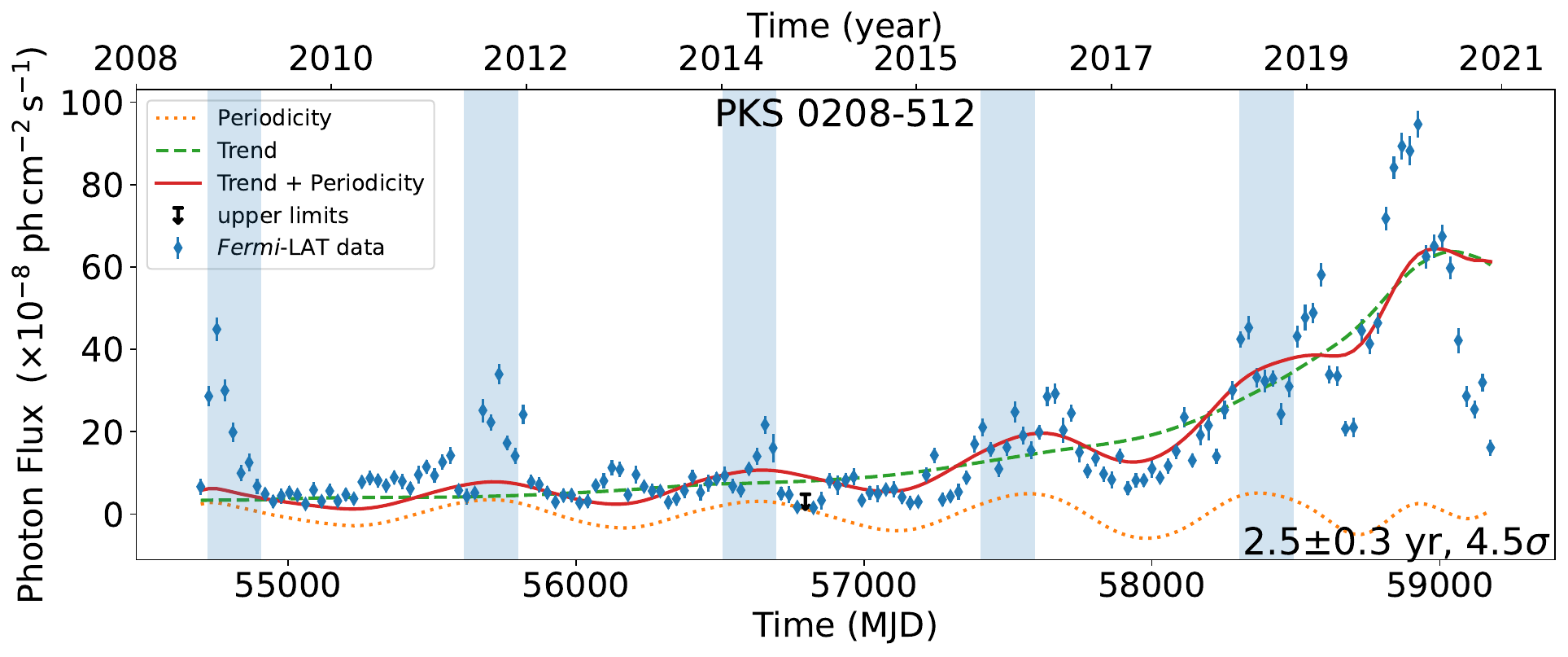}
         \includegraphics[scale=0.262]{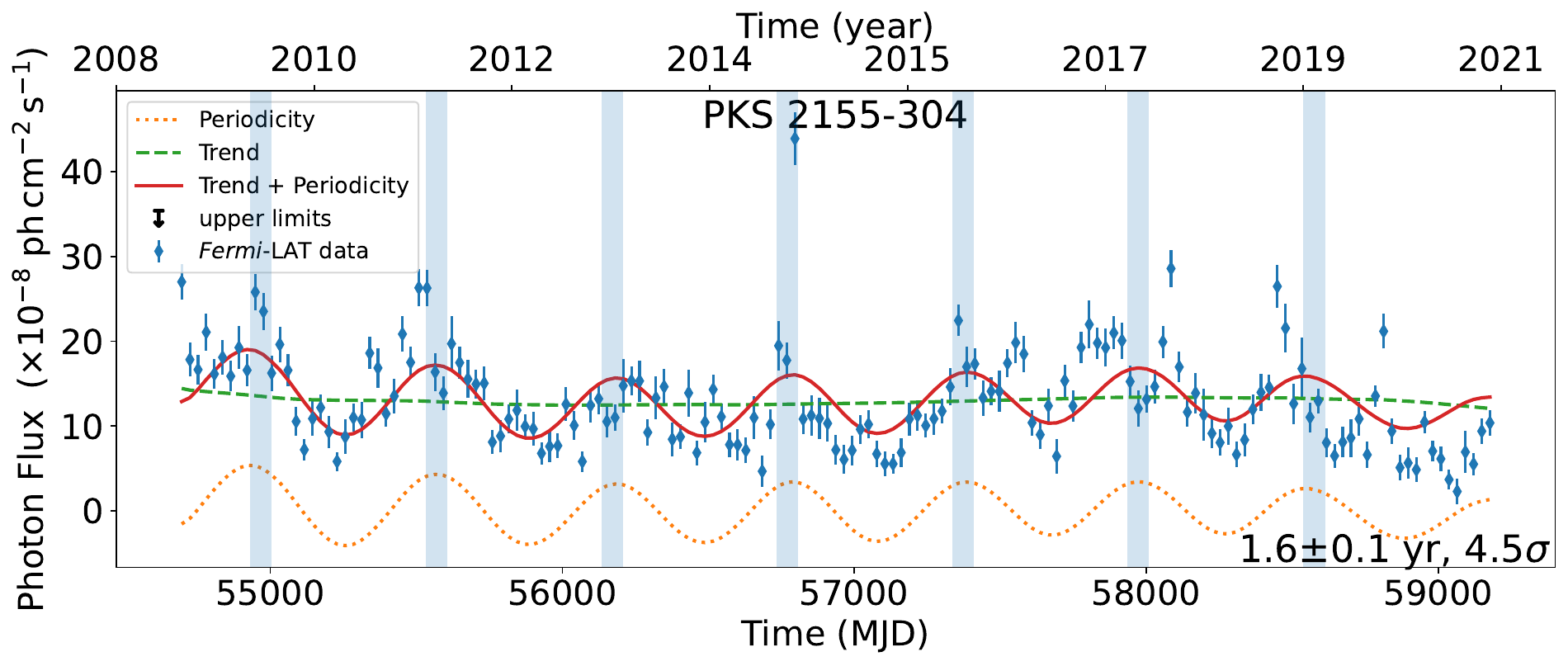}
         
         \includegraphics[scale=0.262]{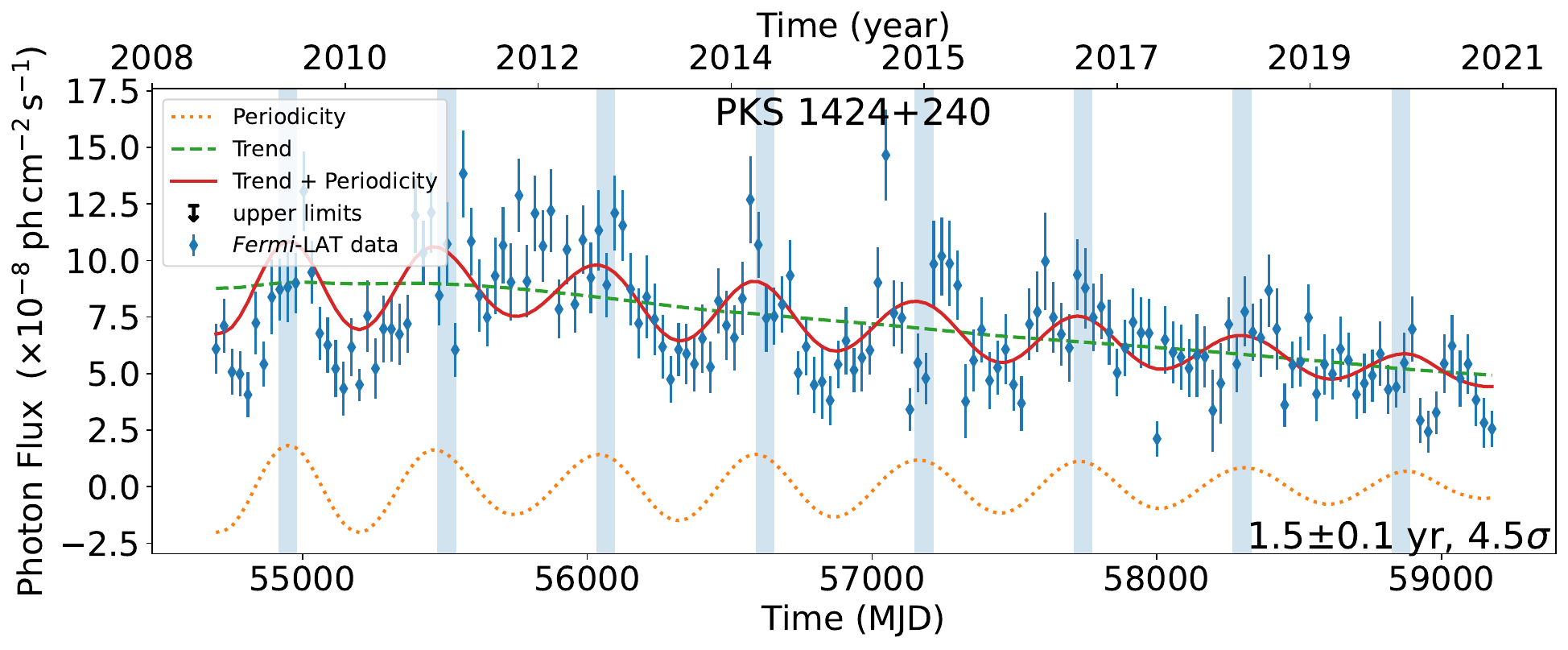}
         \includegraphics[scale=0.262]{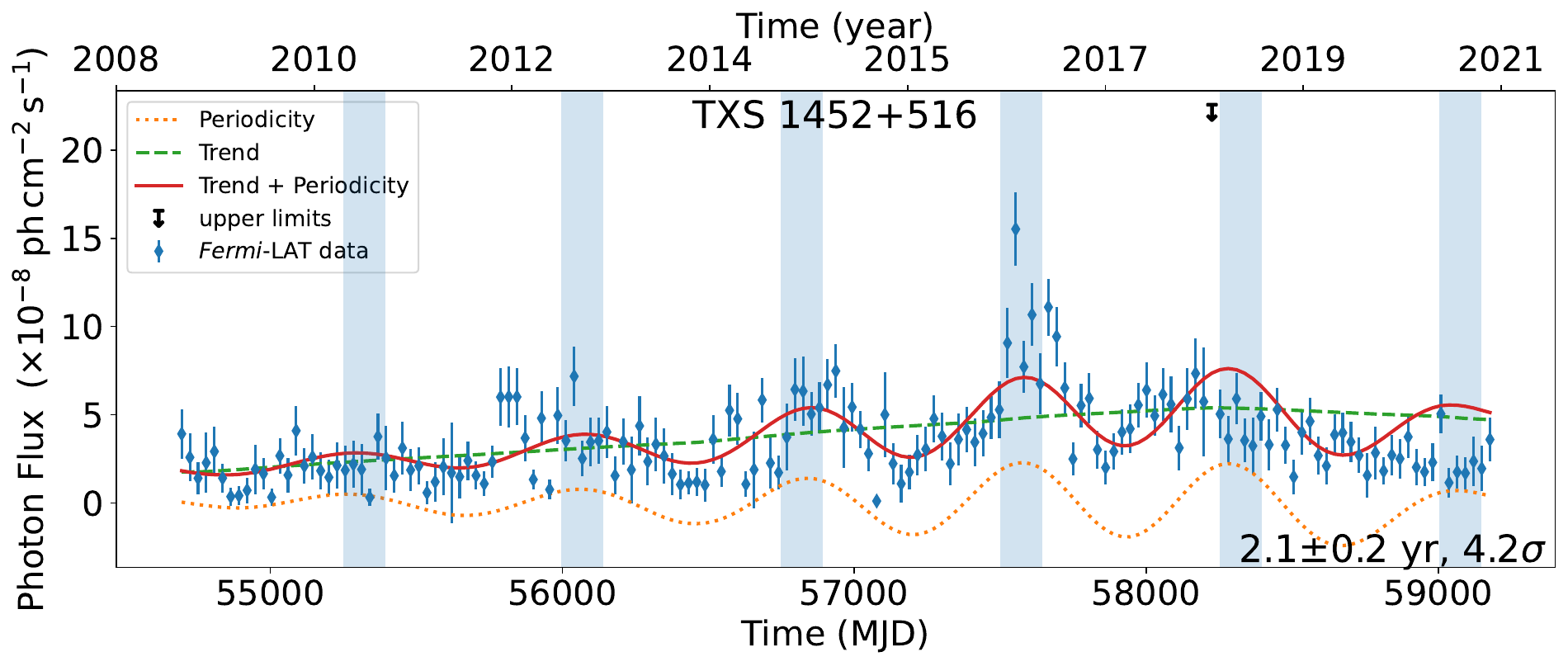}

         \includegraphics[scale=0.262]{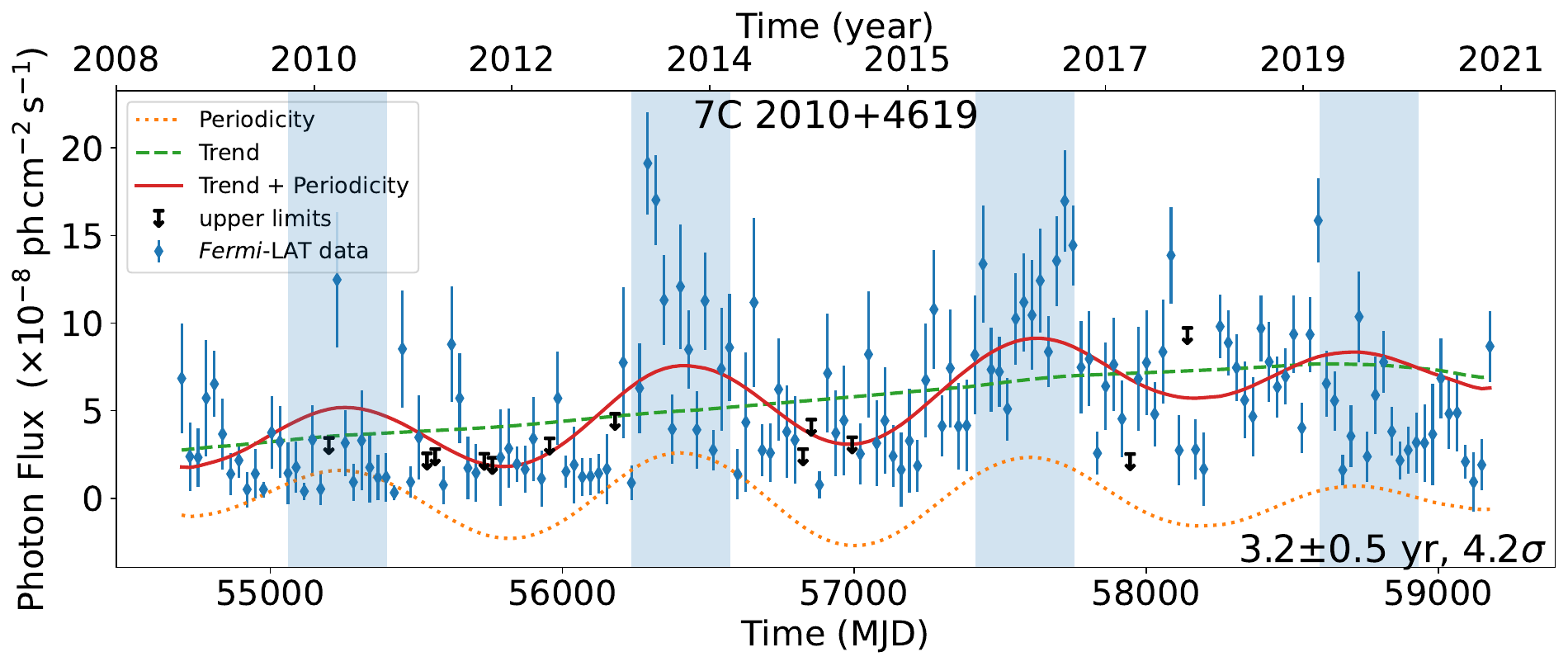} 
         \includegraphics[scale=0.262]{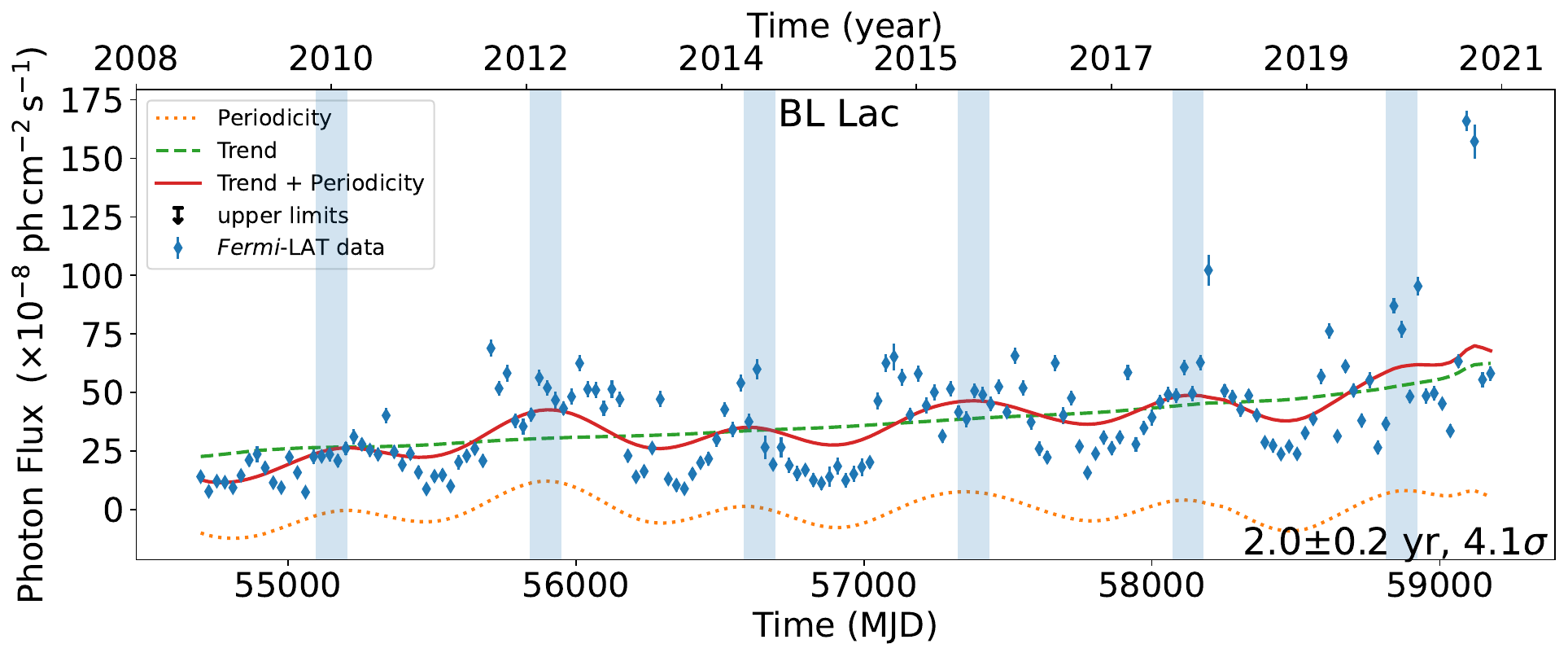}
         \caption{(Continued).}         
\end{figure*}

\begin{figure*}
        \centering
        \ContinuedFloat

         \includegraphics[scale=0.262]{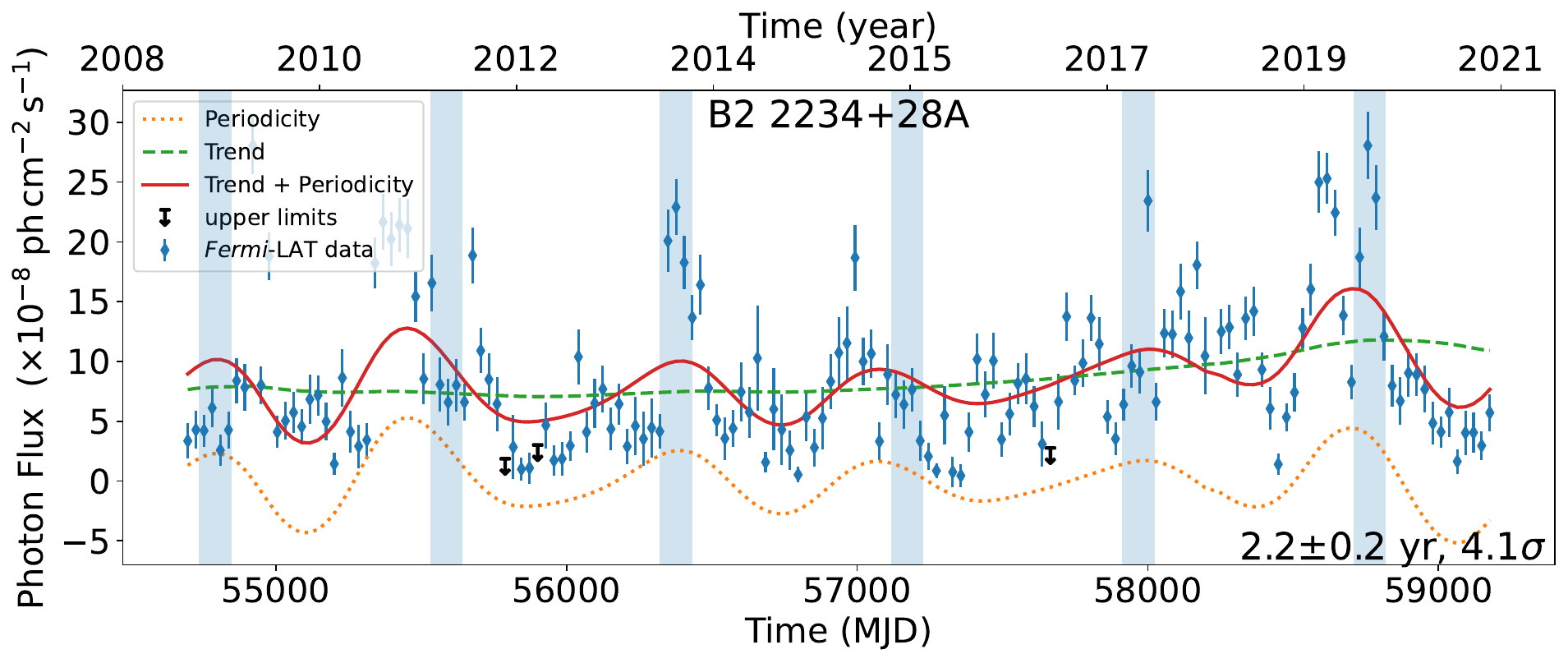}
         \includegraphics[scale=0.262]{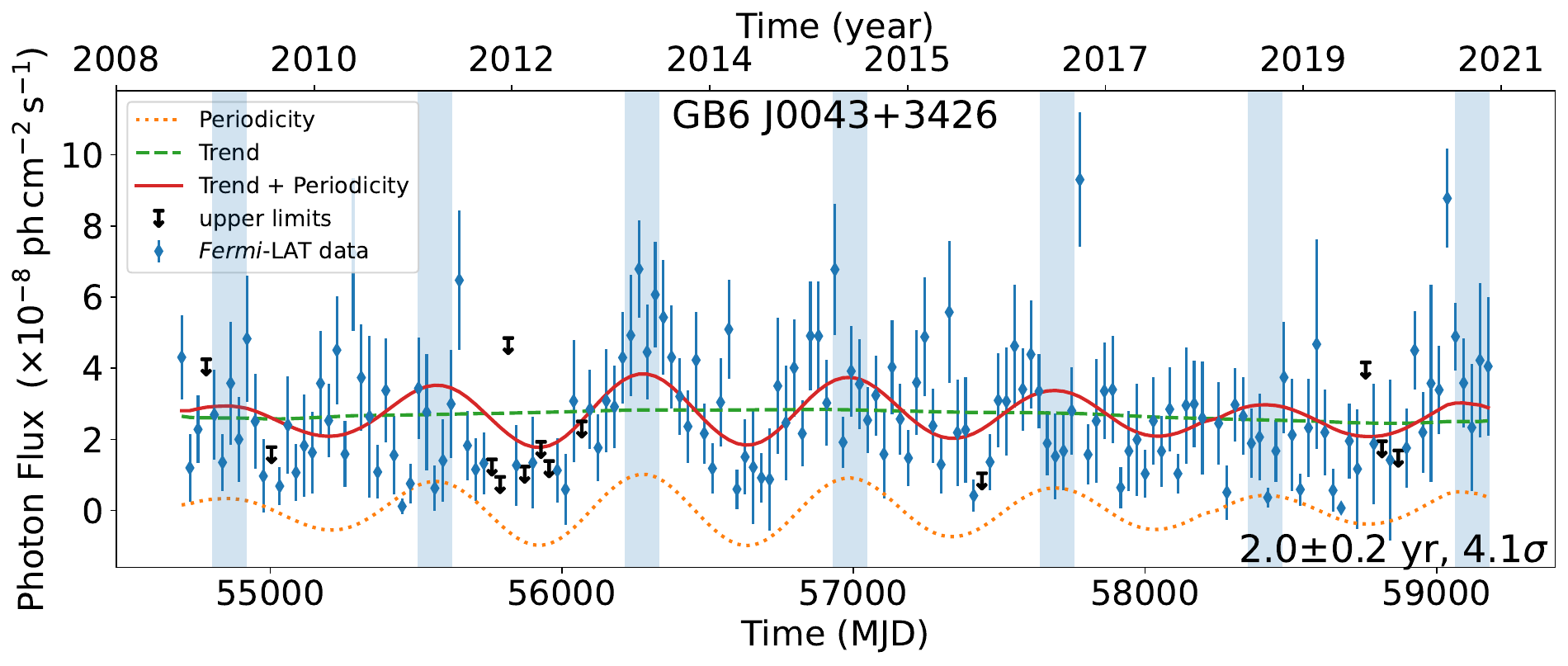}
         
         \includegraphics[scale=0.262]{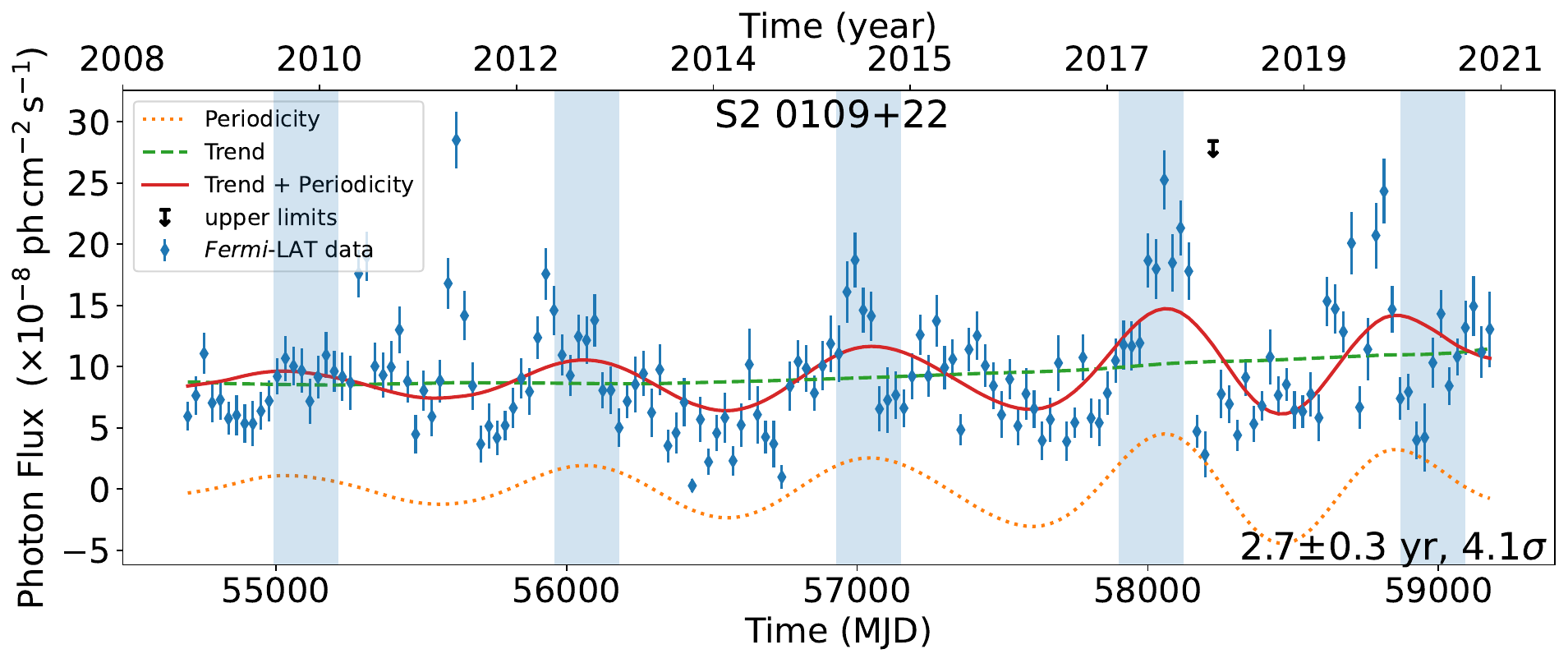}
         \includegraphics[scale=0.262]{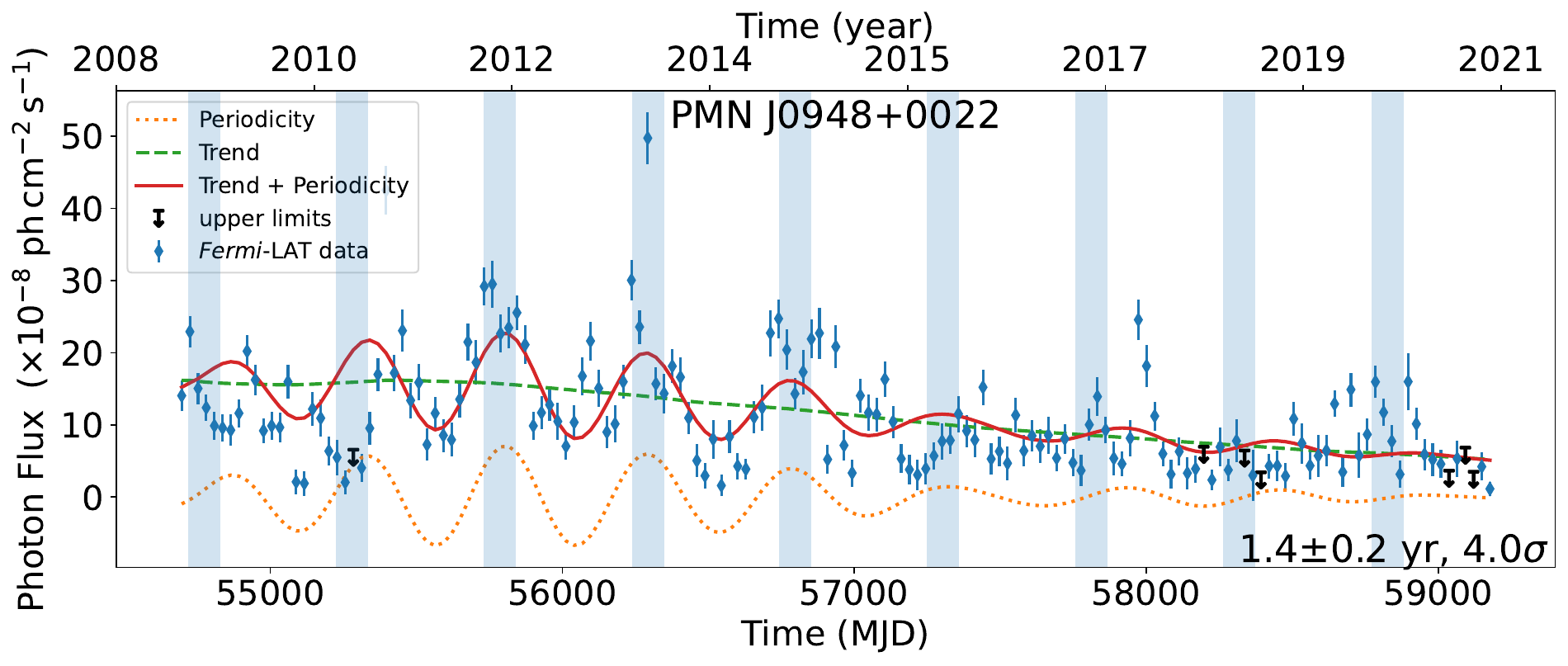}
         
         \includegraphics[scale=0.262]{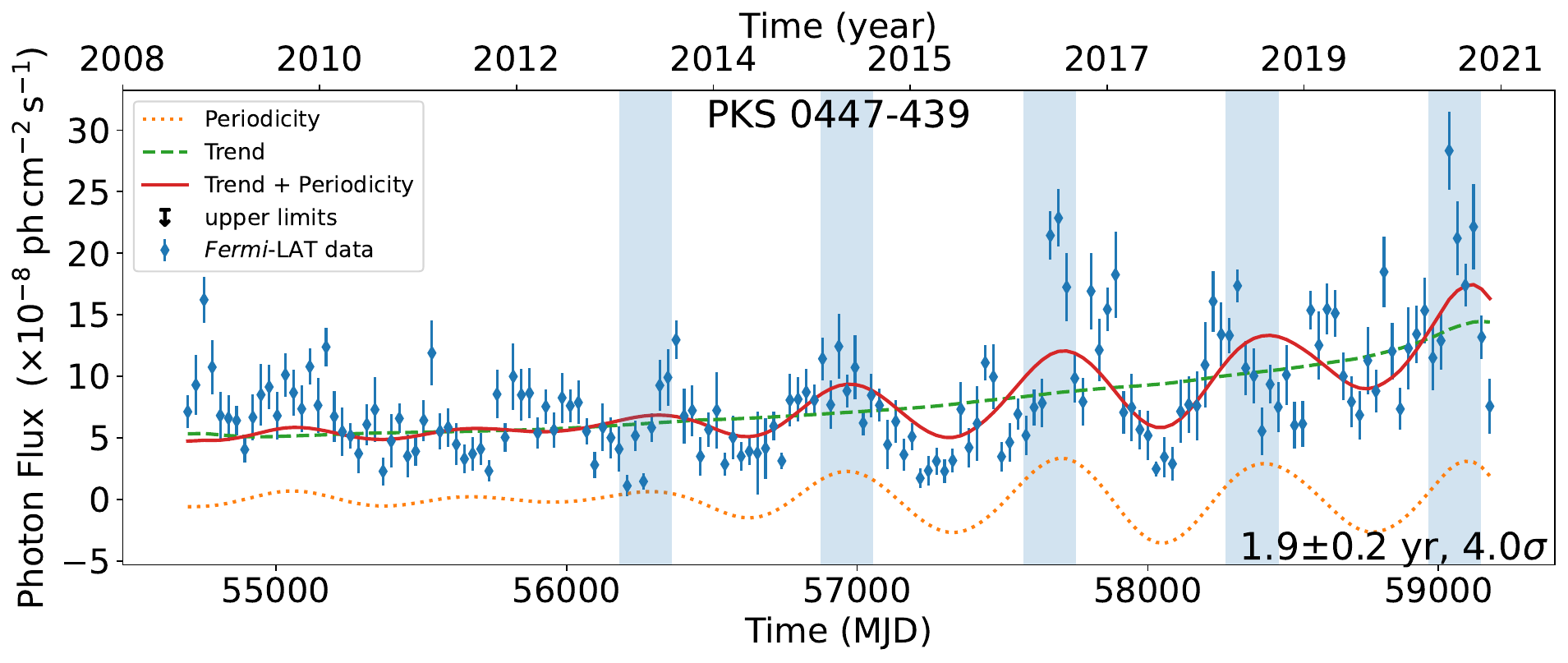}
         \includegraphics[scale=0.262]{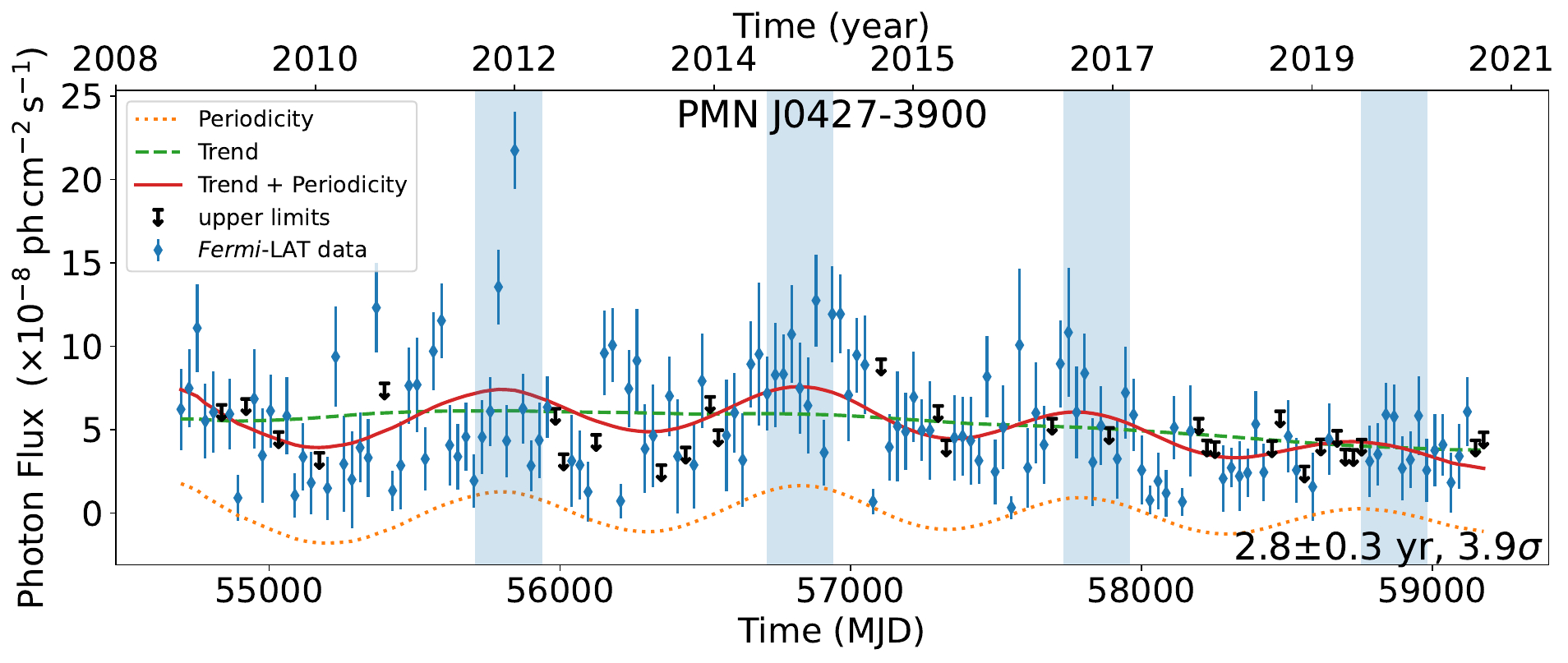}
         
         \includegraphics[scale=0.262]{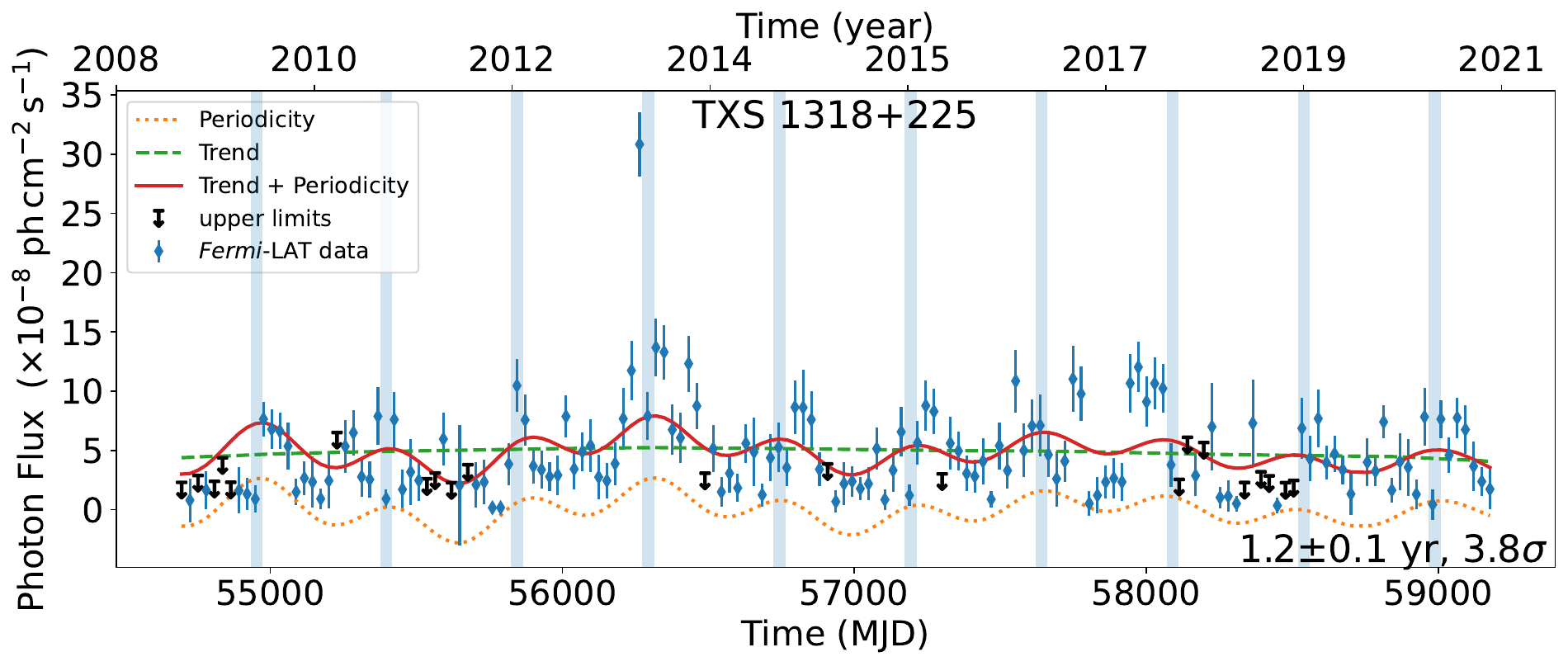} 
         \includegraphics[scale=0.262]{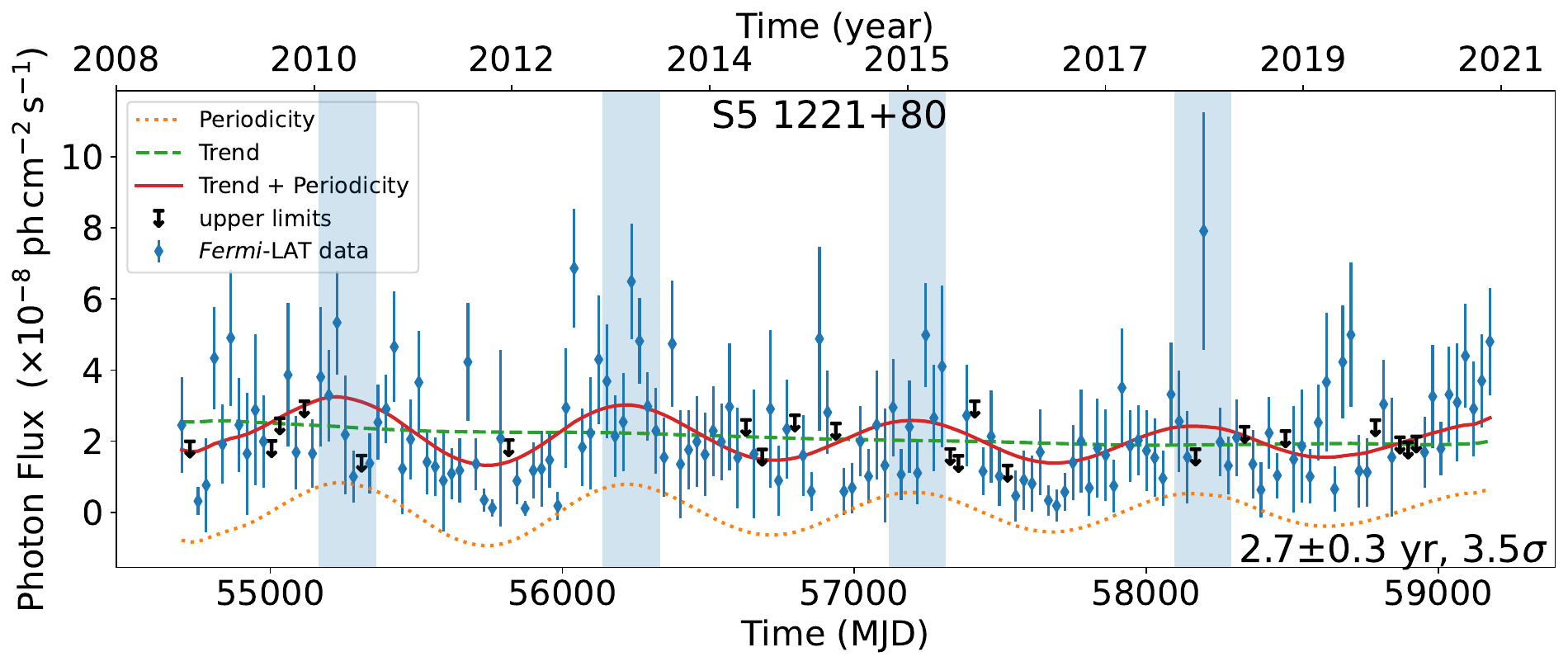}
         
         \includegraphics[scale=0.262]{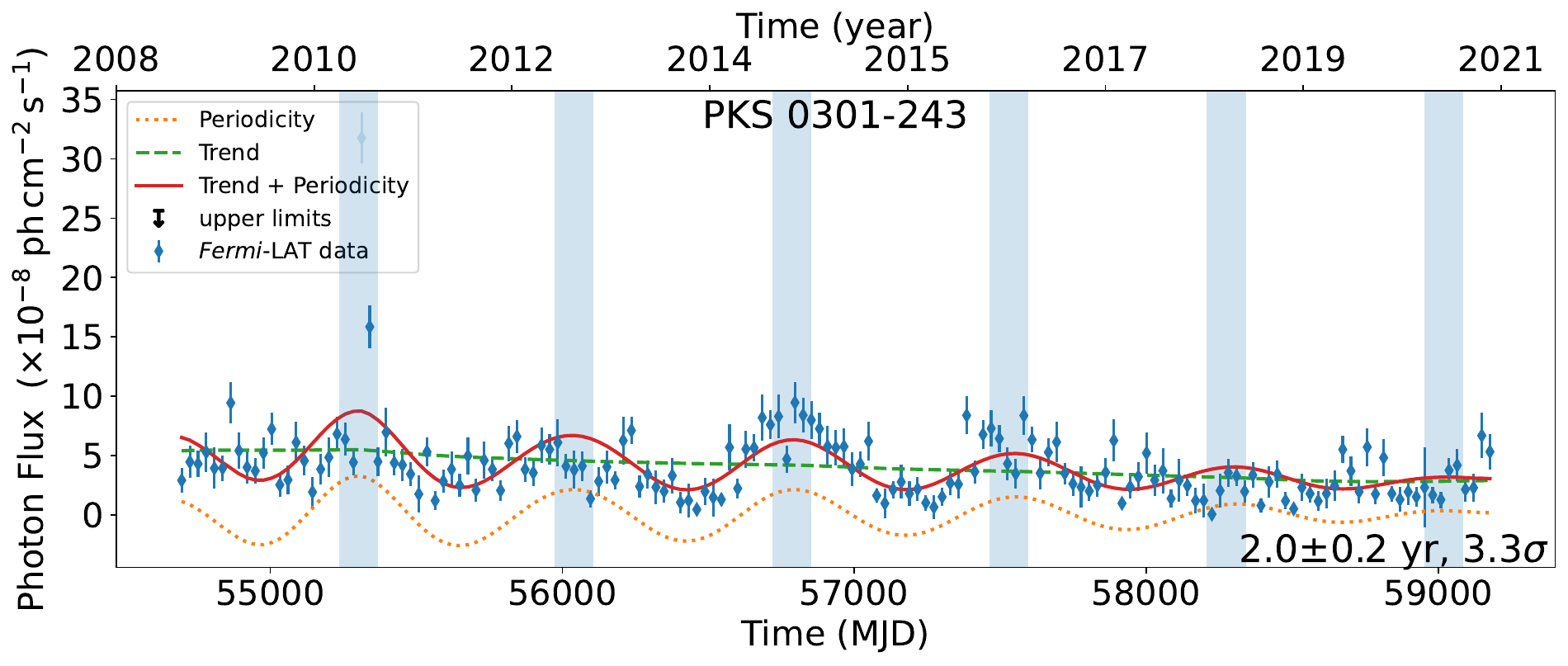} 
         \includegraphics[scale=0.262]{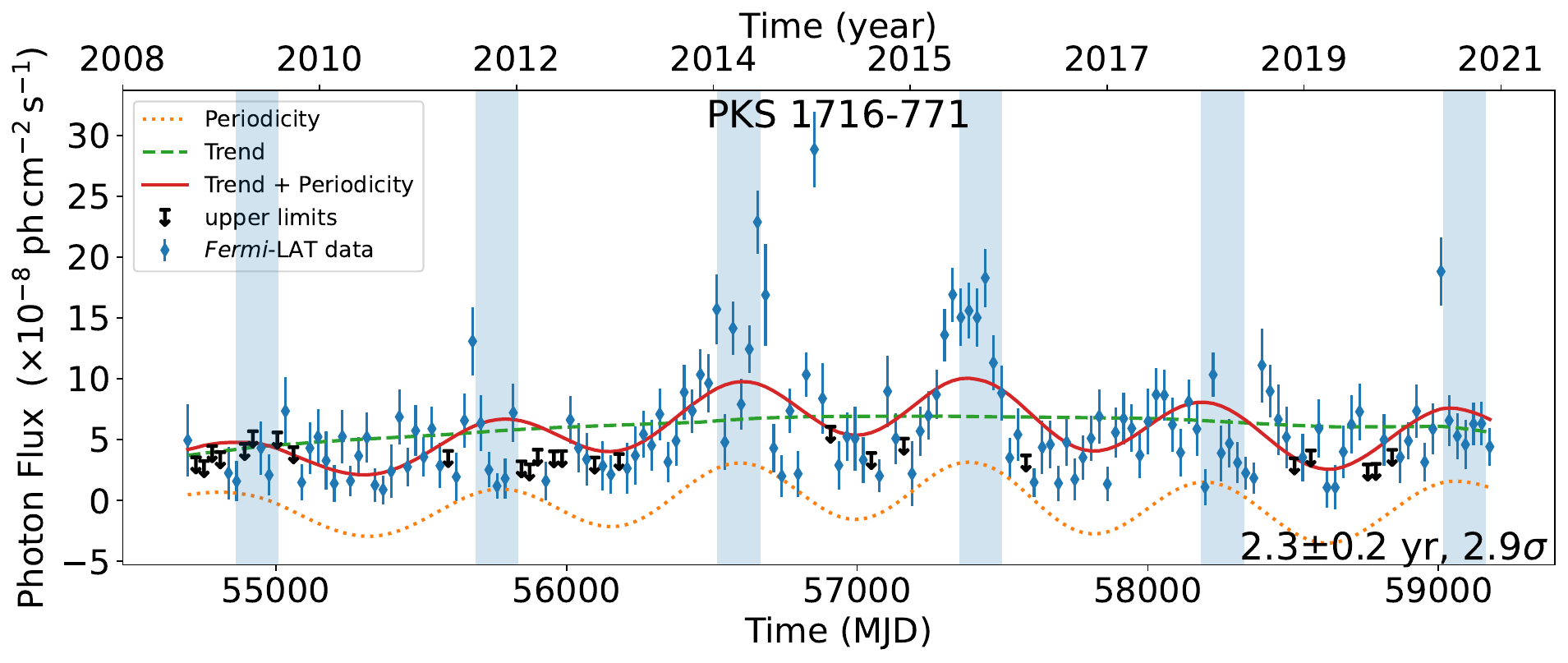}
         
         \caption{(Continued).}         
\end{figure*}

\begin{figure*}
        \centering
        \ContinuedFloat
         \includegraphics[scale=0.262]{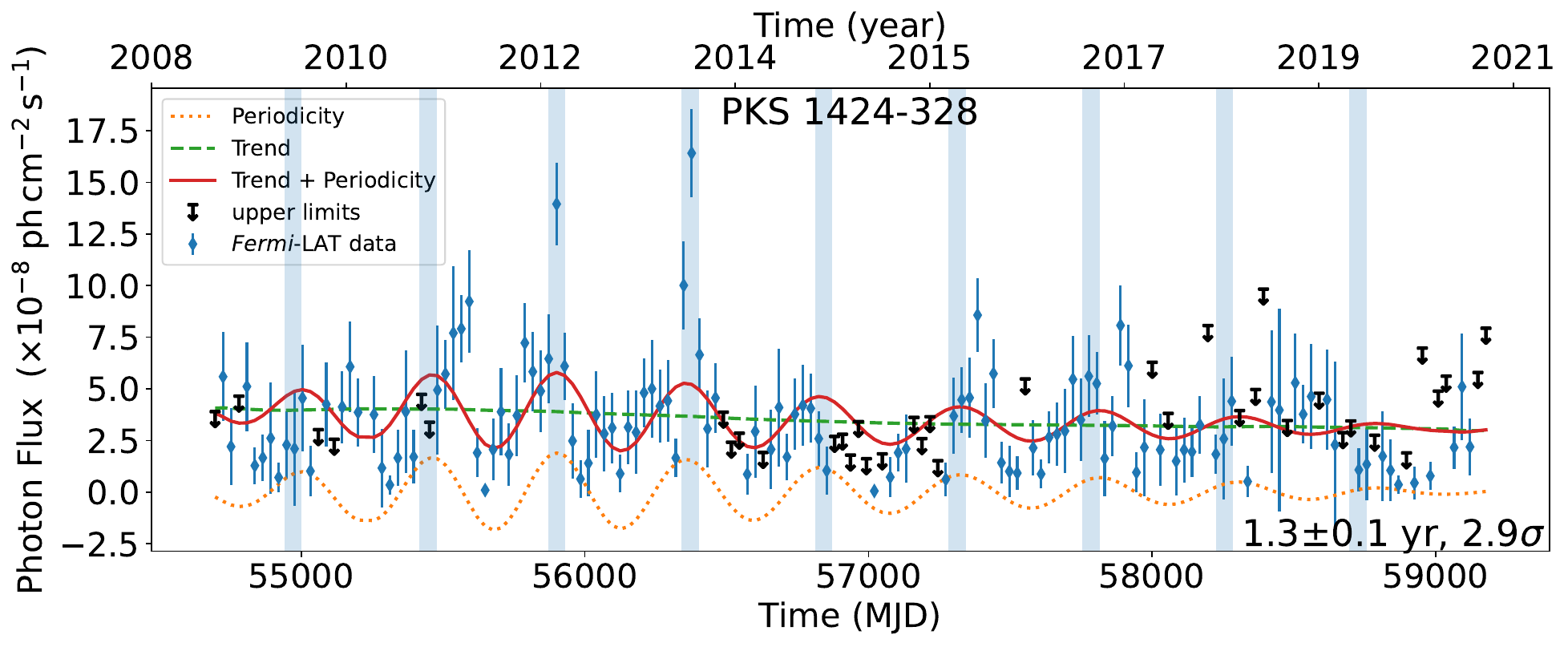}
         \includegraphics[scale=0.262]{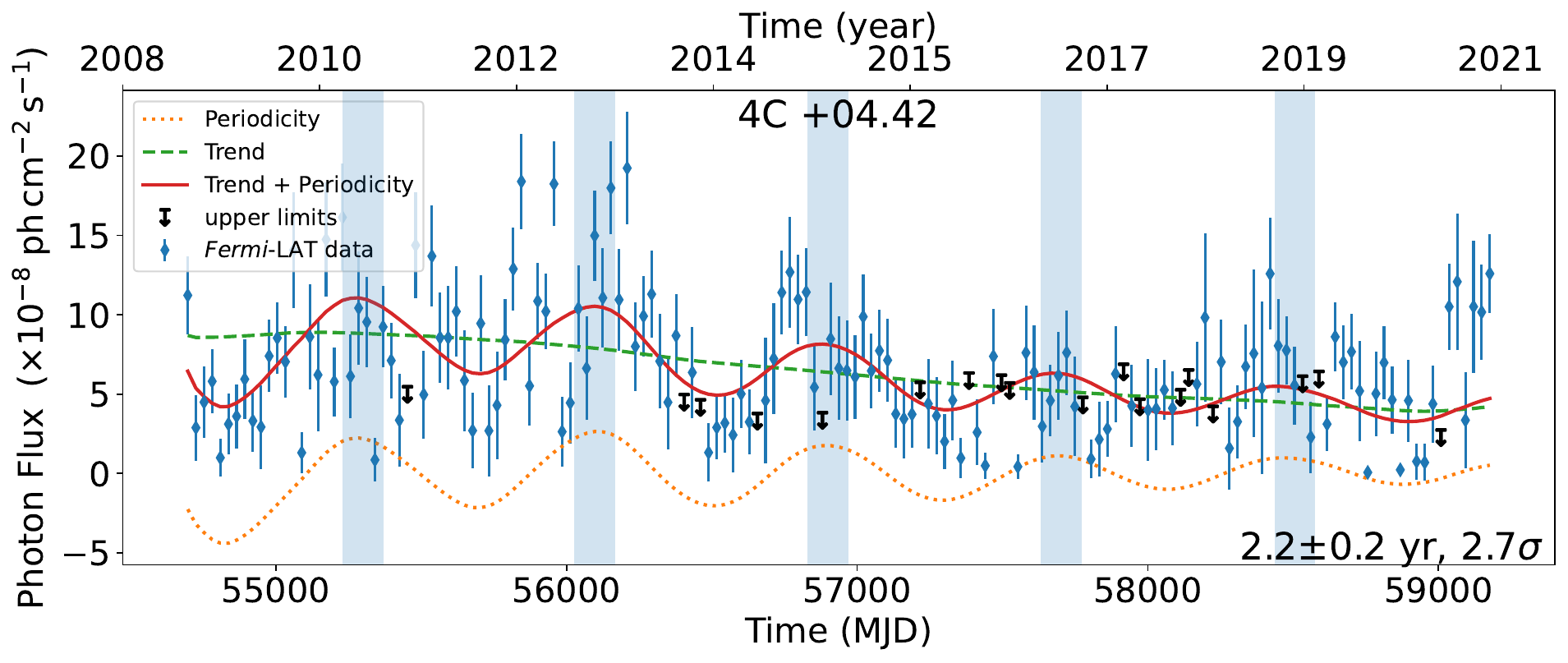}
         
         \includegraphics[scale=0.262]{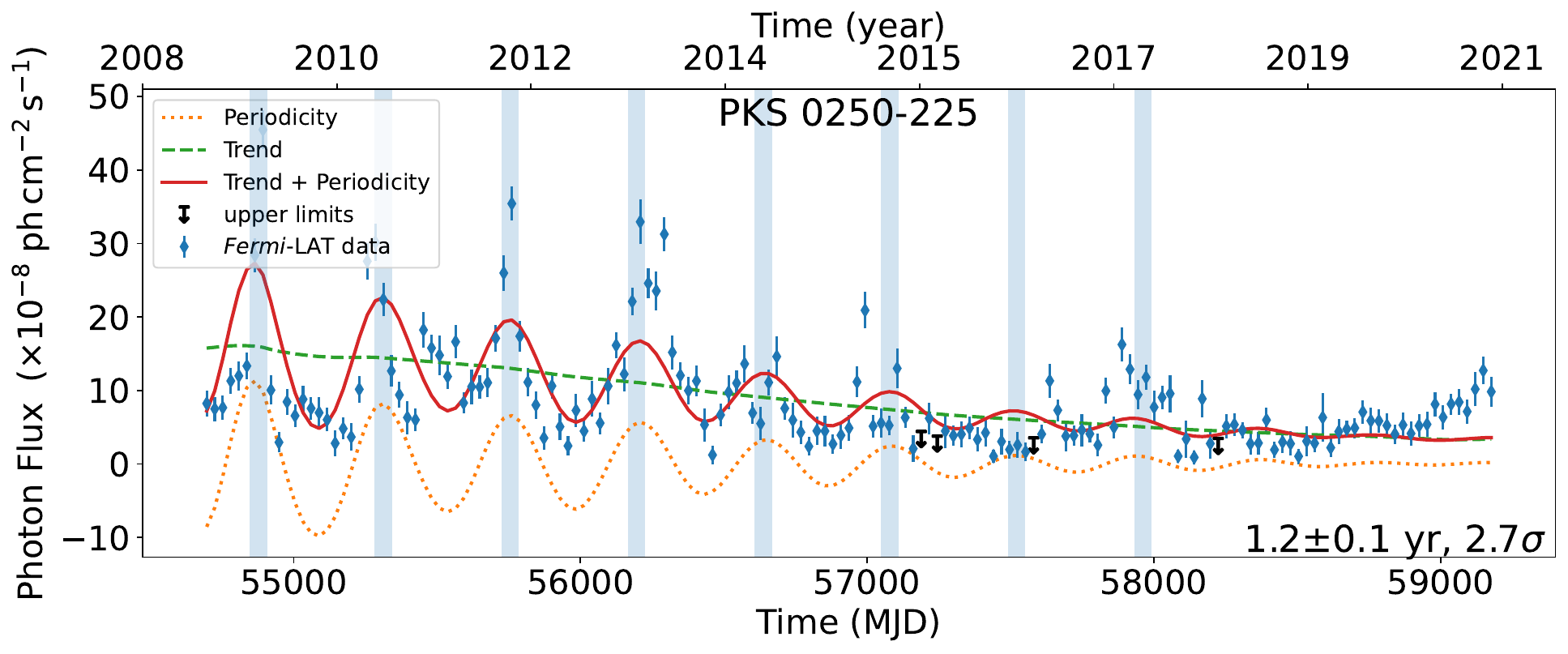}
         \includegraphics[scale=0.262]{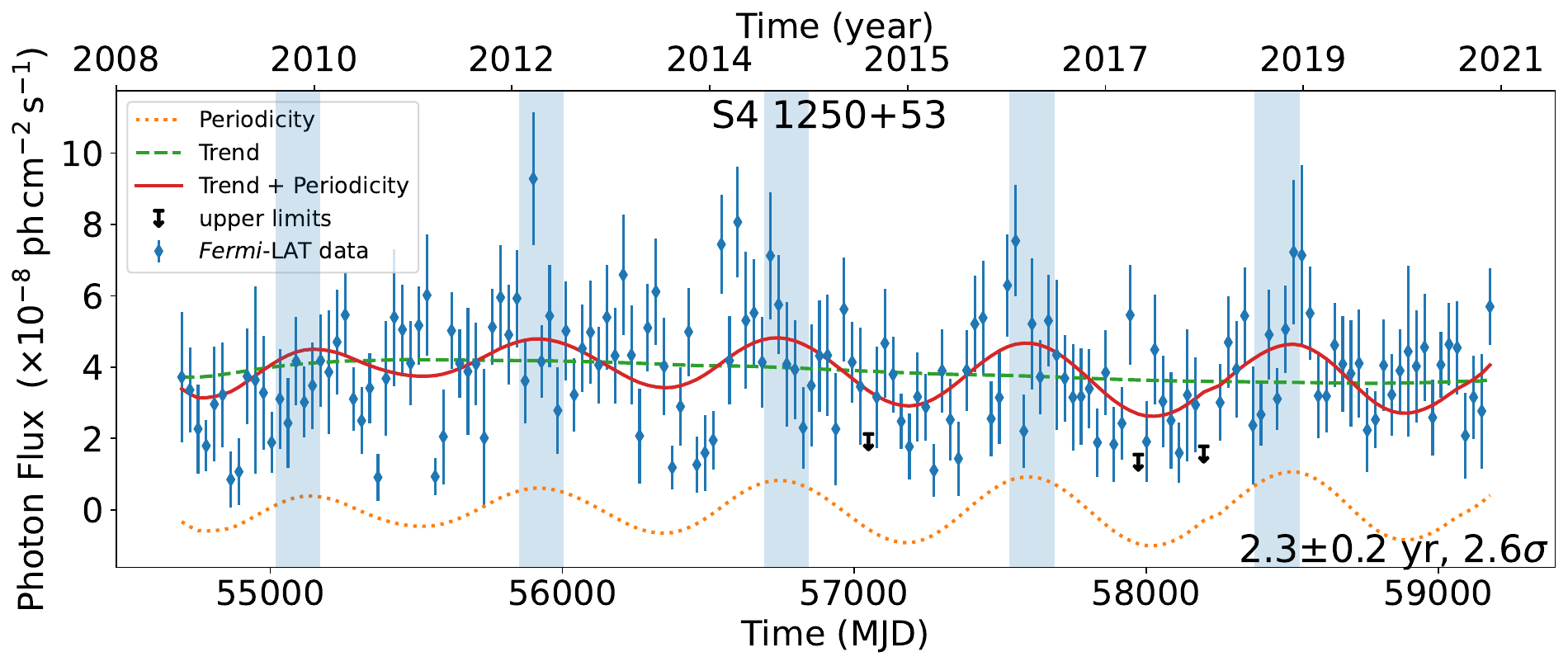}
         
         \includegraphics[scale=0.262]{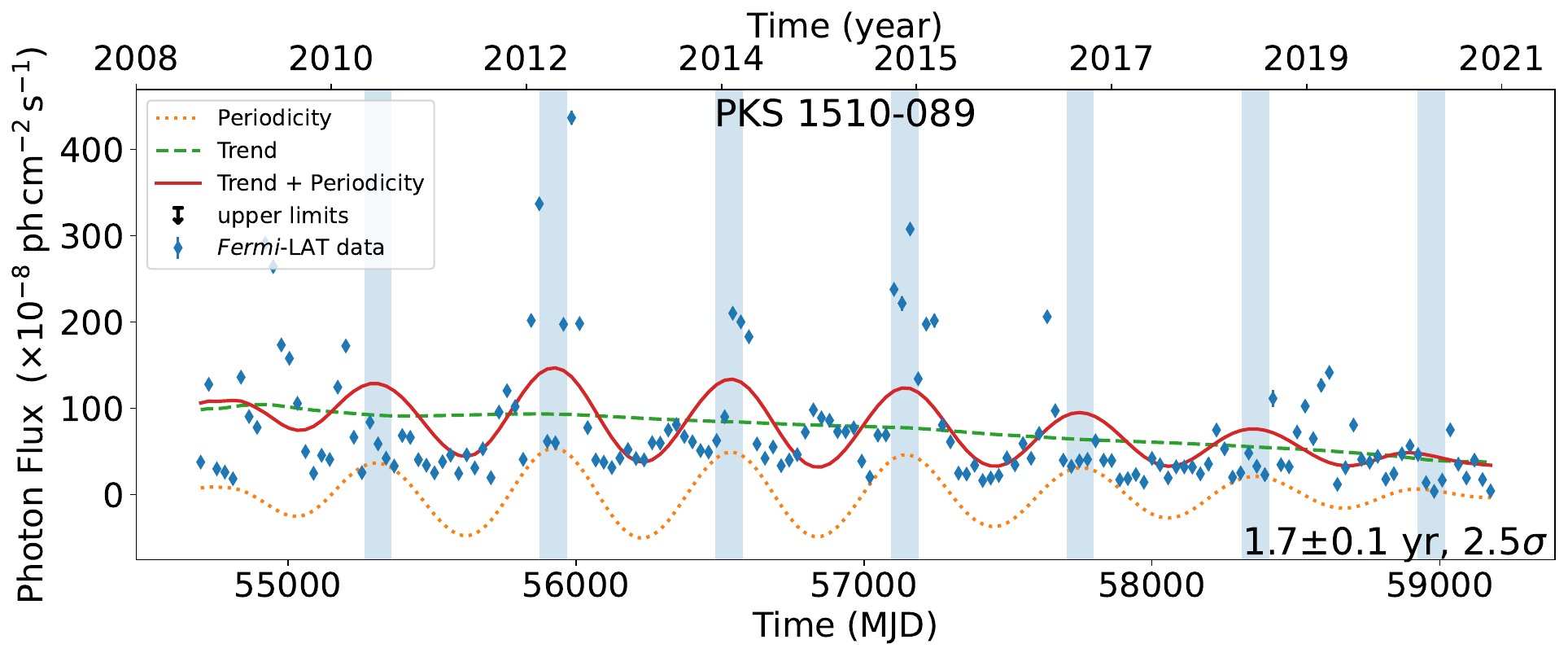}
         \includegraphics[scale=0.262]{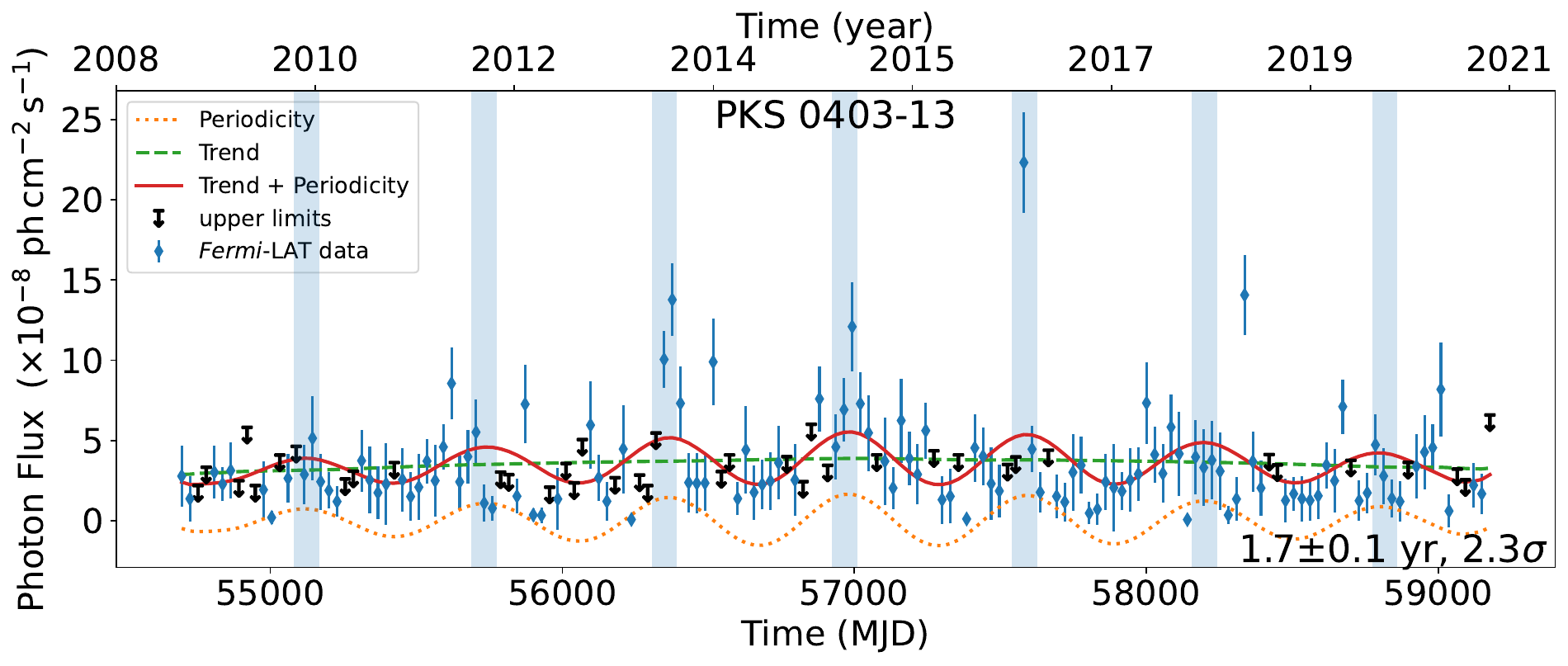}

         \includegraphics[scale=0.262]{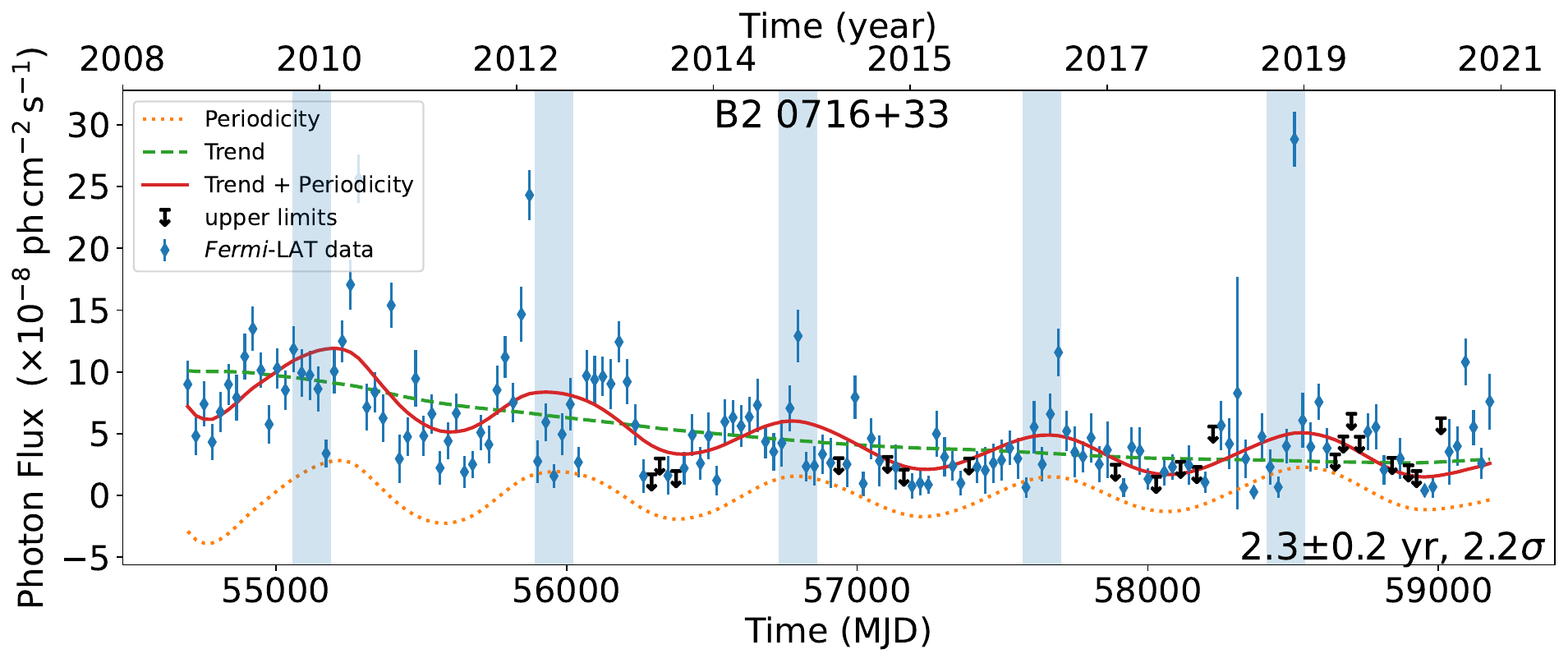}
         \includegraphics[scale=0.262]{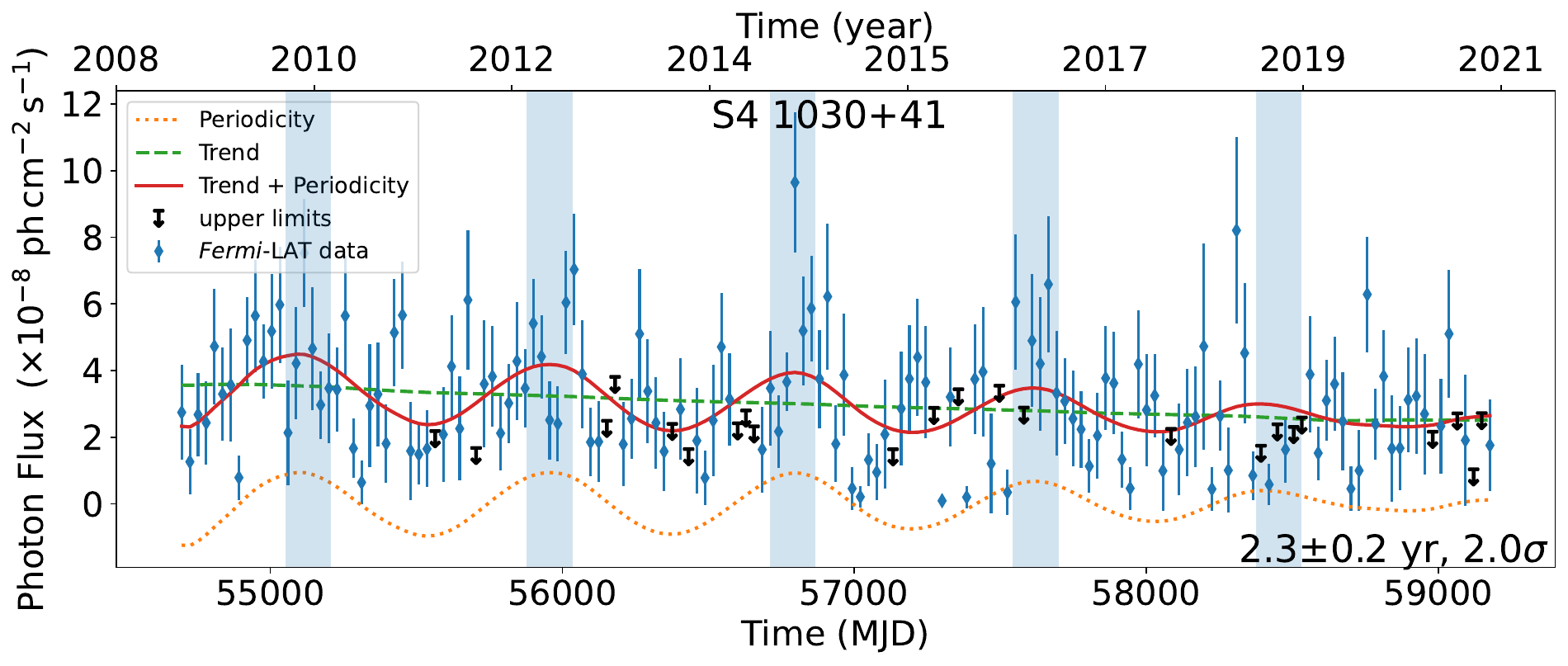}

         \caption{(Continued).}         
\end{figure*}

\newpage
\onecolumn
\section{Forecasting plots}\label{plots_forecast}
This section presents the forecasting plots for our periodicity candidates. The blue dots represent the $\gamma$-ray data from \textit{Fermi}-LAT. Black arrows indicate the upper limits. The orange and green solid lines represent the periodicity + trend and its forecast, respectively. The gray area shows the 95$\%$ confidence interval of the forecast.

\begin{figure*}[h]
	\centering
         \includegraphics[scale=0.262]{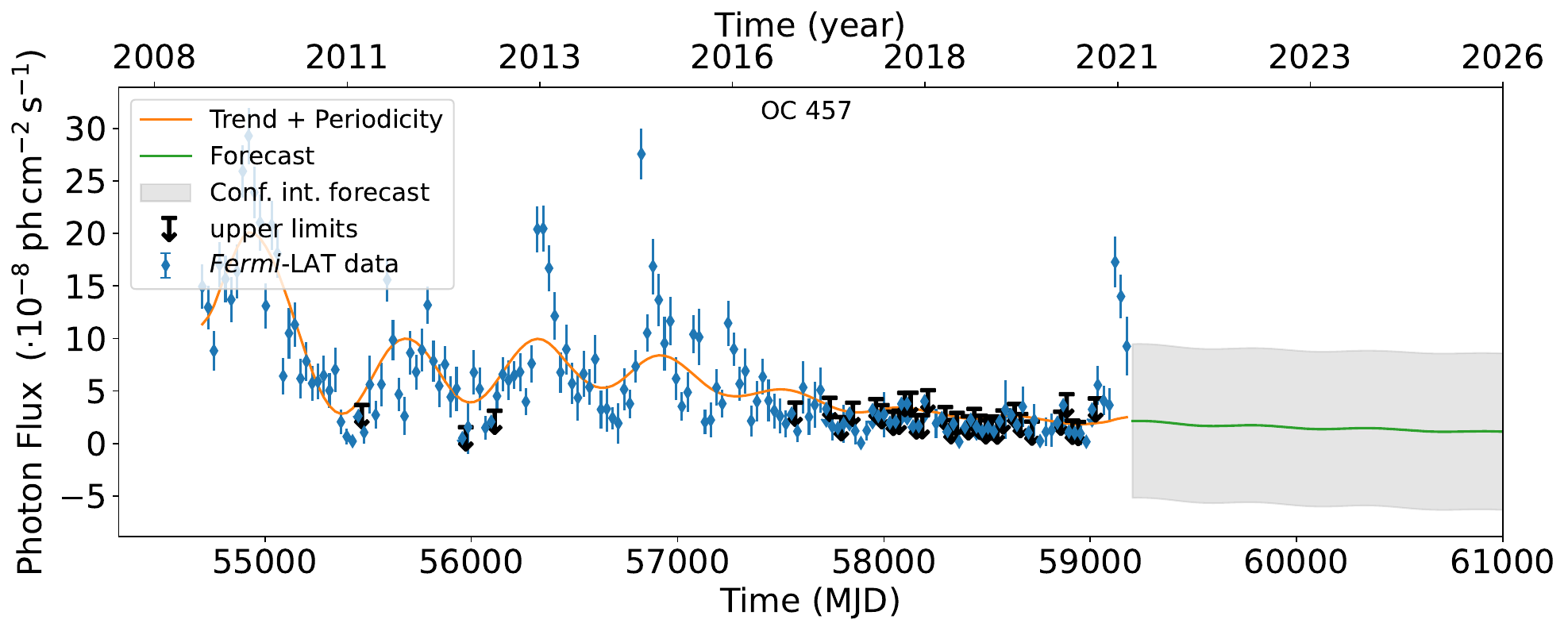}
         \includegraphics[scale=0.262]{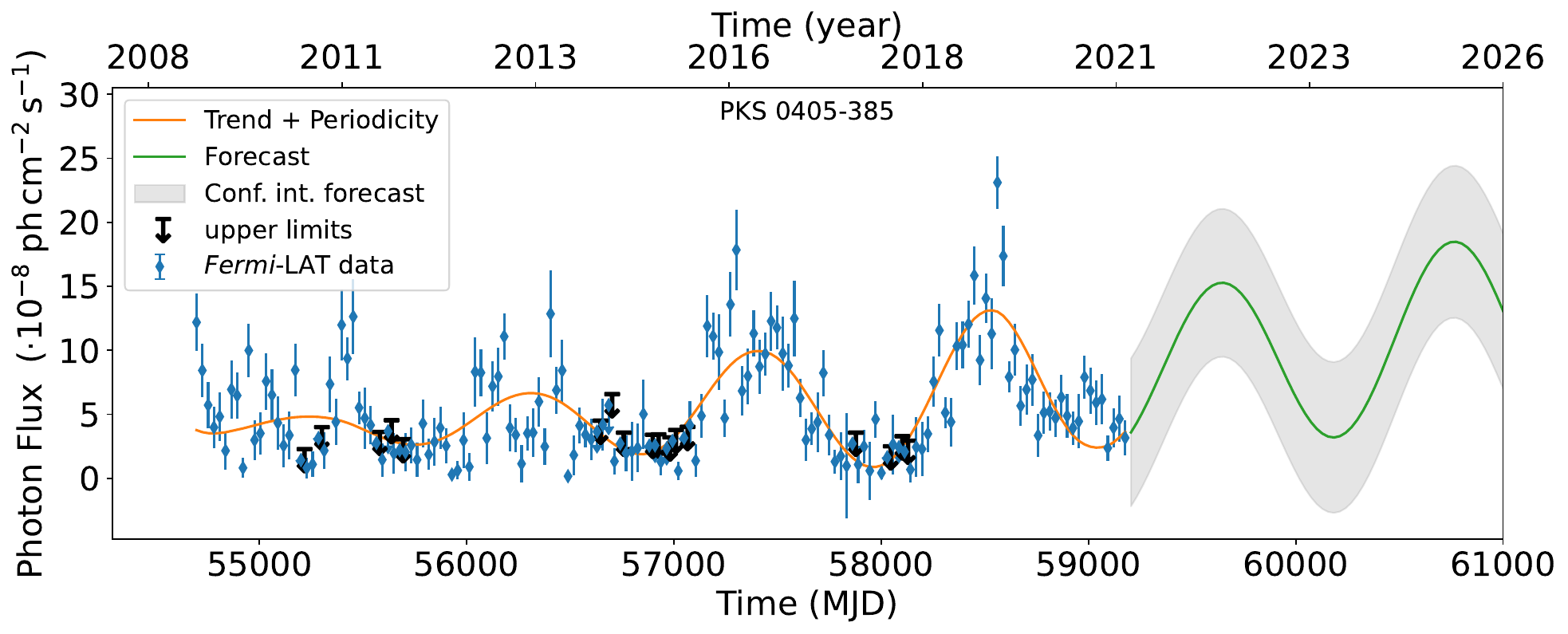}
         
         \includegraphics[scale=0.262]{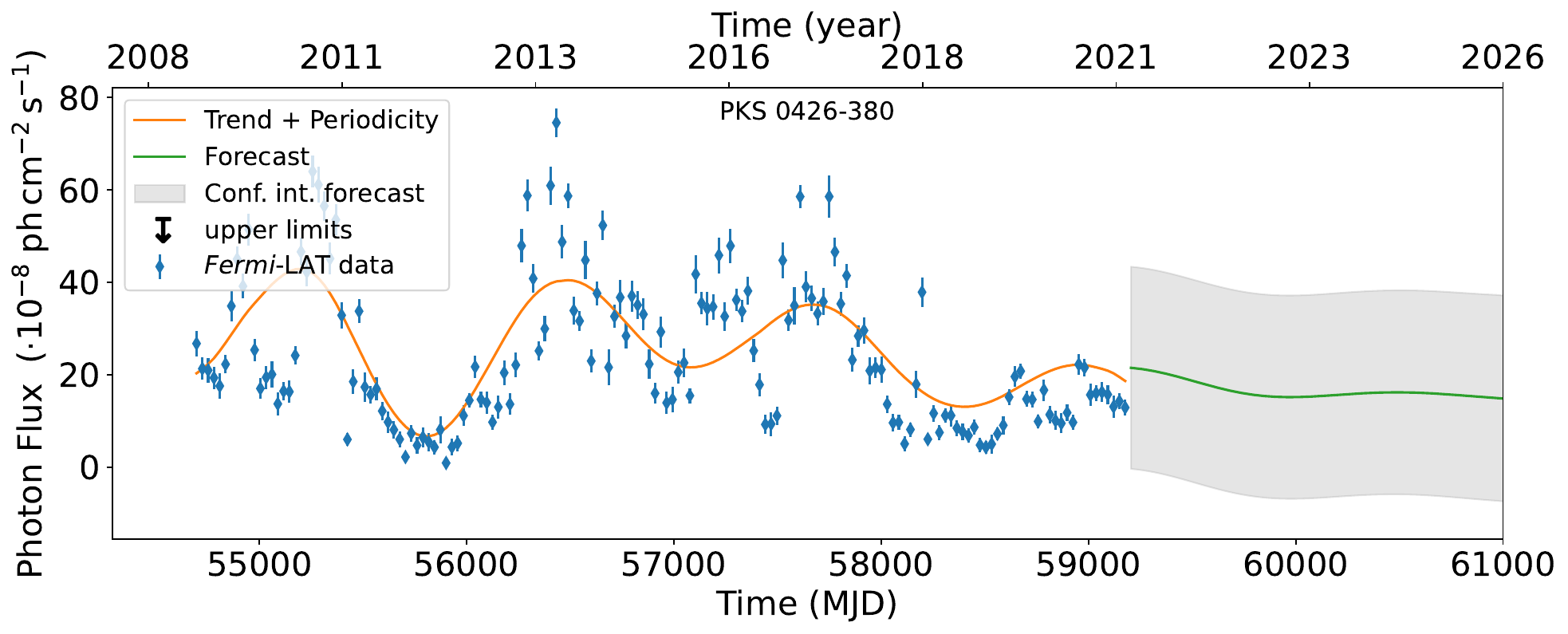}
         \includegraphics[scale=0.262]{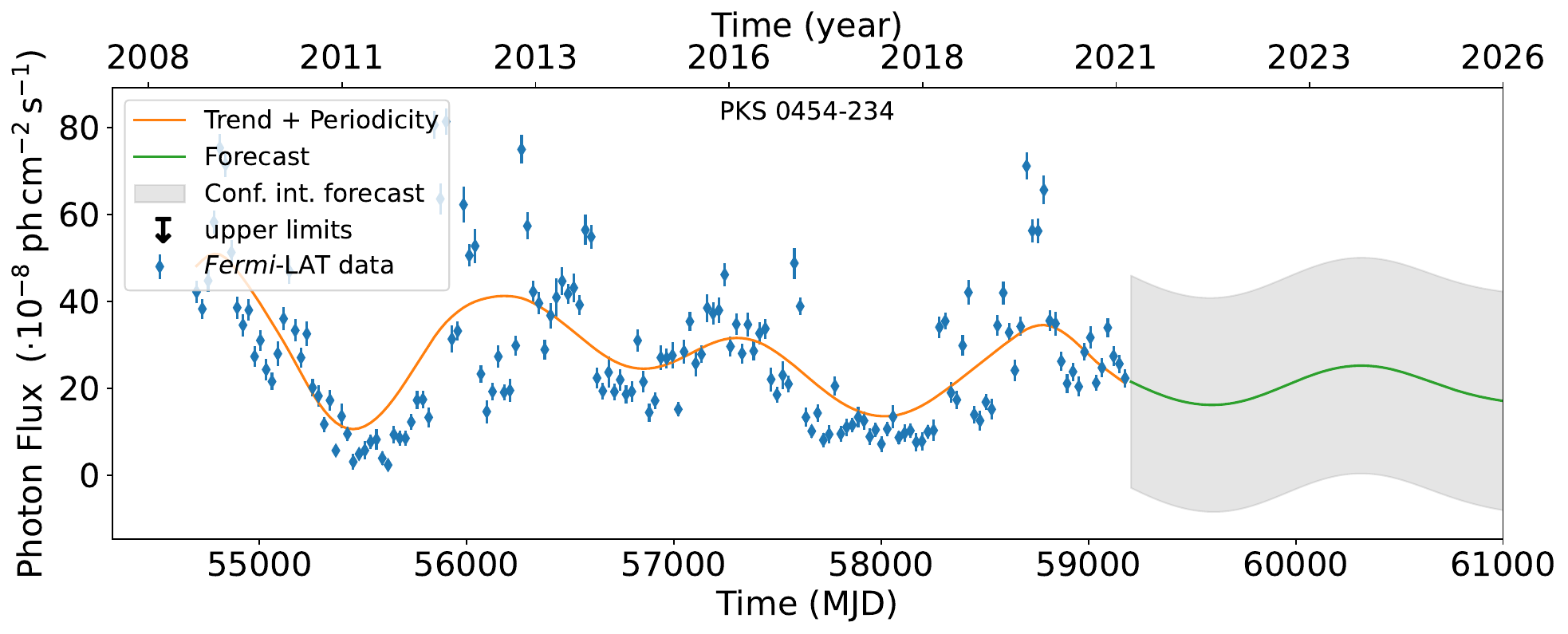}

         \includegraphics[scale=0.26]{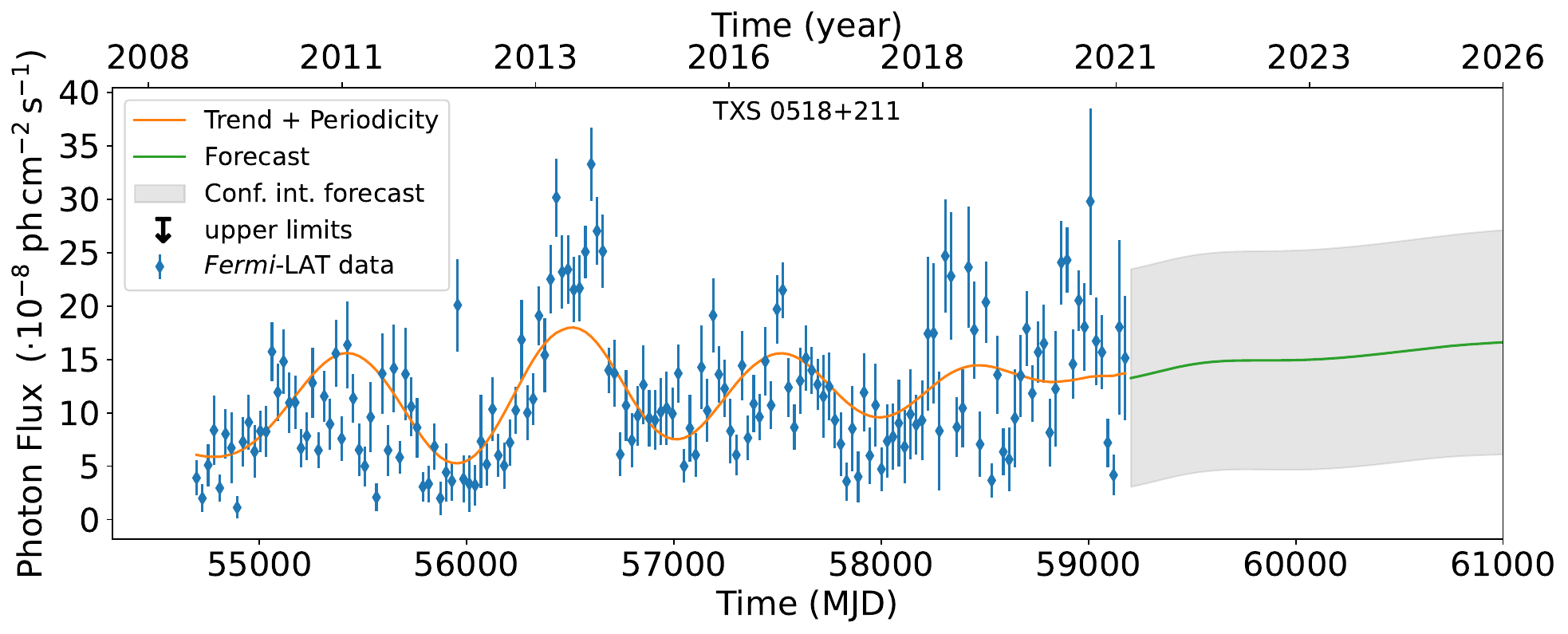}
         \includegraphics[scale=0.262]{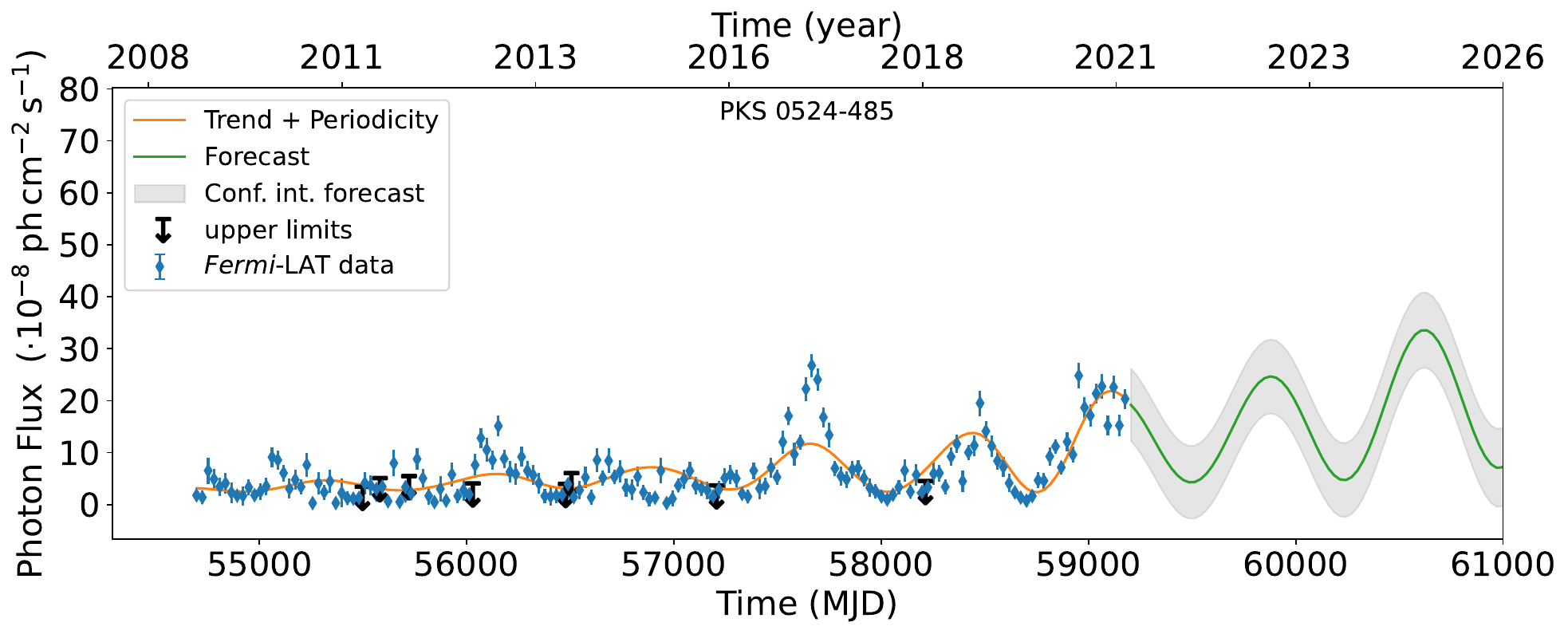}
         
         \includegraphics[scale=0.262]{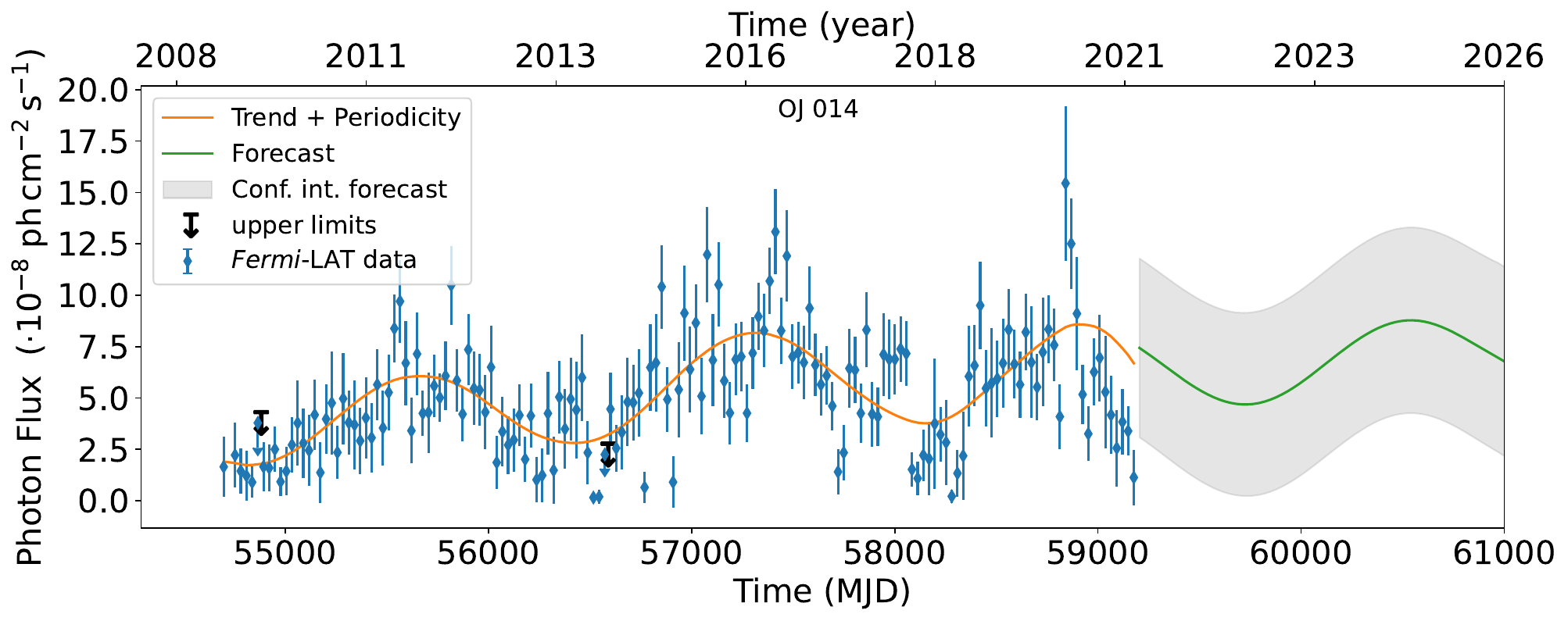}
         \includegraphics[scale=0.262]{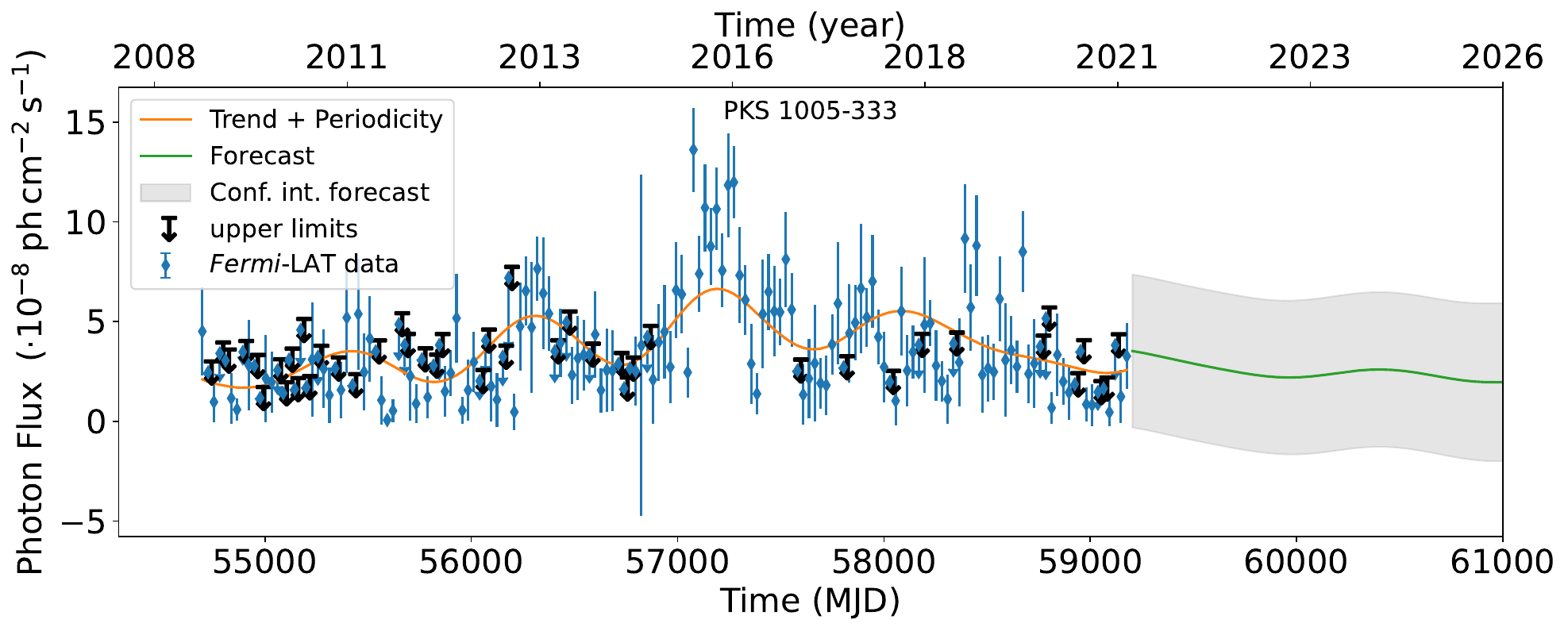}

	\caption{Forecasting LCs of quasi-periodic candidates. Predicted cycles are reported in Table \ref{table4}. }
	\label{fig: plots}
\end{figure*}

\onecolumn
\begin{figure*}
        \centering
        \ContinuedFloat

        \includegraphics[scale=0.262]{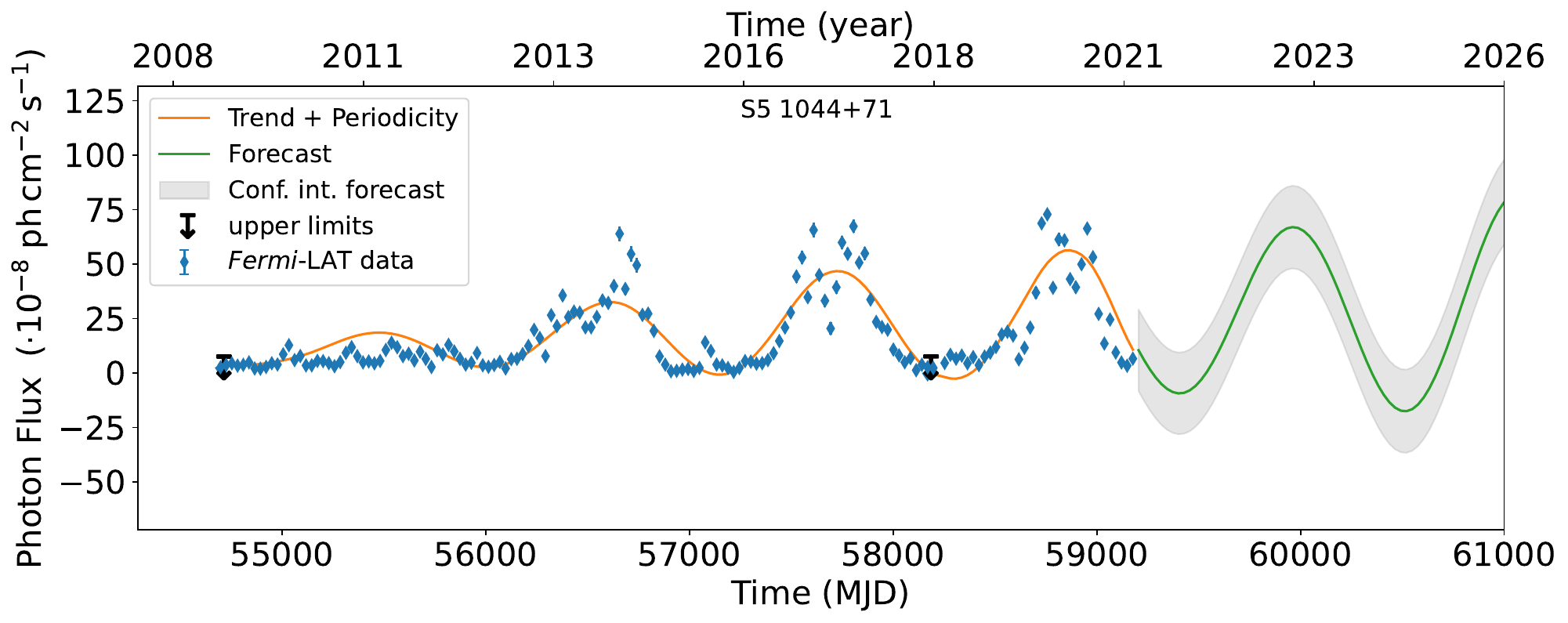}
        \includegraphics[scale=0.262]{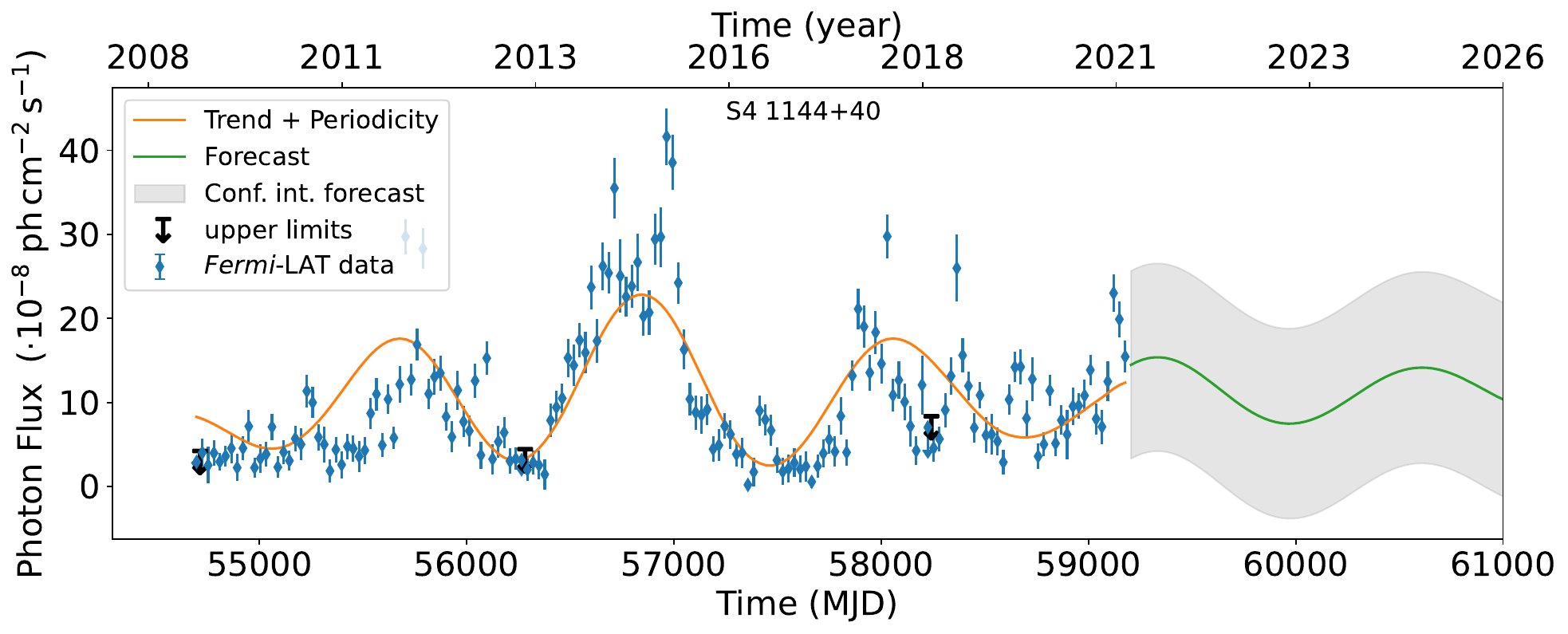}
        
        \includegraphics[scale=0.262]{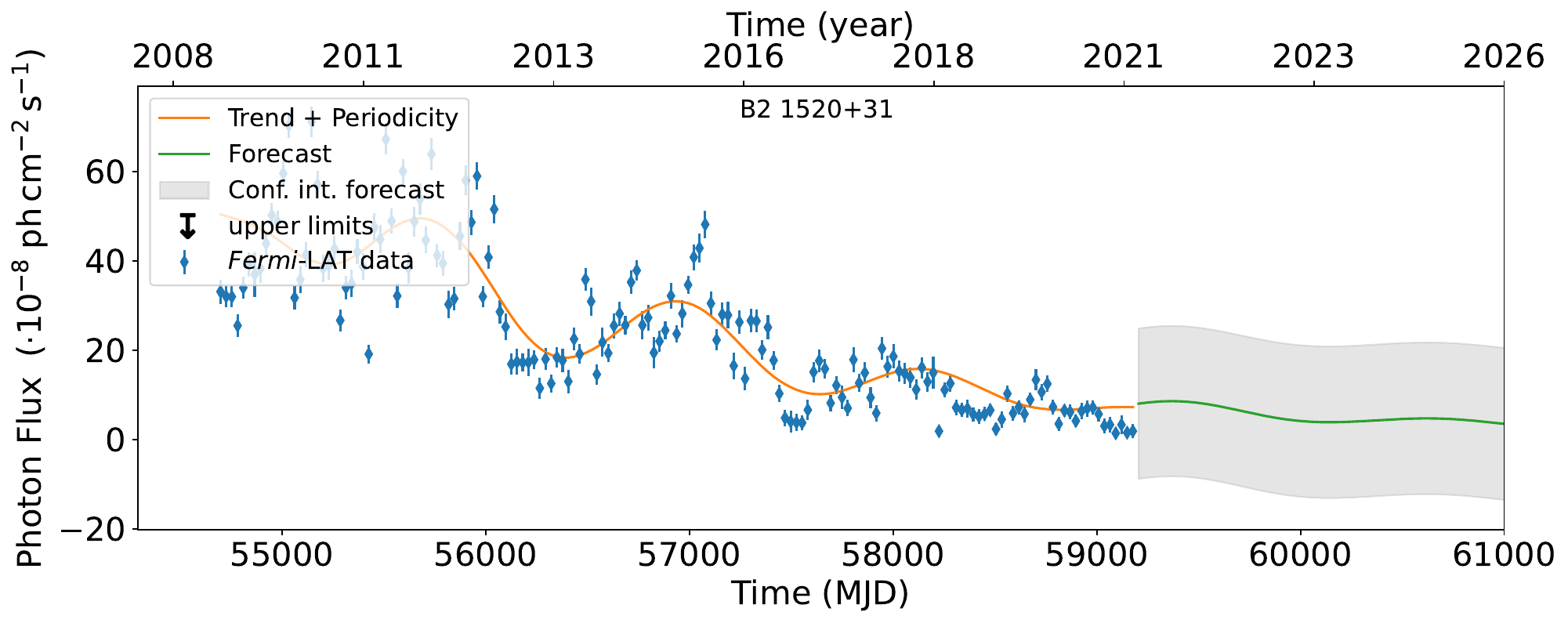}
        \includegraphics[scale=0.262]{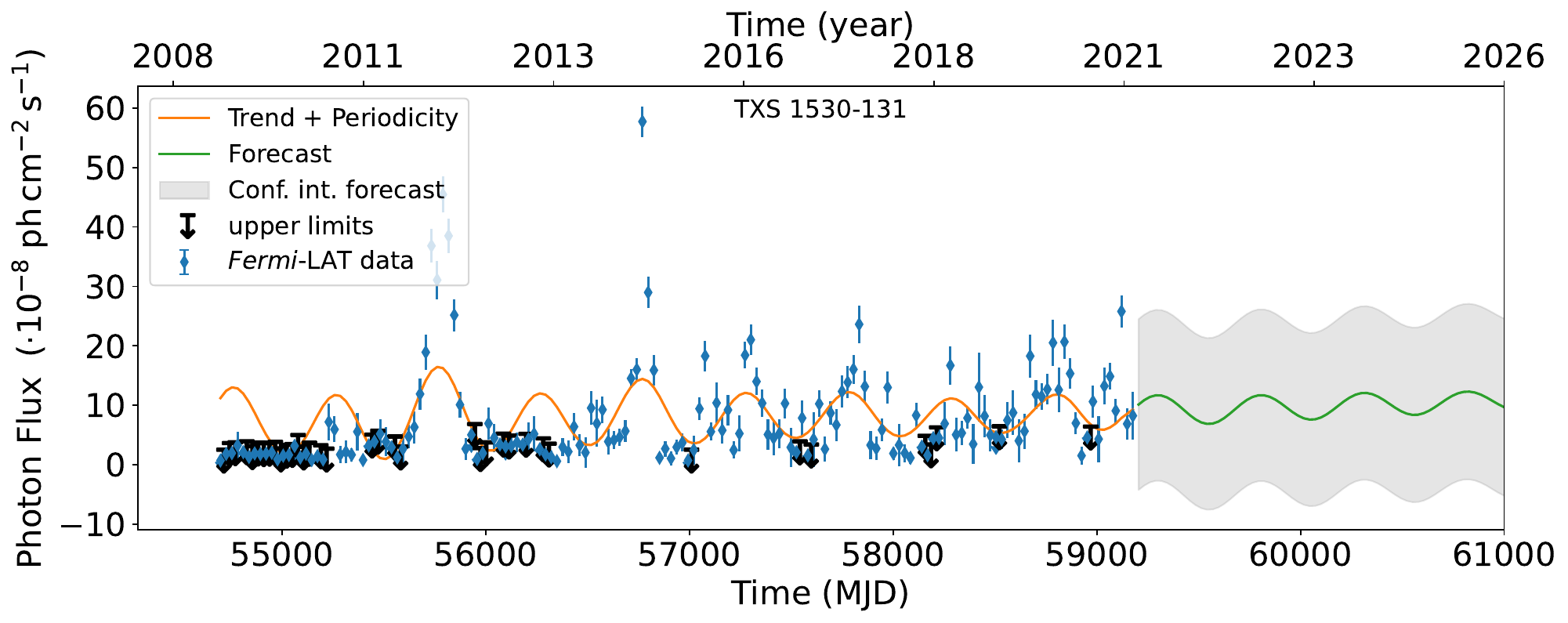}
        
        \includegraphics[scale=0.262]{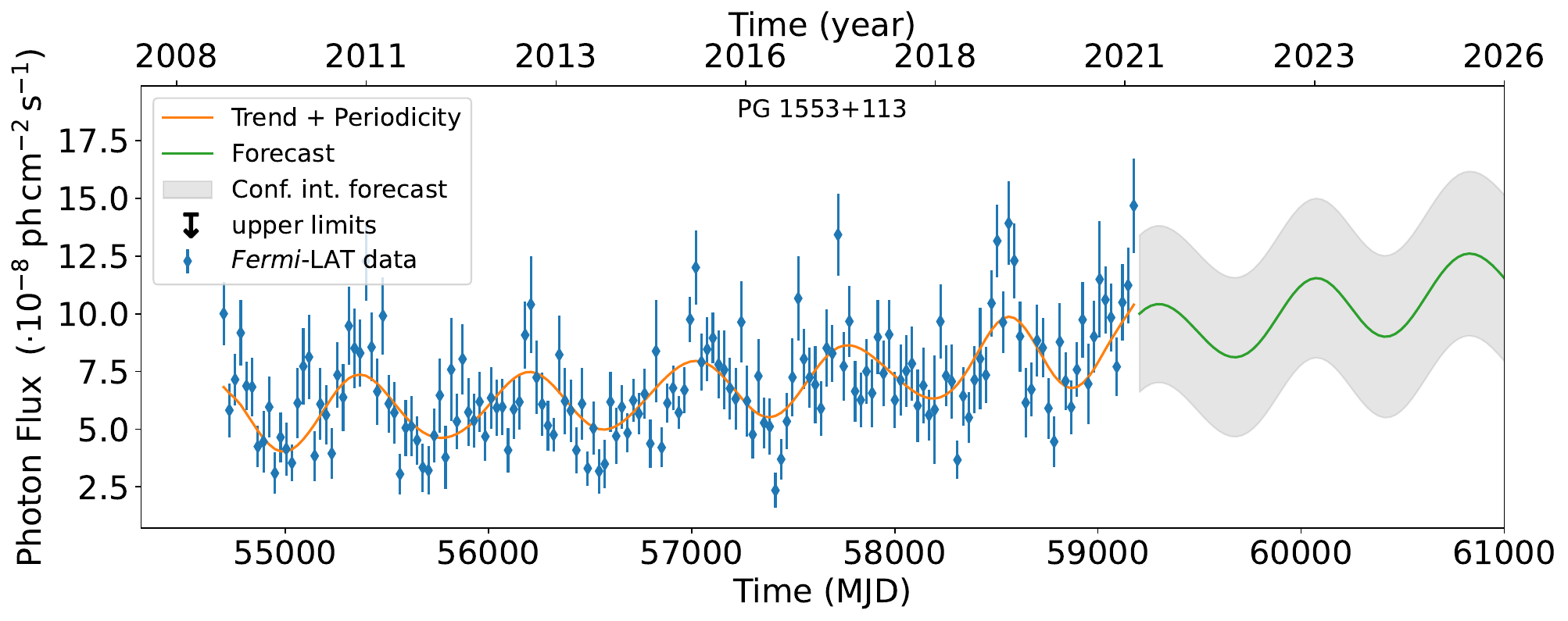}
        \includegraphics[scale=0.262]{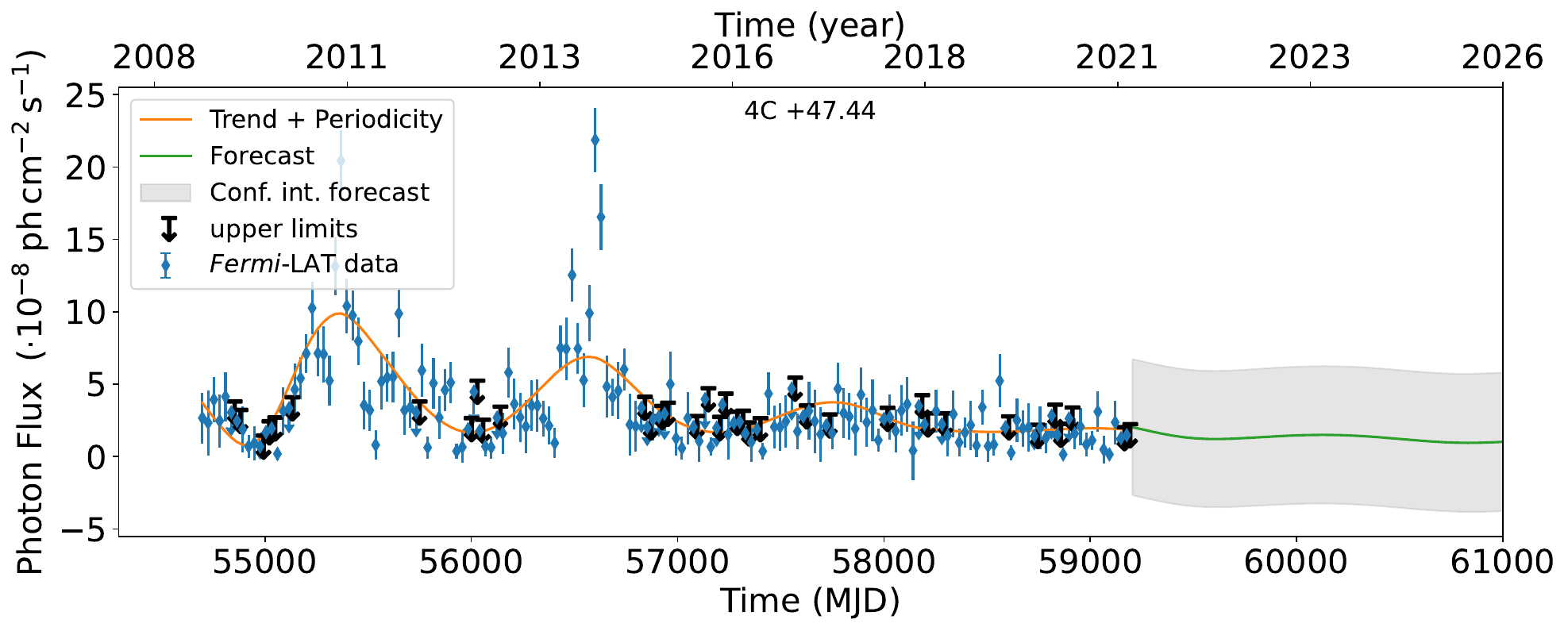}
        
        \includegraphics[scale=0.262]{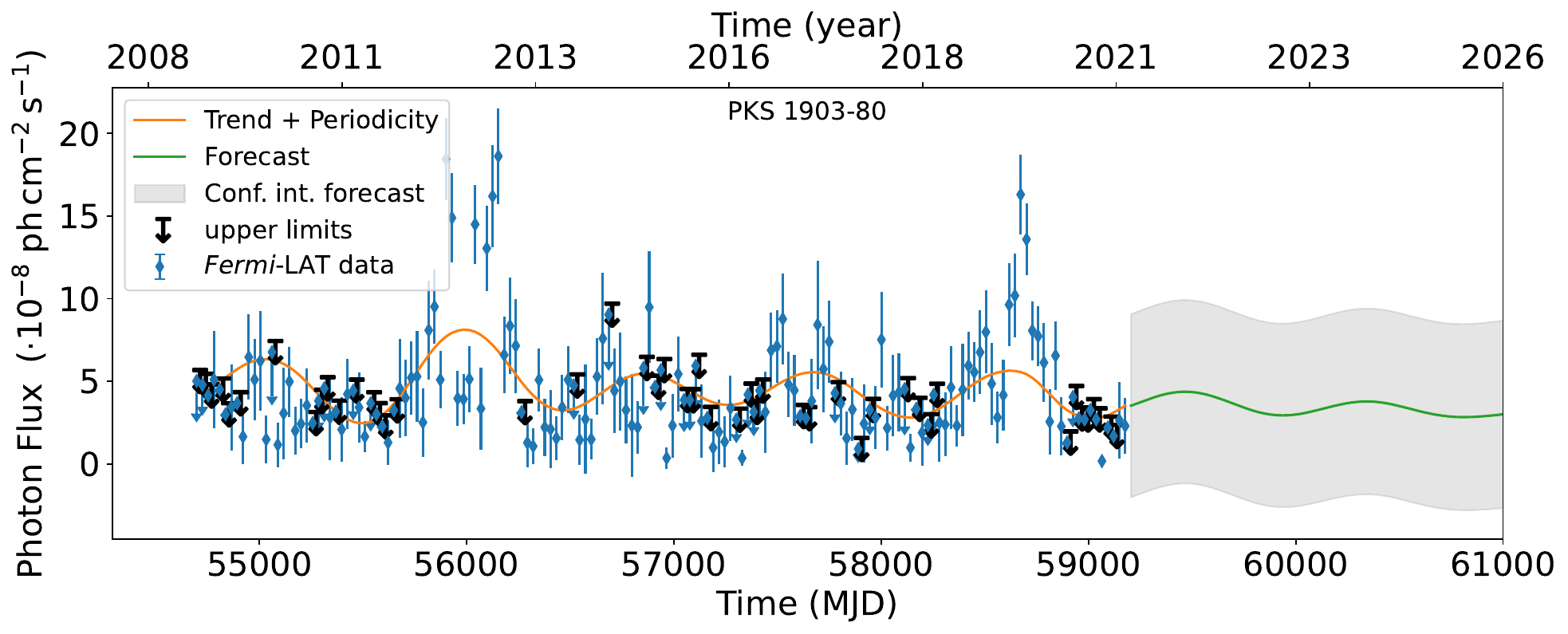}
        \includegraphics[scale=0.262]{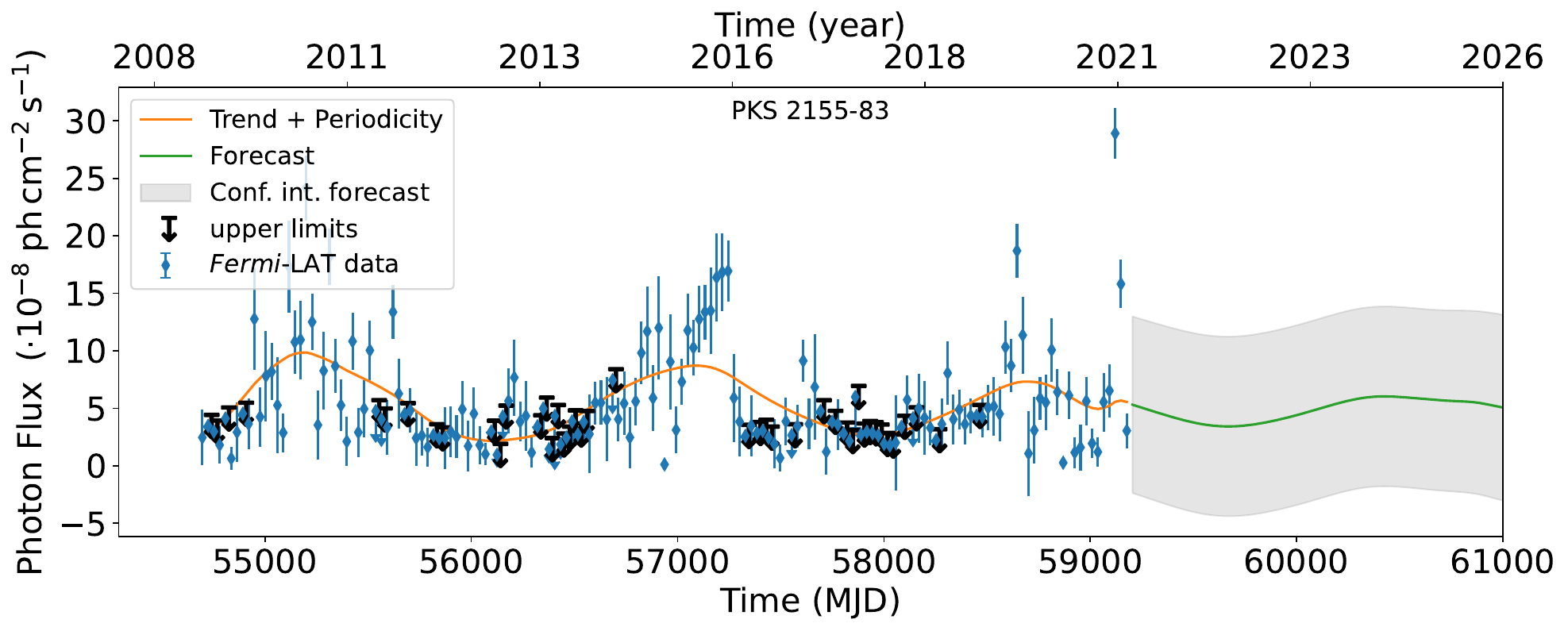}
        
        \includegraphics[scale=0.262]{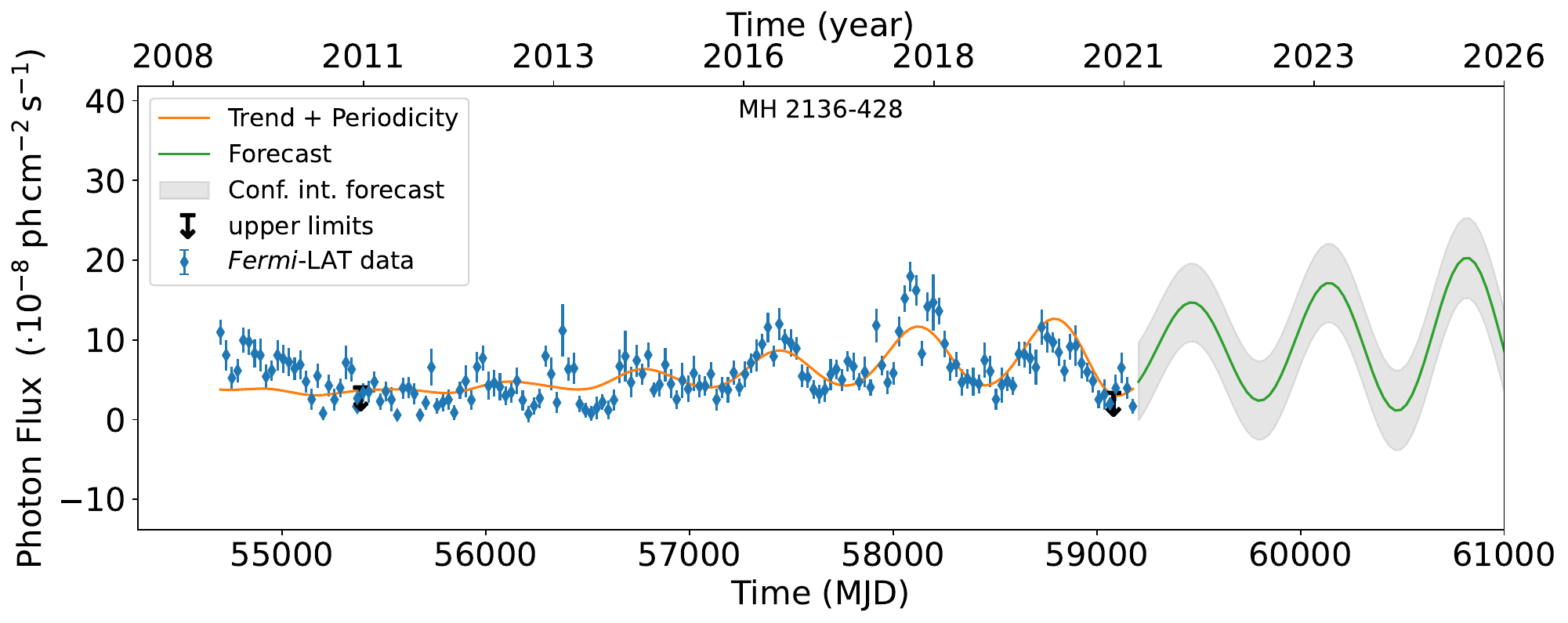}
         \includegraphics[scale=0.262]{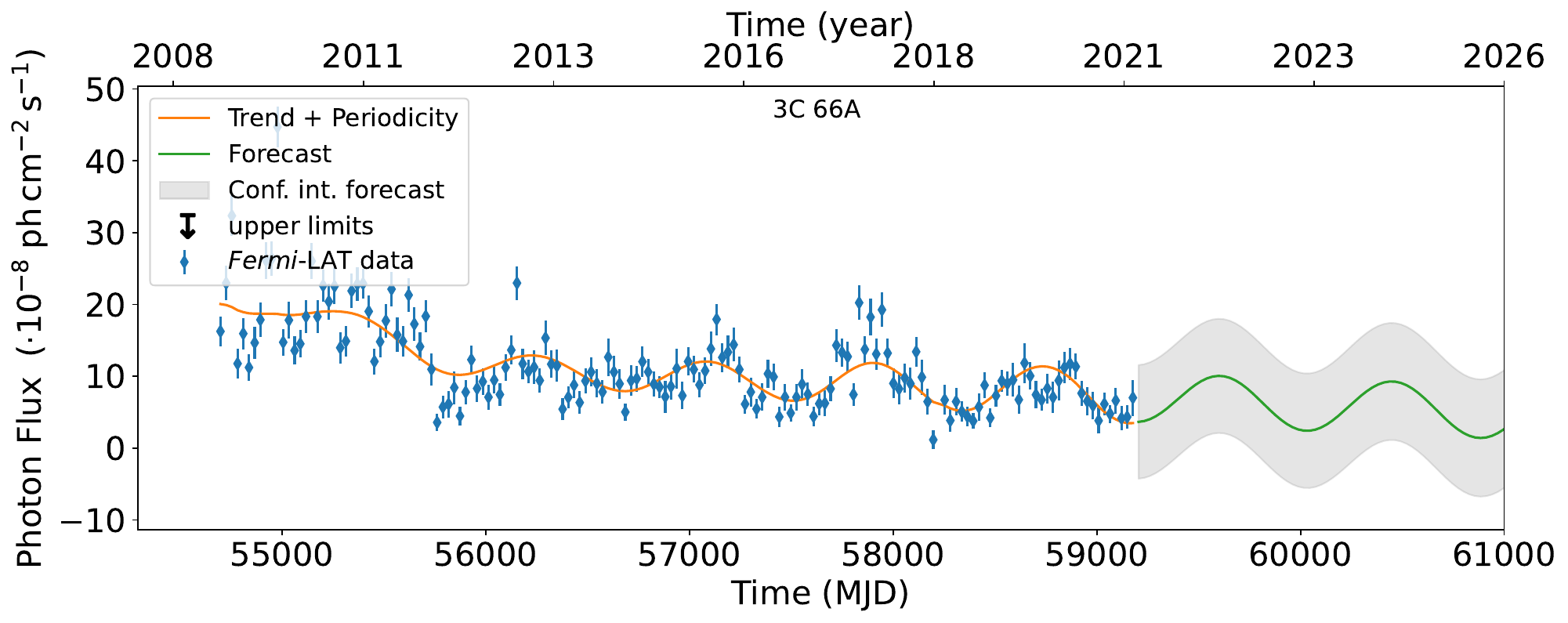}

         \caption{(Continued).}
    
\end{figure*}

\begin{figure*}
        \centering
        \ContinuedFloat

         \includegraphics[scale=0.262]{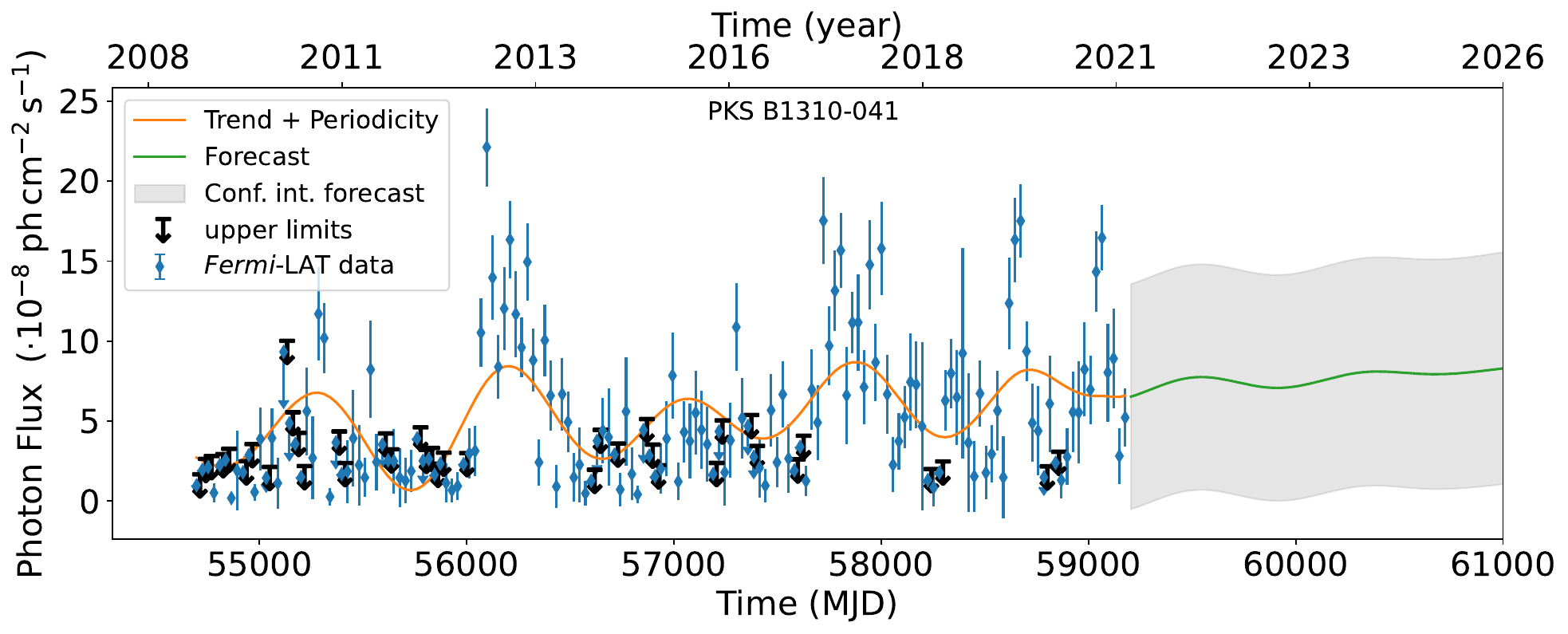}
         \includegraphics[scale=0.262]{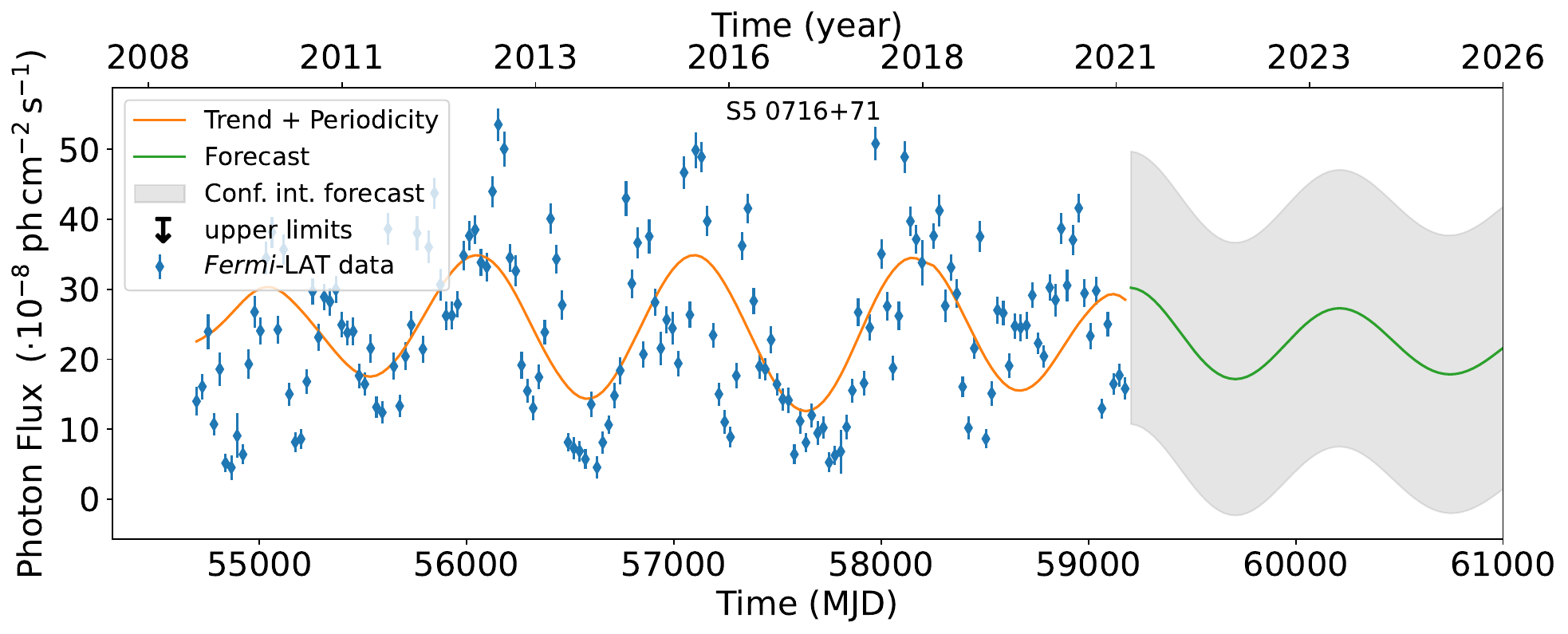}
         
         \includegraphics[scale=0.262]{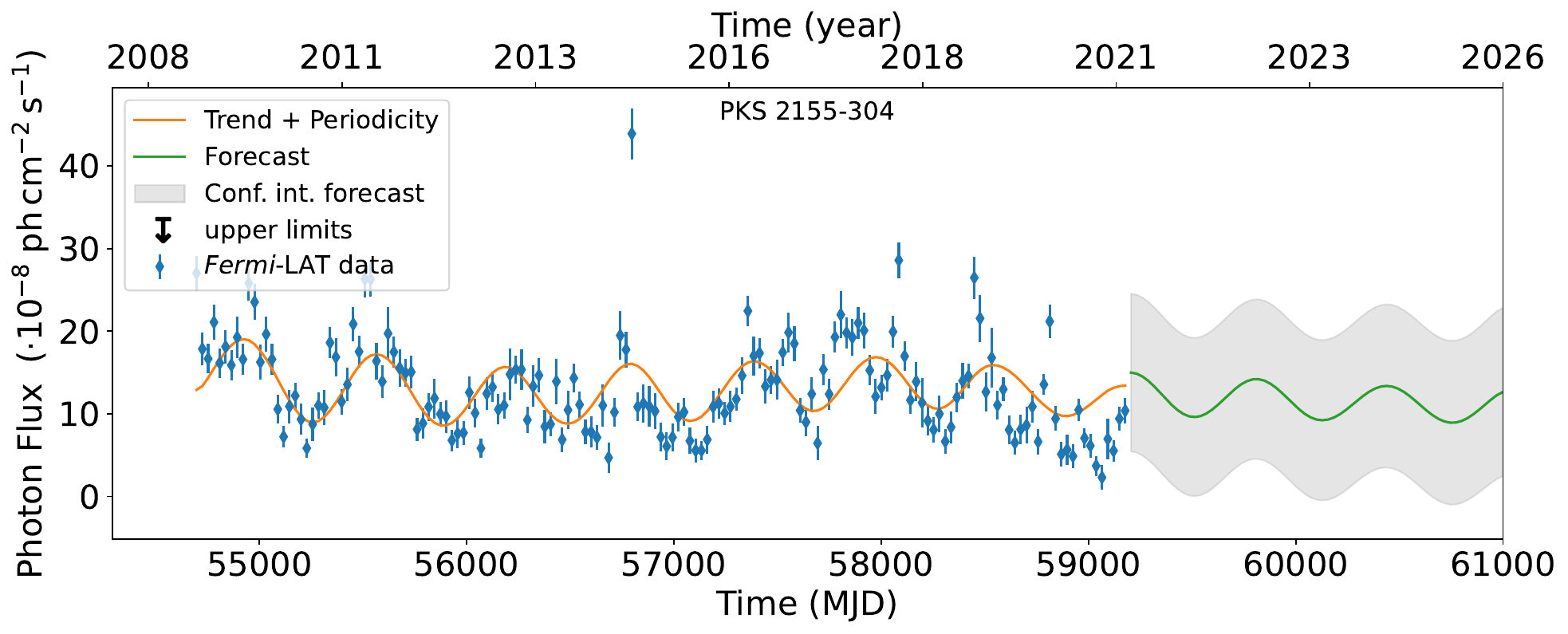}
         \includegraphics[scale=0.262]{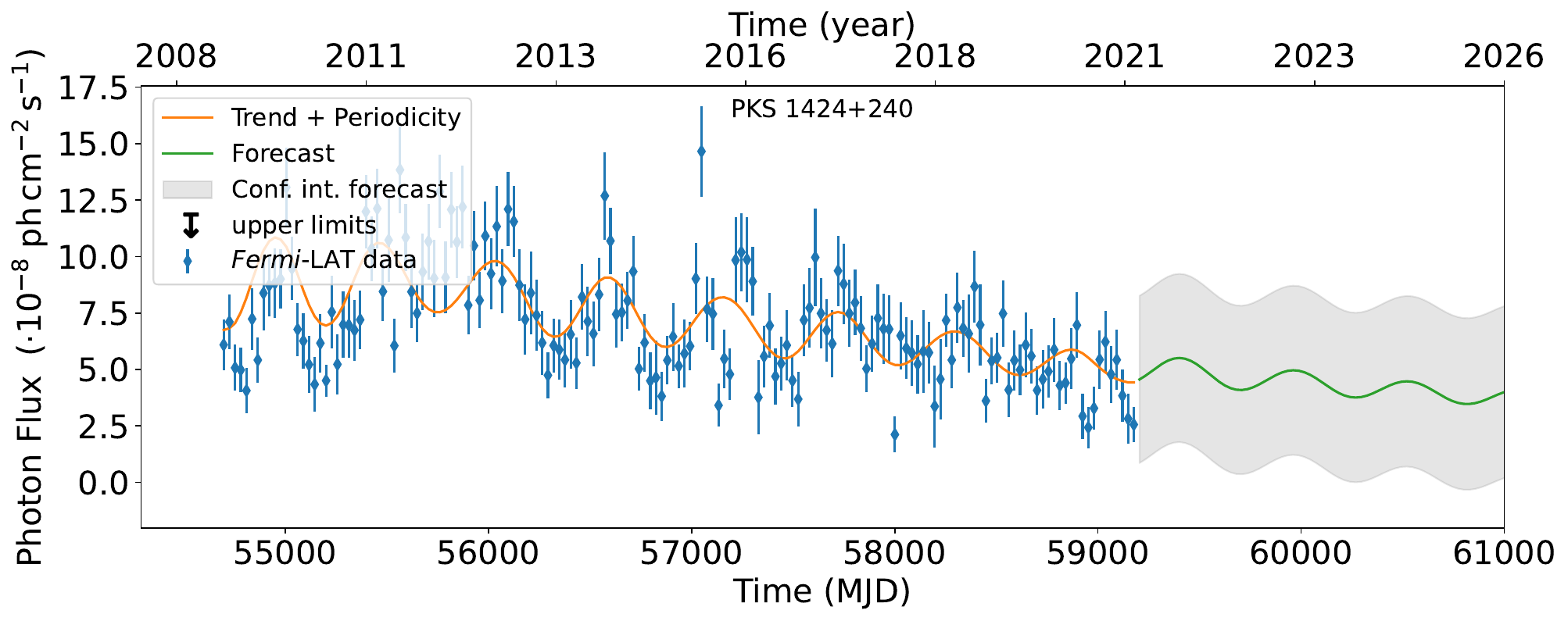}
         
         \includegraphics[scale=0.262]{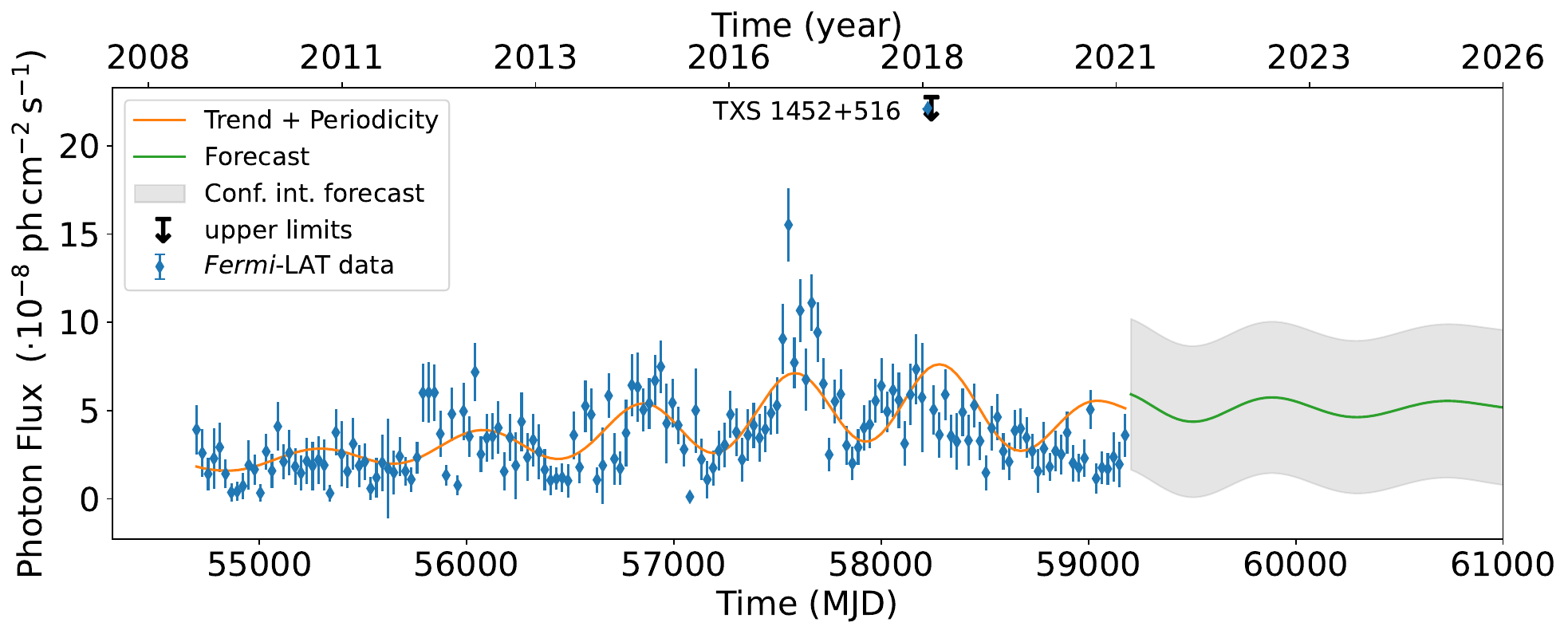}         
         \includegraphics[scale=0.262]{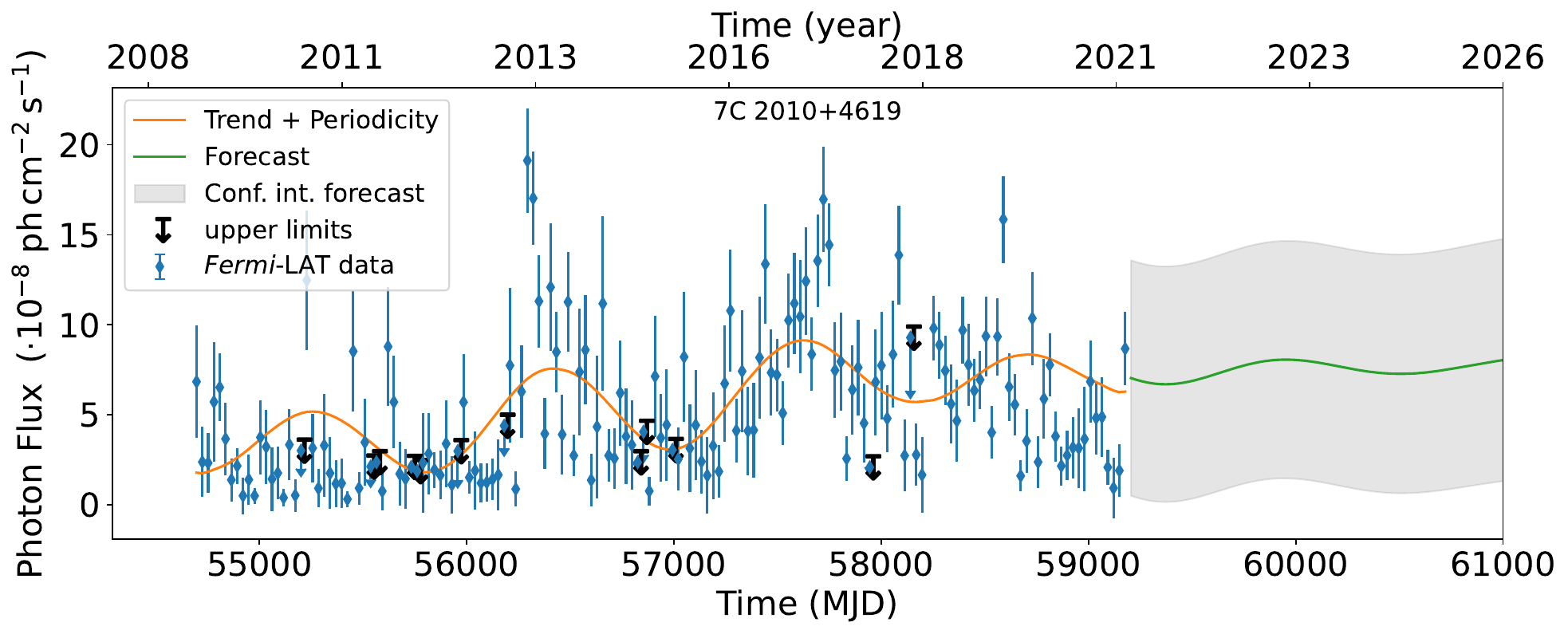}

         \includegraphics[scale=0.262]{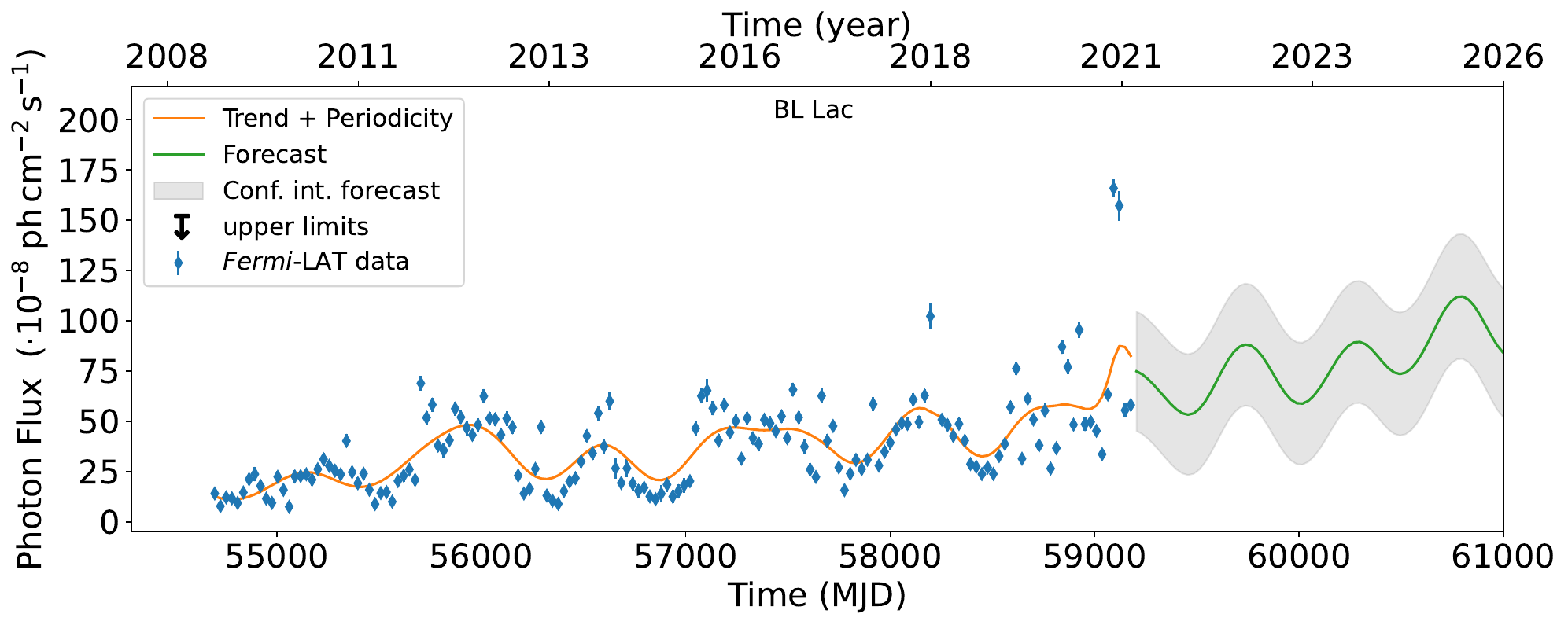}
         \includegraphics[scale=0.261]{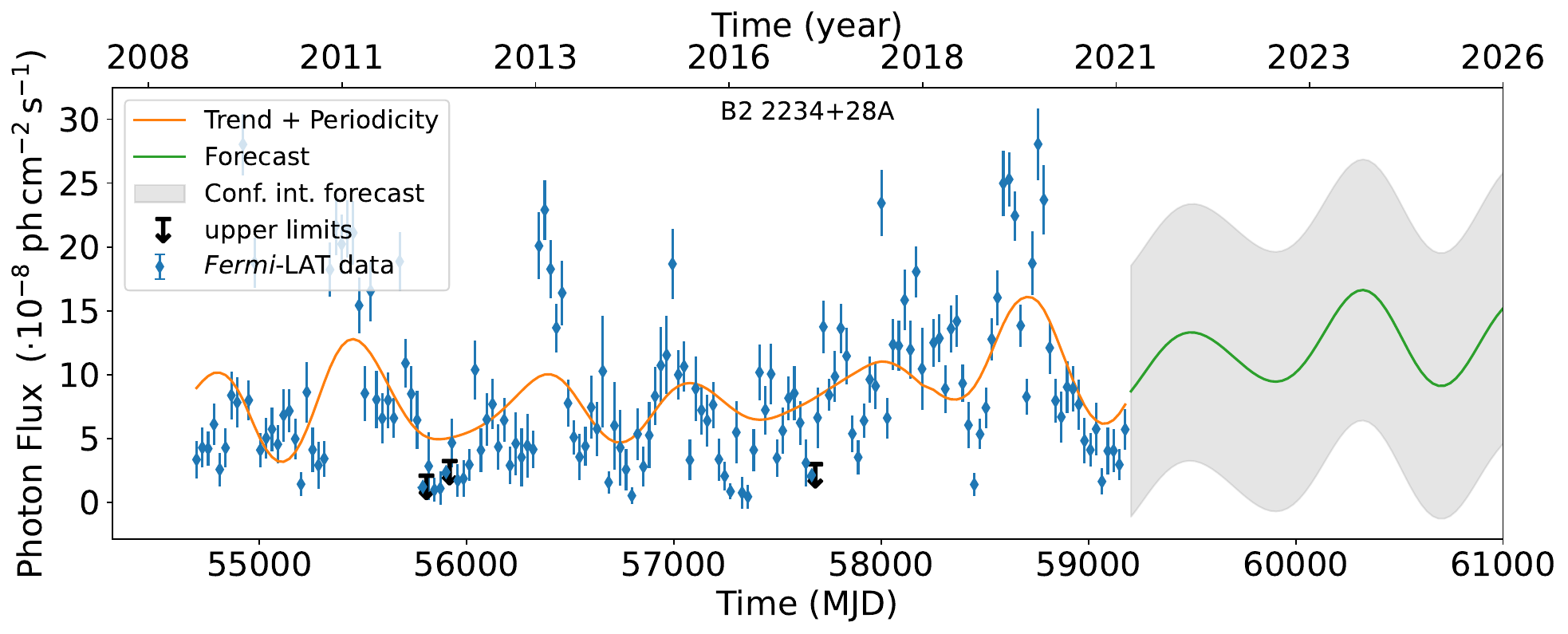}

         \includegraphics[scale=0.261]{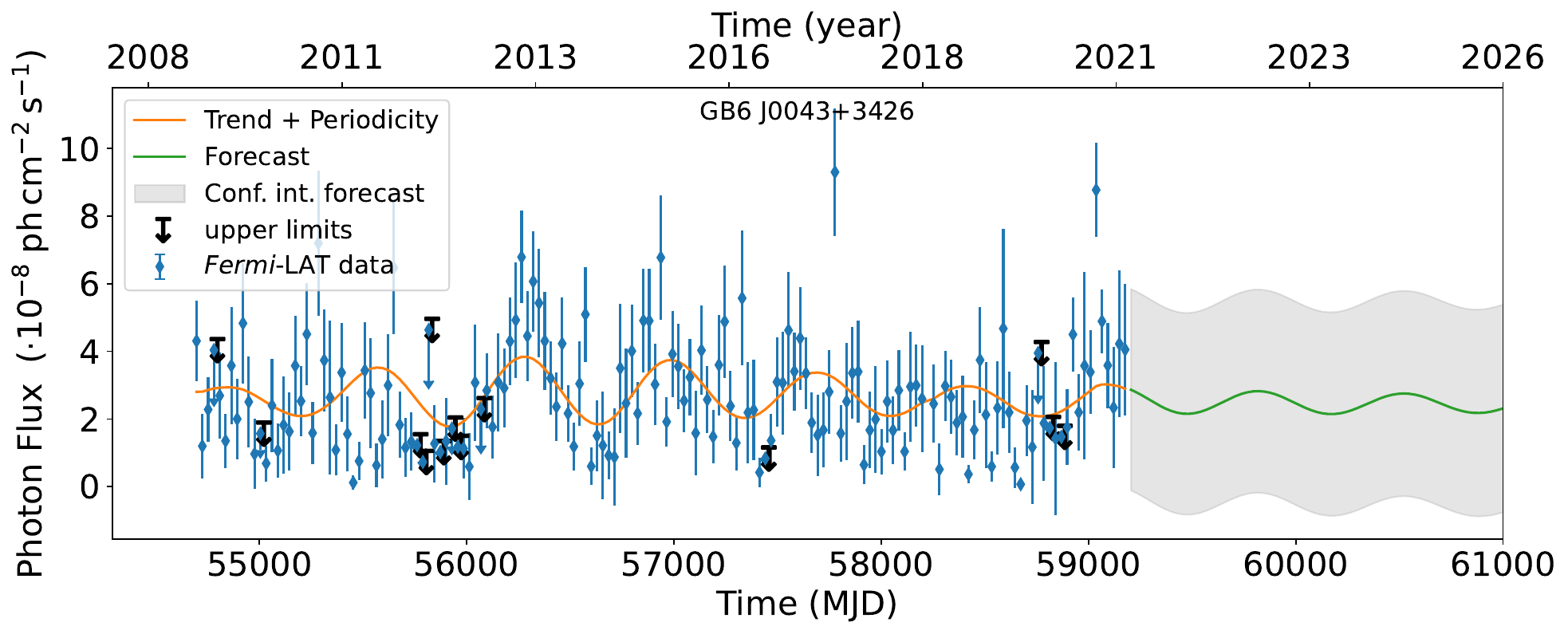}
         \includegraphics[scale=0.262]{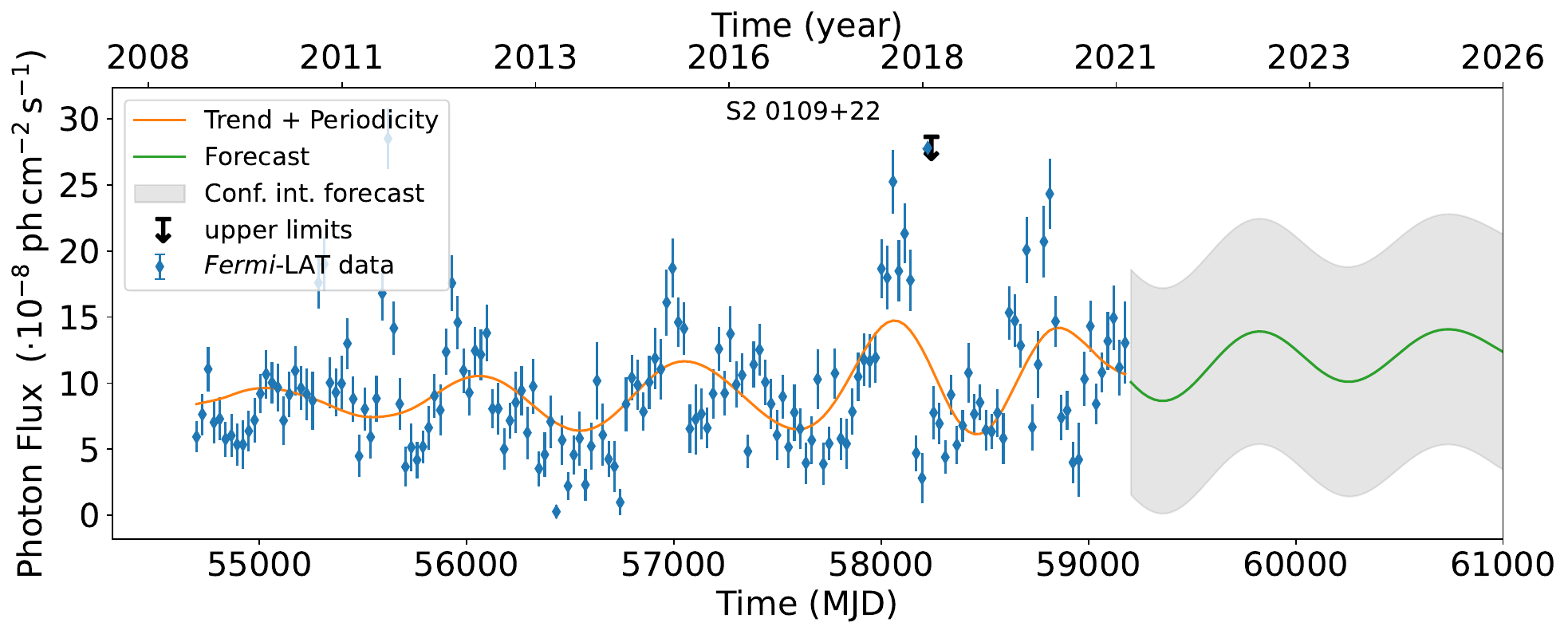}

         \caption{(Continued).}         
\end{figure*}

\onecolumn

\begin{figure*}
        \centering
        \ContinuedFloat

         \includegraphics[scale=0.262]{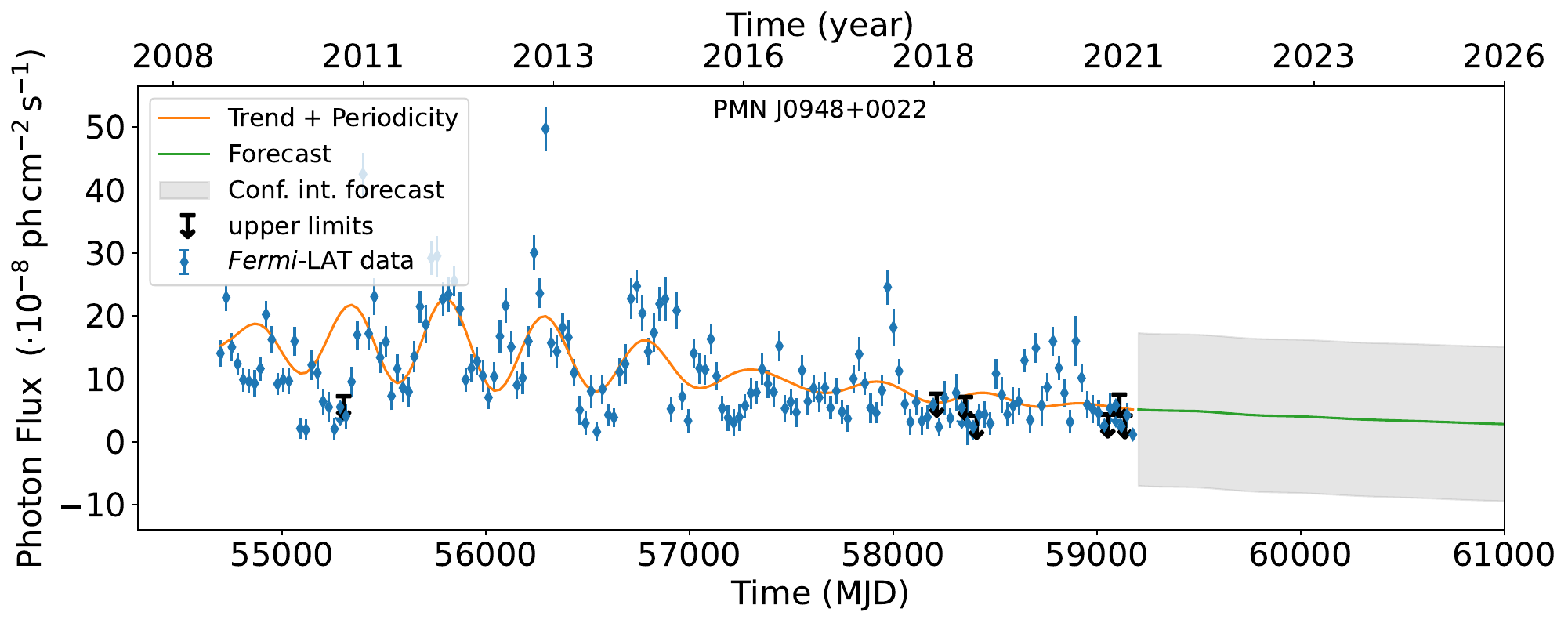}
         \includegraphics[scale=0.262]{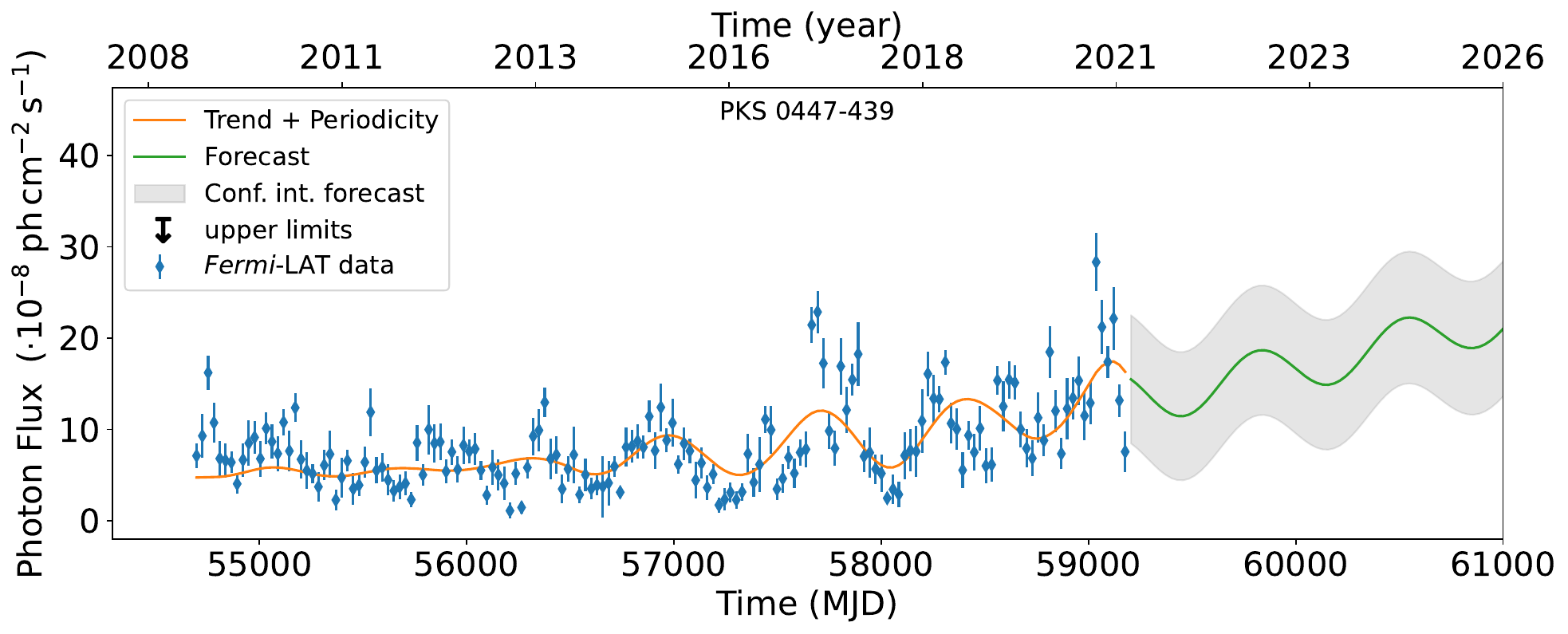}

         \includegraphics[scale=0.262]{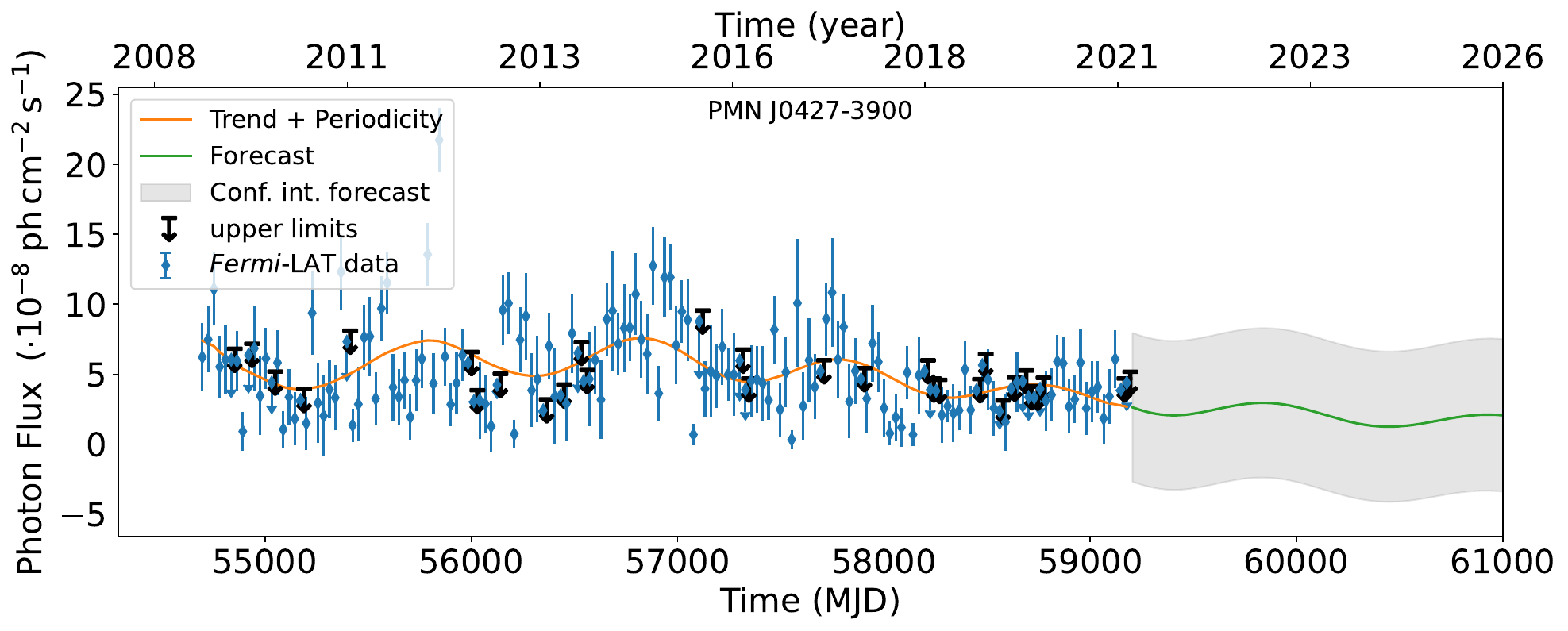}
         \includegraphics[scale=0.262]{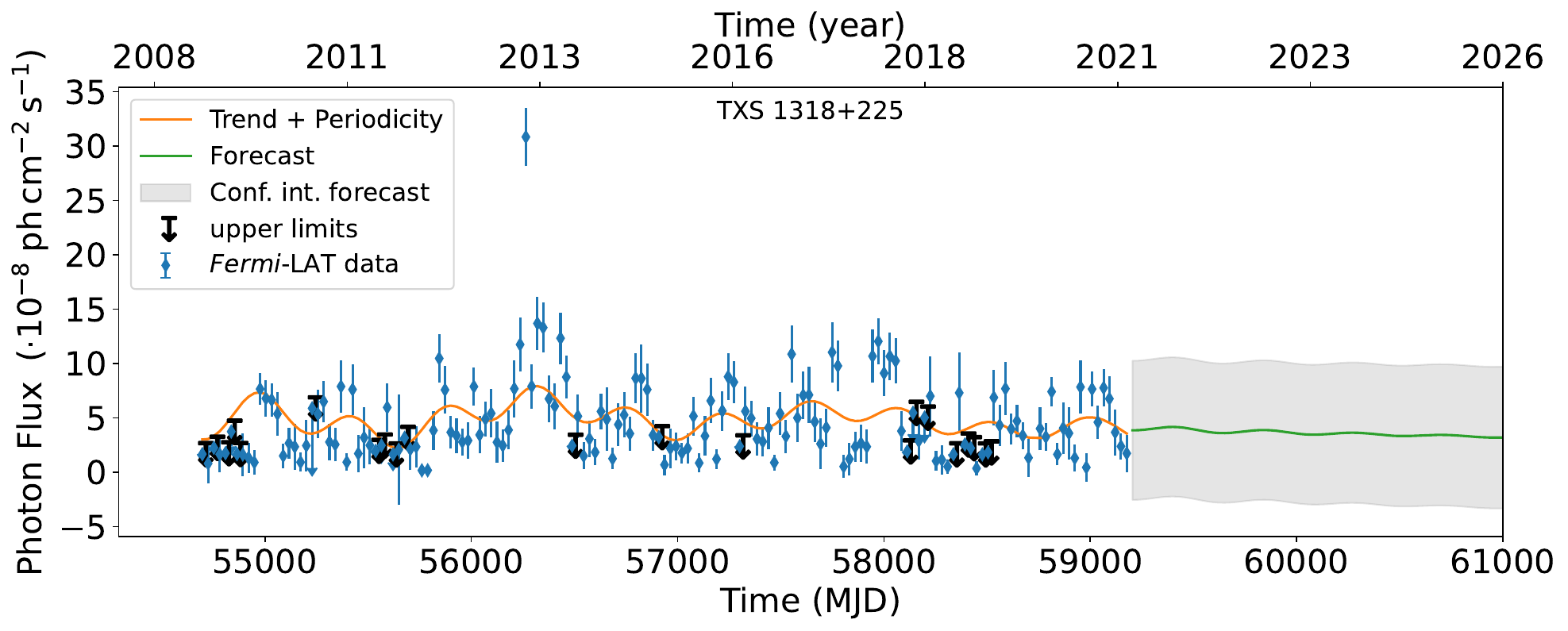}

         \includegraphics[scale=0.262]{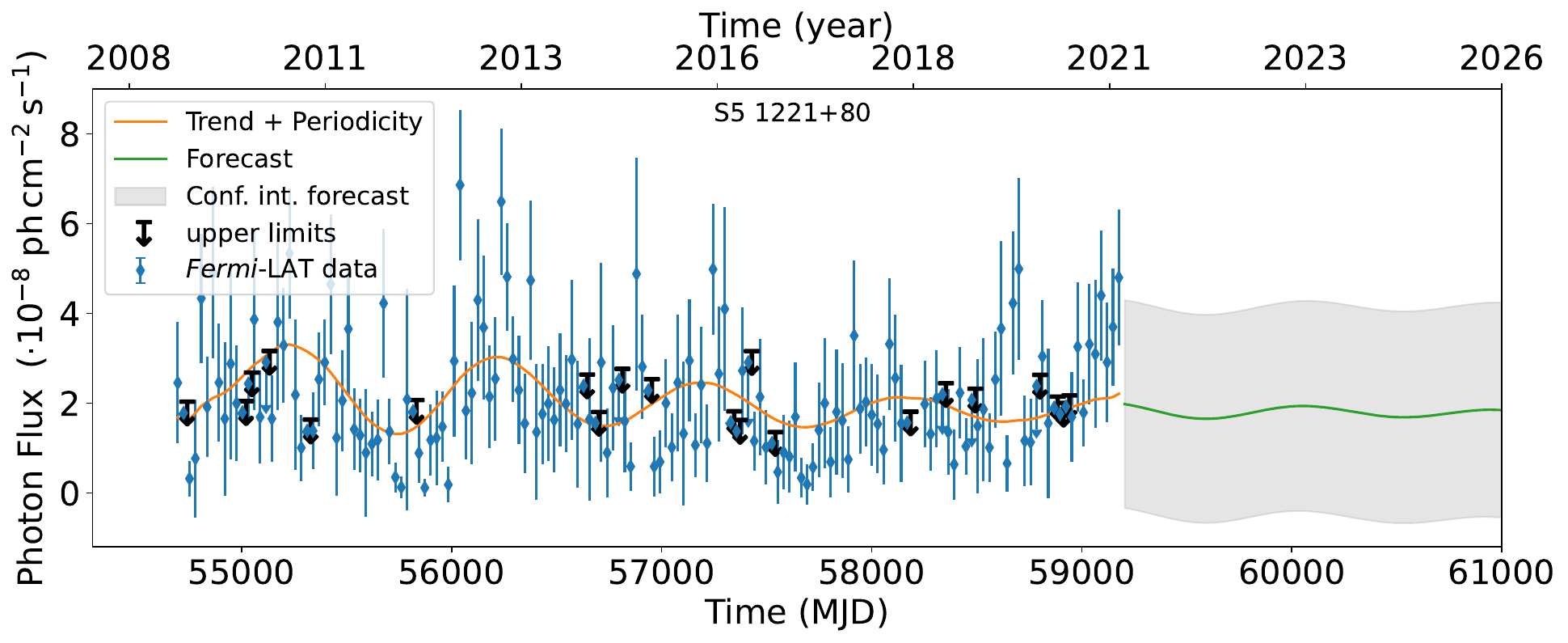}
         \includegraphics[scale=0.262]{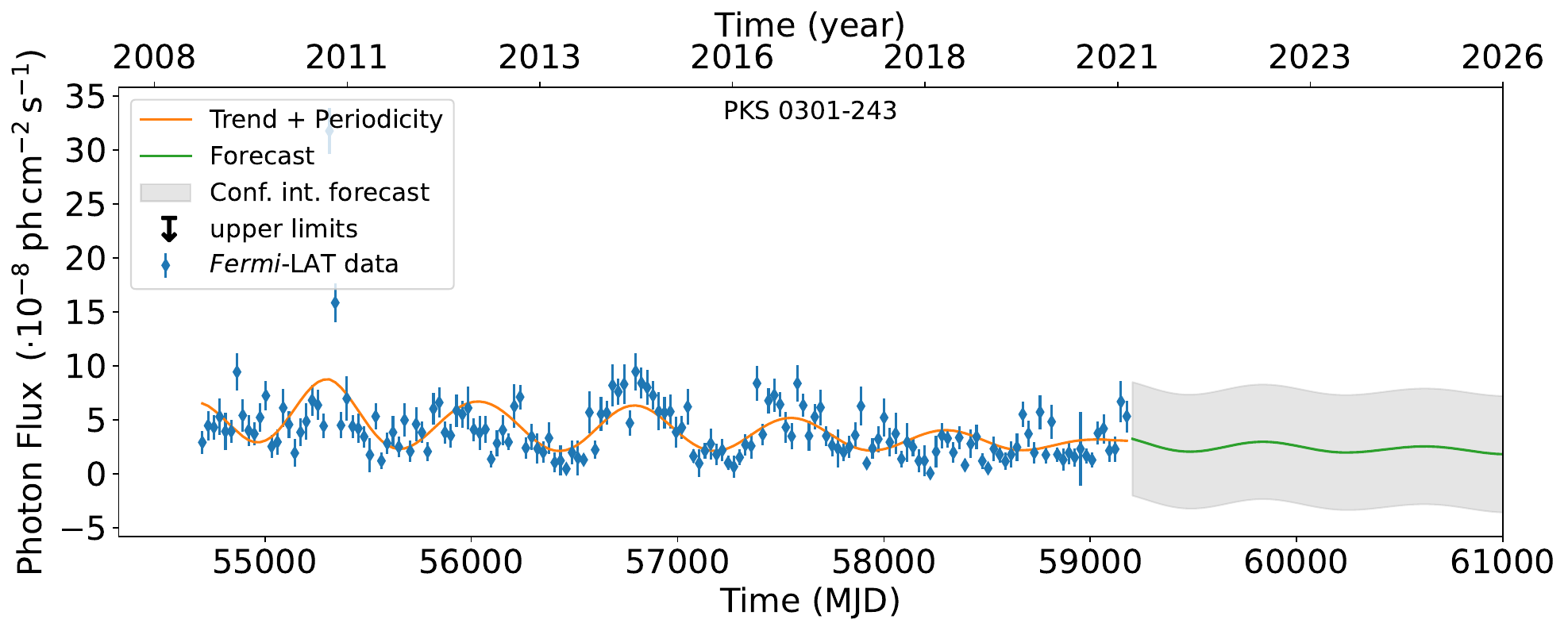}

        \includegraphics[scale=0.262]{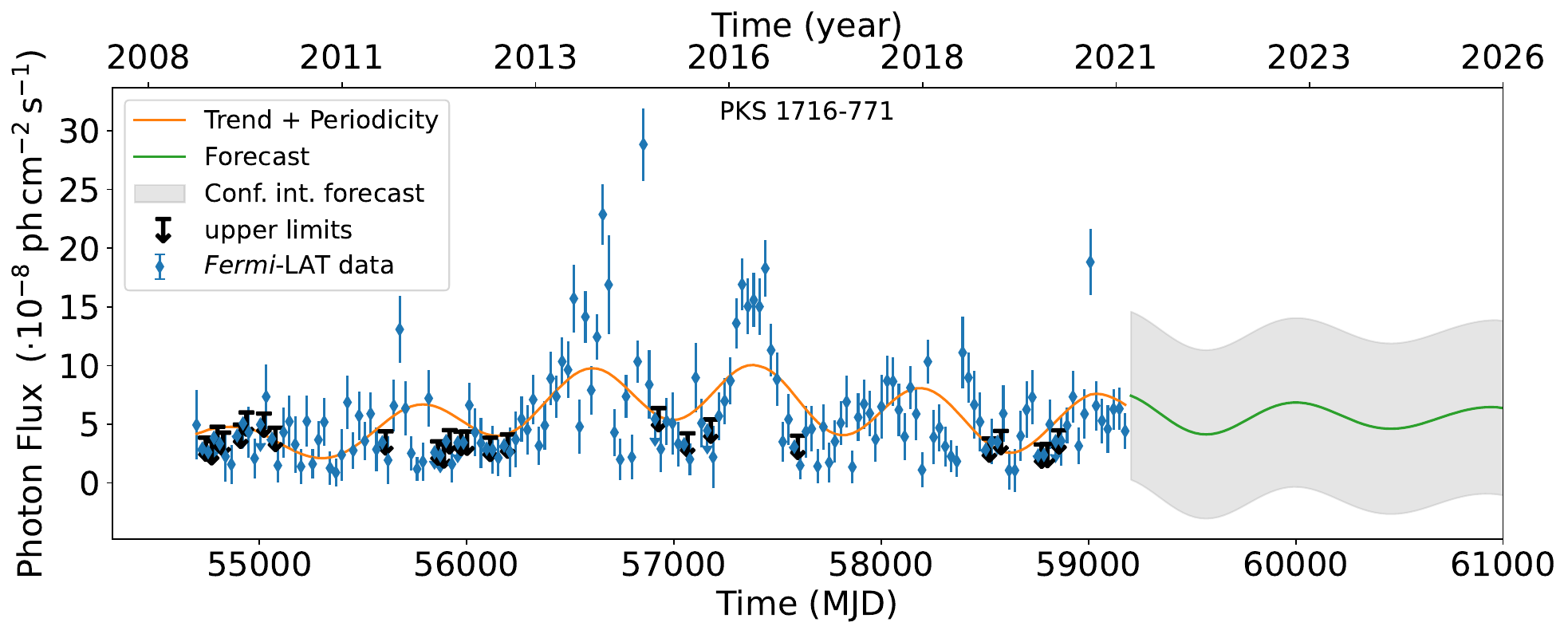}
        \includegraphics[scale=0.262]{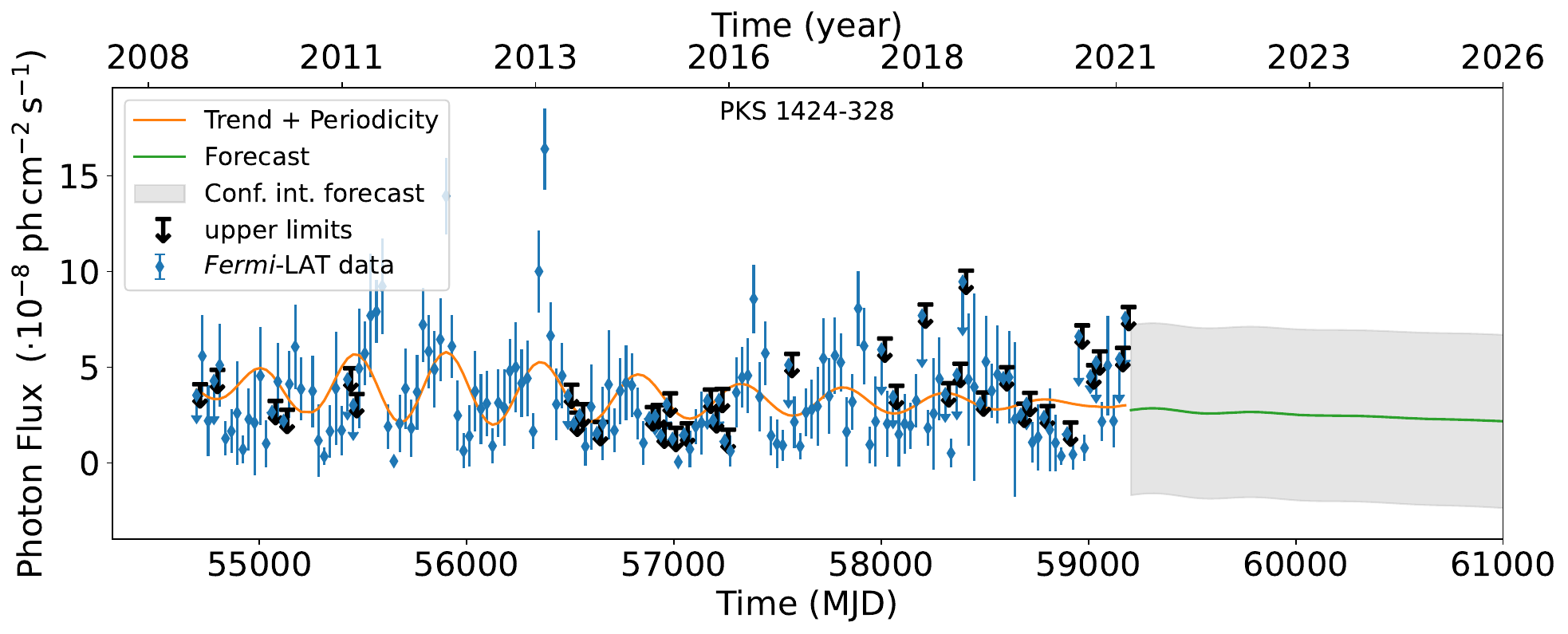}

        \includegraphics[scale=0.262]{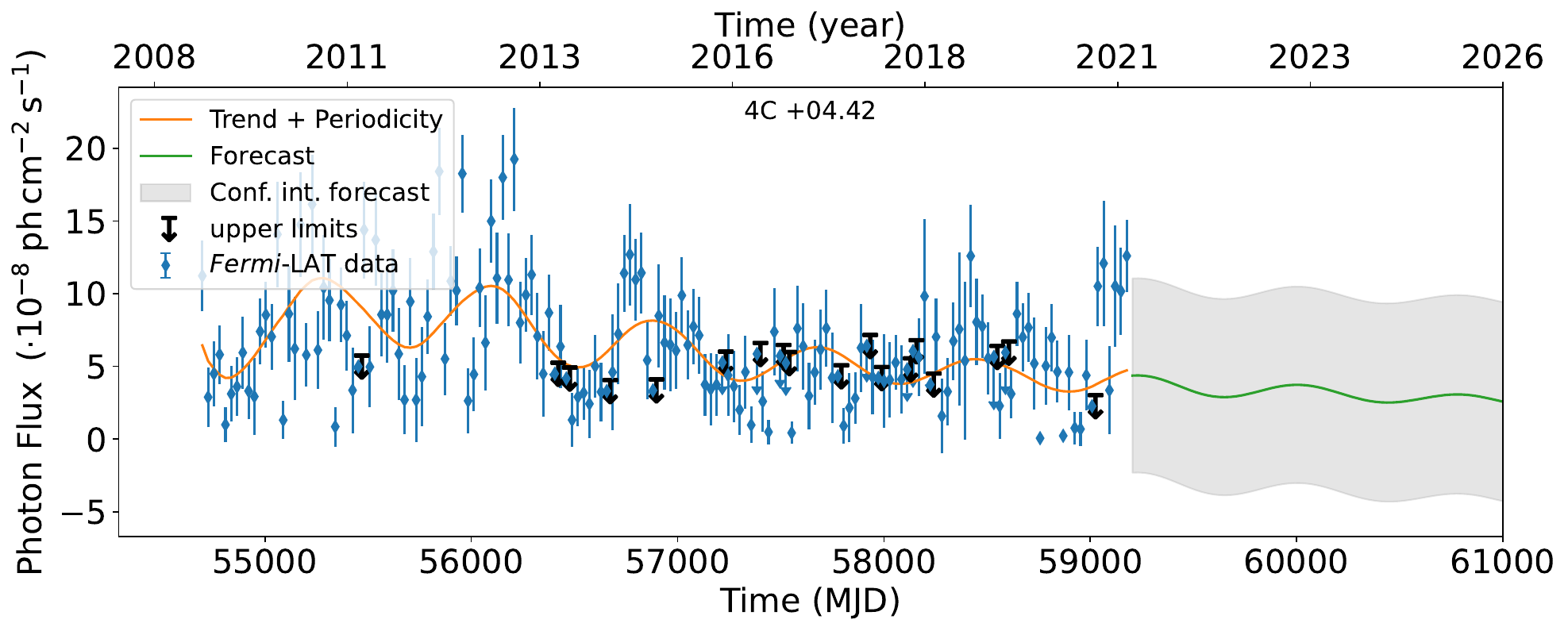}
        \includegraphics[scale=0.262]{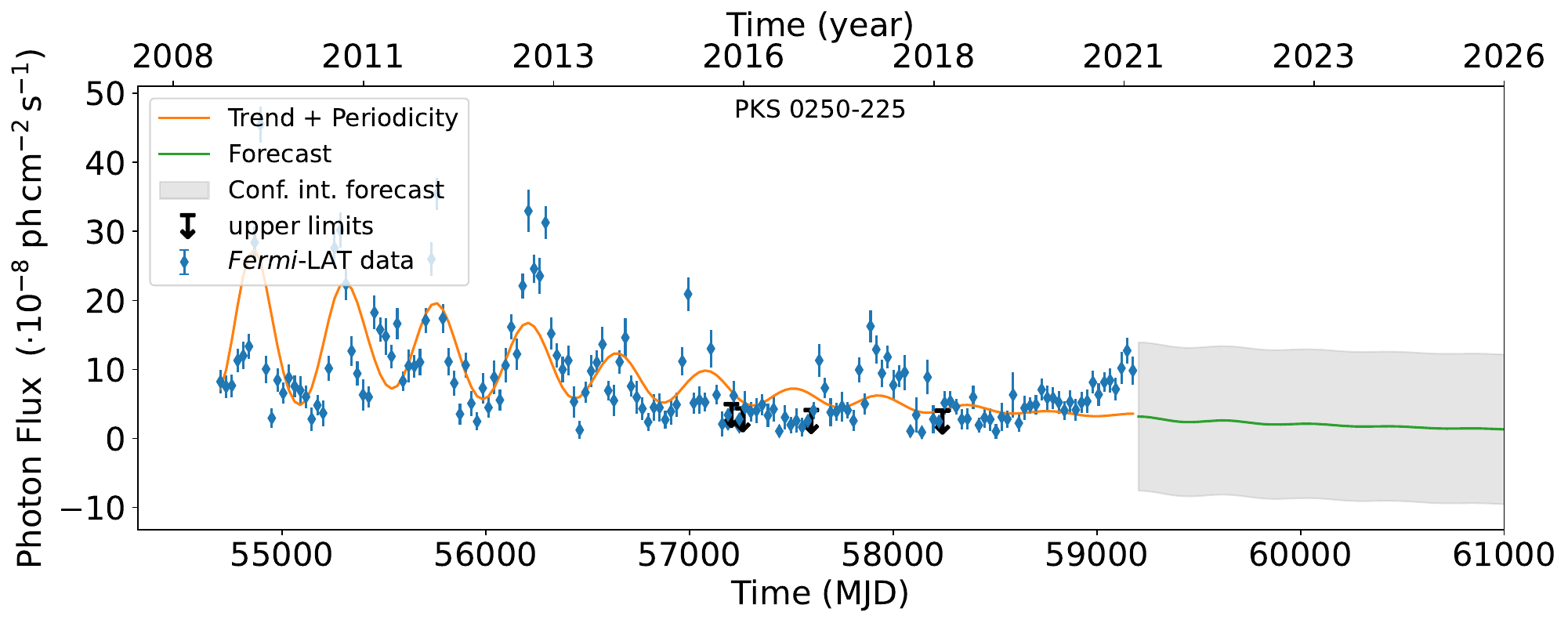}

         \caption{(Continued).}         
\end{figure*}

\begin{figure*}
        \centering
        \ContinuedFloat
 
         \includegraphics[scale=0.262]{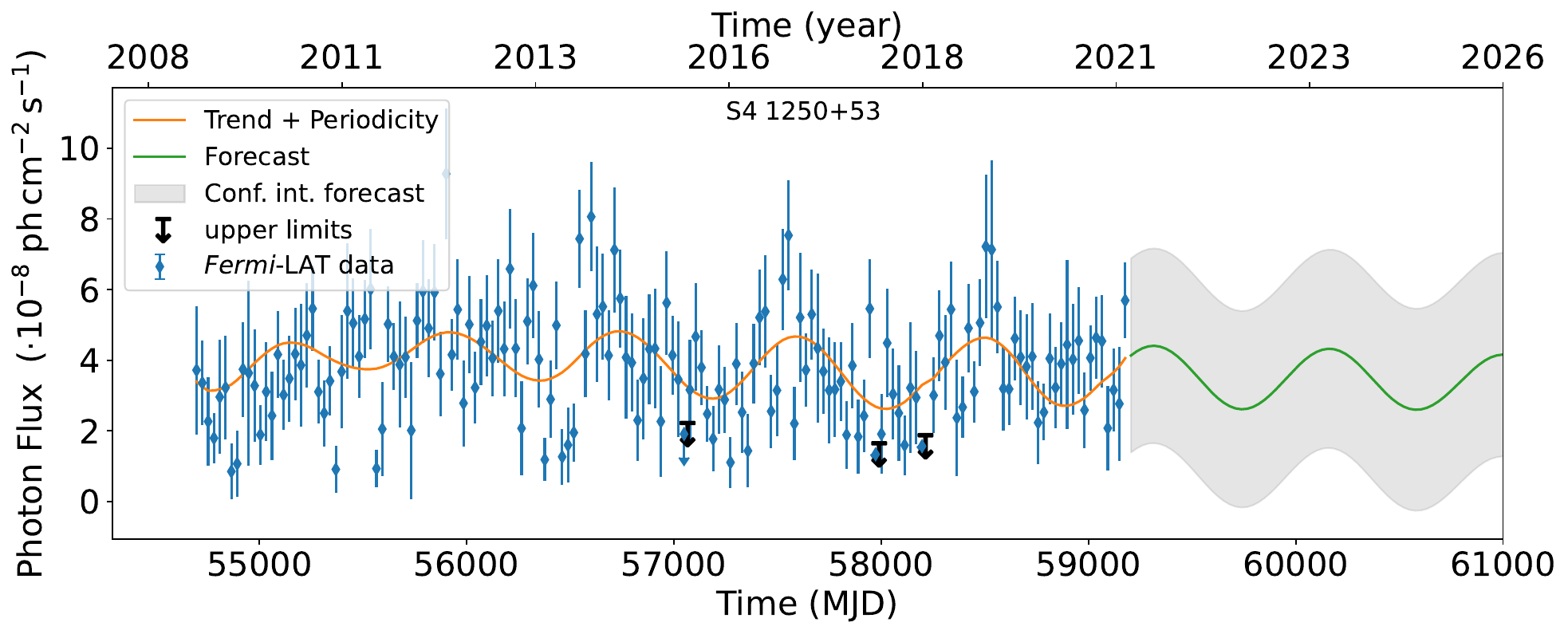}
         \includegraphics[scale=0.262]{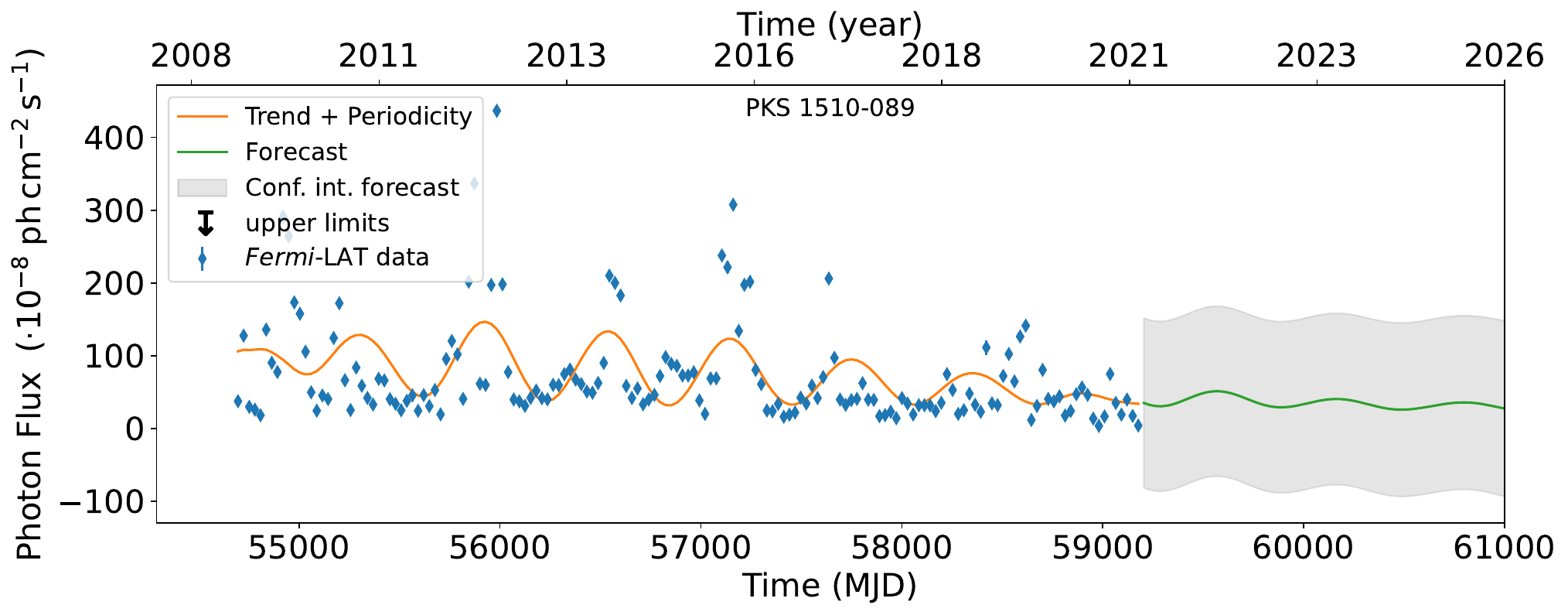}
         
         \includegraphics[scale=0.262]{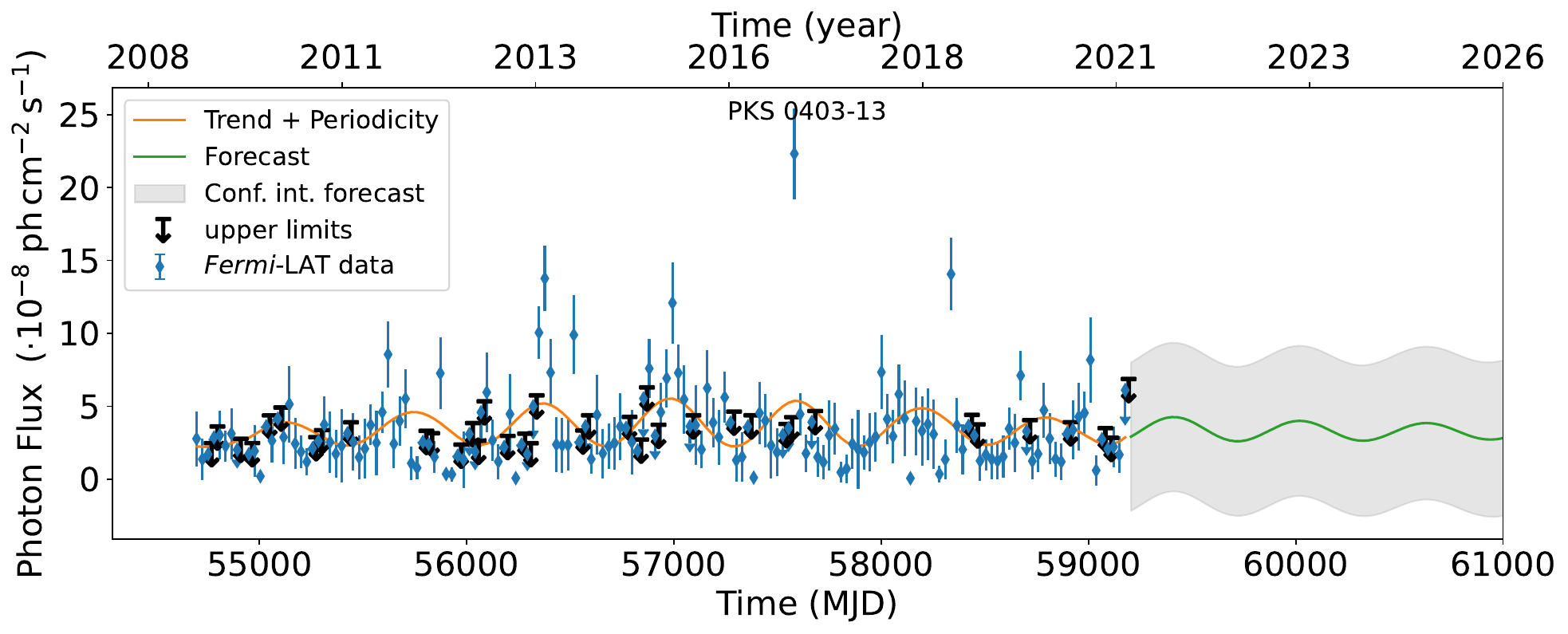}
         \includegraphics[scale=0.262]{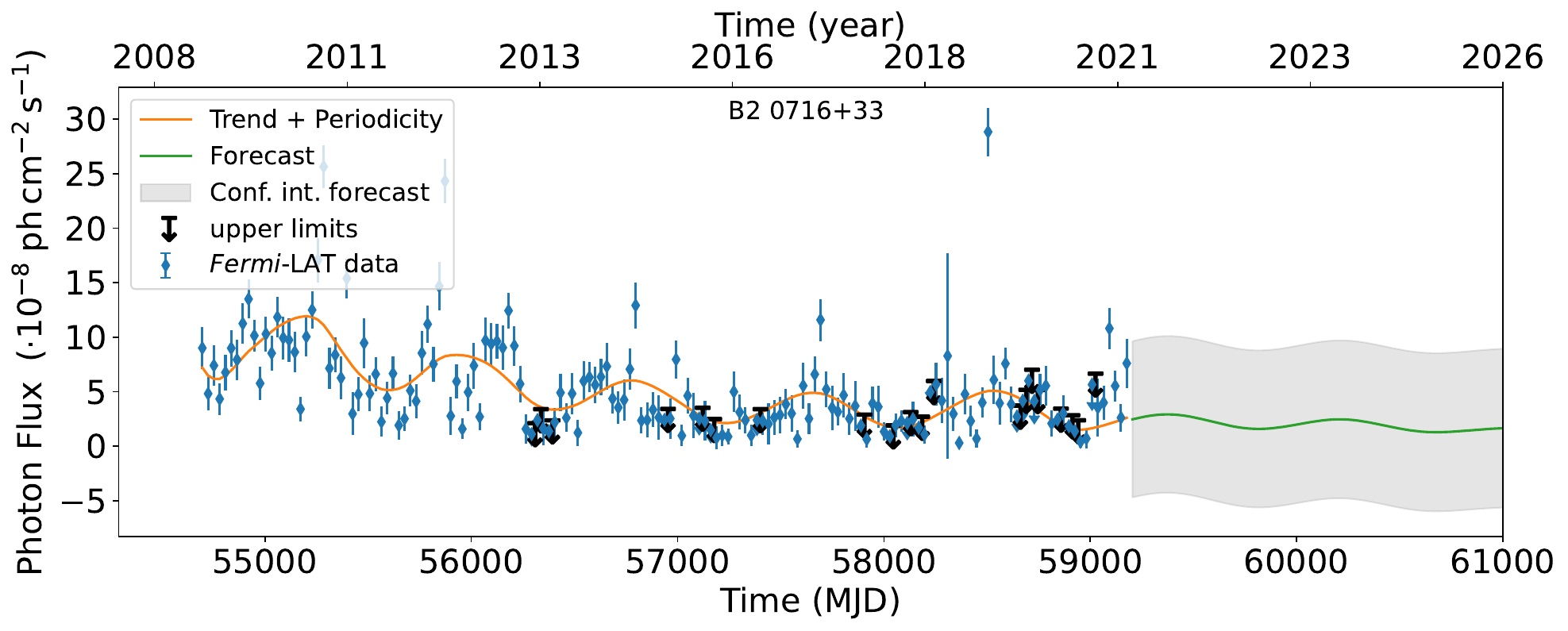}
         
         \includegraphics[scale=0.262]{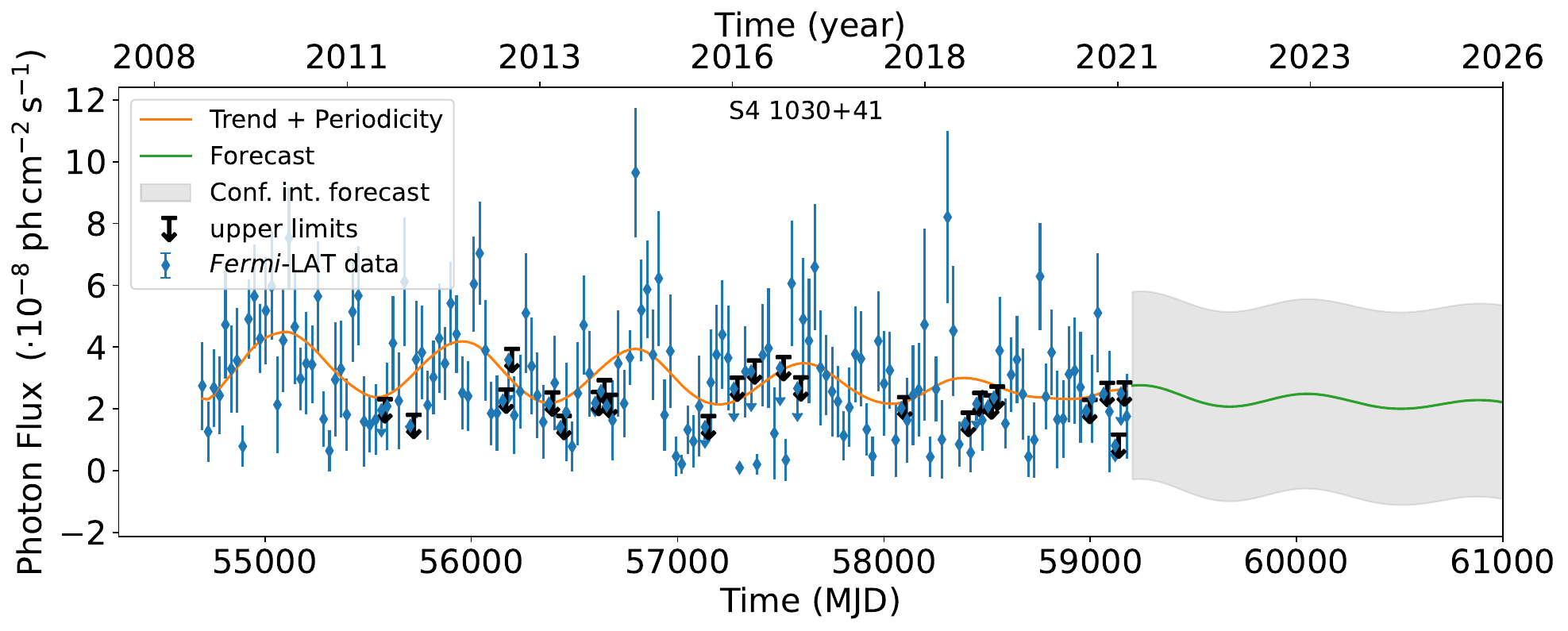} 

         \caption{(Continued).}         
\end{figure*}
\end{appendix}
\end{document}